\pdfoutput=1
\documentclass[%
 reprint,
 superscriptaddress,
showpacs,preprintnumbers,
 amsmath,amssymb,
 aps,
floatfix,
]{revtex4-1}

\usepackage{graphicx}
\usepackage{dcolumn}
\usepackage{bm}
\usepackage{color}
\usepackage{subfig}
\usepackage{xspace}
\usepackage{lineno}
\usepackage{hyperref}
\usepackage{natbib}
\input symbols
\begin{document}



\title{Evidence of Electron Neutrino Appearance in a Muon Neutrino Beam}

\newcommand{\INSTC}{\affiliation{University of Alberta, Centre for Particle Physics, Department of Physics, Edmonton, Alberta, Canada}}
\newcommand{\INSTEE}{\affiliation{University of Bern, Albert Einstein Center for Fundamental Physics, Laboratory for High Energy Physics (LHEP), Bern, Switzerland}}
\newcommand{\INSTFE}{\affiliation{Boston University, Department of Physics, Boston, Massachusetts, U.S.A.}}
\newcommand{\INSTD}{\affiliation{University of British Columbia, Department of Physics and Astronomy, Vancouver, British Columbia, Canada}}
\newcommand{\INSTGA}{\affiliation{University of California, Irvine, Department of Physics and Astronomy, Irvine, California, U.S.A.}}
\newcommand{\INSTI}{\affiliation{IRFU, CEA Saclay, Gif-sur-Yvette, France}}
\newcommand{\INSTCI}{\affiliation{Chonnam National University, Institute for Universe \& Elementary Particles, Gwangju, Korea}}
\newcommand{\INSTGB}{\affiliation{University of Colorado at Boulder, Department of Physics, Boulder, Colorado, U.S.A.}}
\newcommand{\INSTFG}{\affiliation{Colorado State University, Department of Physics, Fort Collins, Colorado, U.S.A.}}
\newcommand{\INSTCJ}{\affiliation{Dongshin University, Department of Physics, Naju, Korea}}
\newcommand{\INSTFH}{\affiliation{Duke University, Department of Physics, Durham, North Carolina, U.S.A.}}
\newcommand{\INSTBA}{\affiliation{Ecole Polytechnique, IN2P3-CNRS, Laboratoire Leprince-Ringuet, Palaiseau, France }}
\newcommand{\INSTEF}{\affiliation{ETH Zurich, Institute for Particle Physics, Zurich, Switzerland}}
\newcommand{\INSTEG}{\affiliation{University of Geneva, Section de Physique, DPNC, Geneva, Switzerland}}
\newcommand{\INSTDG}{\affiliation{H. Niewodniczanski Institute of Nuclear Physics PAN, Cracow, Poland}}
\newcommand{\INSTCB}{\affiliation{High Energy Accelerator Research Organization (KEK), Tsukuba, Ibaraki, Japan}}
\newcommand{\INSTED}{\affiliation{Institut de Fisica d'Altes Energies (IFAE), Bellaterra (Barcelona), Spain}}
\newcommand{\INSTEC}{\affiliation{IFIC (CSIC \& University of Valencia), Valencia, Spain}}
\newcommand{\INSTEI}{\affiliation{Imperial College London, Department of Physics, London, United Kingdom}}
\newcommand{\INSTGF}{\affiliation{INFN Sezione di Bari and Universit\`a e Politecnico di Bari, Dipartimento Interuniversitario di Fisica, Bari, Italy}}
\newcommand{\INSTBE}{\affiliation{INFN Sezione di Napoli and Universit\`a di Napoli, Dipartimento di Fisica, Napoli, Italy}}
\newcommand{\INSTBF}{\affiliation{INFN Sezione di Padova and Universit\`a di Padova, Dipartimento di Fisica, Padova, Italy}}
\newcommand{\INSTBD}{\affiliation{INFN Sezione di Roma and Universit\`a di Roma ``La Sapienza'', Roma, Italy}}
\newcommand{\INSTEB}{\affiliation{Institute for Nuclear Research of the Russian Academy of Sciences, Moscow, Russia}}
\newcommand{\INSTCC}{\affiliation{Kobe University, Kobe, Japan}}
\newcommand{\INSTCD}{\affiliation{Kyoto University, Department of Physics, Kyoto, Japan}}
\newcommand{\INSTEJ}{\affiliation{Lancaster University, Physics Department, Lancaster, United Kingdom}}
\newcommand{\INSTFC}{\affiliation{University of Liverpool, Department of Physics, Liverpool, United Kingdom}}
\newcommand{\INSTFI}{\affiliation{Louisiana State University, Department of Physics and Astronomy, Baton Rouge, Louisiana, U.S.A.}}
\newcommand{\INSTJ}{\affiliation{Universit\'e de Lyon, Universit\'e Claude Bernard Lyon 1, IPN Lyon (IN2P3), Villeurbanne, France}}
\newcommand{\INSTCE}{\affiliation{Miyagi University of Education, Department of Physics, Sendai, Japan}}
\newcommand{\INSTDF}{\affiliation{National Centre for Nuclear Research, Warsaw, Poland}}
\newcommand{\INSTFJ}{\affiliation{State University of New York at Stony Brook, Department of Physics and Astronomy, Stony Brook, New York, U.S.A.}}
\newcommand{\INSTCF}{\affiliation{Osaka City University, Department of Physics, Osaka,  Japan}}
\newcommand{\INSTGG}{\affiliation{Oxford University, Department of Physics, Oxford, United Kingdom}}
\newcommand{\INSTBB}{\affiliation{UPMC, Universit\'e Paris Diderot, CNRS/IN2P3, Laboratoire de Physique Nucl\'eaire et de Hautes Energies (LPNHE), Paris, France}}
\newcommand{\INSTGC}{\affiliation{University of Pittsburgh, Department of Physics and Astronomy, Pittsburgh, Pennsylvania, U.S.A.}}
\newcommand{\INSTFA}{\affiliation{Queen Mary University of London, School of Physics and Astronomy, London, United Kingdom}}
\newcommand{\INSTE}{\affiliation{University of Regina, Department of Physics, Regina, Saskatchewan, Canada}}
\newcommand{\INSTGD}{\affiliation{University of Rochester, Department of Physics and Astronomy, Rochester, New York, U.S.A.}}
\newcommand{\INSTBC}{\affiliation{RWTH Aachen University, III. Physikalisches Institut, Aachen, Germany}}
\newcommand{\INSTDD}{\affiliation{Seoul National University, Department of Physics and Astronomy, Seoul, Korea}}
\newcommand{\INSTFB}{\affiliation{University of Sheffield, Department of Physics and Astronomy, Sheffield, United Kingdom}}
\newcommand{\INSTDI}{\affiliation{University of Silesia, Institute of Physics, Katowice, Poland}}
\newcommand{\INSTEH}{\affiliation{STFC, Rutherford Appleton Laboratory, Harwell Oxford,  and  Daresbury Laboratory, Warrington, United Kingdom}}
\newcommand{\INSTCH}{\affiliation{University of Tokyo, Department of Physics, Tokyo, Japan}}
\newcommand{\INSTBJ}{\affiliation{University of Tokyo, Institute for Cosmic Ray Research, Kamioka Observatory, Kamioka, Japan}}
\newcommand{\INSTCG}{\affiliation{University of Tokyo, Institute for Cosmic Ray Research, Research Center for Cosmic Neutrinos, Kashiwa, Japan}}
\newcommand{\INSTF}{\affiliation{University of Toronto, Department of Physics, Toronto, Ontario, Canada}}
\newcommand{\INSTB}{\affiliation{TRIUMF, Vancouver, British Columbia, Canada}}
\newcommand{\INSTG}{\affiliation{University of Victoria, Department of Physics and Astronomy, Victoria, British Columbia, Canada}}
\newcommand{\INSTDJ}{\affiliation{University of Warsaw, Faculty of Physics, Warsaw, Poland}}
\newcommand{\INSTDH}{\affiliation{Warsaw University of Technology, Institute of Radioelectronics, Warsaw, Poland}}
\newcommand{\INSTFD}{\affiliation{University of Warwick, Department of Physics, Coventry, United Kingdom}}
\newcommand{\INSTGE}{\affiliation{University of Washington, Department of Physics, Seattle, Washington, U.S.A.}}
\newcommand{\INSTGH}{\affiliation{University of Winnipeg, Department of Physics, Winnipeg, Manitoba, Canada}}
\newcommand{\INSTEA}{\affiliation{Wroclaw University, Faculty of Physics and Astronomy, Wroclaw, Poland}}
\newcommand{\INSTH}{\affiliation{York University, Department of Physics and Astronomy, Toronto, Ontario, Canada}}

\INSTC
\INSTEE
\INSTFE
\INSTD
\INSTGA
\INSTI
\INSTCI
\INSTGB
\INSTFG
\INSTCJ
\INSTFH
\INSTBA
\INSTEF
\INSTEG
\INSTDG
\INSTCB
\INSTED
\INSTEC
\INSTEI
\INSTGF
\INSTBE
\INSTBF
\INSTBD
\INSTEB
\INSTCC
\INSTCD
\INSTEJ
\INSTFC
\INSTFI
\INSTJ
\INSTCE
\INSTDF
\INSTFJ
\INSTCF
\INSTGG
\INSTBB
\INSTGC
\INSTFA
\INSTE
\INSTGD
\INSTBC
\INSTDD
\INSTFB
\INSTDI
\INSTEH
\INSTCH
\INSTBJ
\INSTCG
\INSTF
\INSTB
\INSTG
\INSTDJ
\INSTDH
\INSTFD
\INSTGE
\INSTGH
\INSTEA
\INSTH

\author{K.\,Abe}\INSTBJ
\author{N.\,Abgrall}\INSTEG
\author{H.\,Aihara}\thanks{also at Kavli IPMU, U. of Tokyo, Kashiwa, Japan}\INSTCH
\author{T.\,Akiri}\INSTFH
\author{J.B.\,Albert}\INSTFH
\author{C.\,Andreopoulos}\INSTEH
\author{S.\,Aoki}\INSTCC
\author{A.\,Ariga}\INSTEE
\author{T.\,Ariga}\INSTEE
\author{S.\,Assylbekov}\INSTFG
% ...

\collaboration{The T2K Collaboration}\noaffiliation

\date{\today}

\begin{abstract}
The T2K collaboration reports evidence for electron neutrino appearance at the atmospheric mass splitting, $|$\dmsq$|$ $\approx 2.4\times10^{-3}$ \evsq. An excess of electron neutrino interactions over background is observed from a muon neutrino beam with a peak energy of $0.6$ \gev at the Super-Kamiokande (SK) detector 295~km from the beam's origin. 
Signal and background predictions are constrained by data from near detectors located 280~m from the neutrino production target.
We observe 11 electron neutrino candidate events at the SK detector when a background of $3.3\pm0.4$(syst.) events is expected.  
The background-only hypothesis is rejected with a $p$-value of 0.0009 (3.1$\sigma$), and a fit assuming $\nu_{\mu}\rightarrow\nu_{e}$ oscillations with \sttt=1, \dcp=0 and $|\dmsq|=2.4\times10^{-3}$~\evsq~yields \stot={$0.088{}^{+0.049}_{-0.039}$}(stat.+syst.).

\end{abstract}
\pacs{14.60.Pq,14.60.Lm,12.27.-a,29.40.ka}
\maketitle


\section{\label{sec:intro} Introduction}

The phenomena of neutrino oscillations through the mixing of massive neutrinos have been well established by
 experiments observing neutrino interaction rates from
solar~\cite{Cleveland:1998nv,PhysRevLett.63.16,Abdurashitov1994234,Anselmann1994377,PhysRevLett.86.5651,PhysRevLett.89.011301,PhysRevLett.101.091302}, atmospheric~\cite{Fukuda:1998ub,Hirata1988416,PhysRevD.46.3720,Ahlen1995481,Allison:1996yb,PhysRevD.81.092004}, 
reactor~\cite{PhysRevLett.100.221803} and accelerator~\cite{PhysRevD.74.072003,Adamson:2012rm,PhysRevLett.106.181801,Abe:2012gx}
 sources.  
With few exceptions, such as the results from the LSND~\cite{Aguilar:2001ty} and MiniBooNE collaborations~\cite{PhysRevLett.105.181801}, the observations are consistent with the mixing
of three neutrinos, governed by three mixing angles:  $\theta_{12}~\approx34^{\circ}$,
$\theta_{23}\approx45^{\circ}$ and $\theta_{13}$; and an  as-yet-undetermined CP-violating phase, \dcp. Neutrino mixing also depends on three mass states, $m_i$, and therefore two independent mass splittings,  $|\dmsq|\approx 2.4\times10^{-3}$~\evsq~(atmospheric) and $\dmsqso\approx 7.6\times10^{-5}$~\evsq~(solar), where $\Delta m^{2}_{ij}={m_i}^2-{m_j}^2$.
Additional understanding of neutrino mixing can be gained by observing the appearance of one flavor of neutrino interactions in a beam of another flavor through charged current interactions. Recently, T2K~\cite{PhysRevLett.107.041801} has reported on the appearance of electron neutrinos in a beam of muon neutrinos, and the OPERA~\cite{Agafonova2010138} and Super-Kamiokande~\cite{PhysRevLett.110.181802} collaborations have reported on the appearance of tau neutrinos from accelerator-based and atmospheric muon neutrino sources, respectively.

The oscillations of $\nu_{\mu}\rightarrow\nu_{e}$ that T2K searches for are of particular interest since the observation 
of this mode
at a baseline over energy ratio ($L/E$) of $\sim1$~GeV$/500$~km implies a non-zero value for the mixing angle $\theta_{13}$.
Until recently, the mixing angle $\theta_{13}$ had only been constrained to be less than 11$^{\circ}$ by
reactor~\cite{Apollonio:2002gd} and accelerator~\cite{Adamson:2011qu,PhysRevLett.110.171801} neutrino experiments.  With data collected through 2011, the
T2K experiment found the first indication of non-zero $\theta_{13}$ in the oscillation of muon neutrinos
to electron neutrinos~\cite{PhysRevLett.107.041801}.  Since then, a non-zero value of $\theta_{13}=9.1^{\circ}\pm0.6^{\circ}$~\cite{PhysRevD.86.010001} has been confirmed from the disappearance of reactor electron anti-neutrinos observed by the Daya Bay~\cite{PhysRevLett.108.171803},
RENO~\cite{PhysRevLett.108.191802} and Double Chooz~\cite{PhysRevLett.108.131801} experiments. 
 In this paper,
T2K updates its measurement of electron neutrino appearance using additional data collected through 2012 and 
improved analysis methods.

The probability for electron neutrino appearance in a muon neutrino beam with energy $E_{\nu}$ of $\mathcal{O}(1)$~GeV 
propagating over a baseline $L$ of $\mathcal{O}(100)$~km
is dominated by the term (in units of $c,\hbar=1$):
\begin{linenomath*}
\begin{equation}
P_{\nu_{\mu}\rightarrow\nu_{e}} \approx \textrm{sin}^2\theta_{23}~\textrm{sin}^22\theta_{13}~\textrm{sin}^{2}\frac{\dmsq L}{4E_{\nu}}.
\end{equation}
\end{linenomath*}
This leading term is identical for neutrino and antineutrino oscillations.
Since the probability depends on \stt, a precise determination of $\theta_{13}$ requires measurements
of $\theta_{23}$.  
The dependence on \stt can lift the degeneracy of solutions with $\theta_{23}>\pi/4$ and $\theta_{23}<\pi/4$ 
that are present when $\theta_{23}$ is measured from muon neutrino
survival, which depends \sttt.

The electron neutrino appearance probability also includes sub-leading terms which depend on \dcp~and terms that 
describe matter interactions~\cite{Freund:2001pn}: 

\begin{linenomath*}
\begin{align}
\label{eq:subleading}
P_{\nu_{\mu}\rightarrow\nu_{e}}  = &\frac{1}{(A-1)^2}~\textrm{sin}^22\theta_{13}~\textrm{sin}^2\theta_{23}~\textrm{sin}^2[(A-1)\Delta] \nonumber \\
  -(+) &\frac{\alpha}{A(1-A)}~\textrm{cos}\theta_{13}~\textrm{sin}2\theta_{12}~\textrm{sin}2\theta_{23}~\textrm{sin}2\theta_{13}\times \nonumber \\
 & \textrm{sin}\dcp~\textrm{sin}\Delta~\textrm{sin}A\Delta~\textrm{sin}[(1-A)\Delta] \nonumber \\
  + &\frac{\alpha}{A(1-A)}~\textrm{cos}\theta_{13}~\textrm{sin}2\theta_{12}~\textrm{sin}2\theta_{23}~\textrm{sin}2\theta_{13}\times \nonumber \\
 & \textrm{cos}\dcp~\textrm{cos}\Delta~\textrm{sin}A\Delta~\textrm{sin}[(1-A)\Delta] \nonumber \\
  + &\frac{\alpha^2}{A^2}~\textrm{cos}^2\theta_{23}~\textrm{sin}^22\theta_{12}~\textrm{sin}^2A\Delta \nonumber \\
\end{align}
\end{linenomath*}
Here $\alpha=\frac{\dmsqso}{\Delta m^{2}_{32}}<<1$, $\Delta = \frac{\Delta m^{2}_{32} L}{4E_{\nu}}$ and $A = 2\sqrt{2}G_FN_e\frac{E_{\nu}}{\Delta m^{2}_{32}}$, where $N_e$ is the electron density of the Earth's crust.
In the three-neutrino paradigm CP violation can only occur when all three mixing angles, 
including $\theta_{13}$, have non-zero values.
The second term has a negative sign for neutrinos and a positive sign for antineutrinos and violates CP, which 
suggests the
possibility of observing CP violation by measuring the difference in the electron neutrino appearance probabilities 
for neutrinos and 
antineutrinos.  Since the CP-violating term can only appear in an appearance probability, 
a measurement of \nue appearance, such as the one described in this paper, 
is an important milestone towards future searches for CP violation. 
The $A$ dependence in the oscillation probability arises from matter effects and introduces a dependence on 
the sign of \dmsq.  We refer to $\dmsq>0$ as the normal mass hierarchy and $\dmsq<0$ as the inverted mass hierarchy.

This paper is organized as follows. Section~\ref{sec:experiment} is a brief overview of the T2K experiment and the data-taking periods. Section~\ref{sec:ana_overview} summarizes the analysis method and components, including the flux (Section~\ref{sec:flux}), neutrino interaction model (Section~\ref{sec:neut_int}) and near detector and far detector data samples (Section~\ref{sec:nd280} and Section~\ref{sec:sk_selection} respectively). The fit to near detector data, described in Section~\ref{sec:extrapolation}, is used to constrain the far detector rate and associated uncertainties. Finally, Section~\ref{sec:sk_fit_method} describes how the far detector \nue sample is used to estimate \stot.

\section{\label{sec:experiment} Experimental Overview and Data Collection}
The T2K experiment~\cite{Abe:2011ks} is optimized to observe electron neutrino appearance in a muon neutrino beam.
We sample a beam of muon neutrinos generated at the J-PARC accelerator facility in Tokai-mura, Japan, at baselines
of 280~m and 295~km from the neutrino production target. 
The T2K neutrino beam line accepts a 31~GeV/$c$ proton beam from the J-PARC accelerator complex.  The proton beam is
delivered in 5~$\mu$s long spills with a period that has been decreased from 3.64~s to 2.56~s over the data-taking periods described in this paper.  Each spill consists of 8 equally spaced bunches (a significant subset of the data
was collected with 6 bunches per spill) that are $\sim15$~ns wide.
The protons strike a 91.4~cm long graphite target,
producing hadrons including pions and kaons,
and positively charged particles are focused by a series of three magnetic horns operating at 250~kA. 
The pions, kaons and some muons decay
 in a 96~m long volume to produce a predominantly muon neutrino beam.  The remaining protons and particles which have
not decayed are stopped in a beam dump.  A muon monitor situated downstream of the beam dump measures the profile of 
muons from hadron decay and monitors the beam direction and intensity.

\begin{figure}[h]
  \begin{center}
    \includegraphics[keepaspectratio=true,width=0.48\textwidth]{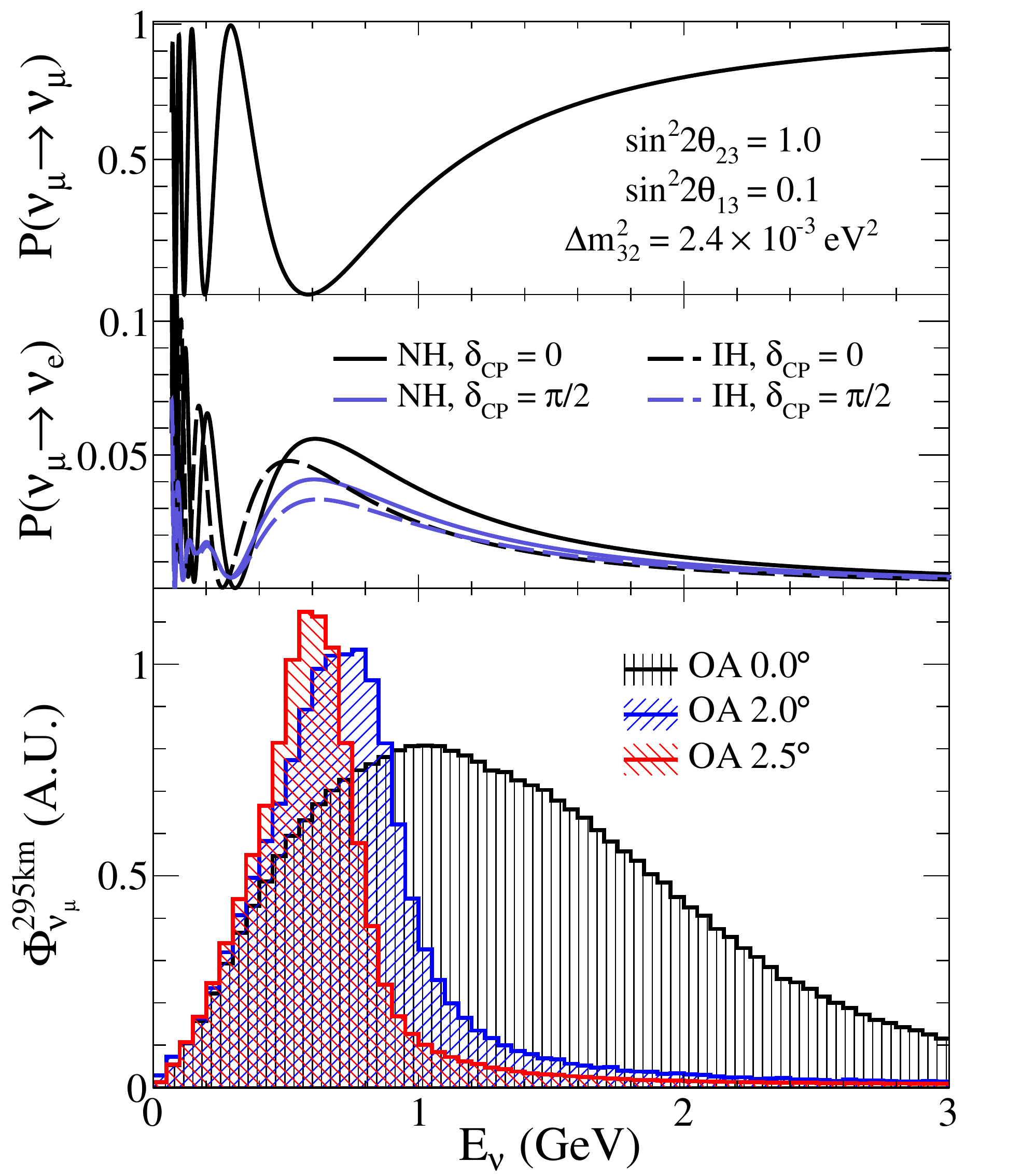}
    \caption{ The muon neutrino survival probability (top) and electron neutrino appearance probabilities (middle)
             at 295 km, and the unoscillated neutrino fluxes for different values of the off-axis angle (OA) (bottom).
             The appearance probability is shown for two values of the phase $\dcp$, and for normal (NH) and inverted (IH) mass hierarchies.}
    \label{fig:oaeffect}
  \end{center}
\end{figure}

We detect neutrinos at both near (280~m from the target)
and far (295~km from the target) detectors.  The far
detector is the Super-Kamiokande (SK) water Cherenkov detector.
The beam
is aimed 2.5$^{\circ}$ (44~mrad) away from the target-to-SK axis to optimize the neutrino energy spectrum for the oscillation measurements.  
The off-axis
configuration~\cite{bnl_offaxis,Mann:1993zk,Helmer:1994ac}
takes advantage of the kinematics of pion decays to 
produce a narrow band beam.  The angle is chosen so that the spectrum peaks at the first oscillation maximum, 
as shown in Fig.~\ref{fig:oaeffect},
maximizing the signal in the oscillation region and minimizing feed-down backgrounds from high energy neutrino
interactions.  This optimization is possible because the value of $|$\dmsq$|$~is already relatively well known.

The near detectors measure the properties of the beam at a baseline where oscillation effects are 
negligible.  The on-axis INGRID detector~\cite{Otani2010368,Abe2012} consists of 16 modules of interleaved scintillator/iron layers 
in a cross configuration centered on the nominal neutrino beam axis, covering $\pm5$~m transverse to the beam direction along the horizontal and vertical axes. The INGRID detector monitors the neutrino event rate stability
at each module, and the neutrino beam direction using the profile of event rates across the modules.

\begin{figure}[h]
  \begin{center}
    \includegraphics[keepaspectratio=true,width=0.48\textwidth]{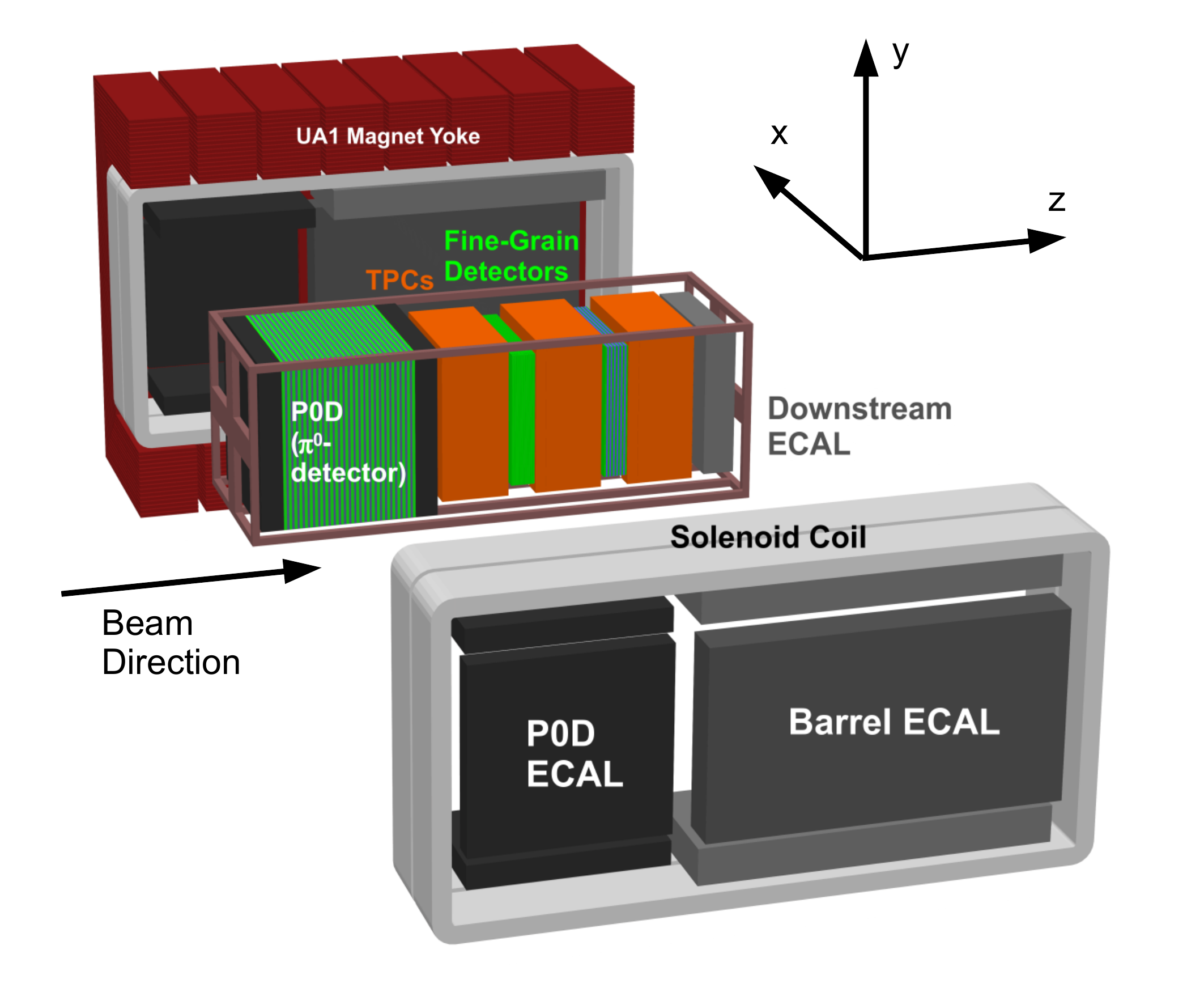}
    \caption{An exploded illustration of the ND280 detector.  The description of the component detectors
can be found in the text.}
    \label{fig:nd280_det}
  \end{center}
\end{figure}

The off-axis ND280 detector is a magnetized multi-purpose detector that is situated along the same direction as 
SK.  It measures the neutrino beam composition and energy spectrum prior to oscillations and is used to study neutrino interactions.
The ND280 detector utilizes a 0.2~T magnetic field generated by the refurbished UA1/NOMAD magnet and consists of a number of 
sub-detectors: side muon range detectors (SMRDs~\cite{Aoki:2012mf}), electromagnetic calorimeters (ECALs), a \piz detector (\pod~\cite{Assylbekov201248})
and a tracking detector.  The tracking detector is composed of two fine-grained scintillator bar
detectors (FGDs~\cite{Amaudruz:2012pe}) sandwiched between three gaseous time projection chambers (TPCs~\cite{Abgrall:2010hi}).
The first FGD primarily consists of polystyrene scintillator and acts as the target for most of the near detector neutrino interactions that are
treated in this paper.  Hence, neutrino interactions in the first FGD are predominantly on carbon nuclei.
The ND280 detector is illustrated in Fig.~\ref{fig:nd280_det}, where the coordinate convention is also indicated.
The $x$ and $z$ axes are in the horizontal plane, and the $y$ axis is vertical. The origin is at the center of the magnet, 
and the magnetic field is along the $x$ direction. The $z$ axis is the direction to the far detector projected to the horizontal plane.

\begin{figure}[h]
  \begin{center}
    \includegraphics[keepaspectratio=true,width=0.48\textwidth]{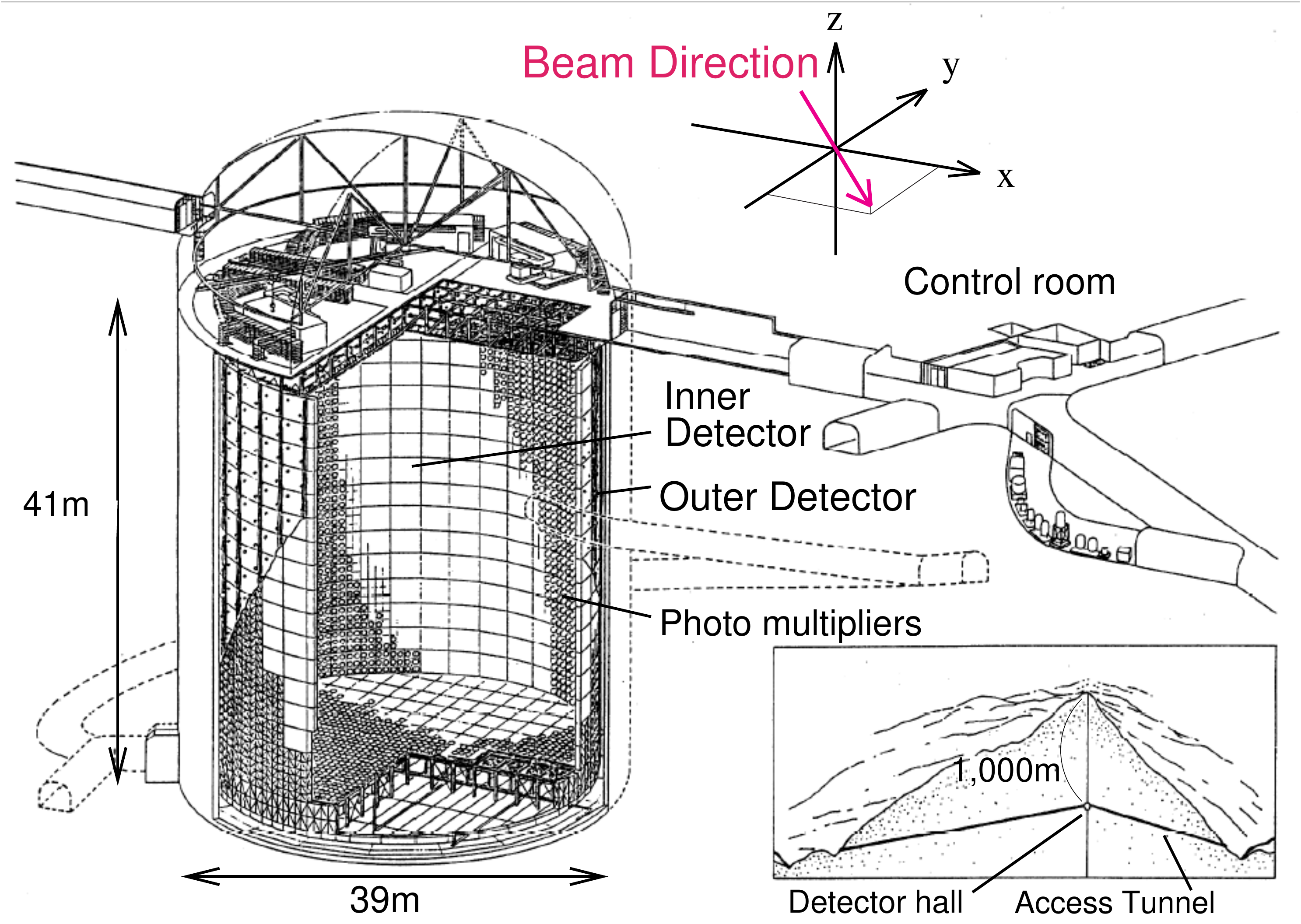}
    \caption{An illustration of the SK detector.}
    \label{fig:sk_det}
  \end{center}
\end{figure}

The SK far detector~\cite{Fukuda:2002uc}, as illustrated in Fig.~\ref{fig:sk_det}, is a 50~kt water Cherenkov detector
located in the Kamioka Observatory.
The cylindrically-shaped water tank is optically separated
to make two concentric detectors :
an inner detector (ID) viewed by 11129 inward-looking 20 inch photomultipliers,
and an outer detector (OD) with 1885 outward-facing 8 inch photomultipliers.
The fiducial volume is defined to be a cylinder whose surface is 2~m away from the ID wall, providing a fiducial mass of 22.5~kt.
Cherenkov photons from charged particles produced in neutrino interactions
form ring-shaped patterns on the detector walls, and are detected by the
photomultipliers.
The ring topology can be used to identify the type of particle and, for charged current interactions, the flavor of the neutrino that interacted. For example, electrons from electron neutrino interactions undergo large multiple scattering and induce electromagnetic showers,
resulting in fuzzy ring patterns.  In contrast,
the heavier muons from muon neutrino interactions produce Cherenkov rings with sharp edges.

The T2K experiment uses a special software trigger to associate neutrino interactions in SK to neutrinos produced in the T2K beam. The T2K trigger records
all the photomultiplier hits within $\pm500$~$\mu$s of the beam
arrival time at SK.
Beam timing information is measured spill-by-spill
at J-PARC and immediately passed to the online computing system at SK.
The time synchronization between the two sites is done using
the Global Positioning System (GPS) with $<150$~ns precision
and is monitored with the Common-View method~\cite{CommonView}.
Spill events recorded by the T2K triggers are processed offline to apply 
the usual SK software triggers used to
search for neutrino events, and any candidate events found are extracted
for further T2K data analysis.
Spills used for the far detector data analysis are selected by beam and
SK quality cuts.
The primary reason spills are rejected at SK is due to the requirement that there are no events in the $100$~$\mu$s before the beam window,  
which is necessary to reject decay electrons from cosmic-ray muons.

In this paper we present neutrino data collected during
the three run periods listed in Table~\ref{tab:t2k_runs}.
The total SK
data set corresponds to $3.01\times10^{20}$ protons on target (POT) or $4\%$ of the T2K design exposure.
About 50\% of the data, the Run 3 data, were collected after T2K and J-PARC recovered from the 2011 Tohoku earthquake. 
 A subset of data
corresponding to $0.21\times10^{20}$ POT from Run 3 was collected with the magnetic horns operating at 
205~kA instead of the nominal
value of 250~kA.  The size of the total data set is approximately two times that of T2K's previously published
electron neutrino appearance result~\cite{PhysRevLett.107.041801}.

\begin{table}
\begin{center}
\caption{T2K data-taking periods and the integrated protons on target (POT) for SK
data collected in those periods. }
\label{tab:t2k_runs}
\begin{tabular}{lcc}
\\ \hline
Run Period & Dates & Integrated POT by SK  \\ \hline
Run 1 & Jan. 2010-Jun. 2010 &  $0.32\times10^{20}$ \\
Run 2 & Nov. 2010-Mar. 2011 &  $1.11\times10^{20}$ \\
Run 3 & Mar. 2012-Jun. 2012 &  $1.58\times10^{20}$ \\
\hline
\end{tabular}
\end{center}
\end{table}

\begin{figure*}
  \begin{center}
    \includegraphics[keepaspectratio=true,width=0.95\textwidth]{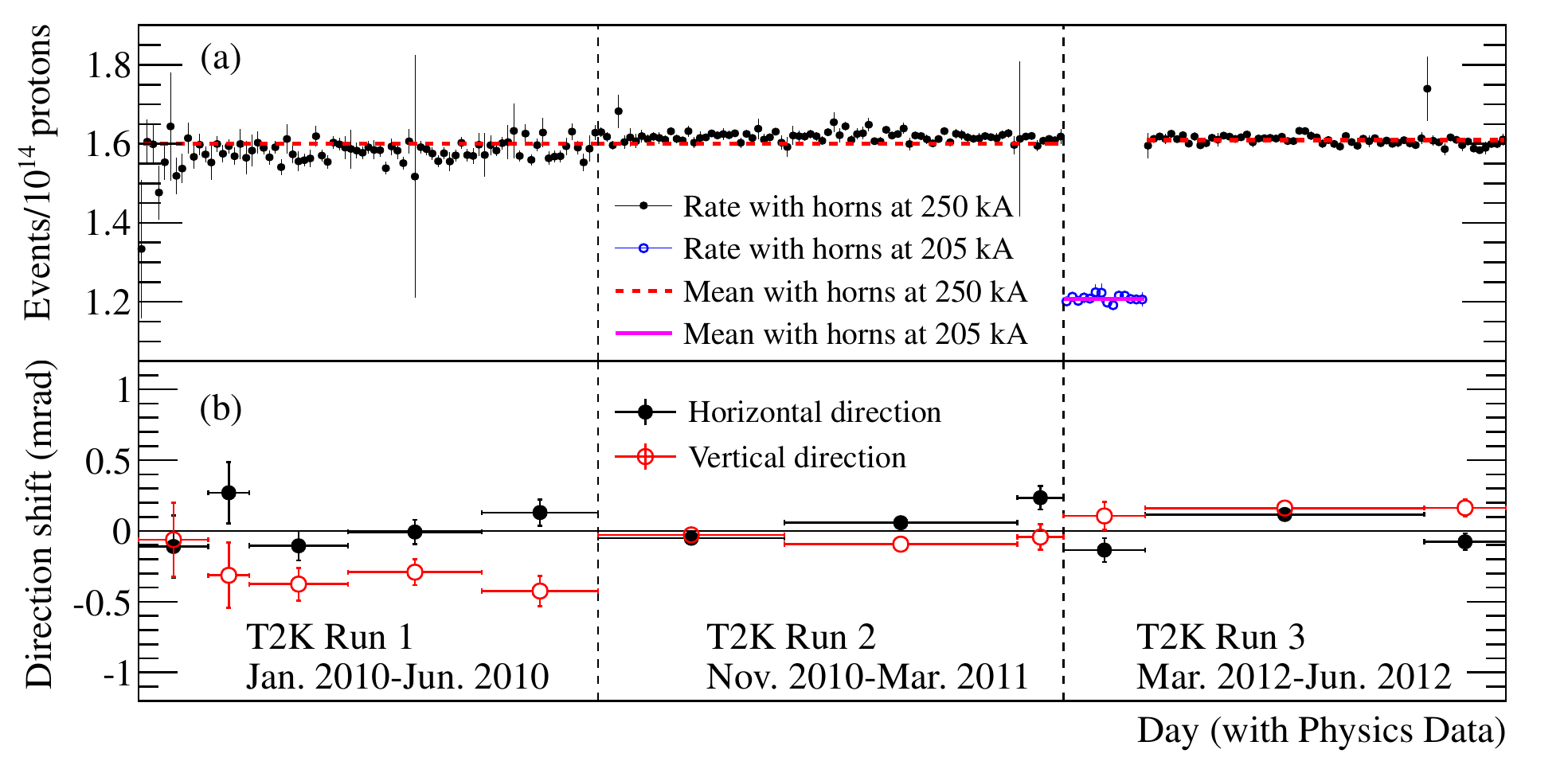}
    \caption{The time dependence of the POT-normalized reconstructed neutrino event rate (a) and the beam direction (b) measured by INGRID.  The error bars show the statistical uncertainty only. The points shown for the direction measurement include sequential data grouped in periods of stable beam conditions. }
    \label{fig:ingrid_meas}
  \end{center}
\end{figure*}

We monitor the rate and direction of the neutrino beam over the full data-taking period with the INGRID detector.
As illustrated in Fig.~\ref{fig:ingrid_meas}, the POT-normalized neutrino event rate is stable to within 1\%, and the beam
direction is controlled well within the design requirement of 1~mrad, which corresponds to a 2\% shift in the peak energy of the neutrino spectrum.

\section{\label{sec:ana_overview} Analysis Overview}

We search for $\nu_{\mu} \rightarrow \nu_{e}$ oscillations via charged current quasi-elastic (CCQE) interactions of $\nu_e$ 
at SK.  Since the recoil proton from the target nucleus is typically below Cherenkov threshold, these events are 
characterized by a single electron-like ring and no other activity.  The most significant background sources are 
$\nu_e$ from muon and kaon decays that are intrinsic to the neutrino beam, 
and neutral current $\pi^0$ (NC$\pi^0$) events where the detector response to the photons from the $\pi^0$ decay is 
consistent with a single electron-like ring.  The selection of $\nu_e$ candidates is described in 
Section~\ref{sec:sk_selection}.

We estimate the oscillation parameters and produce confidence intervals using a model that 
describes the probabilities to observe $\nu_e$ candidate events at SK in bins of electron momentum (magnitude and direction),
as described in Section~\ref{sec:sk_fit_method}. 
The probabilities depend on the values of the oscillation parameters as well as many nuisance parameters that arise 
from uncertainties in neutrino fluxes, neutrino interactions, and detector response. The point where the likelihood is 
maximum for the observed data sample gives the oscillation parameter estimates, and the likelihood ratio at other 
points is used to construct confidence intervals on the parameters. 

We model the neutrino flux with a data-driven simulation that takes as inputs measurements of the proton beam, 
hadron interactions and the horn fields~\cite{PhysRevD.87.012001}.  The uncertainties on the flux model parameters
arise largely from the uncertainties on these measurements. The flux model and its uncertainties are described in 
Section~\ref{sec:flux}. 

We model the interactions of neutrinos in the detectors assuming interactions on a quasi-free nucleon
using a dipole parametrization for vector and axial form factors.
 The nuclei are treated as a relativistic Fermi gas, and outgoing hadrons are subject to 
interactions in the nucleus, so-called ``final state interactions''.
We validate the neutrino interaction model with comparisons to
independent neutrino cross section measurements at  $\mathcal{O}(1)$~GeV
and pion
scattering data.  We set the uncertainties on the interaction 
model with comparisons of the model to data and
alternate models.  The neutrino interaction model and its uncertainties are described in Section~\ref{sec:neut_int}.

We further constrain the flux and interaction model parameters with a fit to samples of neutrino interaction candidates
in the ND280 detector.  Selections containing a negative muon-like particle provide high purity samples 
of $\nu_{\mu}$ interactions, which constrain both the $\nu_{\mu}$ flux that determines 
signal and NC$\pi^0$ backgrounds at SK,  and the intrinsic $\nu_e$ flux.  In the energy range of interest, the intrinsic 
$\nu_e$ are predominantly produced from the decay chain $\pi^+ \rightarrow \mu^++\nu_{\mu}$, 
$\mu^+ \rightarrow e^++\nu_e+\bar{\nu}_{\mu}$, and to a lesser extent by three-body kaon decays. 
Hence, the $\nu_e$ flux is correlated with the $\nu_{\mu}$ 
flux through the production of pions and kaons in the T2K beam line.  
The charged current interactions that make up most of the 
ND280 samples constrain the charged current interaction model.
While \nue interactions are indirectly constrained by \num interactions, we also include uncertainties which account for differences between the \num and \nue cross section model.
The ND280 neutrino interaction sample selection is described in Section~\ref{sec:nd280}, and the fit of the 
neutrino flux and interaction models to
this data is described in Section~\ref{sec:extrapolation}.

\section{\label{sec:flux} Neutrino Flux Model}

We simulate the T2K beam line to calculate the neutrino flux at the near and far detectors in the absence 
of neutrino oscillations,
and this flux model is used as an input to predict neutrino interaction event rates at the detectors.

The flux simulation begins with the primary proton beam upstream of the collimator that sits in front of the T2K target.  
The interactions of particles in the target, beam line components, decay volume walls and 
beam dump, and their decays, are simulated.
The simulation and its associated uncertainties are driven by measurements of the
primary proton beam profile, measurements of the magnetic fields of the T2K horns, and hadron
production data, including NA61/SHINE measurements~\cite{Abgrall:2011ae,Abgrall:2011ts}.
First, we model the interactions of the primary beam protons and subsequently produced particles in the
graphite target with a FLUKA 2008~\cite{Ferrari:2005zk,Battistoni:2007zzb} simulation. 
We pass any particles that exit the target into a GEANT3~\cite{GEANT3}
simulation that tracks particles through the magnetic horns
and decay region, and decays hadrons and muons to neutrinos.  The hadron interactions in the
GEANT3 simulation are modeled with GCALOR~\cite{GCALOR}.  
To improve agreement between selected hadron interaction measurements and the simulation, we 
weight simulated events based on the stored information of the true initial and final state hadron 
kinematics for hadron interactions in events producing neutrinos.  

\begin{figure}
\centering
\includegraphics[width=0.48\textwidth]{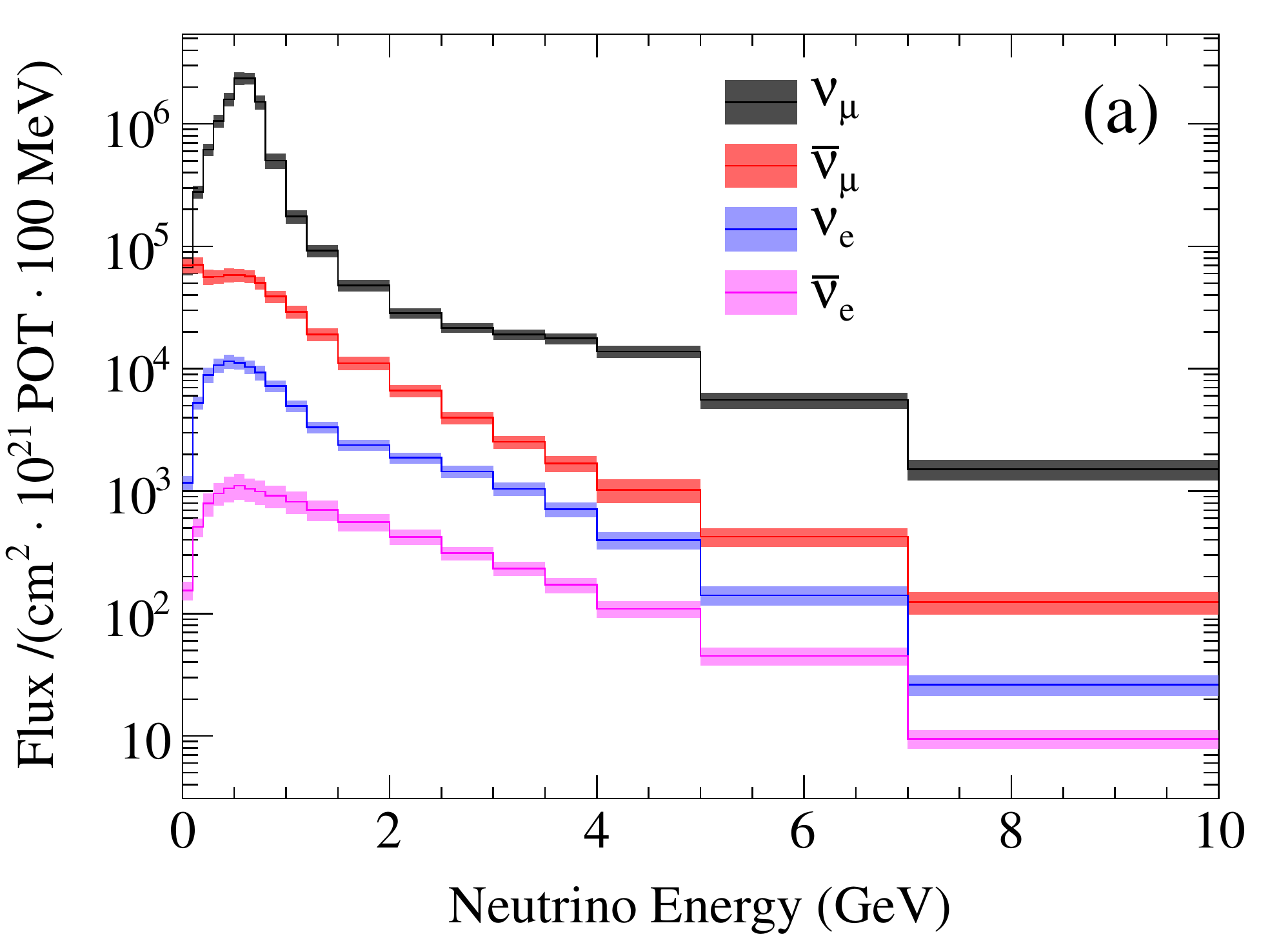}
\includegraphics[width=0.48\textwidth]{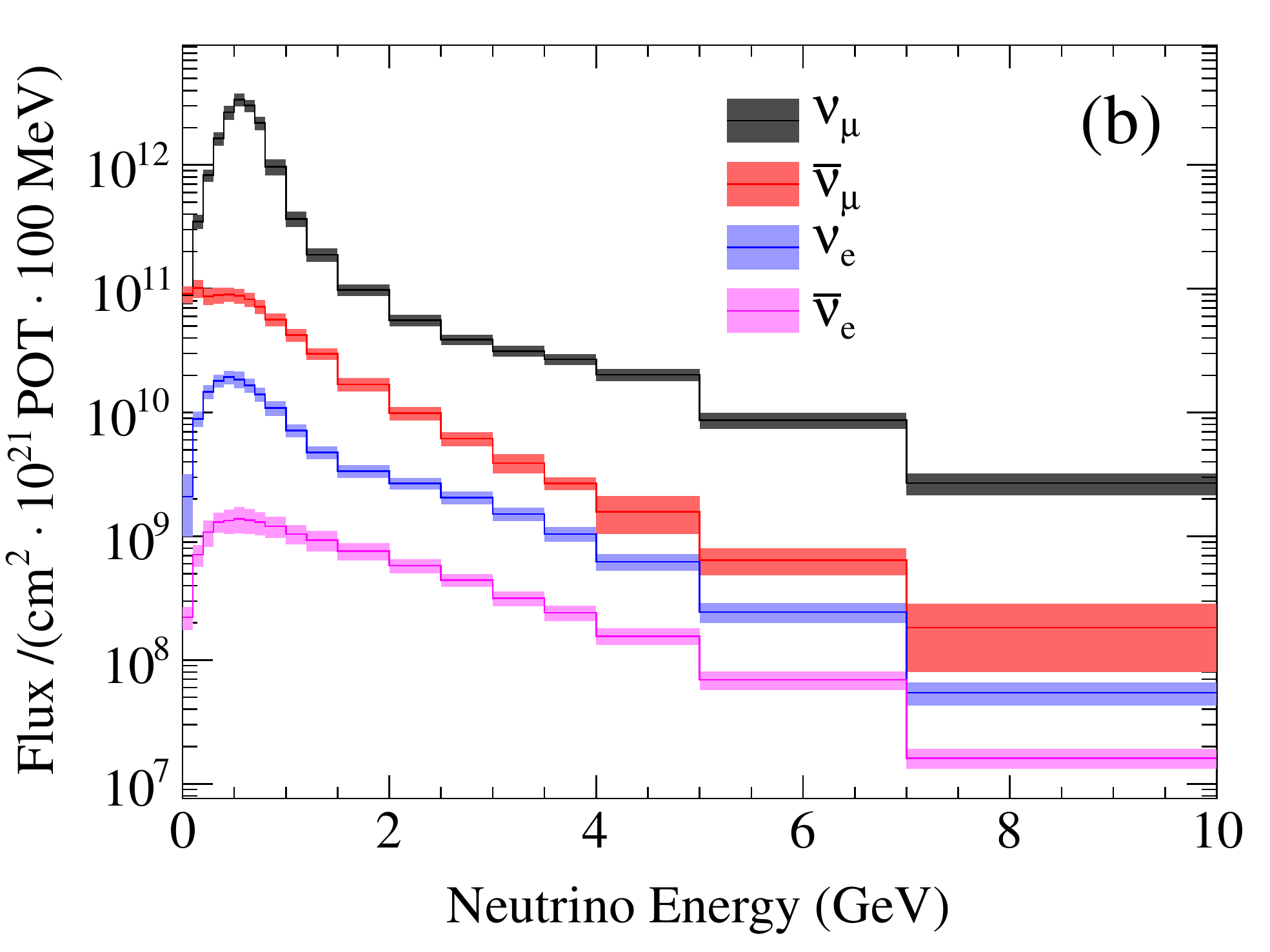}
\caption{The T2K flux prediction at SK (a) and ND280 (b) for neutrinos and antineutrinos with systematic error bars.  The flux above $E_\nu=10$~GeV is not shown; the flux is simulated up to $E_\nu=30$~GeV. }
\label{fig:flux_pred}
\end{figure}

The predicted flux at the SK and ND280 detectors, including systematic errors, is shown
in Fig.~\ref{fig:flux_pred}.  Here we describe the methods for weighting the flux
and evaluating uncertainties based on proton beam measurements, hadron interaction data, alignment measurements, horn current
and field measurements, and the beam direction measurement from the INGRID detector.
More details of the flux calculation are described in Ref.~\cite{PhysRevD.87.012001}.

\subsection{Weighting and systematic error evaluation methods}
\label{sec:flux_reweight}
To tune the flux model and study its uncertainties, adjustments are made by weighting events based on 
kinematics of the hadron interactions or the primary proton.
The sensitivities to nuisance parameters that arise from such uncertainties as the hadron production model, 
proton beam profile, or horn currents, are evaluated by their effect on the predicted neutrino spectrum.

We use one of two approaches for each uncertainty source, depending on whether the uncertainty source 
has correlations that need to be treated.
For error sources described by a number of correlated underlying parameters, 
we use weighting methods when possible. 
The nuisance parameters are sampled according to their covariance and the corresponding flux predictions for the 
$k$ samples, 
$\phi^k$, are calculated. A large number of parameters sets, N (typically 500 or more), are used to calculate the 
fractional covariance using:
\begin{linenomath*}
\begin{equation}
v_{ij} = \frac{1}{N}\sum_{k=1}^{N}\frac{(\phi_{i}^{nom}-\phi_{i}^{k})(\phi^{nom}_{j}-\phi_{j}^{k})}{\phi_{i}^{nom}\phi_{j}^{nom}}.
\end{equation}
\end{linenomath*}
Here $\phi_{i}^{nom}$ is the nominal flux prediction and $i$ specifies a neutrino energy bin, flavor
and detector at which the flux is evaluated.  
We evaluate hadron interaction and proton beam profile uncertainties with this method.

For systematic variations that cannot be treated by weighting simulated events, such as misalignment of beam line
elements or changes to the horn currents, we produce new simulated event samples with $\pm1\sigma$ variations of the
nuisance parameters and calculate the fractional covariance matrix:
\begin{linenomath*}
\begin{align}
v_{ij} & = \frac{1}{2}\frac{(\phi_{i}^{nom}-\phi_{i}^{+})(\phi^{nom}_{j}-\phi_{j}^{+})}{\phi_{i}^{nom}\phi_{j}^{nom}} \notag \\
       & + \frac{1}{2}\frac{(\phi_{i}^{nom}-\phi_{i}^{-})(\phi^{nom}_{j}-\phi_{j}^{-})}{\phi_{i}^{nom}\phi_{j}^{nom}}.
\end{align}
\end{linenomath*}
$\phi_{i}^{+}$ and $\phi_{i}^{-}$ are the flux prediction for $+1\sigma$ and $-1\sigma$ variations of
the nuisance parameter.  We evaluate horn and target alignment and horn current and field uncertainties
with this method.

The total fractional flux covariance matrix is the sum of fractional flux covariance matrices calculated for each
source of uncertainty.  
For the fits to data described in Sections~\ref{sec:extrapolation} and~\ref{sec:sk_fit_method}, variations of the
flux prediction are modeled with parameters $b_{i}$ that scale the normalization of the flux in bins of neutrino
energy and flavor at a given detector.  The covariance matrix of the $b_{i}$, $(V_{b})_{ij}$, is simply the total 
fractional flux covariance matrix described here.  
Since the $b_{i}$ are separated for the near and far detectors, their covariances account for the correlations 
between the flux predictions at the two detectors. The covariances can therefore be used directly in simultaneous 
fits of near and far detector data or to calculate the uncertainty on the ratio of flux spectra at the two detectors.

The following sections describe each source of flux systematic uncertainty.

\subsection{Proton beam monitoring and simulation}
\label{sec:flux_pbeam}
We simulate the proton beam according to the proton orbit and optics parameters measured
by the proton beam position and profile monitors, and the number of protons measured by the 
intensity monitors.  
These monitors are described elsewhere~\cite{Abe:2011ks,Bhadra201345}.
We measure proton beam properties for each run period by reconstructing the
beam profile at the upstream end of the collimator that sits before the T2K target 
for each beam spill.
The sum of profiles for each beam spill, weighted by the number of protons, gives the proton beam profile 
that we input to the flux simulation.
Table~\ref{tab:pbeam_center_xy} summarizes the  
measured mean position, angle, emittance, Twiss $\alpha$ parameter~\cite{McDonald:1988nv} and width of the proton beam at the collimator, and their uncertainties
 for a typical run period. 
The largest contributions to the flux uncertainty from the proton beam simulation arise from 
the alignment uncertainties of the beam monitors.

The effect of the proton beam profile uncertainty on the flux is studied by varying the parameters in 
Table~\ref{tab:pbeam_center_xy} within their uncertainties while accounting for the parameter correlations.
The uncertainties on $Y$ and $Y^{\prime}$ are dominant and are studied on a simulated ``wide beam" flux sample that has
a profile in the $y-y^{\prime}$ (proton vertical position and angle) plane that covers the measured uncertainties.  The wide beam sample is 
weighted for variations of $Y$ and $Y^{\prime}$ and the effect on the flux is studied.  The variations
correspond to shifts in the off-axis angle of $\sim0.35$ mrad, or shifts in the off-axis spectrum peak of 
$\sim 10$~MeV.

\begin{table}[tb]
  \caption{\label{tab:pbeam_center_xy}
Summary of measured proton beam profile parameters and uncertainties at the collimator for a typical run period : mean position ($X$,$Y$)
and angle ($X^{\prime}$,$Y^{\prime}$), width~($\sigma$), emittance~($\epsilon$), and Twiss parameter~($\alpha$).}
\begin{center}
\begin{tabular} {lcc|cc}
\hline
            &\multicolumn{2}{c|}{X Profile} & \multicolumn{2}{c}{Y Profile} \\
Parameter                       & Central Value & Error & Central Value & Error \\ \hline
$X,Y$ (mm)                     & 0.00          & 0.35  &  -0.37        & 0.38  \\
$X^{\prime},Y^{\prime}$ (mrad) & 0.03          & 0.07  &  0.07         & 0.28  \\
$\sigma$ (mm)                   & 4.03          & 0.14  & 4.22          & 0.12  \\
$\epsilon$ ($\pi$ mm mrad)      & 4.94          & 0.54  & 6.02          & 3.42  \\
$\alpha$                        & 0.33          & 0.08  & 0.34          & 0.41  \\
\hline
\end{tabular}
\end{center}
\end{table}

\subsection{Hadron production data, weighting and uncertainties}
\label{sec:flux_had}
The pion and kaon differential production measurements we use to weight the
T2K flux predictions are summarized in Table~\ref{tab:haddatat2k}.

\begin{table}
\centering
\caption{Differential hadron production data relevant for the T2K neutrino flux predictions.}
\begin{tabular}{llll}
\hline
Experiment & Beam Mom. & Target & Particles \\
\hline
NA61/SHINE \cite{Abgrall:2011ae,Abgrall:2011ts} & 31~GeV/$c$  & C & $\pi^\pm$, $K^+$ \\
Eichten \textit{et  al.}~\cite{eichten} & 24~GeV/$c$ & Be, Al, ... & $p$, $\pi^\pm$, $K^\pm$\\
Allaby \textit{et al.}~\cite{allaby} & 19.2~GeV/$c$ & Be, Al, ... & $p$, $\pi^\pm$, $K^\pm$\\
BNL-E910 \cite{e910} & 6.4-17.5~GeV/$c$ & Be & $\pi^\pm$ \\
\hline
\end{tabular}
\label{tab:haddatat2k}
\end{table}

We weight charged meson differential production multiplicities to the NA61/SHINE $\pi^{+}/\pi^{-}$~\cite{Abgrall:2011ae} 
and $K^{+}$~\cite{Abgrall:2011ts} thin target production data, which covers most of phase space
relevant for the off-axis flux.  We use additional kaon differential production data from Eichten {\it et al.}~\cite{eichten} 
and Allaby {\it et al.}~\cite{allaby} to weight $K^{+}$ multiplicities in the phase space not covered by the
NA61/SHINE measurements, and for $K^{-}$ multiplicities.
To estimate the uncertainty of pion production by 
secondary protons, we use differential pion production data from the BNL-E910
experiment~\cite{e910} that were collected in interactions with proton beam energies less than the
T2K primary proton beam energy.

We use measurements of the inelastic cross sections for proton, pion, and
kaon beams with carbon and aluminum targets~\cite{Abrams,Allaby1969500,Allardyce:1973ce,Bellettini:1966zz,Bobchenko,Carroll,Cronin,Chen,Denisov,Longo,Vlasov} to weight based on
particle interaction and absorption rates in the flux prediction. 
In particular, NA61/SHINE measures the inclusive ``production'' cross section of 31~GeV/$c$ protons
on carbon: $\sigma_{prod} = 229.3\pm 9.2$ mb~\cite{Abgrall:2011ae}.
The production cross section is defined as:
\begin{linenomath*}
\begin{equation}
\sigma_{prod} = \sigma_{inel} - \sigma_{qe}.
\label{eq:prod_xsec}
\end{equation}
\end{linenomath*}
Here, $\sigma_{qe}$ is the quasi-elastic scattering cross section, {\it i.e.} scattering off of individual 
bound nucleons that breaks up or excites the nucleus, but does not produce additional hadrons. The inclusive 
production cross section is used in the weighting of
the flux prediction, and the quasi-elastic cross section is subtracted from measurements where 
necessary.

We apply hadron interaction-based weights to simulated events in two steps.  
The multiplicity of pions and kaons
produced in interactions of nucleons on the target nuclei is defined as:
\begin{linenomath*}
\begin{equation}
\frac{dn}{dp}(p,\theta) = \frac{1}{\sigma_{prod}}\frac{d\sigma}{dp}(p,\theta).
\label{eq:diffprod}
\end{equation}
\end{linenomath*}
Here $p$ and $\theta$ are the momentum and angle relative to the incident particle of the 
produced particle in the lab frame.
We apply multiplicity weights that are the ratio of the measured and simulated 
differential multiplicities: 
\begin{linenomath*}
\begin{equation}
W(p,\theta) = \frac{[\frac{dn}{dp}(p,\theta)]_{data}}{{[\frac{dn}{dp}(p,\theta)]_{MC}}}.
\label{eq:prodweight}
\end{equation}
\end{linenomath*}

We adjust the interaction rates of protons, charged pions and charged kaons as well, with
weights that account for attenuation in the target:
\begin{linenomath*}
\begin{eqnarray}
W &=& \frac{\sigma_{prod}'}{\sigma_{prod}} e^{-x(\sigma_{prod}'-\sigma_{prod})\rho}.
\label{eq:xsecweight}
\end{eqnarray}
\end{linenomath*}
Here $\rho$ is the number density of nuclear targets in the material, $\sigma_{prod}$ is the original
inclusive production cross section in the simulation, $\sigma_{prod}'$ is the inclusive production cross section
to which the simulation is being weighted, and $x$ is the distance traversed by the particle through
the material.  The total
weight is the product of weights from all materials through which the particle propagates.

For pion and kaon production in secondary nucleon interactions, or in the phase space covered by the alternative
kaon production data sets, we converted weights to an $x_{F}-p_{T}$
dependence, where $p_{T}$ is the transverse momentum of the produced particle and $x_{F}$ is
the Feynman $x$~\cite{Feynman69} defined as:
\begin{linenomath*}
\begin{equation}
x_{F} = p_{L}/p_{max}.
\end{equation}
\end{linenomath*}
Here $p_{L}$ is the longitudinal momentum of the produced particle in the center of mass frame, and
$p_{max}$ is the maximum momentum the produced particle can have.  We apply the $x_{F}-p_{T}$ dependent weights
after converting simulated hadron interactions to the $x_{F}-p_{T}$ basis.  This method
assumes that the pion and kaon multiplicities expressed in the $x_{F}-p_{T}$ basis are independent of 
the collision center of mass energy.

\begin{figure}[ht]
\centering
\includegraphics[width=0.48\textwidth]{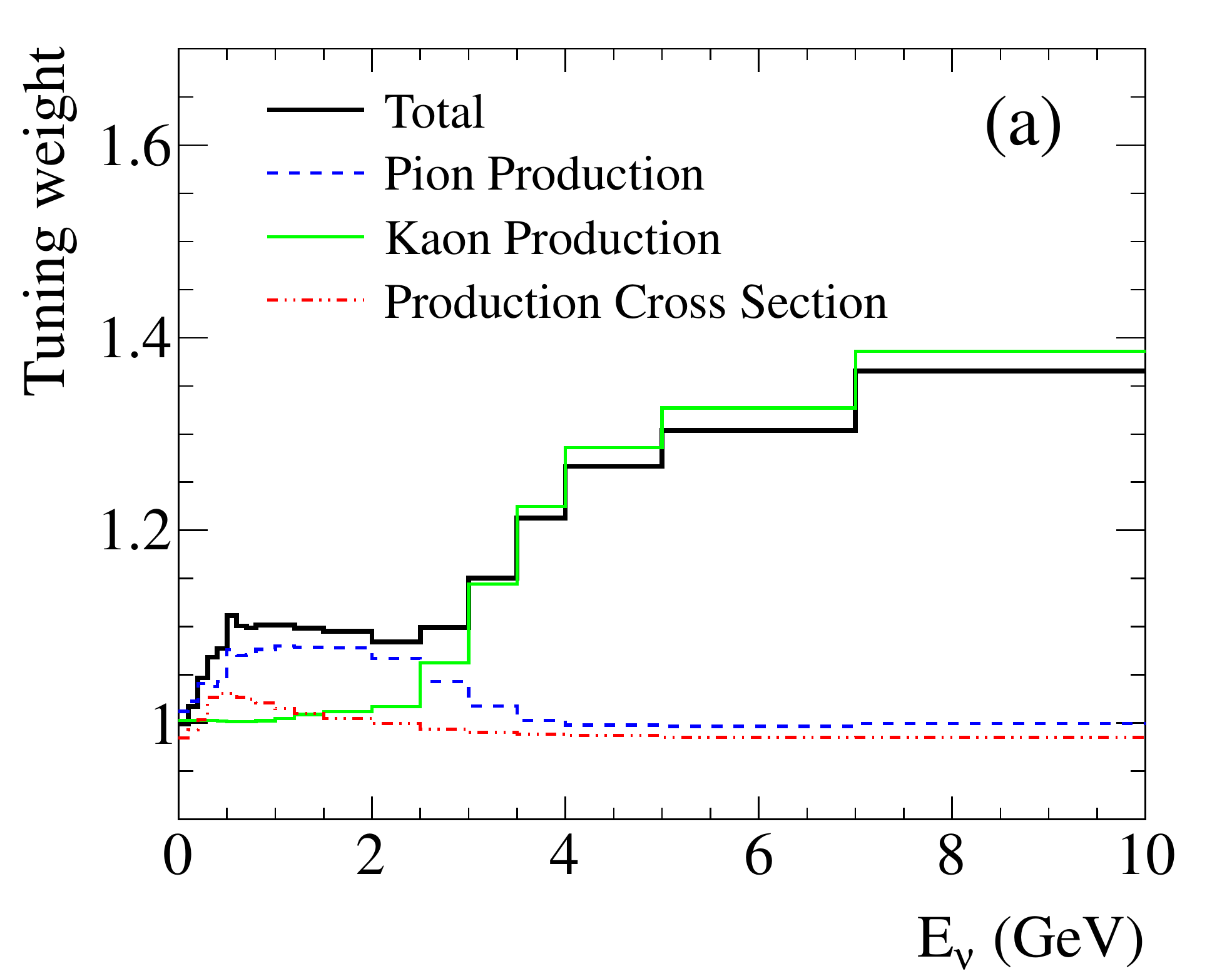}
\includegraphics[width=0.48\textwidth]{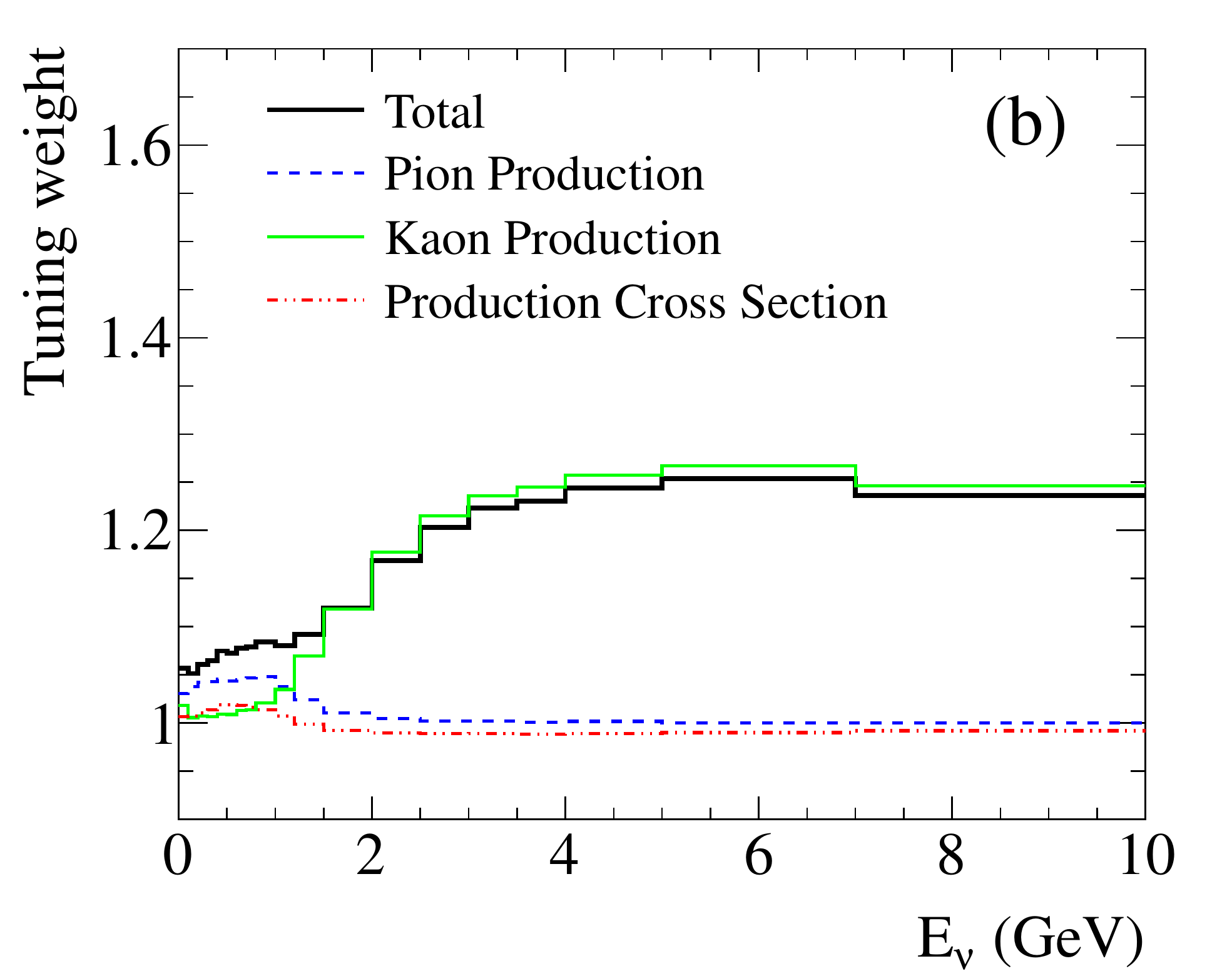}
\caption{Ratio of the hadron interaction weighted flux to the nominal flux for $\nu_{\mu}$ (a),
$\nu_{e}$ (b) flux predictions at SK.  The effects of the pion production, kaon production and 
inclusive production cross section weighting are shown separately and in total. }
\label{fig:flux_had_tune}
\end{figure}

The effect of the hadron interaction
weighting on the SK $\nu_{\mu}$ and $\nu_{e}$ flux are shown as the ratios of weighted to nominal flux in 
Fig.~\ref{fig:flux_had_tune}. The weighting of pion multiplicities is a $10\%$ effect at low energy, while the
weighting of kaon multiplicities affects the flux by as much as $40\%$ in the high energy tail.  The large
weighting effect for kaons is due to the underestimation of kaon production above kaon momenta of 3~GeV/$c$ in
the simulation.  The effect of the inclusive production cross section weighting on the flux prediction is less than 4\% for all energies.

The uncertainties on the hadron multiplicity measurements contribute to the total uncertainty on the flux.
Typical NA61/SHINE $\pi^{\pm}$ data points have $\sim7\%$ systematic error, corresponding to a maximum uncertainty
of 6$\%$ on the flux.  In addition, we evaluate uncertainties on the $x_{F}$ scaling assumption (less than 3\%),
and regions of the pion phase space not covered by data (less than 2\%).
The dominant source of uncertainty on the kaon
production is the statistical uncertainty on the NA61/SHINE measurements. 

The uncertainties on the inclusive production cross section measurements reflect the discrepancies that are seen
between different measurements at similar incident particle energies.  These discrepancies are similar in size
to $\sigma_{qe}$ and may arise from ambiguities in the actual quantity being measured by each experiment.
We apply an uncertainty equal to the $\sigma_{qe}$ component to the inclusive production cross section measurements 
(typically larger than the individual measurement errors), and the uncertainty propagated to the flux is less than 8\% for all energies.

We apply an additional uncertainty to the production of secondary nucleons,
for which no adjustments are made in the current flux prediction 
The uncertainty is based on the discrepancy between the FLUKA
modeling of secondary nucleon production and measurements by Eichten et. al.~\cite{eichten} and Allaby et. al.~\cite{allaby}.
The uncertainty propagated to the flux is less than 10\% for all energies.  

\begin{figure}
\centering
\includegraphics[width=0.48\textwidth]{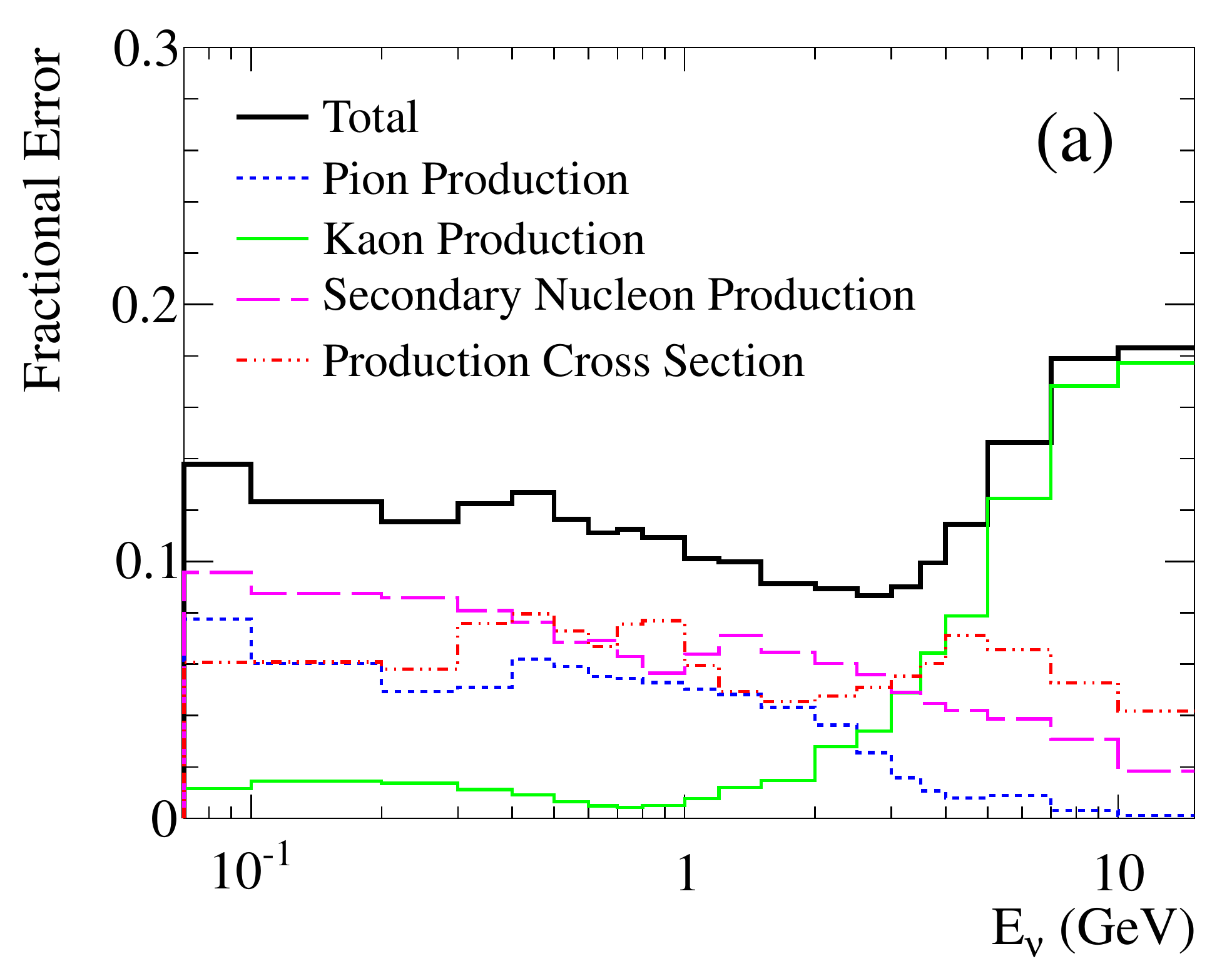}
\includegraphics[width=0.48\textwidth]{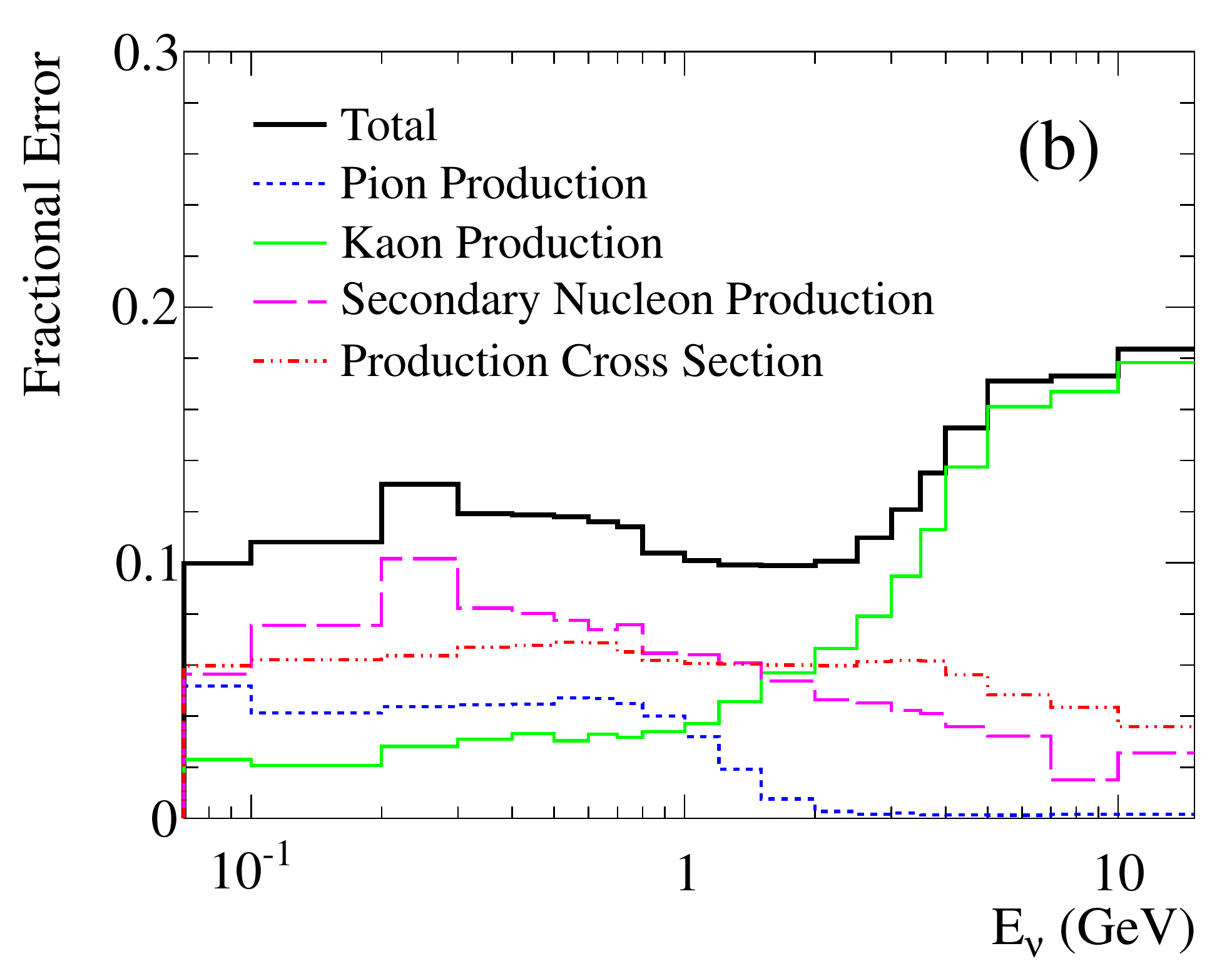}
\caption{The fractional hadron interaction errors on $\nu_{\mu}$ (a),
$\nu_{e}$ (b) flux predictions at SK.}
\label{fig:flux_had_unc}
\end{figure}

The neutrino energy-dependent hadron interaction uncertainties on the SK $\nu_{\mu}$ and $\nu_{e}$ 
flux predictions are summarized in Fig.~\ref{fig:flux_had_unc}, and represent the dominant source
of uncertainty on the flux prediction.

\subsection{Horn and target alignment and uncertainties}
\label{sec:flux_align}
The horns are aligned relative to the primary beam line with uncertainties of 0.3~mm in the 
transverse x direction and 1.0~mm in 
the transverse y direction and beam direction.  The precision of the horn angular alignment is 0.2 mrad. 
After installation in the first horn, both ends of the target were surveyed, and the target was found to be 
tilted from its intended orientation by 1.3 mrad.
We have not included this misalignment in the nominal flux calculation, but the effect is simulated and included as an uncertainty.  
We also simulate linear and angular displacements of the
horns within their alignment uncertainties and evaluate the effect on the flux.  The total alignment uncertainty on the 
flux is less than 3\% near the flux peak.

\subsection{Horn current, field and uncertainties}
\label{sec:flux_horn}
We assume a $1/r$ dependence of the magnetic field in the flux simulation. 
The validity of this assumption is confirmed by measuring the horn field 
using a Hall probe. The maximum deviation from the calculated values is 2\% 
for the first horn and less than 1\% for the second and third horns. 
Inside the inner conductor of a spare first horn, we observe an anomalous field transverse to the horn axis with a maximum strength of 
0.065~T. 
Flux simulations including the
anomalous field show deviations from the nominal flux of up to 4\%, but only for
energies greater than $1$~GeV.

The absolute horn current measurement uncertainty is 2\% and arises from the uncertainty
in the horn current monitoring.  We simulate the flux
with $\pm$5~kA variations of the horn current, and the
effect on the flux is 2\% near the peak.

\subsection{Off-axis angle constraint from INGRID}
\label{sec:flux_ingrid}
The muon monitor indirectly measures the neutrino beam direction by detecting the muons from meson decays,
while the INGRID on-axis neutrino detector directly measures the neutrino beam direction.
The dominant source of uncertainty on the beam direction constraint is the systematic uncertainty
on the INGRID beam profile measurement, corresponding to a $0.35$ mrad uncertainty.
We evaluate the effect on the flux when the SK or ND280 off-axis detectors are shifted in the simulation by 
0.35 mrad.

\subsection{Summary of flux model and uncertainties}
The T2K flux predictions at the ND280 and SK detectors have been described and are shown in Fig.~\ref{fig:flux_pred}.
We use the flux predictions as inputs to calculate event rates at both the ND280 and SK detectors.  To
evaluate the flux related uncertainties on the event rate predictions, we evaluate the fractional uncertainties on 
the flux prediction in bins of energy for each neutrino flavor.  The bin edges are:
\begin{itemize}
\item $\nu_{\mu}$: 0.0, 0.4, 0.5, 0.6, 0.7, 1.0, 1.5, 2.5, 3.5, 5.0, 7.0, 30.0~GeV
\item $\bar{\nu}_{\mu}$: 0.0, 1.5, 30.0~GeV
\item $\nu_{e}$: 0.0, 0.5, 0.7, 0.8, 1.5, 2.5, 4.0, 30.0~GeV
\item $\bar{\nu}_{e}$: 0.0, 2.5, 30.0~GeV
\end{itemize}
We choose coarse binning for the antineutrino fluxes since they
make a negligible contribution for the event samples described in this paper.  
The neutrino flux has finer bins around the oscillation maximum and coarser bins where the flux prediction uncertainties are strongly correlated.

\begin{figure}
\centering
\includegraphics[width=0.48\textwidth]{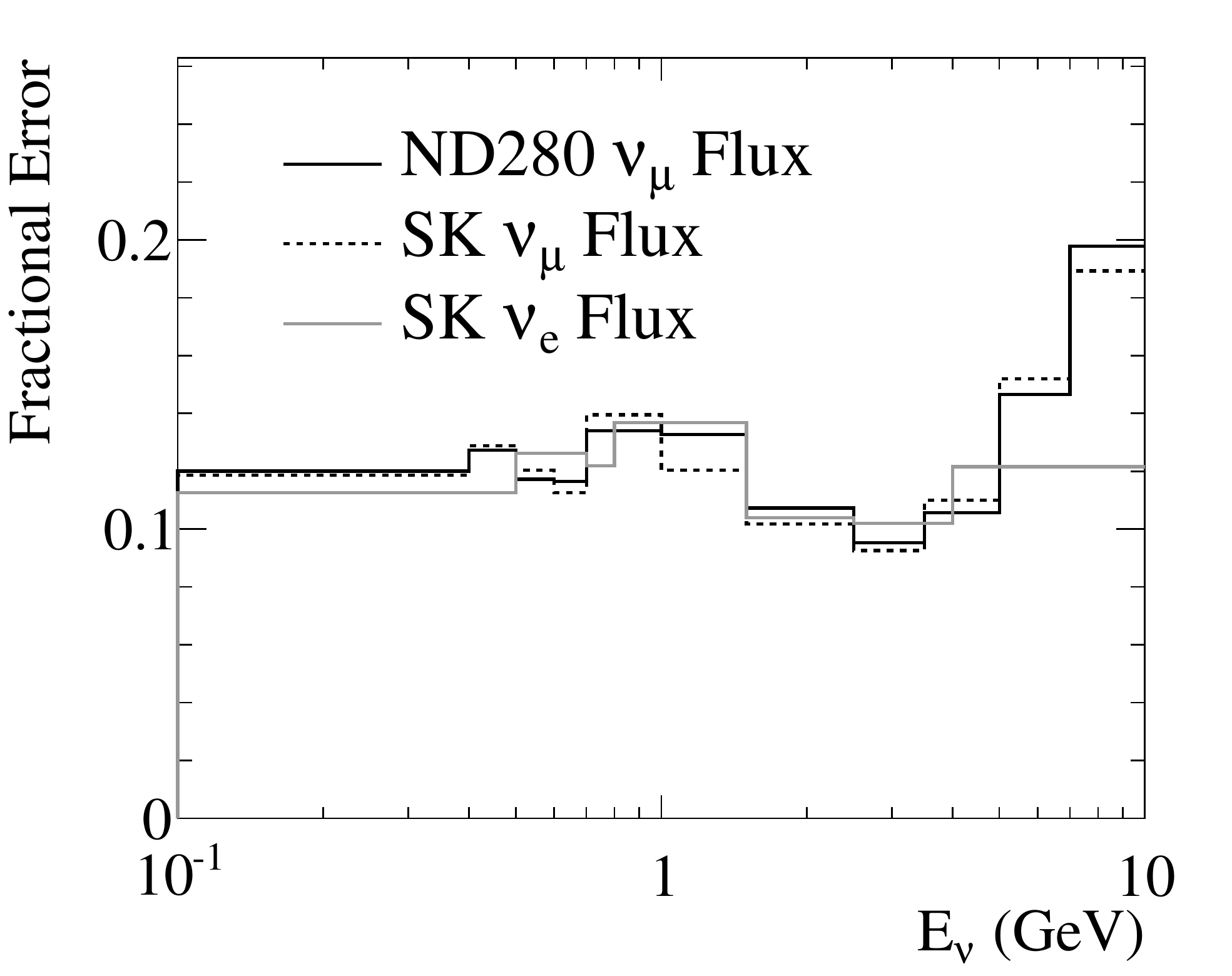}
\caption{The fractional uncertainties on the ND280 $\nu_{\mu}$, SK $\nu_{\mu}$ and SK $\nu_{e}$ flux
evaluated for the binning used in this analysis. This binning is coarser than the binning shown in Fig.~\ref{fig:flux_had_unc} and includes the correlations between merged bins.}
\label{fig:flux_unc_banff}
\end{figure}

\begin{figure*}
\centering
\includegraphics[width=0.85\textwidth]{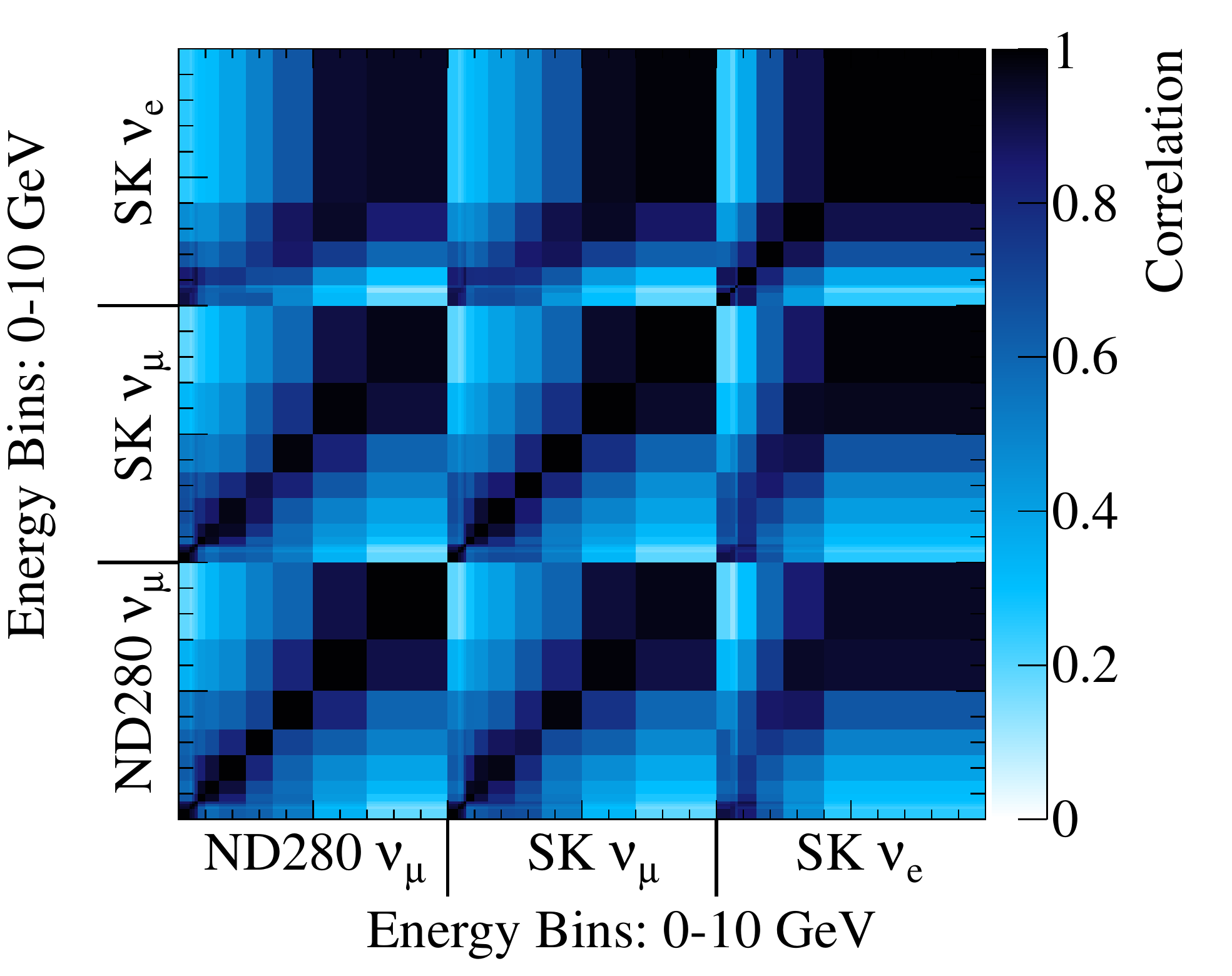}
\caption{The correlations of the flux uncertainties in the
$b_{i}$ bins for the ND280 $\nu_{\mu}$ and SK $\nu_{\mu}$ and $\nu_e$ fluxes.  The axes are the bins in 
neutrino energy for each flavor/detector combination and are proportional to the neutrino energy up to 10~GeV.} 
\label{fig:flux_corr_banff}
\end{figure*}

The uncertainties on the ND280 $\nu_{\mu}$, SK $\nu_{\mu}$ and SK $\nu_{e}$ flux predictions are shown in
Fig.~\ref{fig:flux_unc_banff} and the correlations
are shown in Fig.~\ref{fig:flux_corr_banff}. The correlations shown are evaluated for the binning described above.
The ND280 $\nu_{\mu}$ and SK $\nu_{\mu}$ flux predictions have large correlations, indicating the  
$\nu_{\mu}$ interaction rate at the near detector can constrain the unoscillated $\nu_{\mu}$ interaction
rate at the far detector.  
The SK $\nu_{e}$ 
flux is also correlated with the ND280 $\nu_{\mu}$ flux, since the $\nu_{\mu}$ and $\nu_e$ both originate from 
the $\pi\rightarrow\mu+\nu_{\mu}$ decay chain or kaon decays.  This
correlation also allows us to constrain the expected intrinsic $\nu_e$ rate at the far detector by measuring 
$\nu_{\mu}$ interactions at the near detector.

\section{\label{sec:neut_int} Neutrino Interaction Model}

\begin{figure}[tp]
  \centering
 \includegraphics[width=1.0\linewidth]{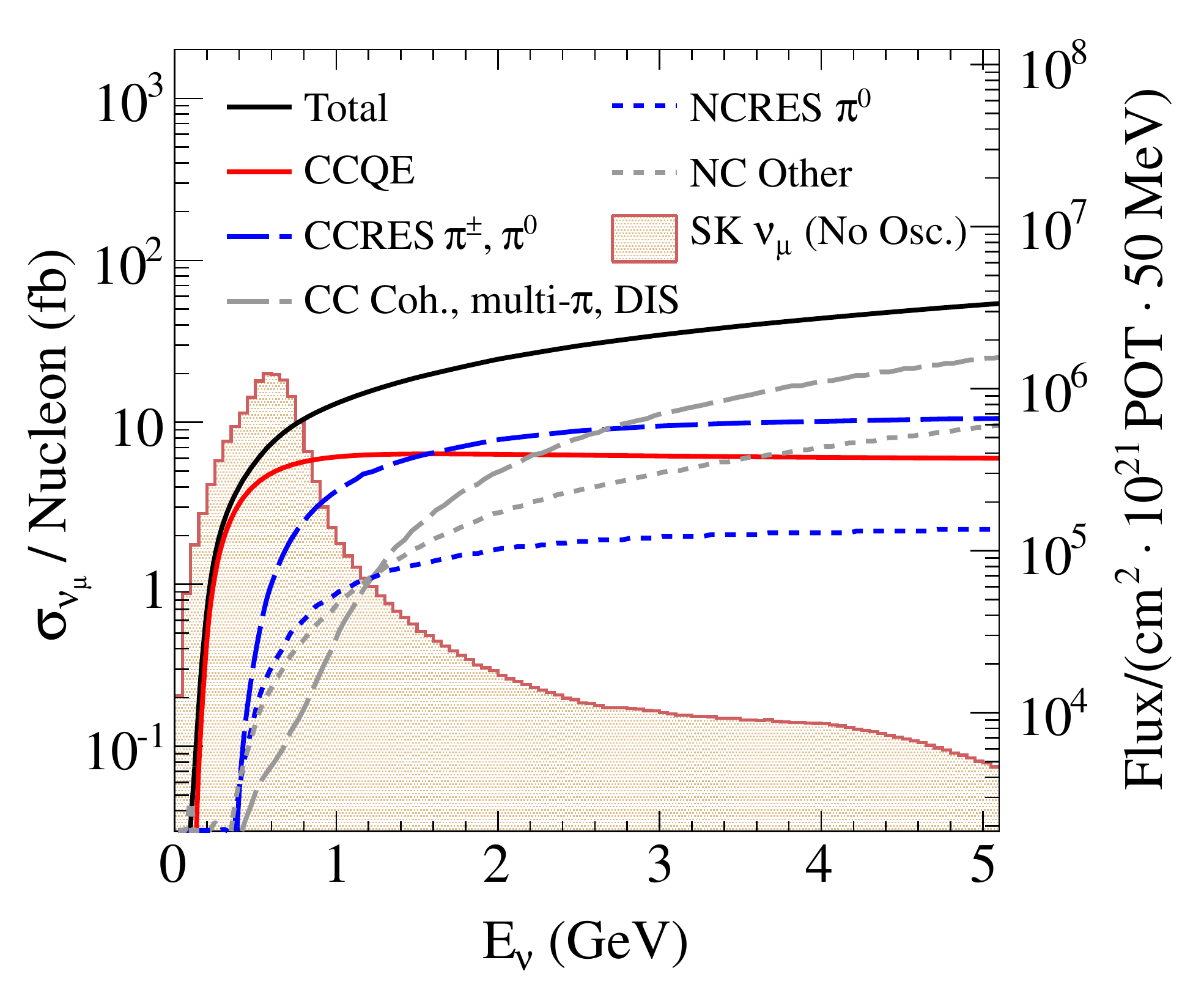} \\
\caption{The NEUT $\nu_{\mu}$ interaction cross section per nucleon on $^{16}$O with a breakdown by interaction process. The ``NC Other" curve includes neutral current coherent pion production, resonant charged pion production, multi-pion production and deep inelastic scattering.
   The predicted $\nu_{\mu}$ flux spectrum at SK with no oscillations is shown for comparison.}

  \label{fig:neut_xsec}
\end{figure}

We input the predicted neutrino flux at the ND280 and SK detectors to the \neut~\cite{Hayato:2009} 
neutrino interaction generator to simulate neutrino interactions in the detectors.  
Fig.~\ref{fig:neut_xsec} illustrates the neutrino-nucleon scattering processes modeled by \neut at the T2K beam 
energies.  The dominant interaction at the T2K beam peak energy is charged 
current quasi-elastic scattering (CCQE):
\begin{linenomath*}
\begin{equation}
\nu_{\ell} + N \rightarrow \ell + N^{\prime},
\end{equation}
\end{linenomath*}
where $\ell$ is the corresponding charged lepton associated with the neutrino's 
flavor (electron or muon), and $N$ and $N^{\prime}$ are the initial and final state nucleons. 
Above the pion production threshold,
single pion production contributes to charged current interactions (CC1$\pi$):
\begin{linenomath*}
\begin{equation}
\nu_{\ell} + N \rightarrow \ell + N^{\prime} + \pi,
\end{equation}
\end{linenomath*}
and neutral current interactions (\ncpis):
\begin{linenomath*}
\begin{equation}
\nu + N \rightarrow \nu + N^{\prime} + \pi.
\end{equation}
\end{linenomath*}
In the high energy tail of the T2K flux, multi-pion and deep inelastic scattering (DIS) processes
become dominant.

\subsection{\label{sec:neut_sim} NEUT simulation models}

CCQE interactions in \neut are simulated using the model of Llewellyn Smith~\cite{LlewellynSmith:1972}, 
with nuclear effects described by the relativistic Fermi gas model of Smith and 
Moniz~\cite{SmithMoniz:1972,SmithMonizErratum}. 
Dipole forms for the vector and
axial-vector form factors in the Llewellyn Smith model are used, with
characteristic masses $M_V = 0.84\, \mathrm{GeV}$ and $M_A = 1.21\,
\mathrm{GeV}$ respectively in the default simulation. 
The Fermi momentum $p_F$ is set to $217\, \mathrm{MeV/c}$ for carbon and $225\,
\mathrm{MeV/c}$ for oxygen, and the binding energy is set to 25~MeV for carbon and 27~MeV for oxygen. 

\neut simulates the production of pions via the excitation of hadronic resonances
using the model of Rein and Sehgal~\cite{ReinSehgal:1981}. The simulation
includes 18 resonances below $2\, \mathrm{GeV}$, along with
interference terms. In the energy
range relevant for T2K, resonance production is dominated by the
$\Delta(1232)$. 
For 20\% of the $\Delta$s produced within a nucleus, \neut also simulates pion-less $\Delta$ decay, in which 
the $\Delta$ de-excites in the nuclear medium without the emission of pions.
\neut includes the production of pions in coherent scattering of the neutrino on the target nucleus
based on the Rein and Sehgal model.  

Multi-pion and DIS interactions in \neut are simulated using the
GRV98 parton distribution functions~\cite{Gluck:1998xa}.  Where the invariant mass of the outgoing
hadronic system ($W$) is in the range $1.3<W<2.0$~GeV/$c^2$, a custom program is used~\cite{Nakahata:1986zp},
and only pion multiplicities of greater than one are considered to avoid double counting with the Rein
and Sehgal model.  For $W>2.0$~GeV/$c^2$ PYTHIA/JETSET~\cite{Sjostrand:1993yb} is used.  Corrections to the
small $Q^2$ region developed by Bodek and Yang are applied~\cite{Bodek:2003wd}.

\neut uses a cascade model to simulate the interactions of hadrons as they propagate
through the nucleus. For pions with momentum below $500 \;
\mathrm{MeV}/c$, the method of Salcedo \emph{et al.}~\cite{Salcedo:1988} is used.
Above pion momentum of $500 \; \mathrm{MeV}/c$ the 
 scattering cross sections are modeled using measurements of $\pi^\pm$ scattering on
free protons~\cite{dePerio:2011zz}.

Additional details on the \neut simulation can be found elsewhere~\cite{Abe:2011ks}.

\subsection{Methods for varying the \neut model}

Uncertainties in modeling neutrino interactions are a significant contribution to the overall systematic uncertainty in the $\nu_e$ appearance analysis reported in this paper. 
In the rest of this section, we describe these uncertainties with nuisance parameters that vary the \neut interaction models.
The parameters, listed in Table~\ref{tab:xsec_params}, are chosen and their central values and uncertainties are set to cover the systematic uncertainties on the interaction models derived from comparisons of \neut to external data or alternative models. 
They are a combination of free parameters in the \neut model and ad-hoc 
empirical parameters. 
The parameter values and uncertainties are
further constrained by the fit to neutrino data from the T2K ND280 detector, as described in
Section~\ref{sec:extrapolation}. 
To tune the \neut model parameters and evaluate the effect of neutrino interaction uncertainties, adjustments are 
carried out by applying weights to simulated \neut event samples from T2K or external experiments, such as \mb.

\begin{table*}
  \caption{\label{tab:xsec_params} The parameters used to vary the \neut cross section model and a brief description
of each parameter.}
\begin{center}
\begin{tabular} {ll}
\hline
\multicolumn{2}{l}{CCQE Cross Section }         \\ \hline
\hspace{0.1in} $M_A^{QE}$ & The mass parameter in the axial dipole form factor for quasi-elastic interactions \\ 
\hspace{0.1in} $x^{QE}_{1}$ & The normalization of the quasi-elastic cross section for $E_{\nu}<1.5$~GeV \\ 
\hspace{0.1in} $x^{QE}_{2}$ & The normalization of the quasi-elastic cross section for $1.5<E_{\nu}<3.5$~GeV \\ 
\hspace{0.1in} $x^{QE}_{3}$ & The normalization of the quasi-elastic cross section for $E_{\nu}>3.5$~GeV \\  \hline
\multicolumn{2}{l}{Nuclear Model for CCQE Interactions (separate parameters for interactions on O and C) }         \\ \hline
\hspace{0.1in} $x_{SF}$ & Smoothly changes from a relativistic Fermi gas nuclear model to a spectral function model  \\
\hspace{0.1in} $p_F$ & The Fermi surface momentum in the relativistic Fermi gas model \\ \hline
\multicolumn{2}{l}{Resonant Pion Production Cross Section }         \\ \hline
\hspace{0.1in} $M_A^{RES}$ & The mass parameter in the axial dipole form factor for resonant pion production interactions \\ 
\hspace{0.1in} $x^{CC1\pi}_{1}$ & The normalization of the \cc resonant pion production cross section for $E_{\nu}<2.5$~GeV \\ 
\hspace{0.1in} $x^{CC1\pi}_{2}$ & The normalization of the \cc resonant pion production cross section for $E_{\nu}>2.5$~GeV \\ 
\hspace{0.1in} $x^{NC1\pi^0}$ & The normalization of the NC$1\pi^0$ cross section \\
\hspace{0.1in} $x_{1\pi E_{\nu}}$ & Varies the energy dependence of the $1\pi$ cross section for better agreement with \mb data \\
\hspace{0.1in} $W_{\textrm{eff}}$ & Varies the distribution of $N\pi$ invariant mass in resonant production  \\
\hspace{0.1in} $x_{\pi-less}$ & Varies the fraction of $\Delta$ resonances that decay or are absorbed without producing a pion \\ \hline
\multicolumn{2}{l}{Other }         \\ \hline
\hspace{0.1in} $x^{CCcoh.}$ & The normalization of \cc coherent pion production \\
\hspace{0.1in} $x^{NCcoh.}$ & The normalization of NC coherent pion production \\
\hspace{0.1in} $x^{NCother}$ & The normalization of NC interactions other than NC$1\pi^0$ production \\
\hspace{0.1in} $x_{CCother}$ & Varies the \cc multi-$\pi$ cross section normalization, with a larger effect at lower energy  \\
\hspace{0.1in} $\vec{x}_{FSI}$ & Parameters that vary the microscopic pion scattering cross sections used in the FSI model \\
\hspace{0.1in} $x_{\nu_e/\nu_{\mu}}$ & Varies the ratio of the \cc $\nu_e$ and $\nu_{\mu}$ cross sections \\
\hline
\end{tabular}
\end{center}
\end{table*}

\subsection{\label{sec:neut_fits}\neut model comparisons to external data and tuning}
A detailed description of the \neut model tuning using external data comparisons can be found in 
Appendix~\ref{sec:neut_data}.  Here we provide a brief summary.

\subsubsection{\label{sec:neut_fsi_tune} FSI model tuning and uncertainty}
The \neut FSI model includes parameters which alter the microscopic pion interaction probabilities 
in the nuclear medium.
The central values of these parameters and their uncertainties are determined from fits to pion scattering 
data~\cite{Ashery:1981tq,PhysRevC.48.2800,PhysRevC.61.054615}.
We consider variations of the FSI parameters within the uncertainties from the fit of the
pion scattering data, and evaluate the uncertainties on the predicted event rates for ND280 and 
SK selections.

\subsubsection{\label{sec:neut_ccqe}CCQE model uncertainty }
\begin{figure}
  \centering
 \includegraphics[width=1.0\linewidth]{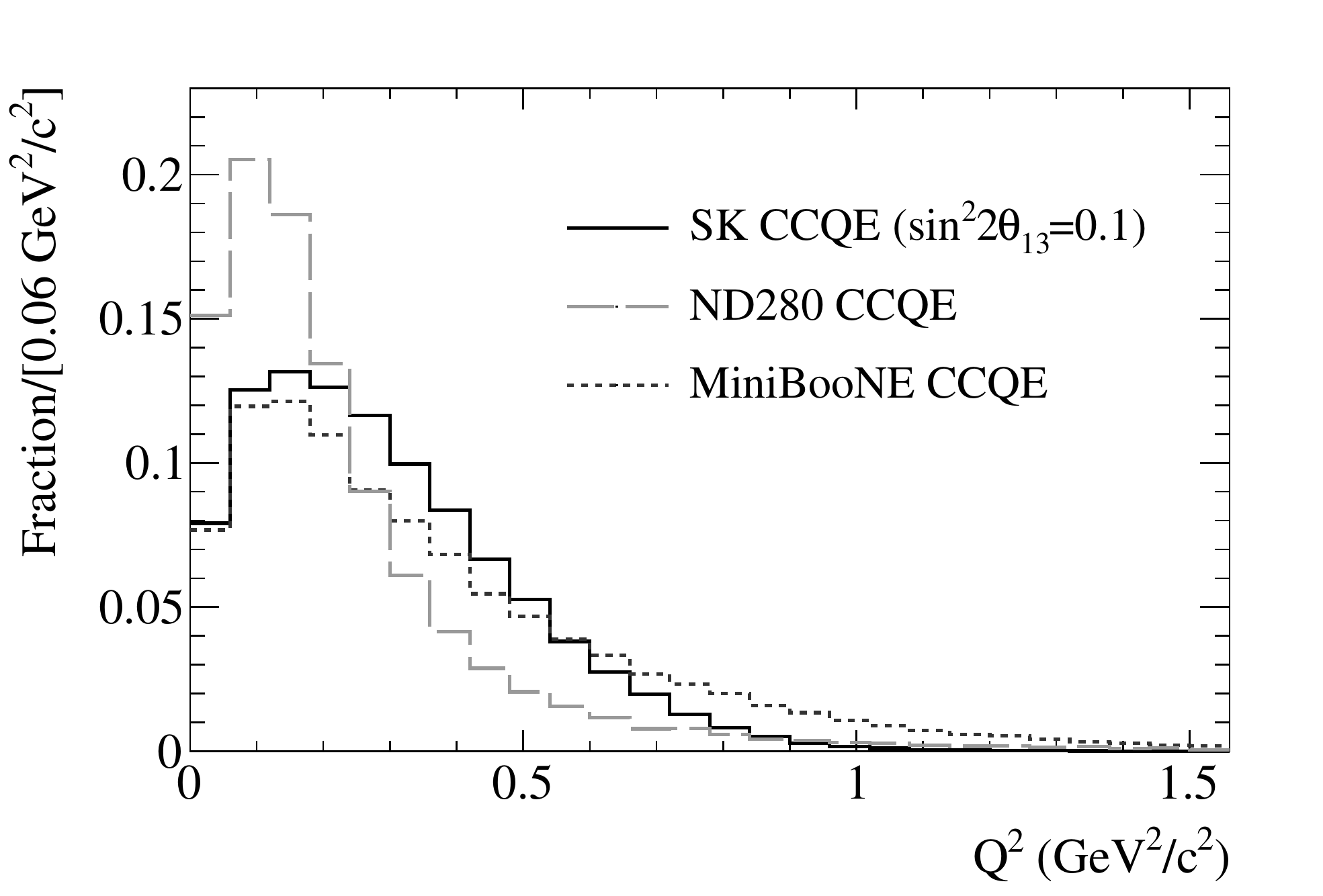} \\
  \caption{The predicted $Q^2$ distributions for CCQE interactions in the
ND280 CCQE selection, the \mb CCQE selection, and the SK \nue appearance selection. }
  \label{fig:q2_comp}
\end{figure}

The most detailed measurement of CCQE scattering on light nuclei in
the region of $1\, \mathrm{GeV}$ neutrino energy has been made by \mb,
which has produced double-differential cross sections in the muon
kinetic energy and angle, $(T_\mu, \cos
\theta_\mu)$~\cite{mb-ccqe}. 
We compare the agreement of NEUT to the \mb CCQE data in addition to our own near 
detector measurement of CCQE events
(Section~\ref{sec:nd280}) since the \mb detector has 4$\pi$ acceptance, providing
a kinematic acceptance of the leptons that more closely matches the SK acceptance
for the selection described in Section~\ref{sec:sk_selection}.
This is illustrated in Fig.~\ref{fig:q2_comp}, which compares the predicted true $Q^2$
distributions for CCQE events in the ND280 CCQE selection, the \mb CCQE 
selection, and the SK selection for \nue appearance candidates. 

In order to allow the ND280 data to constrain the CCQE model, we use the difference of the \neut 
nominal value and the best-fit value from fit to MiniBooNE data to set the uncertainty on $M^{QE}_A$, 
$\sigma_{M_{A}^{QE}} =  0.43\, \mathrm{GeV}$. We also set the uncertainty on the low energy
 CCQE normalization, $x_{1}^{QE}$, to the size of the MiniBooNE flux uncertainty, 11\%.  The results 
of the MiniBooNE fit are discussed in more detail in Appendix~\ref{sec:neut_data}.

To allow for the discrepancy in CCQE cross section at $\mathcal{O}(1)$~GeV measured 
by \mb and at $\mathcal{O}(10)$~GeV measured by NOMAD~\cite{Lyubushkin:2008pe}, we employ independent
CCQE normalization factors for $(1.5 < E_{\nu} < 3.5)$~GeV ($x_2^{QE}$) and
$E_{\nu} > 3.5$~GeV ($x_3^{QE}$), each with a prior uncertainty of 30\% and a nominal value
of unity. 

Alternate explanations have been proposed to reconcile the MiniBooNE data with a $M^{QE}_A\approx 1.0$~GeV derived from electron scattering and NOMAD data~\cite{Martini:2009,Martini:2010,Amaro:2011,Bodek:2011,Nieves:2012}. These models typically modify the cross section either by enhancing the transverse component of the cross section, or by adding an additional multi-nucleon process to the existing cross section, where the neutrino interacts on a correlated pair of nucleons.
Future improvements to the \neut generator may include a full implementation of alternate CCQE models. However, these models would also require modifications to the kinematics of the exiting nucleons, but no consensus has been reached yet in the field as to how the nucleons should be treated. We consider two possible effects of alternate CCQE models on the \nue appearance analysis.
First, the effect in $Q^2$ for these models is often similar to increasing $M^{QE}_A$ and \cite{Nieves:2012} shows that other improvements to the CCQE cross section can  be represented by an experiment-specific $M^{QE}_A$(effective), so the increase to the overall cross section from these models is approximately covered by the uncertainty on $M^{QE}_A$.  
Second, a multi-nucleon process would appear as a CCQE-like interaction in the SK detector, but the relationship between the neutrino energy and the lepton kinematics is different than for quasi-elastic scatters, which may affect the determination of oscillation parameters~\cite{Meloni:2012fq,Lalakulich:2012hs}.
Other processes also appear CCQE-like and have a different relationship between lepton kinematics and neutrino energy, such as non-QE events with no pions in the final state (pion-less $\Delta$ decay). The uncertainty on these events indirectly accounts for the effect of multi-nucleon models as these events affect the extracted oscillation parameters in a way similar to how multi-nucleon models would.

\subsubsection{\label{sec:neut_singlepi}Single pion production model tuning and uncertainty}

Measurements of single pion production cross sections on light 
nuclei in the T2K energy range have been made by \mb~\cite{mb-cc1pip,mb-cc1pi0,mb-nc1pi0},
and K2K, which used a 1000 ton water Cherenkov detector~\cite{k2k-nc1pi0}.
We perform a joint fit to the \mb measurements of charged current
single $\pi^+$ production (\ccpip), charged current single $\pi^0$
production (\ccpi) and neutral current single $\pi^0$ production
(\ncpi).
As shown in  Appendix~\ref{sec:neut_data}, we compare the \neut best-fit derived from the \mb single pion data with the K2K measurement,
which is of particular interest since it is the same nuclear target as SK.

\begin{table}[tp]
  \caption{Parameters used in the single pion fits, and their best-fit 
    values and uncertainties. The $1\sigma$ value of the
    penalty term is shown for parameters which are penalized in the
    fit.
    Where parameters are defined in a manner consistent with the T2K data fits,
the same parameter name is used.
}
  \centering
  \begin{tabular}{lcccc}
    \hline
                     & Nominal value & Penalty & best-fit & Error \\
    \hline
    \mares (GeV)     & 1.21          &         & 1.16     & 0.10 \\
    $W_{\textrm{eff}}$         & 1             &         & 0.48     & 0.14 \\
    $x_{CCother}$    & 0             & 0.40    & 0.36     & 0.39 \\
    Normalizations:        & & & & \\
    $x^{CCcoh}$      & 1             &         & 0.66     & 0.70 \\
    $x^{CC1\pi}_1$        & 1             &         & 1.63     & 0.32 \\

    $x^{NCcoh}$       & 1             & 0.30    & 0.96     & 0.30 \\
    $x^{NC1\pi^0}$      & 1             &         & 1.19     & 0.36 \\
    NC 1$\pi^\pm$    & 1             & 0.30    & 0.98     & 0.30 \\
    NC multi-pion/DIS & 1             & 0.30    & 0.99     & 0.30 \\
    \hline
  \end{tabular}
  \label{tab:singlepi-fitparams}
\end{table}

\begin{figure}[htbp]
  \centering
  \includegraphics[width=0.8\linewidth]{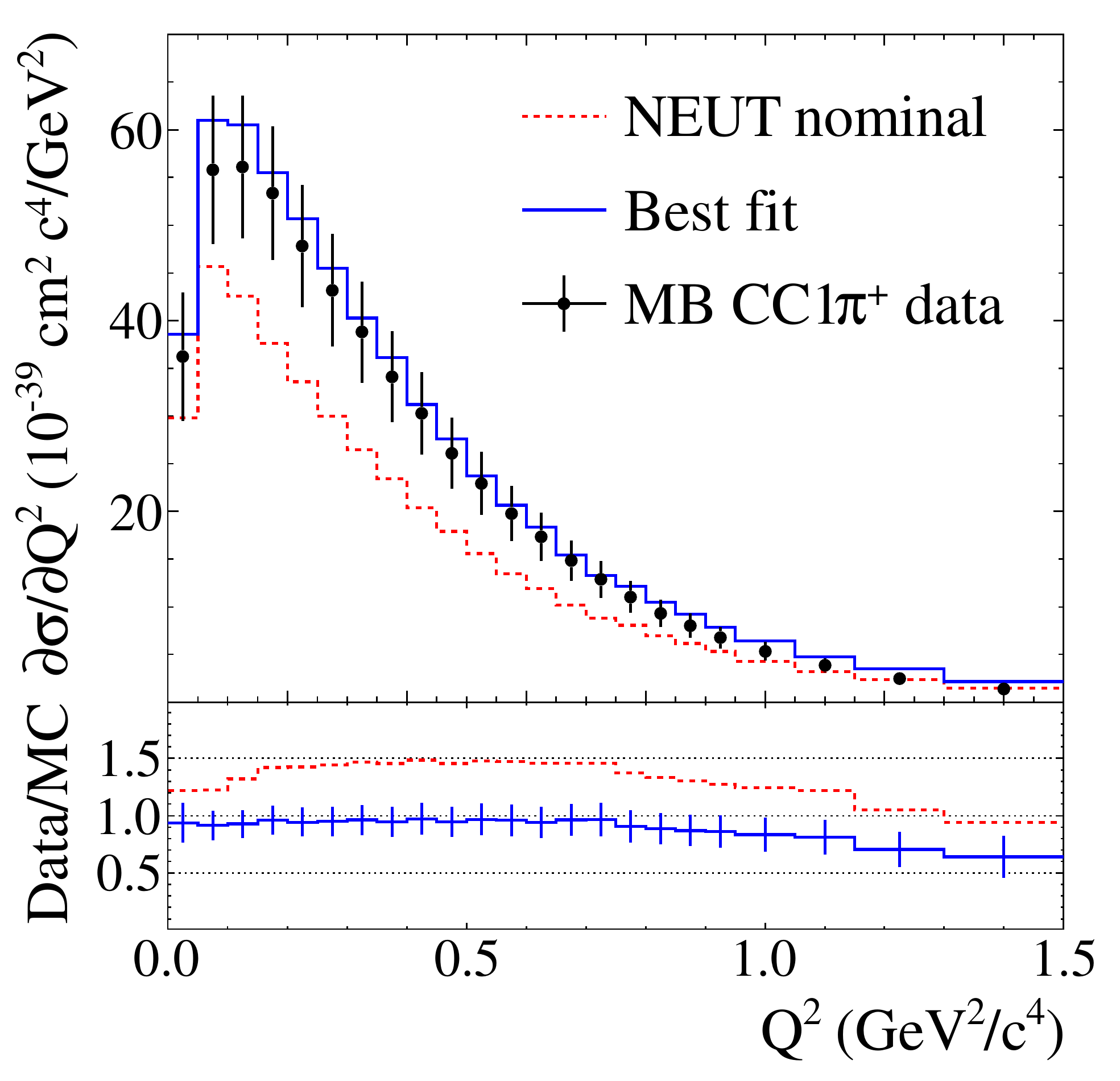} \\
  \includegraphics[width=0.8\linewidth]{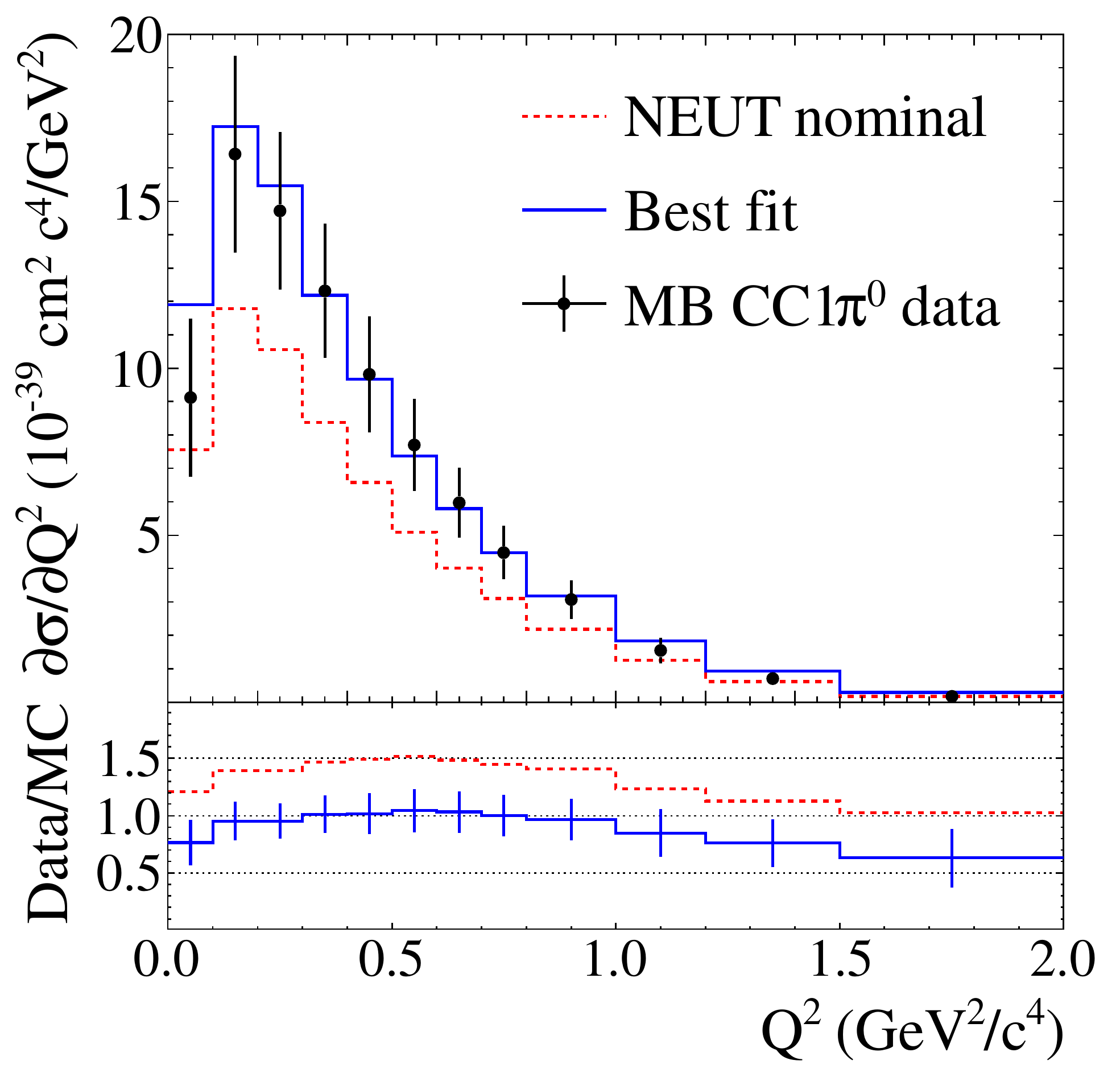} \\
  \includegraphics[width=0.8\linewidth]{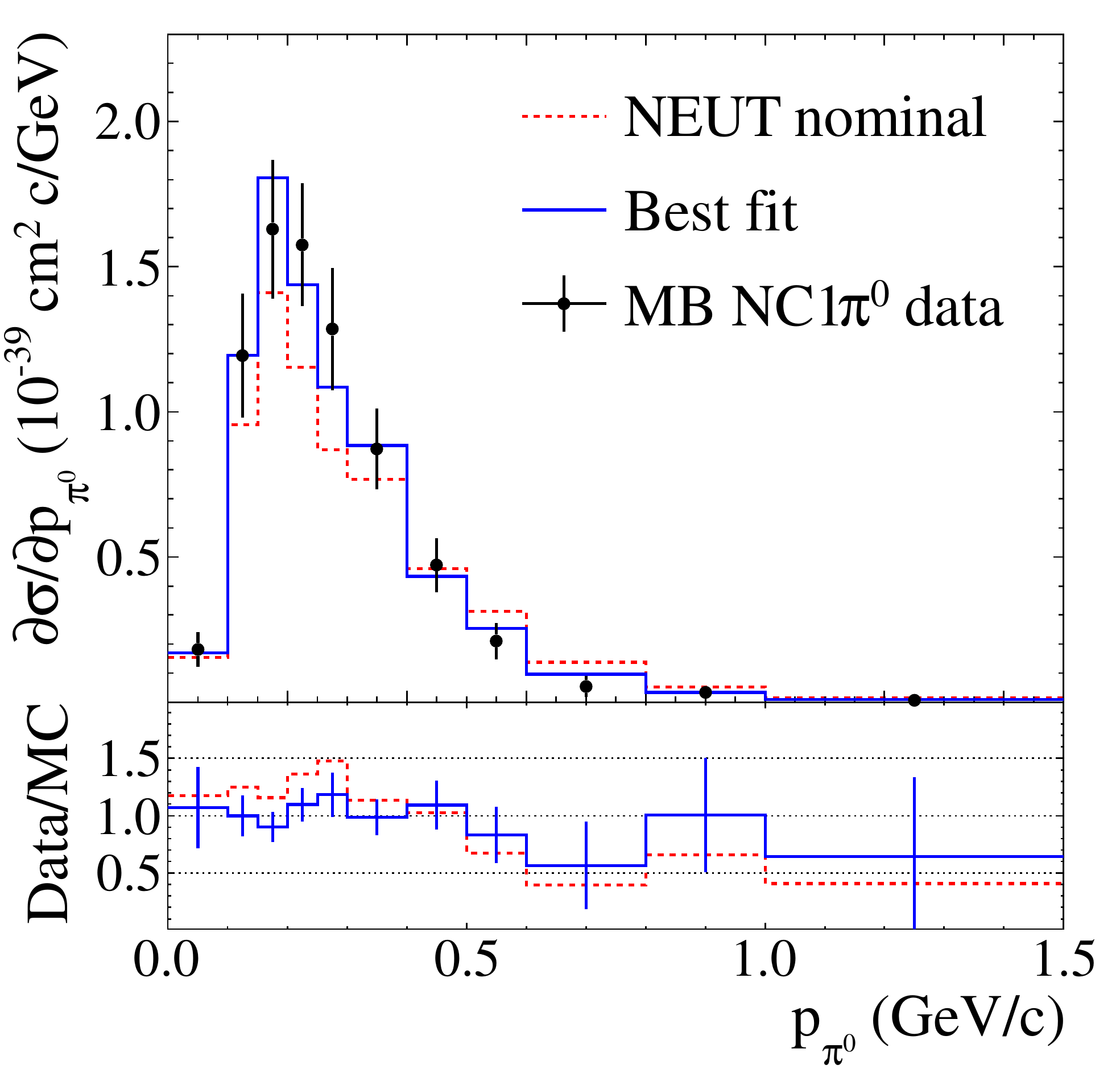} \\
  \caption{Differential cross sections for \ccpip $Q^2$ (top), \ccpi $Q^2$ (middle)
    and \ncpi $p_{\pi^0}$ (bottom)
    used in the single-pion fits to \mb data, and the NEUT nominal and best-fit
    predictions. 
    The MiniBooNE data point errors are statistical+systematic.}
  \label{fig:single-pi-results}
\end{figure}

The parameters listed in Table~\ref{tab:singlepi-fitparams} are varied in the fit to the 
\mb single pion data and their best-fit values and uncertainties are listed.
The parameters include $M_{A}^{RES}$, the axial mass in the Rein and Sehgal model,
 the empirical parameter, $W_{\textrm{eff}}$, discussed in the next paragraph,
and parameters that vary the normalization of various interaction modes.  
Contributions to the samples from CC multi-pion/DIS ($x_{CCother}$)
interactions, NC coherent interactions, NC1$\pi^{\pm}$ interactions and NC multi-pion/DIS interactions are relatively small, so the \mb samples have little power to constrain the associated parameters which are discussed in Section~\ref{sec:neut_other}.
Penalty terms for these parameters are applied using the prior uncertainties listed in 
Table~\ref{tab:singlepi-fitparams}. 

The $W_{\textrm{eff}}$ parameter alters the single pion differential cross section as a function of pion-nucleon
invariant mass $W$, providing a means to change the shape of the \neut prediction for 
\ncpi $\mathrm{d}\sigma / \mathrm{d}p_{\pi^0}$ differential cross section.
Uncertainties in the \ncpi pion momentum distribution enter into the \nue appearance analysis, as the 
momentum and angular distributions of \nue candidates from \ncpi interactions depend on the 
kinematic distribution of the $\pi^{0}$. The NEUT predicted $p_{\pi^{0}}$ spectrum, shown in the bottom
plot of Fig.~\ref{fig:single-pi-results} is broader than the observed MiniBooNE data. 
A decrease to the $W_{\textrm{eff}}$ parameter results in a more sharply-peaked $p_{\pi^{0}}$ spectrum, and achieves agreement between the \neut prediction and the measured cross section; $W_{\textrm{eff}}$ does not alter the total cross section. Future changes to the \neut model that may eliminate the need for $W_{\textrm{eff}}$  include refinements of the treatment of  formation time effects, which have been shown to affect the pion momentum distribution~\cite{Golan:2012}, or modifications to the contribution of higher order resonances relative to $\Delta(1232)$. 

The fitted data and \neut model are shown in  Fig.~\ref{fig:single-pi-results}.
We propagate the fitted parameter values for $M_{A}^{RES}$, $x^{CC1\pi}_{1}$ and $x^{NC1\pi^0}$ and their
correlated uncertainties to the fits of ND280 and SK data. 
The remaining parameters from the fit to \mb data
are marginalized.  We evaluate additional uncertainties on these parameters by re-running the fit to 
\mb data with variations of the FSI model and pion-less $\Delta$ decay turned off.  The deviations of the
fitted parameter values due to these FSI or pion-less $\Delta$ decay variations are applied as parameter
uncertainties, increasing the
uncertainties on $M_{A}^{RES}$, $x^{CC1\pi}_{1}$ and $x^{NC1\pi^0}$ to 0.11~GeV, 0.43 and 0.43 respectively.
The fitted $W_{\textrm{eff}}$ parameter value is not applied to the T2K predictions, but the difference between the nominal value of $W_{\textrm{eff}}$ and the best-fit value from the \mb data fit is treated as an uncertainty.

\begin{figure}
  \centering
  \includegraphics[width=0.95\linewidth]{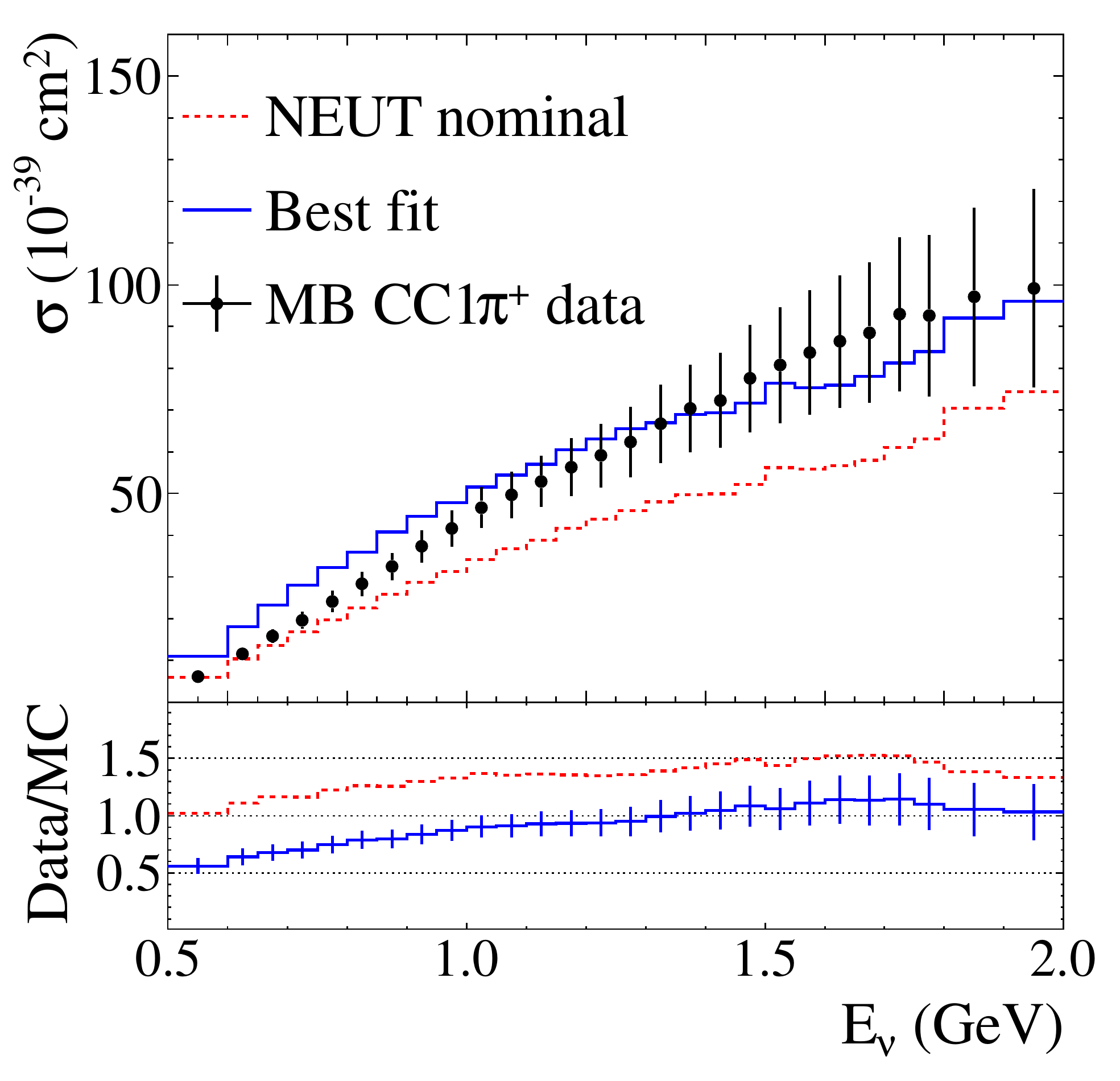}
  \caption{The \ccpip cross section as a function of energy as measured by \mb, with the \neut nominal and best-fit models. The treatment in the analysis of the disagreement between the best-fit \neut and data is discussed in the text.}
  \label{fig:cc1pip_edep}
\end{figure}

An additional uncertainty in the energy-dependent pion production cross section is considered since 
we observe a discrepancy between the fitted \neut model and the \mb \ccpip data, as shown in 
Fig.~\ref{fig:cc1pip_edep}.  We introduce a parameter $x_{1\pi E_{\nu}}$
that represents the energy-dependent tuning which brings the \neut prediction into agreement with the \mb data. 
Uncertainties on the ND280 and SK predictions include the difference between the resonant pion production with
and without this energy-dependent tuning.

The fits to \mb data constrain the normalization of CC$1\pi$ resonant production below 2.5~GeV.  Above
2.5~GeV, we apply a separate normalization uncertainty of 40\% on the parameter $x_2^{CC1\pi}$.  This uncertainty
covers the maximum difference between \mb \ccpip data and \neut at $E_\nu\approx 2$~GeV and is conservative given the CC inclusive cross section measurements~\cite{Wu:2007ab} made at higher energies.

\subsubsection{\label{sec:neut_other} Other interaction channels}

We evaluate the uncertainty on CC coherent pion production based on measurements by 
K2K~\cite{Hasegawa:2005} and SciBooNE~\cite{Hiraide:2008} which place upper limits on the CC coherent
production that are significantly less than the Rein and Sehgal model prediction.  Since no clear
CC coherent signal has been observed at $\mathcal{O}(1)$~GeV
, we apply a 100\% normalization
uncertainty to the \neut CC coherent pion production ($x^{CCcoh}$).

SciBooNE's measurement of the NC coherent pion cross section at $\mathcal{O}(1)$~GeV~\cite{PhysRevD.81.111102} is in good
agreement with the Rein and Sehgal model prediction; the uncertainty on this channel is set to 30\% based on the SciBooNE measurement and is represented by a normalization parameter,  $x^{NCcoh}$.  We define a single parameter $x^{NCother}$ that varies the
normalization of the NC resonant $\pi^{\pm}$, NC elastic and NC multi-pion/DIS/other resonant modes.  The uncertainty on this
normalization parameter is set to 30\%. As there is little NC resonant $\pi^{\pm}$ data, the uncertainty on the NC resonant $\pi^{\pm}$ processes is set to be the same size as the agreement shown in Section~\ref{sec:neut_singlepi} for the NC resonant 1$\pi^0$ cross section (30\%). The NC multi-pion and DIS model was tuned to agree with the CC/NC data using the NEUT predicted CC DIS cross section; the uncertainties on this phenomenological model are set to cover the size of the uncertainties of the CC/NC data~\cite{Musset19781,RevModPhys.53.211} (30\%).

The CC multi-pion/DIS interactions contribute to the ND280 samples discussed in Section~\ref{sec:nd280}.  At
energies greater than 4~GeV, these modes dominate the inclusive cross section and are constrained by 
measurements of the inclusive cross section~\cite{Adamson:2009ju} with $\approx$10\% uncertainties.  At lower 
energies the constraint from the inclusive cross section measurements is weaker since other interactions modes
are significant.  Hence, we apply an uncertainty that is $10\%$  at high energies and increases to $40\%$ near the
threshold for multi-pion production.  The model is adjusted by applying a weight:
\begin{linenomath*}
 \begin{equation}
   \label{eq:ccother}
   w = 1+\frac{x_{CCother}}{E_{\nu}(\mathrm{GeV})}.
 \end{equation}
\end{linenomath*}
The parameter $x_{CCother}$ is allowed to vary around a nominal value of 0 with a prior uncertainty of 0.4 GeV.

\subsection{\label{sec:nucl_model_unc} Nuclear model uncertainties}
\neut models nuclei with a relativistic Fermi gas model (RFG) using a Fermi momentum $p_{F}$ from electron
scattering data~\cite{Moniz:1971mt}.  We evaluate the uncertainty on the CCQE cross section for 
variations of $p_{F}$ within its uncertainty of $30$ MeV/$c$. This uncertainty covers the uncertainty from
the electron scattering data and has been inflated to cover possible discrepancies in the CCQE cross section
at low $Q^2$.  The uncertainty is applied independently for interactions on carbon and oxygen targets.

We also consider alternatives to the RFG model of the nuclei by making comparisons to a spectral function
nuclear model implemented in the NuWro neutrino interaction generator~\cite{Ankowski:2005wi}.
The discrepancy in CCQE interactions models with the RFG and spectral
function are assigned as uncertainty and represented by the parameter $x_{SF}$ which smoothly varies the 
predicted lepton kinematics between the RFG ($x_{SF}=0$) and spectral function ($x_{SF}=1$) models. We apply 
the uncertainties for the nuclear model independently for carbon and oxygen cross sections.

\subsection{\label{sec:neut_flavor_unc} $\nu_{e}$ cross section uncertainty}
Differences between \num and \nue in the cross section are also considered, as the \cc \num sample at ND280 is 
used to infer the \cc \nue rate at the far detector.  The spectral function uncertainty is calculated separately 
for \num and \nue as well as target material. In addition, an overall 3\% uncertainty on the ratio of \num and 
\nue CC neutrino-nucleon cross sections ($x_{\nue/\num}$) is included, based on calculations~\cite{Day:2012gb} 
over T2K's energy range.

\subsection{\label{sec:xsec_summary} Summary of the neutrino cross section model, tuning and uncertainties}
The cross section model parameters values and uncertainties are listed in Table~\ref{tab:xsec_param_unc}.
These priors are used as inputs to fits to the T2K ND280 and SK data sets, and include the results of the \mb single pion model fit.
For parameters related to the nuclear modeling, such as $x_{SF}$, $p_F(^{12}$C) and $p_F(^{16}$O), we apply separate 
uncorrelated parameters for the modeling of interactions on $^{12}$C and $^{16}$O.  Hence, the fit to ND280 data 
does not constrain the nuclear modeling parameters used when modeling interactions at SK.  Of the remaining parameters, 
we treat them as correlated for ND280 and SK if they are strongly constrained by ND280 data.  These parameters include 
the CCQE cross section parameters, $M_A^{QE}$, $x^{QE}_{1}$, and the CC1$\pi$ cross section parameters, $M_A^{RES}$, 
$x^{CC1\pi}_{1}$.  
To preserve the correlations between NC and CC parameters from the fit to MiniBooNE single pion data, $x^{NC1\pi^0}$ 
is also propagated. All other parameters are not well constrained by the ND280 data and are applied separately for 
ND280 and SK interaction modeling.

\begin{table}
  \caption{\label{tab:xsec_param_unc} The parameters used to vary the \neut cross section model along with the values used in the ND280 fit (input value)
and uncertainties prior to the ND280 and SK data fits.}
\begin{center}
\begin{tabular} {lcc}
\hline
Parameter          & Input Value & Uncertainty \\ \hline
$M_A^{QE}$ (GeV)   & 1.21          & 0.43         \\ 
$x^{QE}_{1}$       & 1.00          & 0.11         \\ 
$x^{QE}_{2}$       & 1.00          & 0.30         \\ 
$x^{QE}_{3}$       & 1.00          & 0.30         \\ 
$x_{SF}$           & 0.0           & 1.0          \\
$p_F(^{12}$C) (MeV/$c$)   & 217      & 30           \\ 
$p_F(^{16}$O) (MeV/$c$)   & 225      & 30           \\ 
$M_A^{RES}$ (GeV)  & 1.16          & 0.11         \\ 
$x^{CC1\pi}_{1}$   & 1.63          & 0.43         \\ 
$x^{CC1\pi}_{2}$   & 1.00          & 0.40         \\ 
$x^{NC1\pi^0}$     & 1.19          & 0.43         \\
$x_{1\pi E_{\nu}}$ & off           & on           \\
$W_{\textrm{eff}}$          & 1.0           & 0.51         \\
$x_{\pi-less}$     & 0.2           & 0.2          \\ 
$x^{CCcoh.}$       & 1.0           & 1.0          \\
$x^{NCcoh.}$       & 1.0           & 0.3          \\
$x^{NCother}$         & 1.0           & 0.3          \\
$x_{CCother}$ (GeV)   & 0.0          & 0.4          \\
$x_{\nu_e/\nu_{\mu}}$ & 1.0        & 0.03         \\
\hline
\end{tabular}
\end{center}
\end{table}

\section{\label{sec:nd280} ND280 Neutrino Data}

We select samples of CC \num interactions in the ND280 detector, which are fitted to constrain the flux and 
cross section models, as described in Section~\ref{sec:extrapolation}.  \cc\ \num interaction candidates are 
divided into two selections, one enhanced in \ccqe-like events, and the second consisting of all other 
CC interactions, which we refer to as the \ccnqe-like selection.  While the $\nu_e$ flux and interaction models
are constrained by the CC \num data, we also select a sample enhanced in \cc \nue interactions to directly 
verify the modeling of the intrinsic \nue rate. 

\subsection{{\label{sec:nd280_sim} ND280 simulation}}

The ND280 detector response is modeled with a GEANT4-based~\cite{Agostinelli2003250,1610988} Monte Carlo (MC) simulation, using the neutrino flux described in Section~\ref{sec:flux} and the NEUT simulation.  The MC predictions presented in this section are not calculated with the cross section parameter tuning described in Table~\ref{tab:singlepi-fitparams}. Neutrino interactions are generated up to 30~GeV for all flavors from the unoscillated flux prediction, with a time distribution matching the beam bunch structure. The ND280 subdetectors and magnet are represented with a detailed geometrical model. To properly represent the neutrino flux across a wider range of off-axis angles, a separate simulation is run to model neutrino interactions in the concrete and sand which surround ND280. The scintillator detectors, including the FGD, use custom models of the scintillator photon yield, photon propagation including reflections and attenuation, and electronics response and noise. The gaseous TPC detector simulation includes the gas ionization, transverse and longitudinal diffusion of the electrons, propagation of the electrons to the readout plane through the magnetic and electric field, and a parametrization of the electronics response. Further details of the simulation of the individual detectors of ND280 can be found in Refs~\cite{Abe:2011ks,Amaudruz:2012pe}.  

\subsection{{\label{sec:nd280_numuselection}$\nu_{\mu}$ candidate selection}}

We select \cc \num interactions by identifying the muons from \ccnumu interactions, which may be accompanied 
by hadronic activity $X$ from the same vertex.  
Of all negatively charged tracks, we identify the highest momentum track in each event that originates in the upstream FGD (FGD1) and enters the middle TPC (TPC2) as the $\mu^{-}$ candidate. 
The negatively charged track is identified using curvature and must start inside the 
FGD1 fiducial volume (FV) that begins 48~mm inward from the edges of FGD1 in $x$ and $y$ and 21~mm inward from the upstream FGD1 edge in $z$.
In this analysis we use only selected tracks with a vertex in FGD1, since it provides a homogeneous target 
for neutrino interactions.
To reduce the contribution from neutrino interactions upstream of the FGD1 FV, any tracks which pass through both the upstream TPC (TPC1) and FGD1 are rejected.
This also has the consequence of vetoing backward-going particles from the \cc interaction vertex, so the resulting selection is predominantly forward-going $\mu^{-}$.

The \mun candidate track energy loss is required to be consistent with a muon.  The identification of particles (PID) is based on a truncated mean of measurements of energy loss 
in the TPC gas, from which a discriminator function is calculated for different particle hypotheses. We apply the discriminator to select muon candidates and reject electron and proton tracks. The TPC PID and TPC performance are described in more detail elsewhere~\cite{Abgrall:2010hi}.  

Events passing the previously described cuts comprise the CC-inclusive sample, and the number of selected events
and the MC predictions are listed in Table~\ref{tab:nd280selection}. 
These data correspond to $2.66\times 10^{20}$ \pot.
The predictions include a correction for the event pile-up that is not directly modeled by the Monte Carlo simulation of the detector. The pile-up correction takes into account the presence of neutrino interactions in the same beam bunch originating in the sand and material surrounding the detector. The size of this correction ranges between 0.5\% and 1\%\ for the different run periods. 
Of \cc\ \num interactions in the FGD1 FV, 47.6\% are accepted by the \cc-inclusive selection, and the resulting selection is 88.1\% pure. The largest inefficiency of the \cc-inclusive selection is from high angle particles which do not traverse a sufficient distance through the TPC to pass the selection criteria.

\begin{figure}
\centering
\includegraphics[trim=3.5cm 0cm 4cm 0.7cm, clip=true, width=0.405\textwidth]{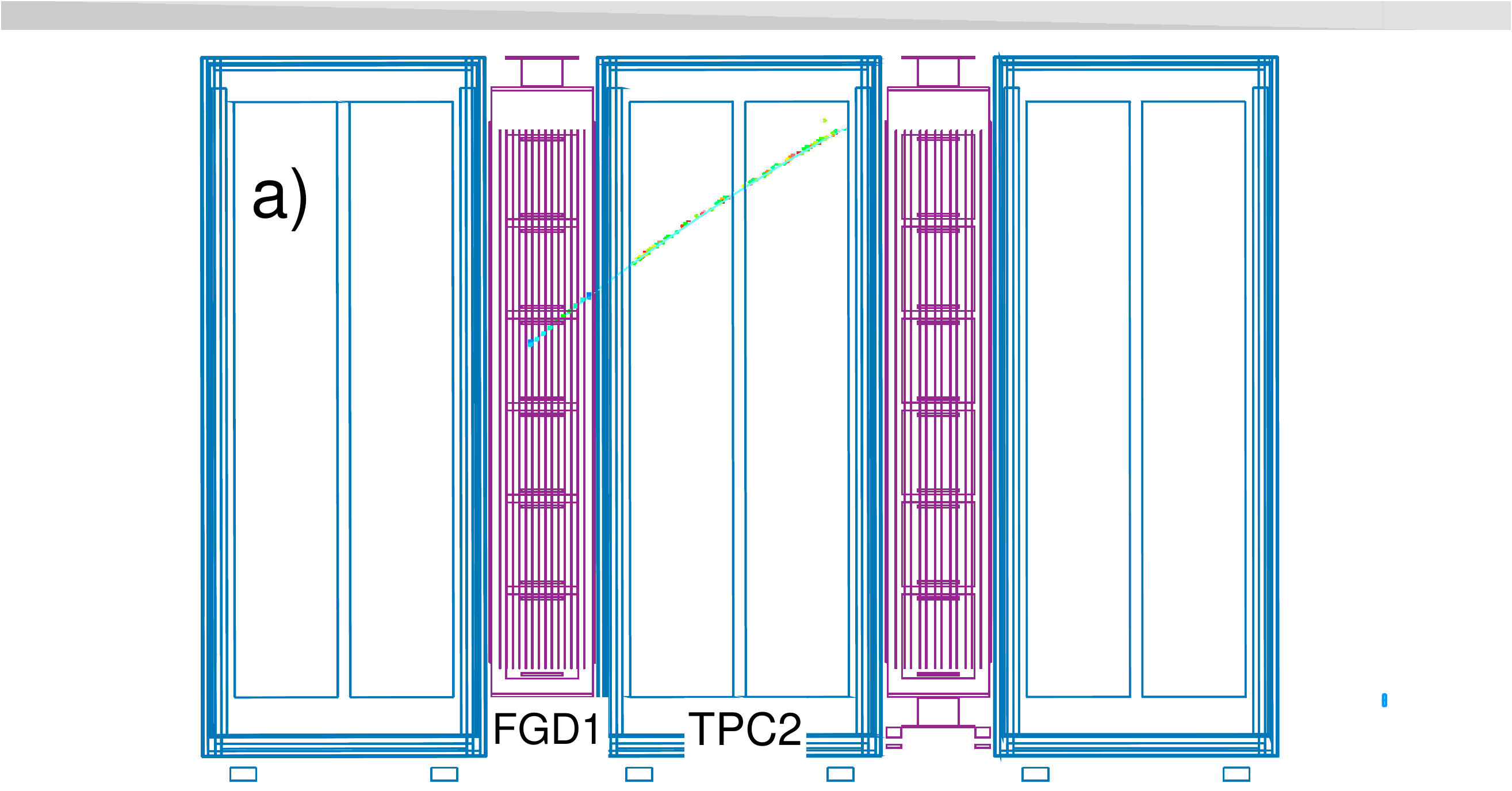}\\
\includegraphics[trim=5cm 1cm 5.5cm 1.7cm, clip=true, width=0.4\textwidth]{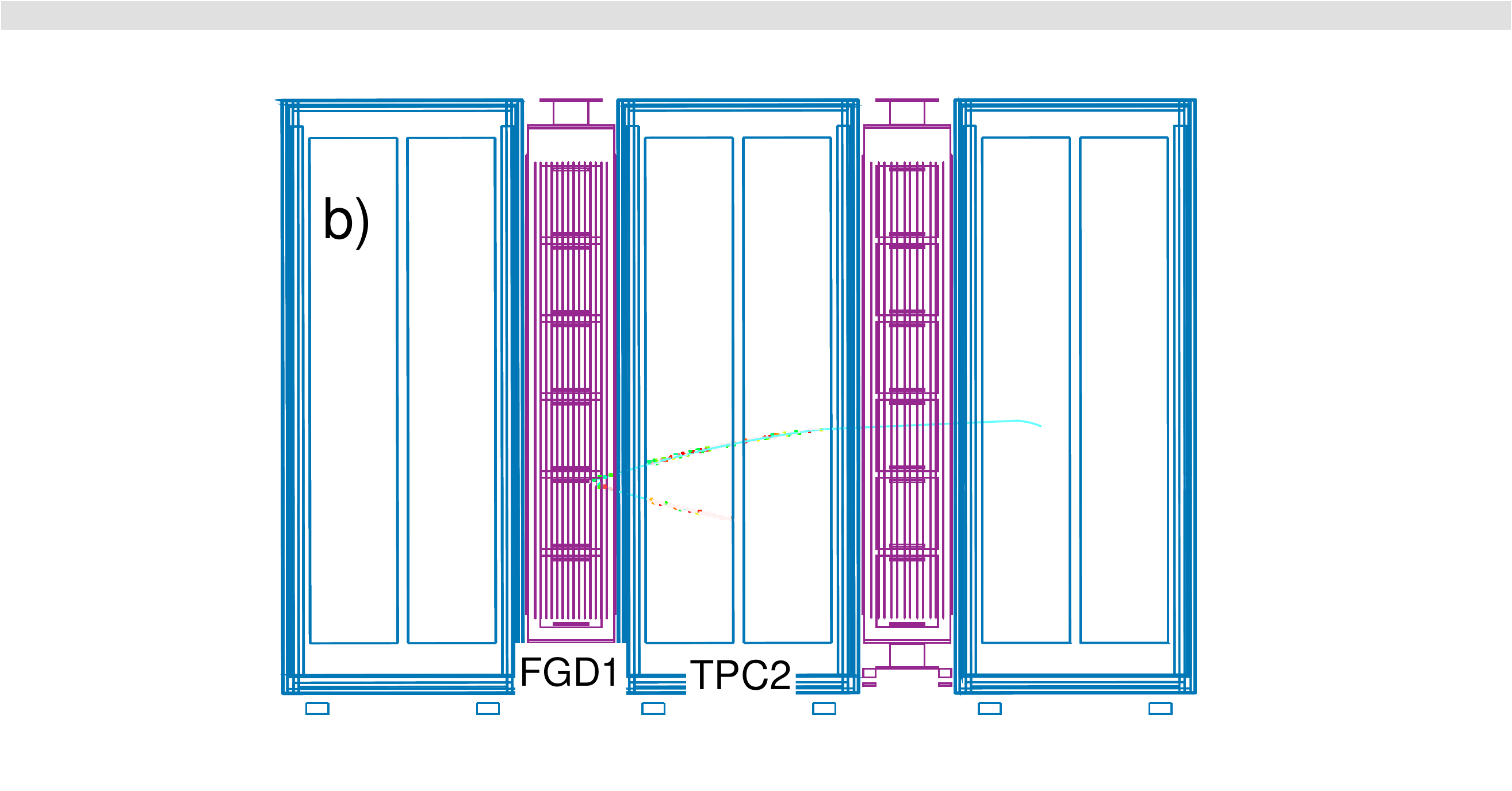}\\
\caption{Event displays of example ND280 \ccqe-like (a) 
and \ccnqe-like (b) selected events.
}
\label{fig:nd280_mc_event_displays}
\end{figure}

We divide the \cc-inclusive \num events into two mutually exclusive samples sensitive to different neutrino interaction types: \ccqe-like and \ccnqe-like.
As the \ccqe neutrino interaction component typically has one muon and no pions in the final state, we separate the two samples by requiring the following for the CCQE-like events:
\begin{itemize}
\item Only one muon-like track in the final state
\item No additional tracks which pass through both FGD1 and TPC2.
\item No electrons from muon decay at rest in FGD1 (Michel electron)
\end{itemize}
A Michel electron  will typically correspond to a stopped or low energy pion that decays to a muon which stops in FGD1, and is identified by looking for a time-delayed series of hits in FGD1. The Michel electron tagging efficiency is 59\%. Events in the \cc-inclusive selection which do not pass the CCQE-like selection comprise the CCnonQE-like sample. 
Example event displays for ND280 events are shown in Fig.~\ref{fig:nd280_mc_event_displays}.

The numbers of selected events in the data and nominal prediction for the \ccqe-like and \ccnqe-like selections are shown in Table~\ref{tab:nd280datareduction_qe}. Table~\ref{tab:nd280fractionalcontribution} shows the composition of the \cc, \ccqe-like and \ccnqe-like selections according to the generated neutrino interaction categories in the Monte Carlo. 
The \ccqe-like sample contains 40.0\% of all \ccqe interactions in the FGD1 FV, and \ccqe interactions comprise 69.5\% of the \ccqe-like sample.

\begin{figure*}
\centering
\begin{tabular}{cc}
\includegraphics[width=0.48\textwidth]{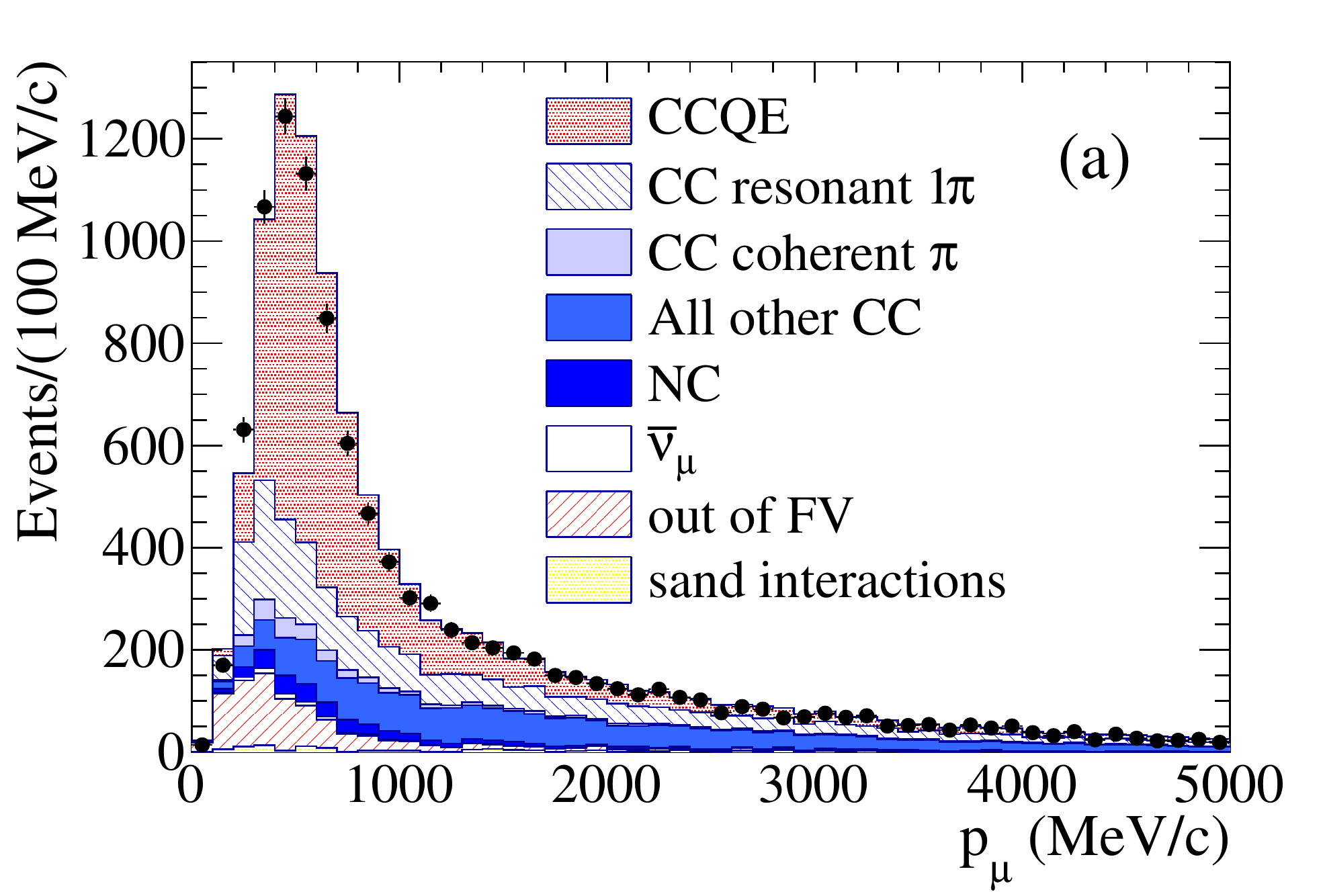}&
\includegraphics[width=0.48\textwidth]{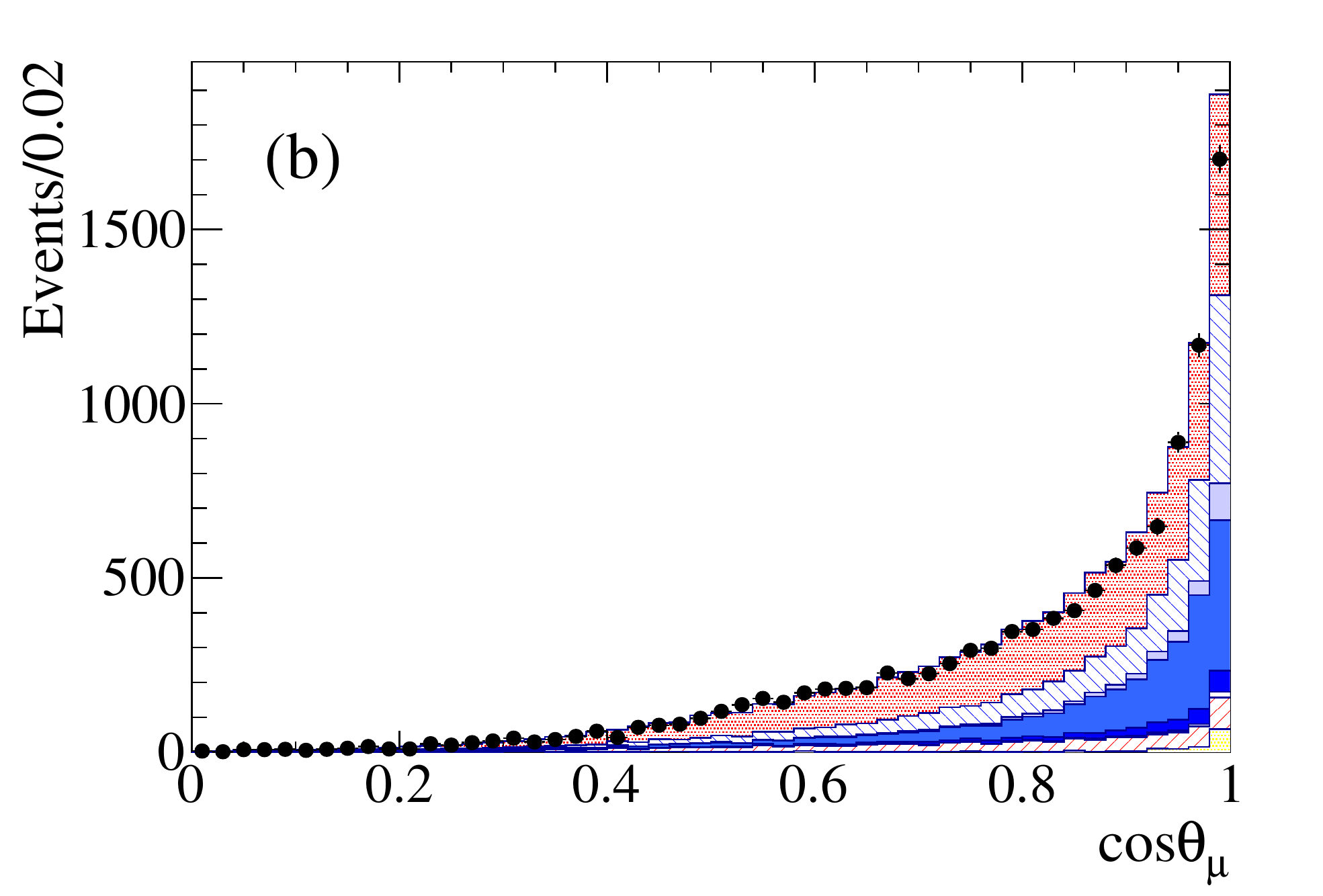} \\
\includegraphics[width=0.48\textwidth]{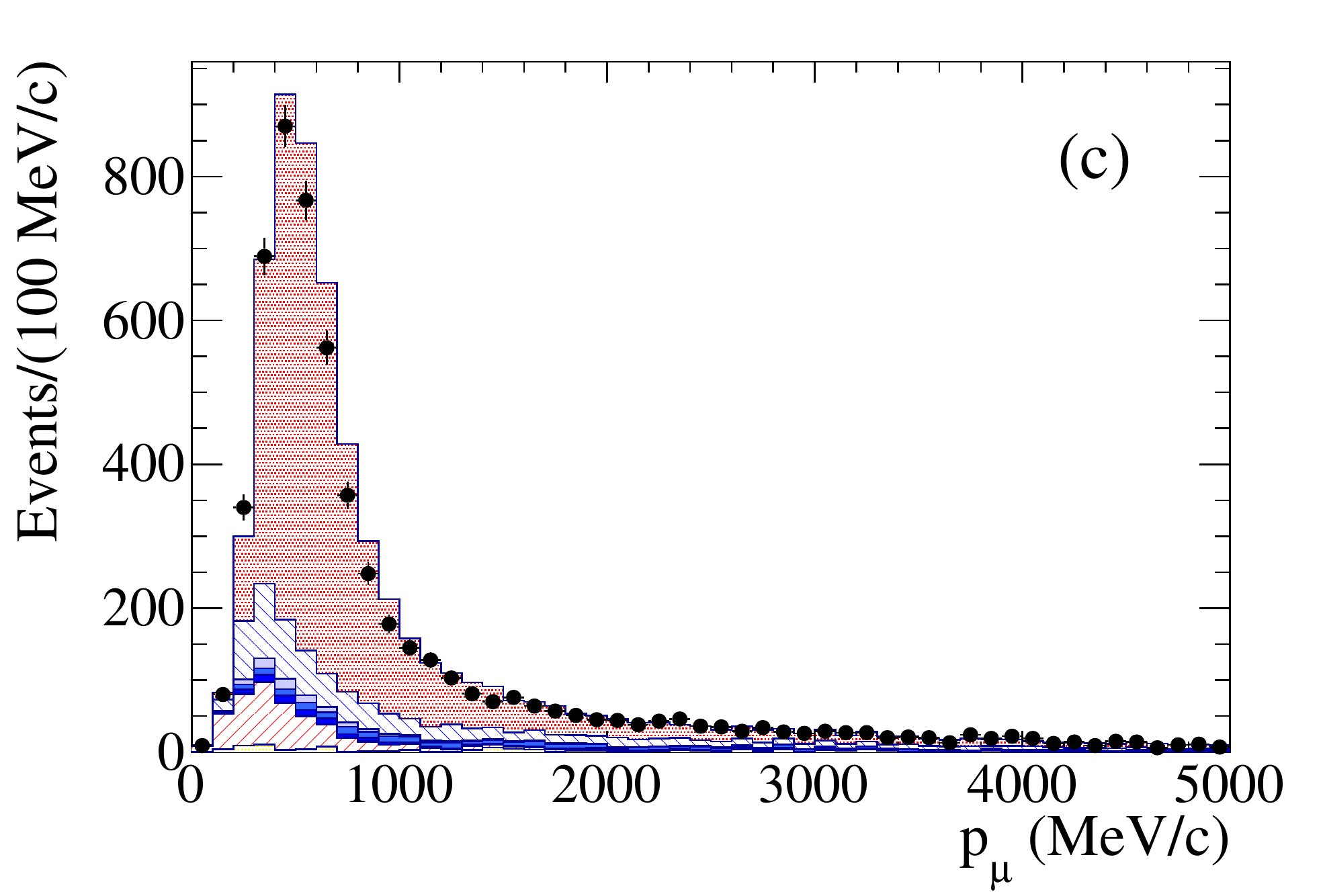} &
\includegraphics[width=0.48\textwidth]{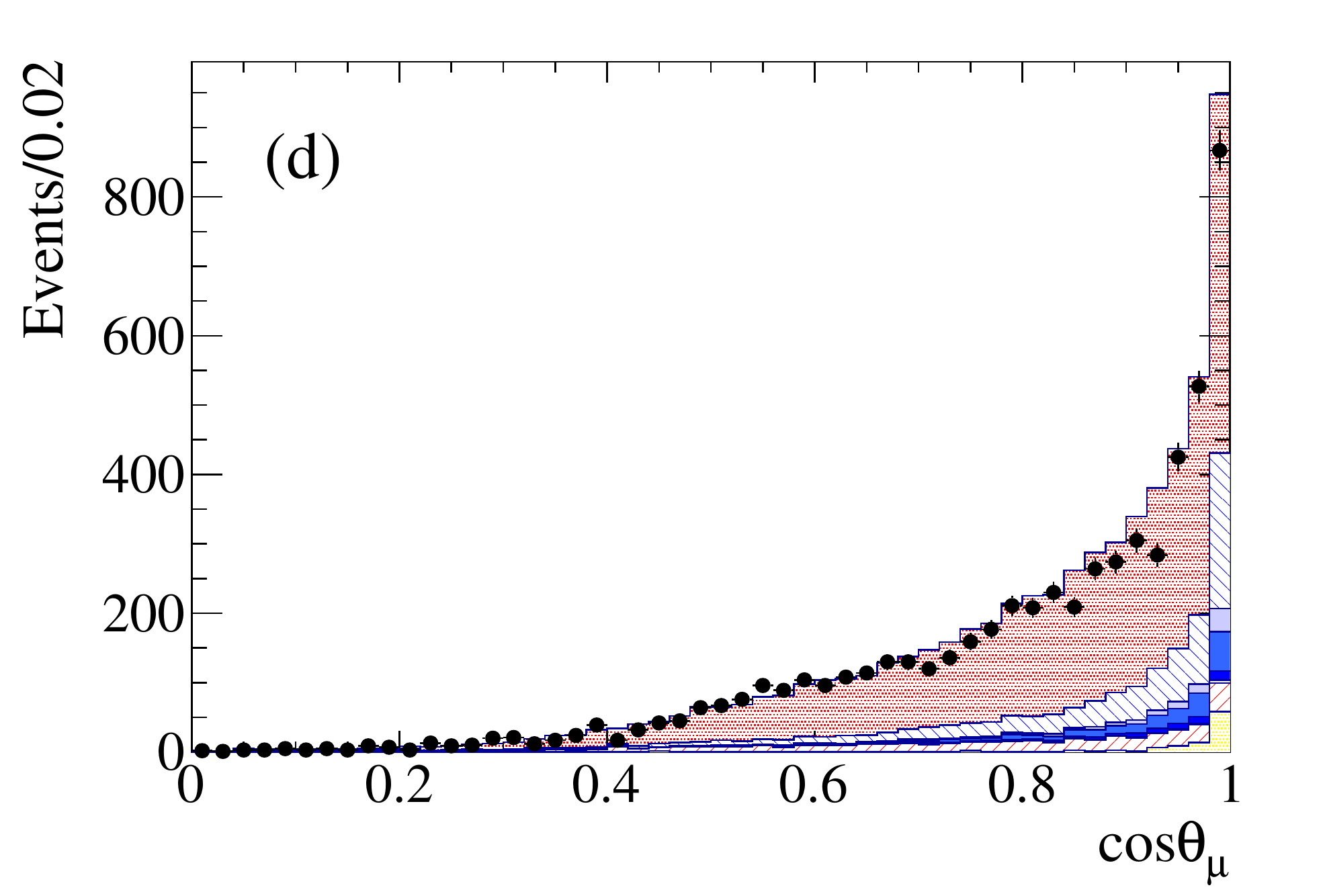}\\
\includegraphics[width=0.48\textwidth]{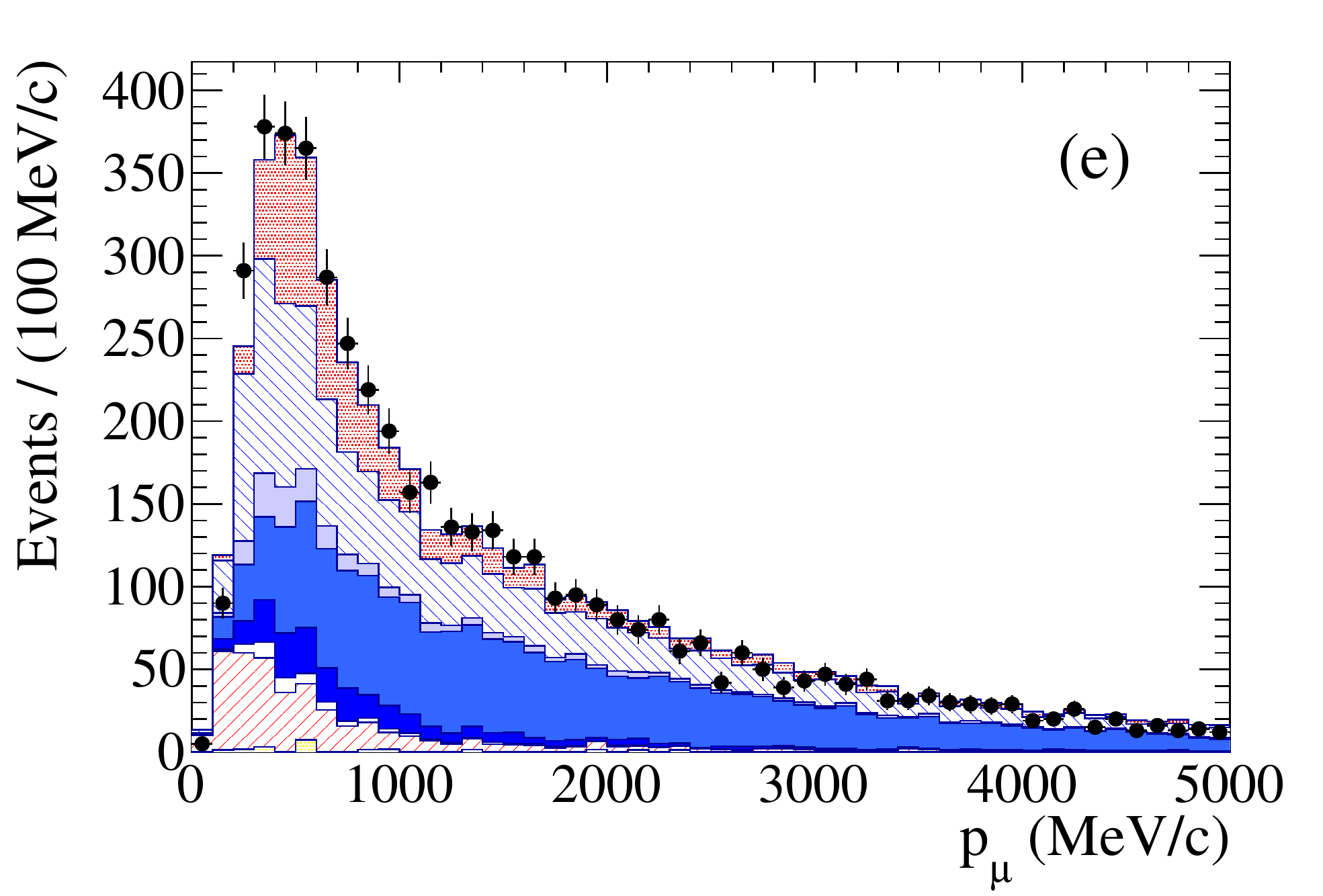} &
\includegraphics[width=0.48\textwidth]{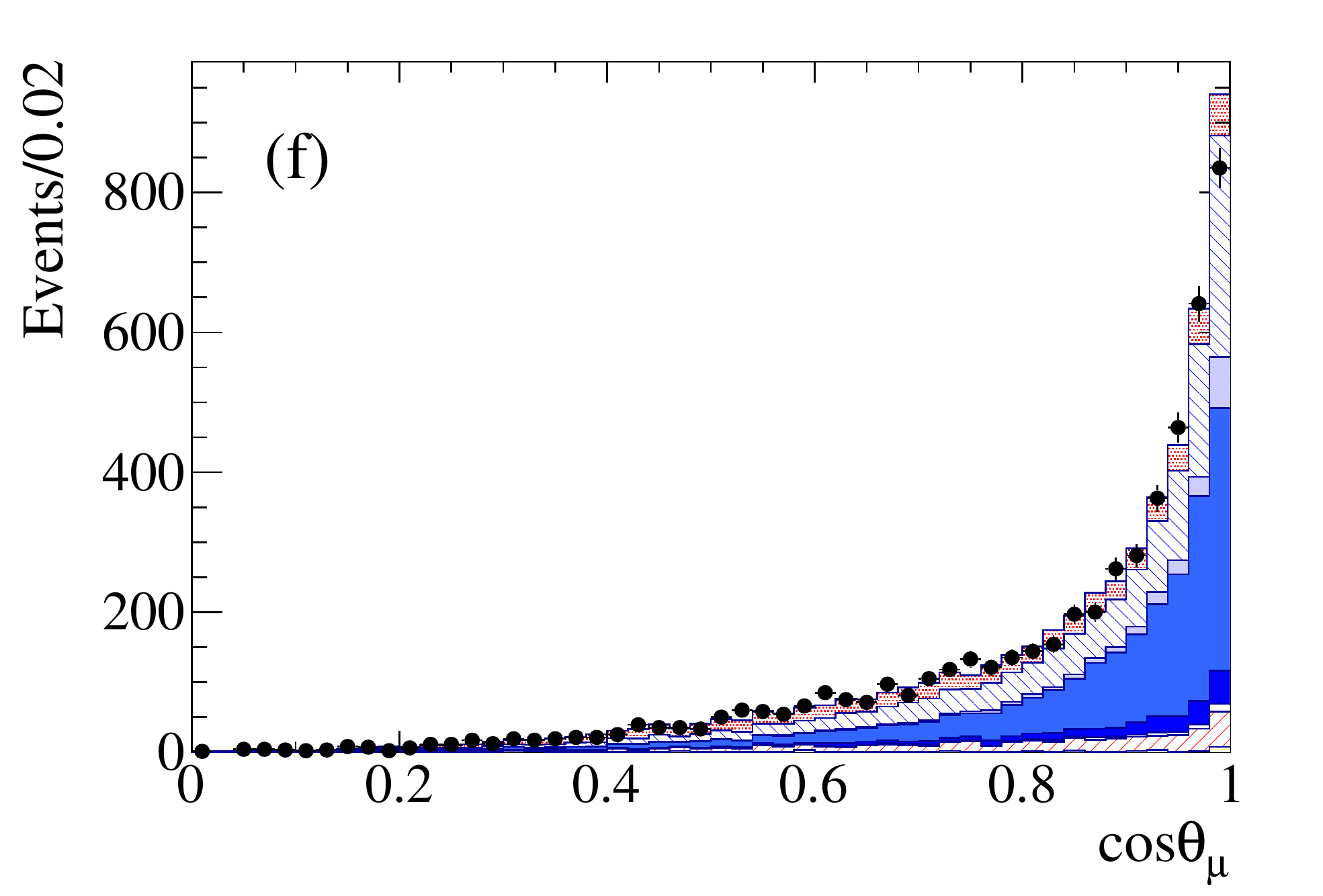}\\
\end{tabular}
\caption{Muon momentum for the \cc-inclusive (a), \ccqe-like (c), and \ccnqe-like (e) samples.
Cosine of the muon angle for the  \cc-inclusive (b), \ccqe-like (d), and \ccnqe-like (f) samples.
 The errors on the data points are the statistical errors.
\label{fig:nd280selection} }
\end{figure*}

Fig.~\ref{fig:nd280selection} shows the distributions of events binned in the muon momentum (\pmu) and cosine of the angle between the muon direction and the $z$-axis (\costmu) for both data and the prediction. In addition, we check the stability of the neutrino interaction rate with a Kolmogorov-Smirnov (KS) test of the accumulated data and find $p$-values of 0.20, 0.12, and 0.79 for the \cc-inclusive, \ccqe-like and \ccnqe-like samples, respectively.

Both \ccqe-like and \ccnqe-like samples provide useful constraints on the neutrino flux and neutrino interaction models.
 The \ccqe-like sample includes the dominant neutrino interaction process at the T2K beam peak energy (CCQE)
and the \ccnqe-like sample is sensitive to the high energy tail of the neutrino flux, where relatively few \ccqe interactions occur. The fit of the flux and cross section models to these data, further described in Section~\ref{sec:extrapolation}, uses two-dimensional \pmu and \costmu distributions for the \ccqe-like and \ccnqe-like samples. 
We use a total of 20 bins per each sample, where \pmu is split into 5 bins and \costmu is split into 4 bins. The data and the expected number of events for this binning are shown in Table~\ref{tab:280numubanff}.

\begin{table}
\caption{Number of data and predicted events for the ND280 \cc-inclusive selection criteria.
\label{tab:nd280selection}}
\centering
\begin{tabular}{l|c|c}
\hline
           & Data &  MC  \\ 
\hline 
Good negative track in FV &  21503  & 21939 \\ 
Upstream TPC veto         &  21479  & 21906 \\ 
\mmu PID                  &  11055  & 11498 \\ 
\hline    
\end{tabular}
\end{table}

\begin{table}
\caption{Number of data and predicted events 
for the ND280 \ccqe-like and \ccnqe-like selection criteria, after the \cc-inclusive selection has been applied.
\label{tab:nd280datareduction_qe}
}
\centering
\begin{tabular}{l|c|c|c|c}
\hline 
                            &\multicolumn{2}{|c|}{\ccqe-like}& \multicolumn{2}{|c}{\ccnqe-like}\\                              
\hline
                            & Data &  MC & Data & MC  \\ 
\hline
TPC-FGD track              &  6238  & 6685  & 4817  & 4813  \\ 
Michel electron            &  5841  & 6244  & 5214  & 5254  \\ 
\hline
\end{tabular}
\end{table}

\begin{table}
\caption{ Breakdown of the three ND280 \cc\ samples by true interaction type as predicted by the MC simulation.
\label{tab:nd280fractionalcontribution}
}
\centering
\begin{tabular}{l|c|c|c}
\hline 
Event type & \cc-inclusive             &  \ccqe-like         &  \ccnqe-like          \\
\hline 
\ccqe                & 44.4  & 69.5  & 14.7  \\ 
\cc resonant $1\pi$  & 21.4  & 14.5  & 29.6 \\
\cc coherent $\pi$   & 2.8   & 1.7   & 4.0 \\
All other \cc        & 18.8  & 3.7   & 36.8 \\
\nc                  & 3.0   & 1.3   & 5.1 \\
\numb                & 0.7   & 0.2   & 1.2 \\
out of FV        & 7.8   & 7.6   & 8.0 \\
sand interactions    & 1.1   & 1.6   & 0.5 \\
\hline
\end{tabular}
\end{table}

\begin{table*}
\centering
\caption{Data (MC) \pmu and \costmu events split in bins as used by the fit described in Section~\ref{sec:extrapolation} at ND280.
\label{tab:280numubanff}}
\begin{tabular}{l|c|c|c|c|c}
\hline
\multicolumn{6}{c}{\ccqe-like sample}\\\hline
& \multicolumn{5}{c}{\pmu (\mevc)}\\ \hline                                   
                          &  0-400       & 400-500 & 500-700 &  700-900 & $>$900 \\\hline
$-1 < \costmu \le 0.84$   & 854 (807.7) & 620 (655.6) & 768 (821.2) & 222 (255.0) & 222 (233.0) \\ 
$0.84 < \costmu \le 0.90$  & 110 (107.2) & 110 (116.3) & 235 (270.6) & 133 (153.5) & 159 (194.7)  \\ 
$0.90 < \costmu \le 0.94$ &  62 (69.1)  &  67 (74.0)  & 142 (179.0) &  90 (121.4) & 228 (274.6) \\ 
$0.94 < \costmu \le 1.0 $ &  92 (95.4)  &  73 (85.4)  & 184 (216.5) & 160 (174.8) &1310 (1339.0)\\ 
\hline\hline
\multicolumn{6}{c}{\ccnqe-like sample}\\
\hline
& \multicolumn{5}{c}{\pmu (\mevc)}\\ \hline                                   
     &    0-400 & 400-500 & 500-700 &  700-900 & $>$900 \\\hline
$-1 < \costmu \le 0.84$  & 560 (517.9) &   262 (272.2) &   418 (400.3) &   256 (237.8) &   475 (515.0) \\
$0.84 < \costmu \le 0.90$ & 83 (80.3) &    42 (35.8) &    83 (80.2) &    86 (74.8) &   365 (389.8) \\
$0.90 < \costmu \le 0.94$& 46 (58.6) &    37 (33.8) &    60 (63.1) &    39 (56.4) &   462 (442.6) \\
$0.94 < \costmu \le 1.00 $& 75 (76.6) &    33 (43.2) &    91 (93.4) &    85 (87.2) &  1656 (1694.7)\\
\hline
\end{tabular}
\end{table*}

\subsection{{\label{sec:nd280_numusystematics}Detector Response Modeling Uncertainties}}

We consider systematic uncertainties on the modeling of the detection efficiency and reconstruction of events which affect:
\begin{itemize}
\item the overall efficiency for selecting \cc interactions
\item the reconstructed track properties (\pmu, \costmu)
\item the sample (either \ccqe-like or \ccnqe-like) in which the event is placed
\end{itemize}
We estimate uncertainties from each category with a variety of control samples that include beam data, cosmic events and 
simulated events.

The uncertainty on the efficiency for selecting \cc \num interactions is propagated from uncertainties on:
the data quality criteria applied to the tracks, track reconstruction and matching efficiencies, PID, and
determination of the track curvature.  We also consider the uncertainty on the detector mass.

The systematic uncertainty on the track momentum determination is from uncertainties on the magnetic field absolute value and field non-uniformity. 
Small imperfections in the magnetic and electric fields can affect the path of the drift electrons, causing a distorted image of the track and a possible bias in the reconstructed momentum. The size of these distortions is constrained from laser calibration data and MC simulations using magnetic field measurements made prior to detector installation. The overall momentum scale is determined from the magnitude of the magnetic field component transverse to the beam direction, $B_x$, which is inferred from the measured magnetic coil current. The momentum resolution is determined in data from studies of tracks which traverse multiple TPCs; the individual momentum calculated for a single TPC can be compared to the momentum determined by nearby TPCs to infer the momentum resolution in data and MC simulation. 

The primary causes of event migration between the \ccqe-like and \ccnqe-like samples are external backgrounds or interactions of pions. External backgrounds in the samples are due to three sources: cosmic rays, neutrino interactions upstream in the surrounding sand and concrete, and neutrino interactions in the ND280 detector outside the FV (out of FV). 
Interactions from the sand or concrete contribute to the number of tracks in the selected event, which can change a \ccqe-like event to a \ccnqe-like event.
Interactions that occur outside of the FGD1 FV are about 7.6\%\ of the total selected \cc-inclusive sample. 
Sources include neutrino interactions in FGD1 outside of the FV, or particles produced in interactions downstream of FGD1 that travel backwards to stop in the FGD1 FV.
Pion absorption and charge exchange interactions in the FGD material can also reduce the probability that a charged pion produces a track in TPC2, affecting the identification of an event as \ccqe-like or \ccnqe-like. The uncertainty on the GEANT4 modeling of pion inelastic scattering is evaluated by comparing the GEANT4 model to pion scattering data. 

\begin{figure}[htbp]
\centering
\begin{tabular}{c}
\includegraphics[angle=0,width=0.5\textwidth]{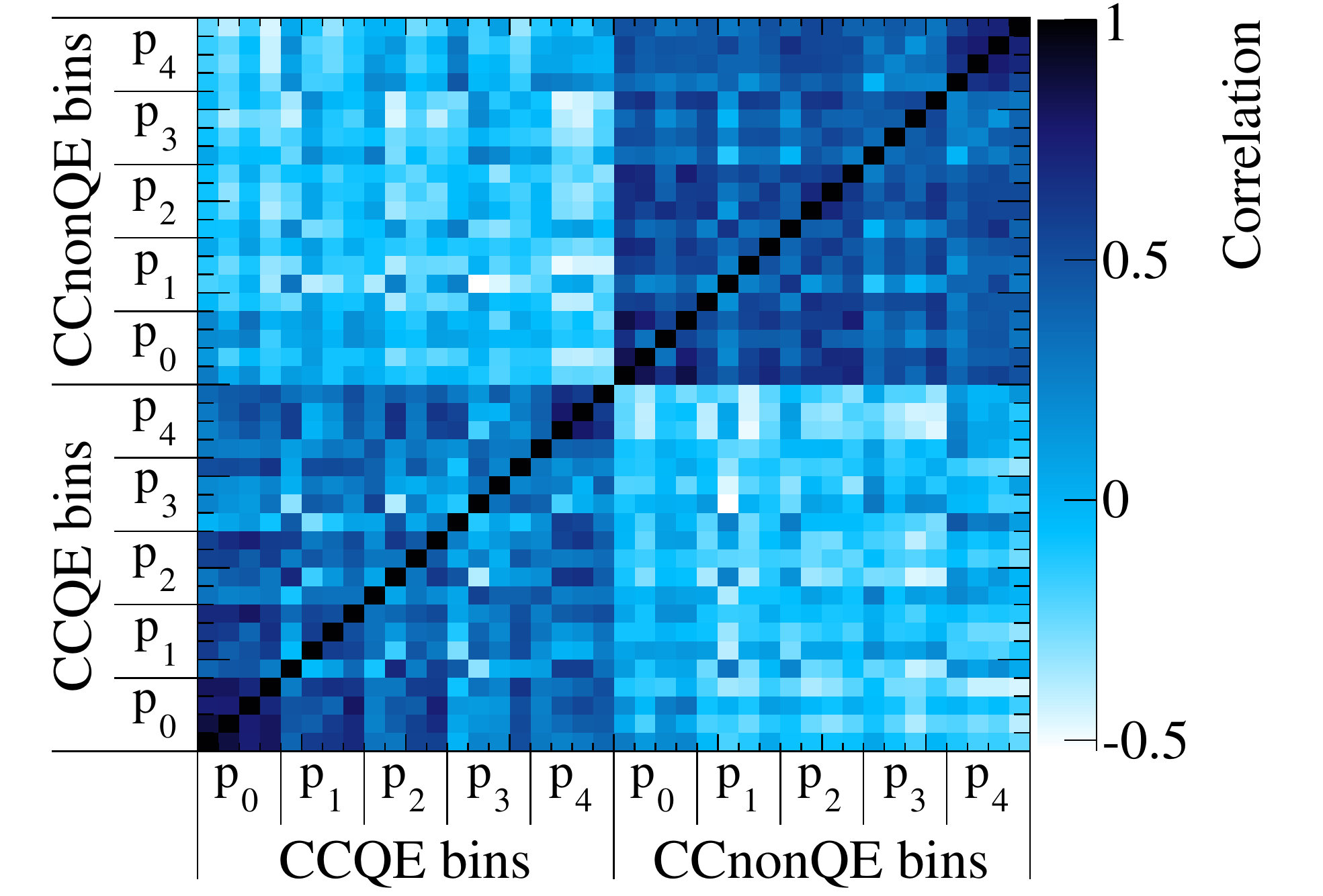}\\
\end{tabular}
\caption{The bin-to-bin correlation matrix from the systematic covariance matrix for the \num selected sample at ND280. The bins are
ordered by increasing \costmu  in groups of increasing muon momentum ($p_{0}$ to $p_4$) for the
two selections.
\label{fig:nd280covariance} }
\end{figure}

For each source of systematic uncertainty, we generate a $40\times40$ covariance matrix with entries for each pair of \pt bins. 
These matrices represent the fractional uncertainty on the predicted numbers of events in each \pt bin for each error source.
The binning used is the same as shown in Table~\ref{tab:280numubanff}, where the first 20 bins correspond to the \ccqe-like sample and the second 20 correspond to the \ccnqe-like sample. The total covariance matrix $V_{d}$ is generated by linearly summing the covariance matrices for each of the systematic uncertainties.
Fig.~\ref{fig:nd280covariance} shows the bin-to-bin correlations from the covariance matrix, which displays the feature of anti-correlations between bins in the \ccqe-like and \ccnqe-like samples arising from systematic error sources, such as the pion absorption uncertainty, that migrate simulated events between samples.  
Table~\ref{tab:nd280syst} summarizes the range of uncertainties across the \pt bins and the uncertainty on the total number of events.

\begin{table}
\caption{Minimum and maximum  fractional errors among all the \pt bins, including the largest error sources.
The last column shows the fractional error on the total number of events, taking into account the correlations 
between the \pt bins.\label{tab:nd280syst} }
\centering
\begin{tabular}{l|c|c}\hline
Systematic error             &\multicolumn{2}{c}{Error Size (\%)} \\\hline
                             & Minimum and &  Total fractional  \\
                             & maximum fractional &  error       \\
                             & error           & \\\hline
B-Field Distortions          & 0.3 - 6.9 & 0.3 \\
Momentum Scale               & 0.1 - 2.1  &  0.1 \\
Out of FV                & 0 - 8.9   &  1.6 \\
Pion Interactions         & 0.5 - 4.7  &  0.5 \\
All Others                   & 1.2 -  3.4 &  0.4  \\\hline
Total                        & 2.1 - 9.7 &  2.5 \\
\hline
\end{tabular}
\end{table}

\subsection{{\label{sec:nd280_nue}Intrinsic \nue candidate selection}}

We also select a sample of \cc \nue interactions to check the consistency of the predicted and measured intrinsic \nue rates. 
The \cc \num selections described earlier provide the strongest constraint on the expected intrinsic \nue rate, through the 
significant correlation of the \num flux to the \nue flux. However, a \cc \nue selection at the near detector 
provides a direct and independent measurement of the intrinsic \nue rate.  

We select \cc \nue interactions by applying the same criteria as described in Section~\ref{sec:nd280_numuselection}, except 
that the energy loss for the highest momentum negatively charged particle is required to be consistent with an electron instead of a muon, and interactions in FGD2 are used to increase the sample size. 
For electrons of momenta relevant to T2K, the energy loss is 30--40\% larger than for muons at the same momenta, and so electrons and muons are well separated since the TPC energy loss resolution is less than $8$\%~\cite{Abgrall:2010hi}. 
In addition, for tracks which reach the downstream ECAL, we use the information from the ECAL to remove events in which the lepton candidate is consistent with a muon. A muon that crosses the ECAL produces a narrow track while an electron releases a large part of its energy, producing an electromagnetic shower. We developed a neural network to distinguish between track-like and shower-like events. For this analysis we select only shower-like events.

 The total number of selected events in the electron candidate sample is 927. The signal efficiency for selecting \cc \nue interactions in the FGD1 and FGD2 FV is 31.9\%\ with an overall 23.7\%\ purity. For higher momenta the relative purity of the selection increases (42.1\% for $p_e>300$\,\mevc).

\begin{figure}[htbp]
\centering
\includegraphics[angle=0,width=0.5\textwidth]{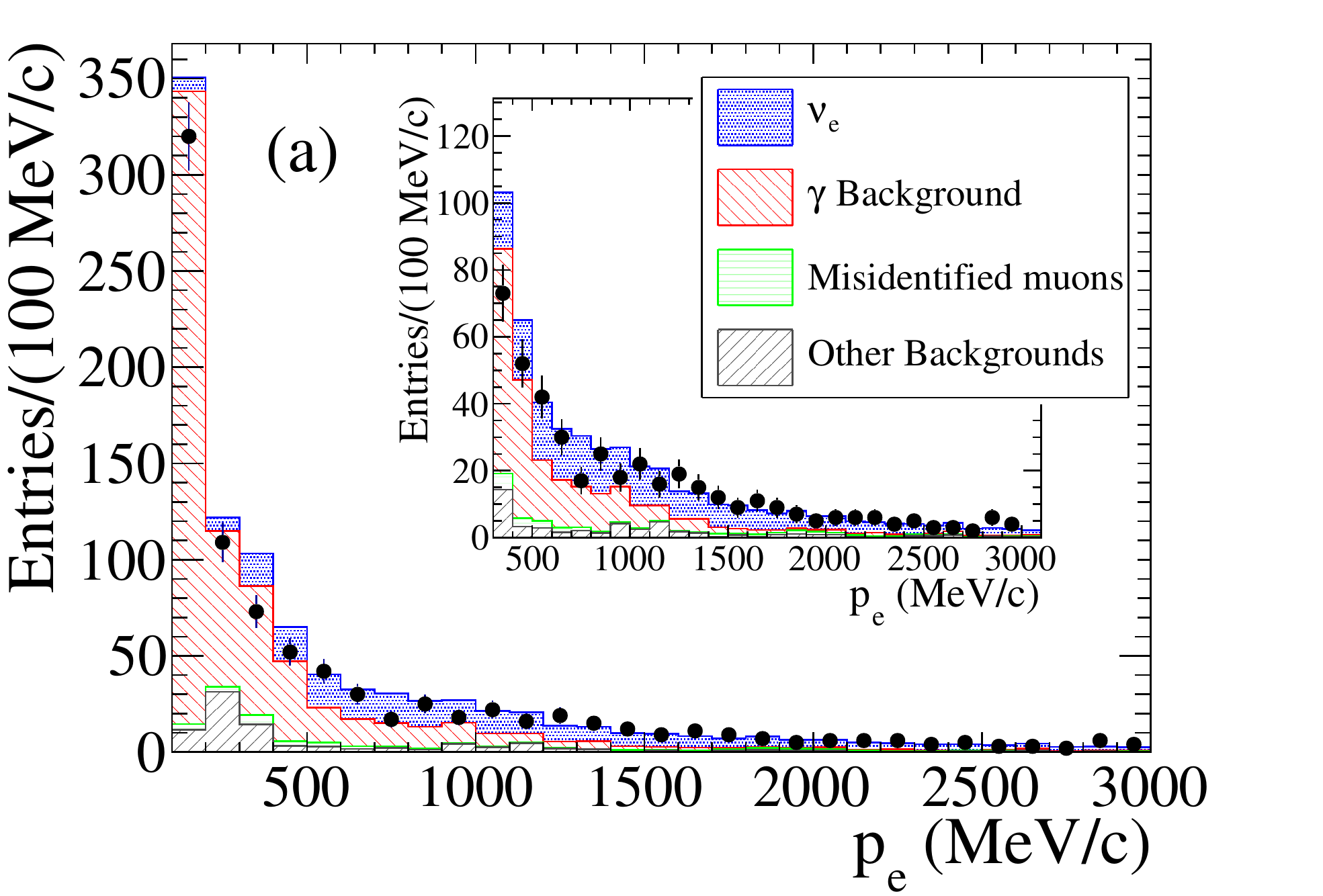}\\ 
\includegraphics[angle=0,width=0.5\textwidth]{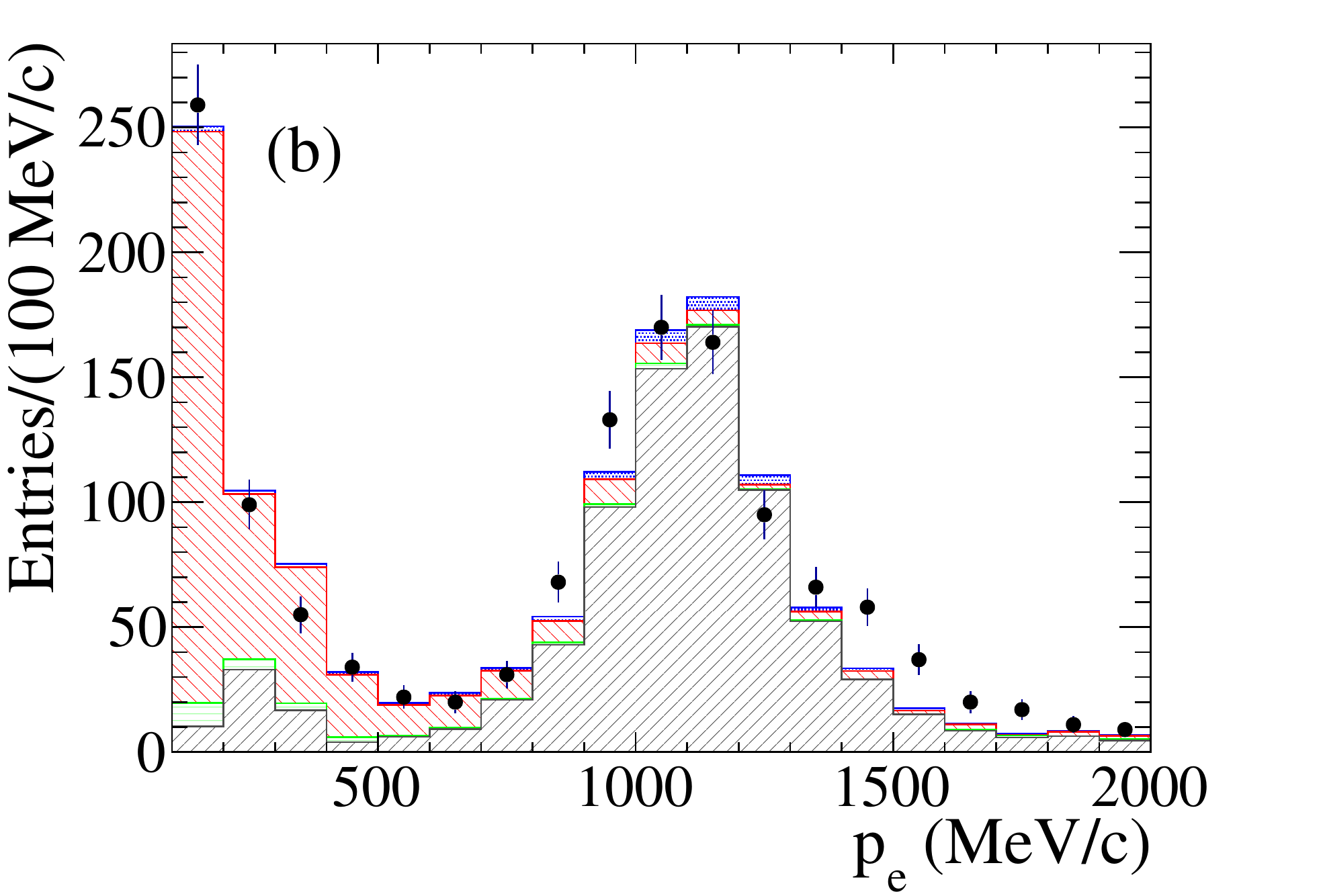} \\ 
\caption{ (a) Momentum distribution of the highest momentum track with negative charge for each event in the electron candidate sample at ND280. The inset shows the region with momentum $\geq 300$\,\mevc. (b) Momentum distribution of the highest momentum track with positive charge for each event of the positron candidate control sample. The ``Other Backgrounds'' component is mainly due to protons and pions from \nc and \cc \num interactions in the FGD.  The energy loss of positrons and protons (pions) is similar at $p\approx$1000\,\mevc (200\,\mevc), resulting in the presence of these particles in the positron candidate sample.
\label{fig:nd280nueselection}}
\end{figure}

 The majority of selected \nue are from kaon decay (80\%). 
The dominant background events ($78\%$ of the total background) are low energy electrons produced by photon conversion in the FGDs, called the $\gamma$ background. The photons come from \piz decays, where the $\piz$s are generated in \num interactions either in the FGD or in the material which surrounds the FGD. A total of $7$\% of the remaining background events are misidentified muons coming from \num interactions.  The probability for a muon to be misidentified as an electron is estimated to be less than 1\% across most of the relevant momentum range. This probability is determined using a highly pure ($>$99\%) sample of muons from neutrino-sand interactions. Finally, background not belonging to the two previous categories is mainly due to protons and pions produced in \nc and \cc \num interactions in the FGD. 
Fig.~\ref{fig:nd280nueselection} (a) shows the momentum distribution of the highest momentum track with negative charge for each event in the selected electron candidate sample. 

We estimate the uncertainties on the detector response modeling for the 
electron candidate sample
in the same manner as described in Section~\ref{sec:nd280_numusystematics}, with additional uncertainties considered for the FGD2 interactions in the selection, and for electron-PID selection. The total detector response systematic uncertainty on the electron candidate sample is 5.7\%, with the TPC PID (3.8\%) uncertainty as the largest.

The rate of intrinsic \nue interactions is determined with a likelihood fit to reconstructed momenta of electron
candidate events. 
To constrain the large background from photons, a control sample of positron (positive charge, electron PID tracks) candidates is used.  
Fig.~\ref{fig:nd280nueselection} (b) shows the momentum distribution of candidate positrons.  The sample is composed of positrons at 
lower energies and protons at higher energies. 
We simultaneously fit the electron and positron candidate 
samples to determine the photon background and \nue signal rate normalizations.
The misidentified muon background component is fixed according to the estimate from the pure muon control sample, and other smaller background sources are fixed according to the nominal predictions. 
Neutrino flux, neutrino cross section, and detector response uncertainties are included in the likelihood fit.  

The inferred rate of \cc \nue events in data from the likelihood fit normalized by the prediction is 
$0.88 \pm 0.10 (stat.) \pm 0.15 (syst.)$.  
The measured \nue rate at the near detector is consistent 
with the prediction within systematic uncertainties.  The neutrino flux and cross section systematic uncertainties 
are the dominant contributions to the total systematic error on the \nue rate. In Section~\ref{sec:extrapolation} we show 
the \nue rate after the flux and cross section parameters are tuned by the fit to the CC \num data. 

\section{\label{sec:extrapolation} ND280 Constraint on the Neutrino Flux and Cross Section Models}

The rate of neutrino interactions measured at the ND280 detector has power to constrain the  neutrino flux 
and interaction models used to predict the $\nu_{e}$ candidate event rate at the SK detector.  
The predicted SK $\nu_e$ signal and neutral current background both depend directly on the
unoscillated $\nu_{\mu}$ flux, while the intrinsic $\nu_e$ background depends on the $\nu_{e}$ flux.  As
shown in Fig.~\ref{fig:flux_corr_banff}, both the SK $\nu_{\mu}$ and $\nu_{e}$ flux predictions are correlated to
the ND280 \num flux prediction through the underlying data and assumptions applied in the flux calculation.
Both the SK $\nu_e$ signal and intrinsic $\nu_e$ background also depend on the charged current interaction model.
Hence, a fit to the CC-inclusive events from ND280 can constrain flux and cross section nuisance parameters 
relevant to the SK prediction.

We fit the near detector \ccqe-like and \ccnqe-like $\nu_\mu$ data to determine tuned values of the $\nu_\mu$ and \nue flux parameters 
and cross section model parameters, described in Sections~\ref{sec:flux} and \ref{sec:neut_int} respectively.  The fit includes
the marginalization of nuisance parameters describing uncertainties in the simulation of the detector response and parameters describing
parts of the neutrino interaction model that are not correlated for ND280 and SK selections. The tuned parameters are then 
applied to predict the $\nu_e$ signal and background interactions at SK. 
The fit also incorporates constraints on the flux and cross section models determined independently from the ND280 data constraint 
to properly propagate all constraints to the SK event rate predictions.

\subsection{\label{subsec:like}ND280 likelihood}

The fit maximizes a likelihood that includes the binned likelihood of the ND280 data and
the prior constraints on the flux model,
the interaction model, and the detector response model:
\begin{linenomath*}
\begin{equation}
\begin{split}
& \mathcal{L}_{ND}(\vec{b},\vec{x},\vec{d}|N^{d}_{i}) = \pi_{flux}(\vec{b})\pi_{xsec}(\vec{x})\pi_{det}(\vec{d})\times \\
& \prod_{i=1}^{N_{bins}}\frac{[N^{p}_{i}(\vec{b},\vec{x},\vec{d})]^{N^{d}_{i}}e^{-N^{p}_{i}(\vec{b},\vec{x},\vec{d})}}{N^{d}_{i}!}.
\end{split}
\end{equation}
\end{linenomath*}
$\pi_{flux}(\vec{b})$, $\pi_{xsec}(\vec{x})$, $\pi_{det}(\vec{d})$ are multivariate normal distributions
that are functions of the flux ($\vec{b}$), neutrino cross section ($\vec{x}$) and detector response ($\vec{d}$)
nuisance parameters.  These functions encode the prior constraints on the nuisance parameters and depend on the
nominal parameter values and the parameter errors or covariance matrices described in previous sections.
 The likelihood includes the product of the Poisson probabilities for the $N_{bins}=40$ bins of the 
\ccqe-like and \ccnqe-like selections.  For
each bin the predicted number of events, $N^{p}_{i}(\vec{b},\vec{x},\vec{d})$, is evaluated based on the values of the nuisance parameters, and compared to 
the measurement, $N^{d}_{i}$. To obtain fit results that more closely follow a $\chi^{2}$ distribution~\cite{Baker1984437}, 
we define the
likelihood ratio:
\begin{linenomath*}
\begin{equation}
\mathcal{L}_{ratio} = \frac{\mathcal{L}_{ND}(\vec{b},\vec{x},\vec{d}|N^{d}_{i})}
{\mathcal{L}_{ND}(\vec{b}^{0},\vec{x}^{0},\vec{d}^{0}, N^{p}_{i} = N^{d}_{i} | N^{d}_{i})}
\end{equation}
\end{linenomath*}
Here the denominator is the likelihood evaluated with $N^p_{i}$ set equal to $N^{d}_{i}$ and the nuisance parameters set to their
nominal values: $\vec{b}^{0}=1$, $\vec{x}^{0}$, $\vec{d}^{0}=1$; both $\vec{b}$ and $\vec{d}$ have nominal values of 1.   The quantity that is minimized is $-2\textrm{ln}(\mathcal{L}_{ratio})$:
\begin{linenomath*}
\begin{equation}
\begin{split}
& -2\textrm{ln}(\mathcal{L}_{ratio}) = \\
&  2 \sum_{i=1}^{N_{bins}}N^{p}_{i}(\vec{b},\vec{x},\vec{d})-N^{d}_{i}
+N^{d}_{i}\textrm{ln}[N^{d}_{i}/N^{p}_{i}(\vec{b},\vec{x},\vec{d})] \\
& +\sum_{i=1}^{N_{b}}\sum_{j=1}^{N_{b}}(1_{i}-b_{i})(V_{b}^{-1})_{i,j}(1-b_{j})  \\
&  +\sum_{i=1}^{N_{x}}\sum_{j=1}^{N_{x}}(x^{0}_{i}-x_{i})(V^{-1}_{x})_{i,j}(x^{0}_j-x_{j}) \\
& +\sum_{i=1}^{N_{bins}}\sum_{j=1}^{N_{bins}}(1-d_{i})(V_{d}(\vec{b},\vec{x})^{-1})_{i,j}(1-d_{j}) \\
& +\textrm{ln}\left(\frac{|V_{d}(\vec{b},\vec{x})|}{|V_{d}(\vec{b}^{0},\vec{x}^{0})|}\right).
\end{split}
\label{eq:delta_chi2}
\end{equation}
\end{linenomath*}
The predicted number of events in each observable bin, $N^{p}(\vec{b},\vec{x},\vec{d})$ depends on the value of the 
$\vec{b}$, $\vec{x}=(\vec{x}^{norm},\vec{x}^{resp})$, and $\vec{d}$ nuisance parameters:
\begin{linenomath*}
\begin{equation}
N^{p}_{i} = d_{i}\sum_{j}^{E_{\nu}bins} \sum_{k}^{Int.modes}{b_{j}x^{norm}_{k}(E_{j})w_{i,j,k}(\vec{x}^{resp})T^{p}_{i,j,k}}.
\end{equation}
\end{linenomath*}
The $T^{p}_{i,j,k}$ are the nominal Monte Carlo templates that predict the event rate for bins in the observables, $i$, 
true neutrino energy, $j$, and neutrino interaction modes, $k$.
The $\vec{b}$ parameters multiply 
the flux prediction in bins of true neutrino energy.   The detector response parameters, $\vec{d}$, 
multiply the expected number of events
in each observable  \pt  bin. 
The $\vec{x}$ are included in the prediction in one of two ways. The $x_{k}^{norm}$ are cross section parameters 
that multiply the neutrino cross section normalization for a given true neutrino energy bin and one of the $k$ interaction 
modes.  We model the effect of the remaining cross section parameters, $\vec{x}^{resp}$,  
with pre-calculated response functions, 
$w_{i,j,k}(\vec{x}^{resp})$, that have a value of 1 for the nominal parameter settings and  
can have a non-linear dependence on the cross section parameters.

The remaining terms in Eq.~\ref{eq:delta_chi2} correspond to the prior constraints on the flux, 
cross section and detector response models discussed in earlier sections. $V_{b}$ is the prior fractional covariance matrix,
corresponding to Figures~\ref{fig:flux_unc_banff} and~\ref{fig:flux_corr_banff}. The covariances of flux predictions at
ND280 and SK are included so that the fit to ND280 data can constrain the SK flux parameters.   The prior covariance matrix 
for the neutrino interaction parameters, $V_{x}$, is diagonal for most parameters with entries corresponding to the errors 
listed in Table~\ref{tab:xsec_param_unc}.  Correlations are included for the parameters constrained by the fit to 
MiniBooNE single pion data.   The $V_{d}$ fractional covariance matrix, with correlations shown in 
Fig.~\ref{fig:nd280covariance}, 
incorporates the simulated detector efficiency and reconstruction uncertainties,  
final state interaction errors and Monte Carlo statistical errors. The final term in the likelihood is present since
the Monte Carlo statistical errors included in $V_{d}$ depend on the $\vec{b}$ and $\vec{x}$ parameters through
the weights applied to the simulated events.   Since
$V_{d}$ is not constant, the determinant from the multivariate normal distribution, $\pi_{det}(\vec{d})$, cannot be dropped
from the $-2\textrm{ln}(\mathcal{L}_{ratio})$.

\subsection{\label{subsec:param_prop}Parameter propagation and marginalization}
This fitting method extrapolates the ND280 constraint on the neutrino flux and interaction model to the 
far detector prediction through
the simultaneous variation of ND280 and SK flux parameters, and the constraint on the
common interaction model parameters.
After the $-2\textrm{ln}(\mathcal{L}_{ratio})$ is minimized, we apply a subset of the fitted parameter values
to the calculation of the expected $\nu_e$ candidate rate at SK.  
The subset of parameters which are substantially constrained by the ND280 data sets and are also relevant to 
the event rate prediction at SK are listed in Table ~\ref{tab:banff_fit_param}. 
Since they are not used to calculate the predicted event rates at SK,
the flux parameters for ND280, nuclear model-dependent cross section parameters, and detector response systematic 
parameters are marginalized by integrating out their dependence in $-2\textrm{ln}(\mathcal{L}_{ratio})$ under the 
assumption of a quadratic dependence near the minimum.
The remaining cross section parameters do not affect the SK event prediction substantially and these are also marginalized.

\subsection{\label{subsec:nd280_fit}ND280 fit results}
\begin{figure*}
\centering
\includegraphics[width=0.95\textwidth]{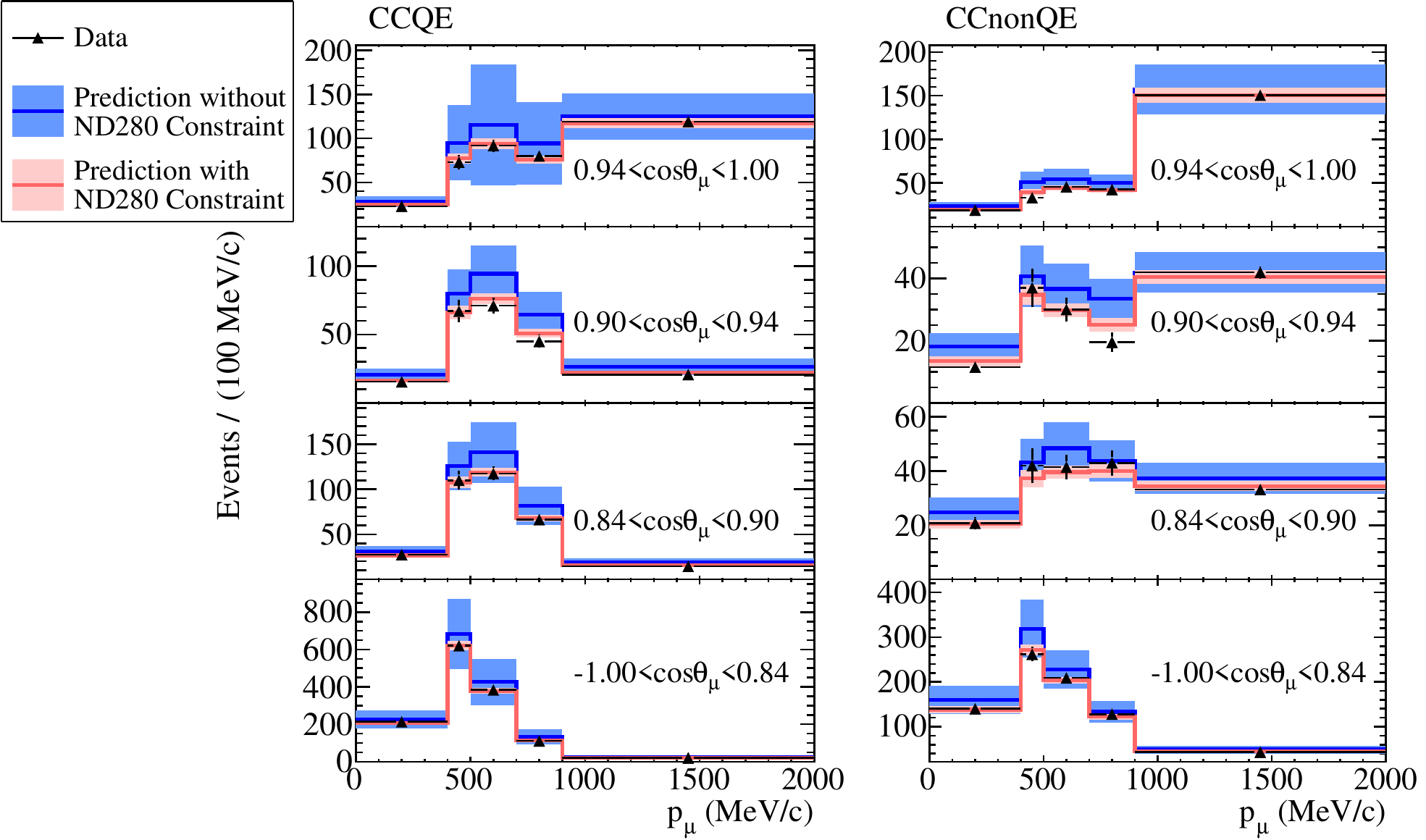}
\caption{The fitted \pmu,\costmu bins from the ND280 \ccqe-like (left) and \ccnqe-like (right) samples. 
All values in the plot are divided by the shown bin width. The $p_{\mu}>900$~MeV/$c$ bin additionally contains the overflow bin, and is normalized by a bin width of 1100~MeV/$c$.
The prediction prior to the fit uses the modifications to the NEUT model parameters derived from fits of the MiniBooNE single pion data. }
\label{fig:nd_fit_ptheta}
\end{figure*}

The resulting \pt distributions from the fit to the ND280 samples
are shown in Fig.~\ref{fig:nd_fit_ptheta}.
We evaluate the post-fit agreement between model and data by generating 2000 pseudo-experiments with statistical 
and systematic variations, and fitting them to obtain the minimum $-2\textrm{ln}(\mathcal{L}_{ratio})$ value for 
each pseudo-experiment.  The distribution of these values resembles a $\chi^2$ distribution of 41 degrees of freedom. 
Thus the value  $[-2\textrm{ln}(\mathcal{L}_{ratio})]_{min}=29.7$ from the fit to data indicates
that the data are consistent with the prediction within the prior uncertainties assigned for the neutrino flux model,
neutrino interaction model, and detector response model.  

\begin{figure}
\centering
\includegraphics[width=0.48\textwidth]{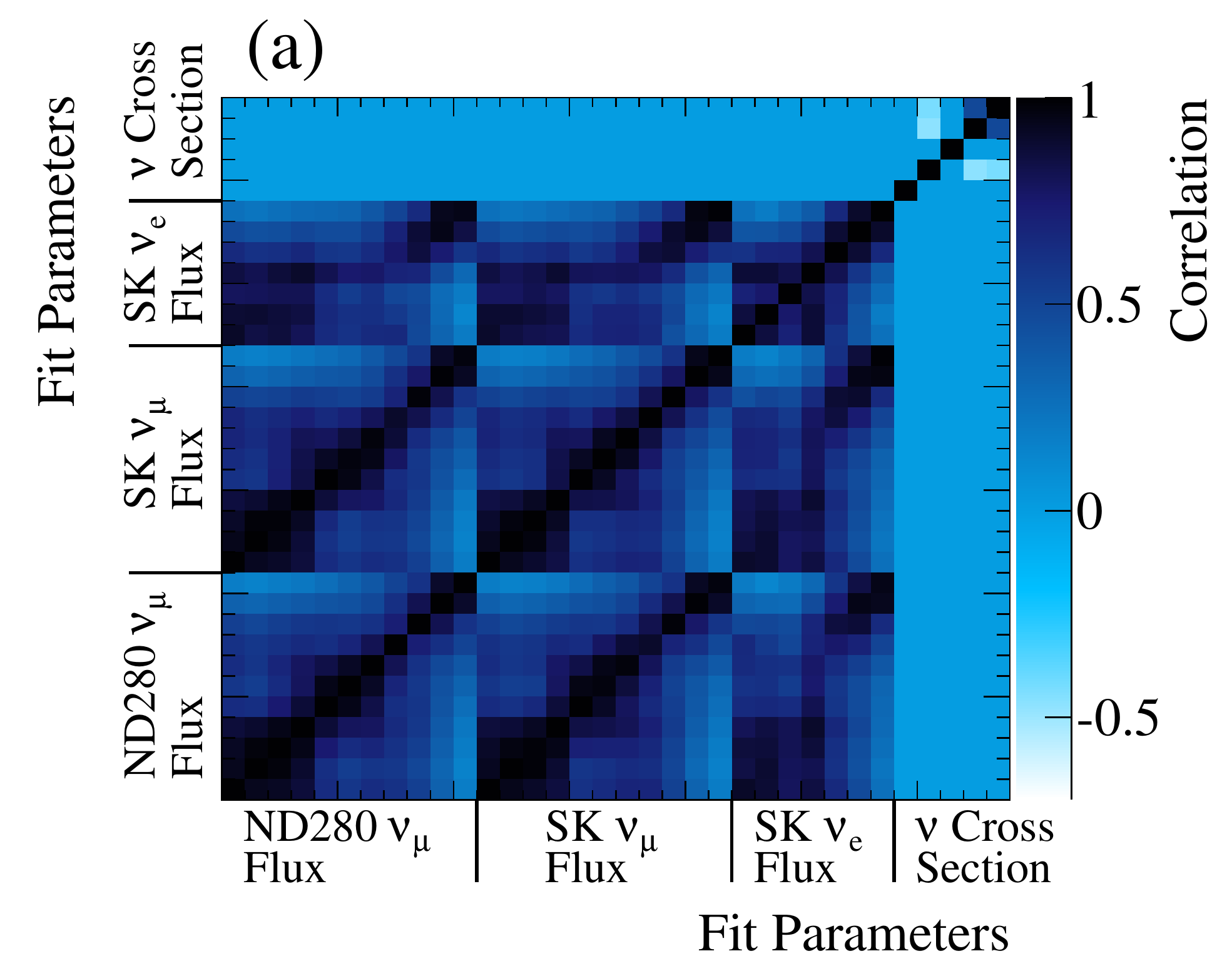}
\includegraphics[width=0.48\textwidth]{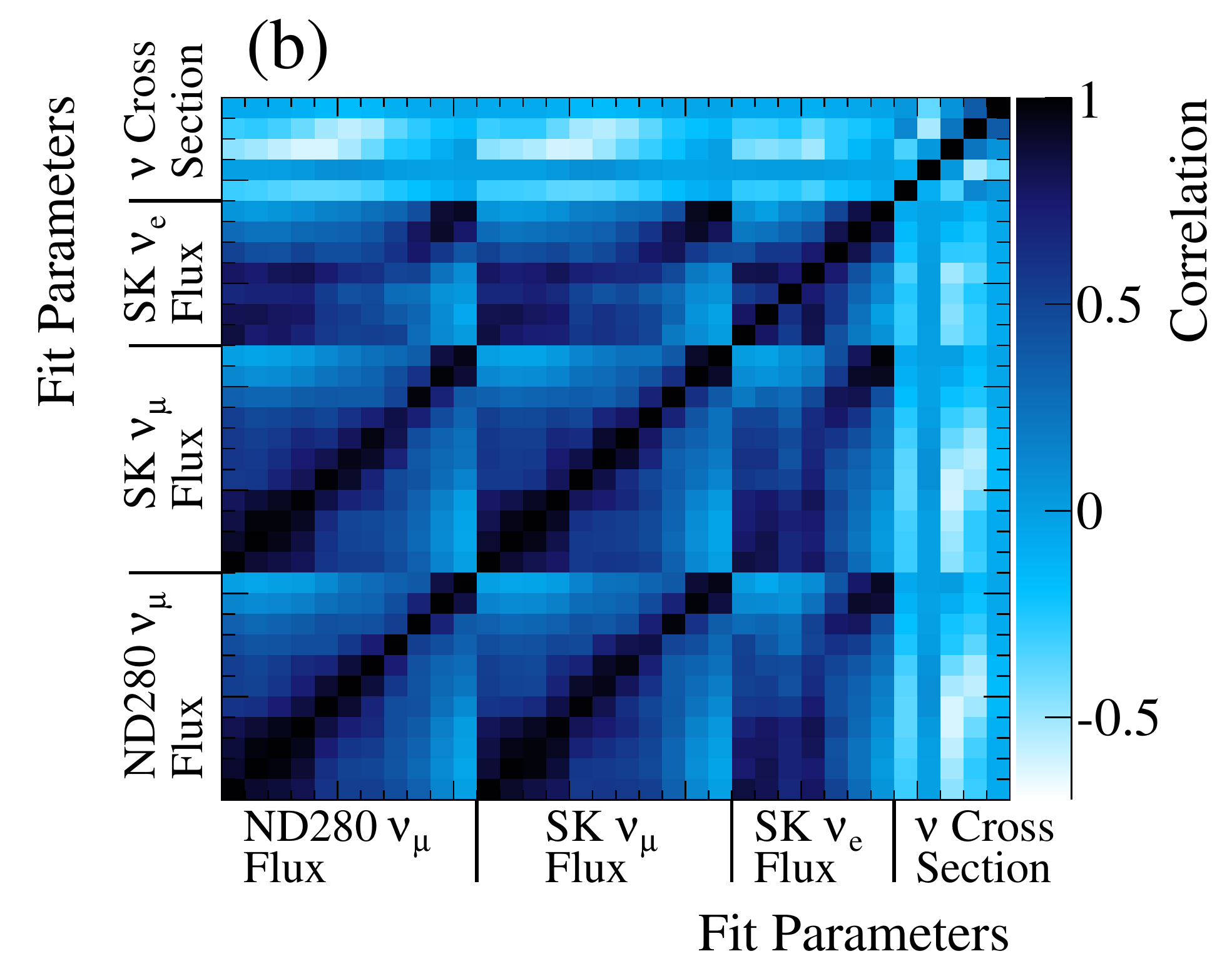}
\caption{The neutrino flux and cross section parameter correlations before (a) and
after (b) the fit to the ND280 data.  The flux parameters are ordered by increasing
energy with the binning listed in Table~\ref{tab:banff_fit_param} (the correlations of the antineutrino flux parameters 
are not shown in this figure).  The cross section parameter ordering is: $M_{A}^{QE}$, $M_{A}^{RES}$, CCQE low
energy normalization, CC1$\pi$ low energy normalization
and NC1$\pi^0$ normalization.}
\label{fig:banff_fit_corr}
\end{figure}

The propagated neutrino flux and cross section parameter values prior to and after the fit
 are listed in Table~\ref{tab:banff_fit_param}.  The fit decreases the flux prediction near the spectrum
peak to improve agreement with the data.  In addition to modifying the parameter central values and
uncertainties, the fit also sets the correlations between parameters.  Prior to the fit, the flux and
cross section model parameters have no correlation, but the fit introduces anti-correlations, as shown in
Fig.~\ref{fig:banff_fit_corr}. The anti-correlations arise because the event rate depends on the 
product of the neutrino flux and the neutrino interaction cross section.

\begin{table}
\begin{center}
\caption{Prior and fitted values and uncertainties of the propagated neutrino flux and
cross section model parameters. }
\label{tab:banff_fit_param}
\begin{tabular}{lcc}
\\ \hline
Parameter & Prior Value & Fitted Value  \\ \hline \hline
$\nu_{\mu}$ 0.0-0.4~GeV & $1.00\pm0.12$ & $0.98\pm0.09$ \\
$\nu_{\mu}$ 0.4-0.5~GeV & $1.00\pm0.13$ & $0.99\pm0.10$  \\
$\nu_{\mu}$ 0.5-0.6~GeV & $1.00\pm0.12$ & $0.98\pm0.09$ \\
$\nu_{\mu}$ 0.6-0.7~GeV & $1.00\pm0.13$ & $0.93\pm0.08$ \\
$\nu_{\mu}$ 0.7-1.0~GeV & $1.00\pm0.14$ & $0.84\pm0.08$ \\
$\nu_{\mu}$ 1.0-1.5~GeV & $1.00\pm0.12$ & $0.86\pm0.08$ \\
$\nu_{\mu}$ 1.5-2.5~GeV & $1.00\pm0.10$ & $0.91\pm0.08$ \\
$\nu_{\mu}$ 2.5-3.5~GeV & $1.00\pm0.09$ & $0.95\pm0.07$ \\
$\nu_{\mu}$ 3.5-5.0~GeV & $1.00\pm0.11$ & $0.98\pm0.08$ \\
$\nu_{\mu}$ 5.0-7.0~GeV & $1.00\pm0.15$ & $0.99\pm0.11$ \\
$\nu_{\mu}$ $>7.0$~GeV  & $1.00\pm0.19$ & $1.01\pm0.15$ \\ \hline
$\bar{\nu}_{\mu}$ 0.0-1.5~GeV & $1.00\pm0.12$ & $0.95\pm0.10$ \\
$\bar{\nu}_{\mu}$ $>1.5$~GeV  & $1.00\pm0.11$ & $0.95\pm0.10$ \\ \hline
$\nu_e$ 0.0-0.5~GeV & $1.00\pm0.13$ & $0.96\pm0.10$ \\
$\nu_e$ 0.5-0.7~GeV & $1.00\pm0.12$ & $0.96\pm0.10$ \\
$\nu_e$ 0.7-0.8~GeV & $1.00\pm0.14$ & $0.96\pm0.11$ \\
$\nu_e$ 0.8-1.5~GeV & $1.00\pm0.10$ & $0.94\pm0.08$ \\ 
$\nu_e$ 1.5-2.5~GeV & $1.00\pm0.10$ & $0.97\pm0.08$ \\
$\nu_e$ 1.5-4.0~GeV & $1.00\pm0.12$ & $0.99\pm0.09$ \\
$\nu_e$ $>4.0$~GeV  & $1.00\pm0.17$ & $1.01\pm0.13$ \\ \hline
$\bar{\nu}_e$ 0.0-2.5~GeV & $1.00\pm0.19$ & $0.97\pm0.18$ \\
$\bar{\nu}_e$ $>2.5$~GeV  & $1.00\pm0.14$ & $1.02\pm0.11$ \\ \hline
$M_{A}^{QE}$ (GeV)   &  $1.21\pm0.45$ & $1.33\pm0.20$      \\
$M_{A}^{RES}$ (GeV)  &  $1.16\pm0.11$ & $1.15\pm0.10$         \\
$x^{QE}_{1}$         &  $1.00\pm0.11$  & $0.96\pm0.09$         \\
$x^{CC1\pi}_1$      &  $1.63\pm0.43$ & $1.61\pm0.29$         \\
$x^{NC1\pi^0}_1$    &  $1.19\pm0.43$ & $1.19\pm0.40$         \\
\hline
\end{tabular}
\end{center}
\end{table}

\subsection{Consistency checks with ND280 data}
\begin{figure*}
\centering
\begin{tabular}{cc}
\includegraphics[angle=0,width=0.45\textwidth]{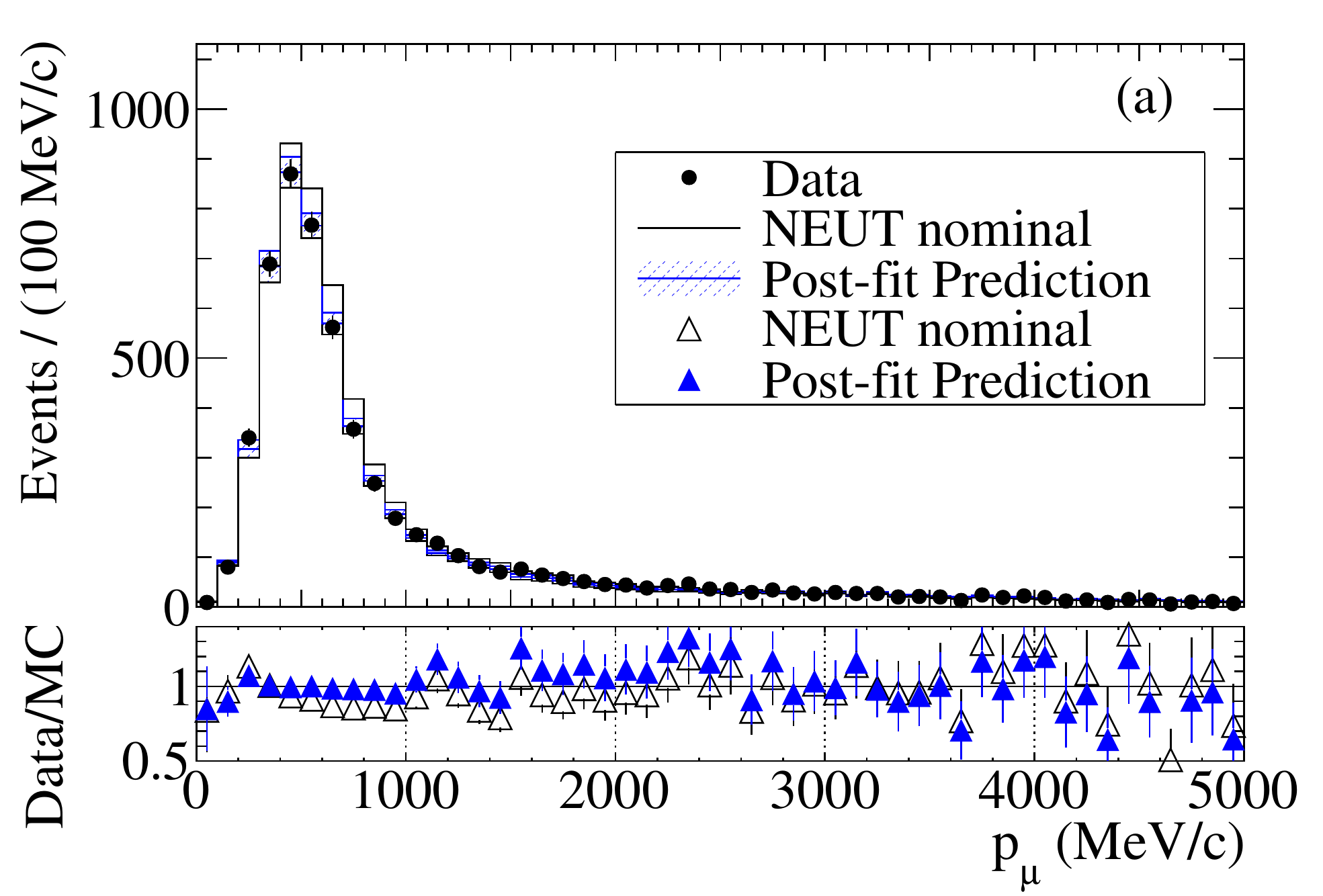}&
\includegraphics[angle=0,width=0.45\textwidth]{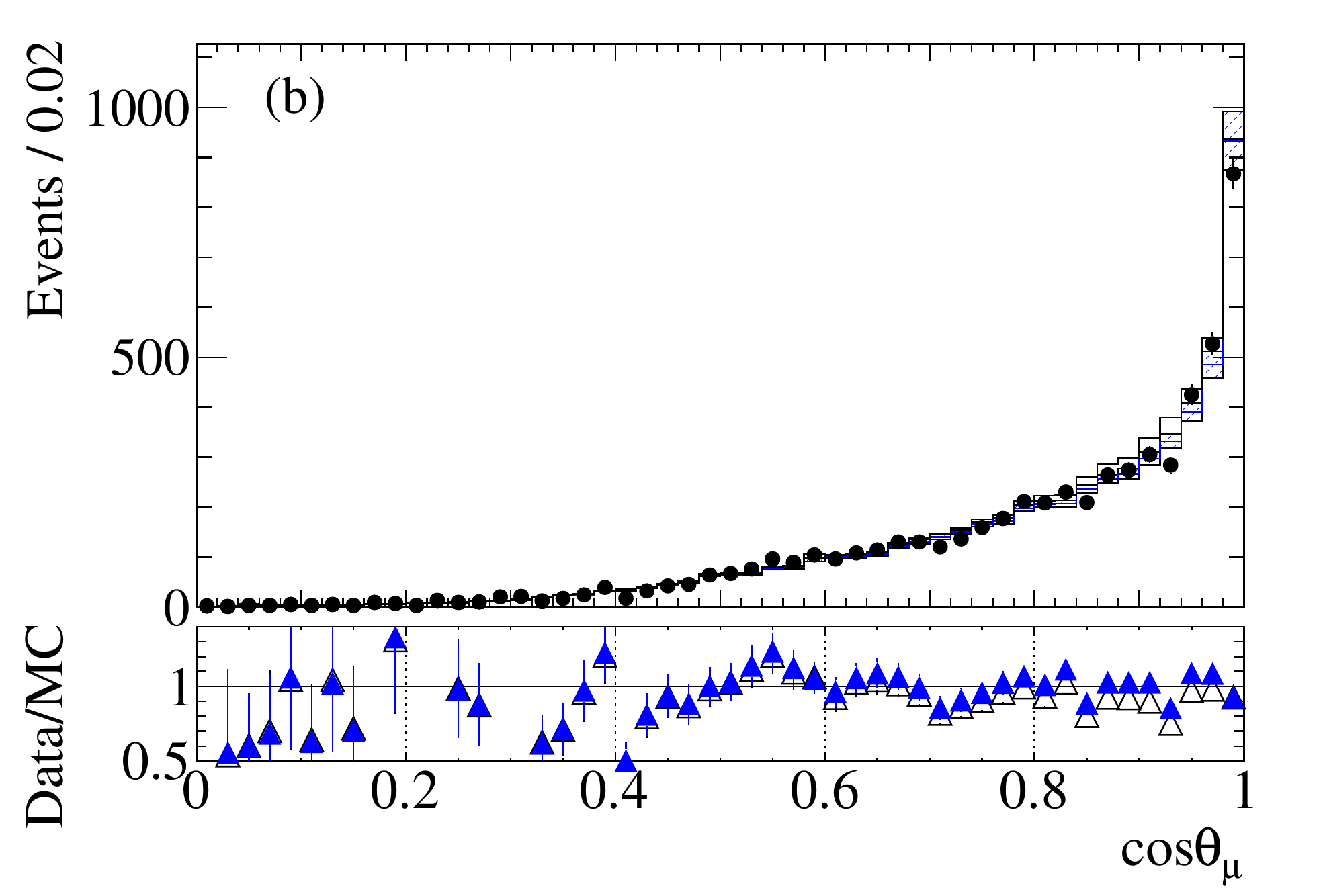}\\
\includegraphics[angle=0,width=0.45\textwidth]{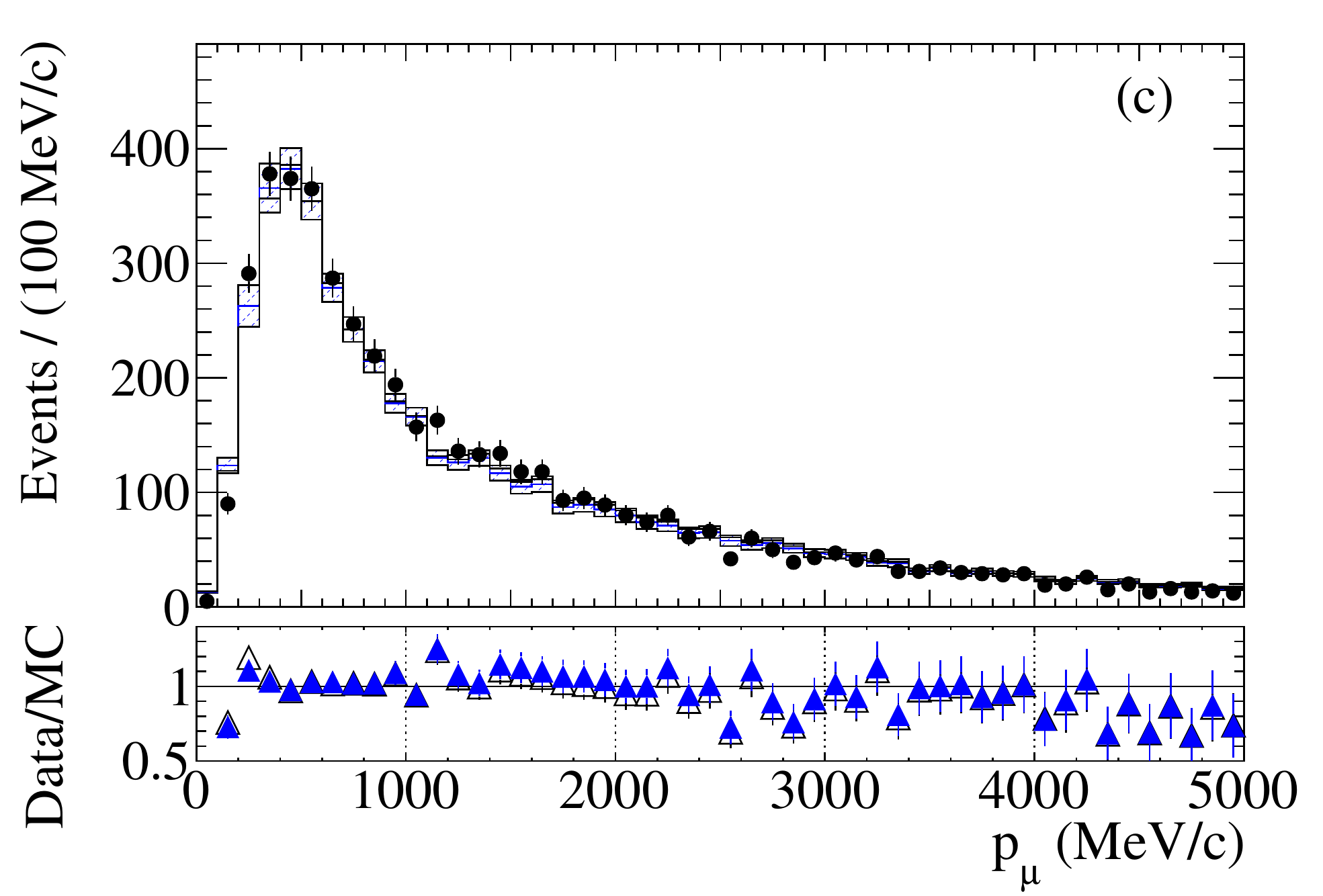}&
\includegraphics[angle=0,width=0.45\textwidth]{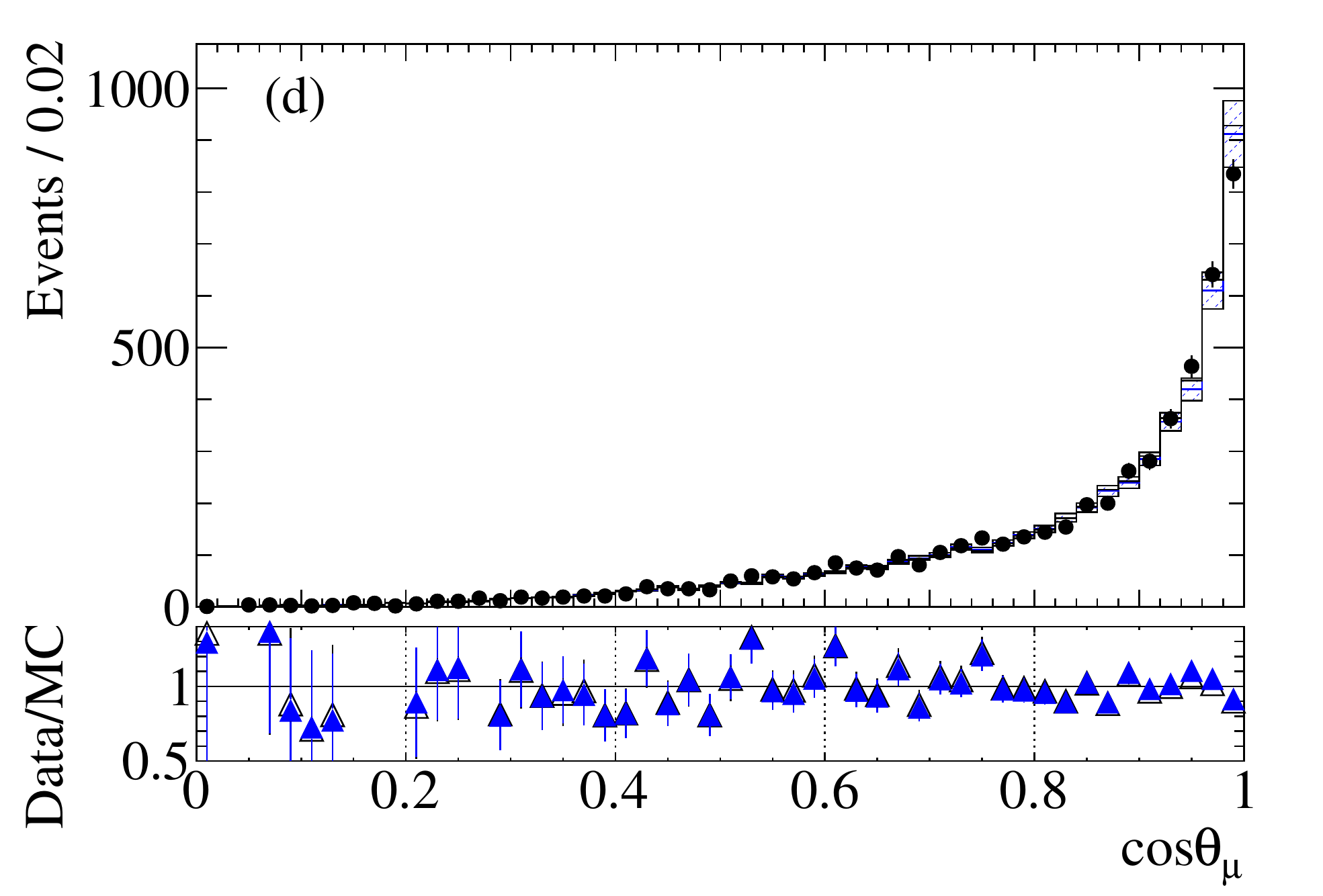}\\
\end{tabular}
\caption{Comparisons of the \pmu (left column) and \costmu (right column) distributions for \ccqe-like \num selected events in (a) and (b) and \ccnqe-like \num selected events in (c) and (d). The solid line represents the NEUT nominal prediction and the hatched region represents the post-fit MC prediction. 
The dots are the data events. Below each graph, the data/MC ratio is shown for both the NEUT nominal prediction (empty triangle) and post-fit MC prediction (full triangle). The error on the points is the statistical error on the data.
\label{fig:nd280postfit}}
\end{figure*}

We perform a consistency check of the fit results by applying the fitted parameters to the ND280 MC simulation and investigating the data and predicted rates in more finely binned kinematic distributions. 
Fig.~\ref{fig:nd280postfit} shows the level of agreement in the muon momentum and angle distributions of the
\ccqe and \ccnqe-like samples before and 
after the fit constraint to the flux and cross section models are applied.  
The fitted flux and cross section models show improved agreement with the data.

We also apply the fitted flux and cross section parameters to the ND280 \cc \nue simulation.  
Adopting the same analysis as in Section~\ref{sec:nd280_nue} while using the fitted cross section and 
flux parameters, we measure the ratio of inferred to predicted CC \nue rate to be
$0.91 \pm{} 0.10\mathrm{(stat.)} \pm{} 0.10\mathrm{(syst.)}$.
The \cc \nue rate remains consistent within the reduced systematic uncertainties after tuning.\\

To check the modeling of NC\piz production, we measure the rate of single \piz with the \pod detector using a 
data set corresponding to $8.55\times10^{19}$ POT. 
The ratio of the measured to the predicted rate is found to be $0.84 \pm{} 0.16 \mathrm{(stat.)} \pm{} 0.18 \mathrm{(syst.)}$.  
When normalized to the corresponding ratio from the
ND280 CC \num selection, we measure a ratio of $0.81 \pm{} 0.15 \mathrm{(stat.)} \pm{} 0.14 \mathrm{(syst.)}$,
indicating that the predicted rate is consistent with the measured rate within errors.

\section{\label{sec:sk_selection} SK Electron Neutrino Selection }

For a non-zero value of $\theta_{13}$, we expect an oscillated $\nu_{\mu} \rightarrow \nu_e$ flux with a peak oscillation
probability near $600$~MeV at the SK detector.  To detect the oscillated $\nu_e$, we select SK events with a single
electron-like Cherenkov light ring, providing a sample that is enhanced in \ccqe \nue interactions.  Additional cuts 
are applied to reduce the backgrounds from intrinsic $\nu_e$ contamination of the beam and $\pi^0$ background.
The selection is described here.

\subsection{The SK detector simulation}
We simulate the predicted event distributions at the far detector with
the neutrino flux prediction up to 30~GeV, the \neut cross section model, and 
a GEANT3-based detector simulation.
The \nue signal events from  $\num\to\nue$ oscillation are produced using the predicted \num spectrum without oscillations, and the \nue cross section; oscillations probabilities are applied after the simulation.
Additionally, the intrinsic \num, \numb, \nue and \nueb components of the beam are generated from the intrinsic flux predictions without oscillations.

\begin{figure}
\centering
\includegraphics[width=0.43\textwidth]{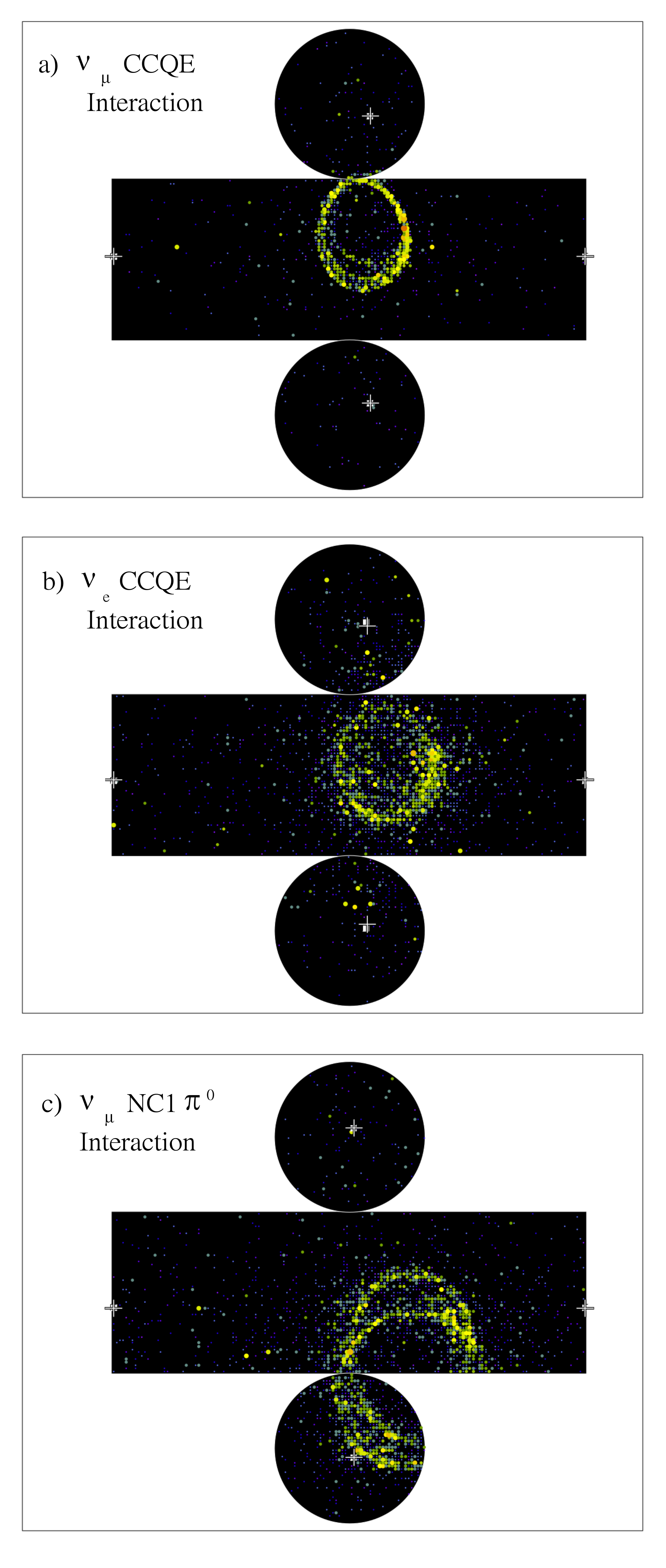}
\caption{Example event displays for the SK simulation of a) $\nu_{\mu}$ CCQE (single well-defined ring from the muon),
b) $\nu_{e}$ CCQE (single diffuse ring from the electron) and c) $\nu_{\mu}$ NC1$\pi^{0}$ interactions (two diffuse rings from the \piz $\rightarrow \gamma \gamma$ decay).  The images show the
detected light pattern at the ID wall, with the cylindrical SK detector shown as a flat projection. The color indicates the amount of charge detected by the PMT, with purple dots corresponding to the least amount of charge, and red the most.
}
\label{fig:sk_mc_event_displays}
\end{figure}

SKDETSIM, a GEANT3-based simulation of the SK detector, simulates the 
propagation of particles produced in the neutrino interactions in the SK detector.
We use the GCALOR physics package to simulate hadronic interactions in water since it
successfully reproduces pion interaction data around
1~GeV. For pions with momentum below 500~MeV, however,
we use custom routines based on the cascade model
used by \neut to simulate interactions of final state hadrons.
SKDETSIM models the propagation of light in water, considering
absorption, Rayleigh scattering,
and Mie scattering as possible interactions.
The parameters employed in the models of these processes have been
tuned using a number of laser calibration sources~\cite{Fukuda:2002uc}.
Example event displays for simulated SK events are shown in Fig.~\ref{fig:sk_mc_event_displays}.

As a final step, we scale the predicted events according to the constrained flux and 
cross section models from the fit to the ND280 \num \cc-inclusive data, and according to the oscillation probability.
The three-neutrino oscillation probability, including matter effects, is calculated for each event with the parameter values shown in Table~\ref{tab:oscparam}, unless otherwise noted.
\begin{table}
\begin{center}
\caption{Default neutrino oscillation parameters and earth matter density
used for the MC prediction.}
\label{tab:oscparam}
\ \\
\begin{tabular}{lc}
\hline \hline
Parameter & Value \\
\hline 
\\
\dmsqso        & 7.6$\times$10$^{-5}$~eV$^{2}$ \\
$|$\dmsq$|$         & 2.4$\times$10$^{-3}$~eV$^{2}$ \\ 
$\sin^2\,\theta_{12}$   & 0.32 \\  
$\sin^2\,2\theta_{23}$   & 1.0 \\  
$\delta_{CP} $           & 0 \\  
Mass hierarchy           & Normal \\ 
$\nu$ travel length      & 295~km \\ 
Earth matter density     & 2.6~g/cm$^{3}$ \\  
\\
\hline \hline
\end{tabular}
\end{center}
\end{table}

\subsection{Neutrino event selection}

We select fully contained (FC) events, which deposit all of their Cherenkov light inside
the SK inner detector (ID), by applying the following selection criteria.
First, any photomultiplier tubes (PMTs) which register sufficient charge, a ``PMT hit'', in the outer detector (OD) are
associated with other nearby PMT hits to form clusters. Events with greater than $15$ hits in the highest charge OD cluster 
are rejected. Second, most of the low energy (LE) events are removed
by requiring that the total charge from ID PMT hits in
a 300\,ns time window must be above 200 photoelectrons (p.e.),
corresponding to visible energy, $E_{vis}$, above 20~MeV.  Visible energy
is defined as the energy of an electromagnetic shower
that produces the observed amount of Cherenkov light. 
In order to remove events caused by radioactivity very close to the PMT, a third cut removes events in which a single ID PMT hit has more than half of the total charge in a 300\,ns time window.

The final FC selection cut rejects events with
ID photomultipliers which produced light because of
a discharge around the dynode, called ``flasher'' events.
The cut identifies flasher events from their timing distribution, which is much broader than neutrino events, and from a repeating pattern of light in the detector. However, neutrino events are sometimes misidentified as flasher events when the neutrino interaction vertex is close to the ID wall.
There have been a total of 8 events that have been rejected
by the flasher cut during all run periods. From event time information and visual inspections, it is clear that all eight events are induced by beam neutrino interactions. The predicted number of rejected beam events from this cut is 3.71 events; the 
probability to observe 8 or more events when 3.71 are expected is 3.6\%.
All eight events have vertices close to the ID wall, and would be rejected by the fiducial cut.

\begin{figure}
\centering
\includegraphics[width=0.4\textwidth]{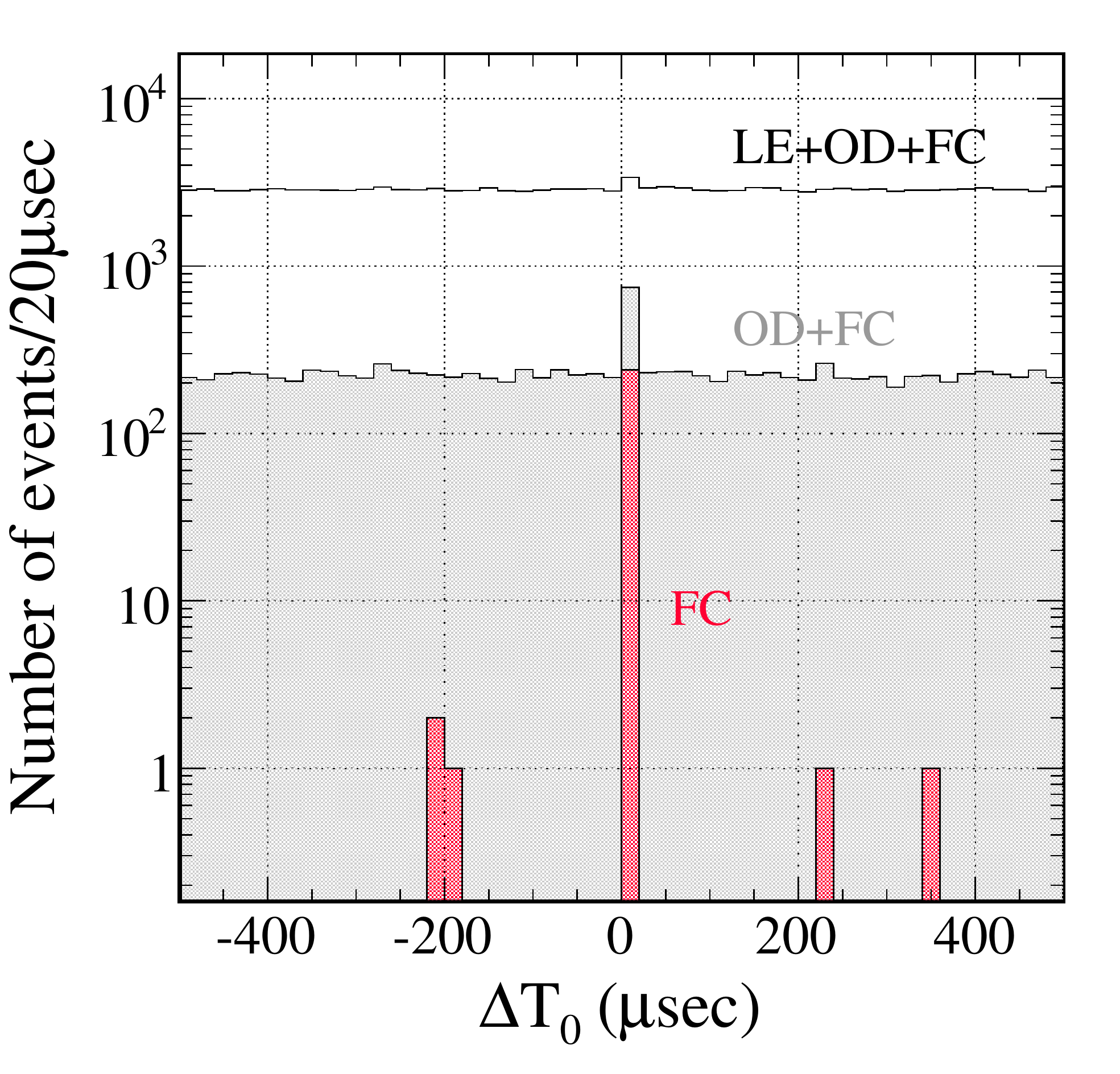}
\caption{
$\Delta T_0$ distribution of all FC, OD and LE events
observed in the $\pm500\,\mu$s T2K windows.
The OD histogram is stacked on the FC histogram,
and the LE histogram is stacked on the OD and FC histograms.
}
\label{fig:sk_dt0}
\end{figure}

\begin{figure}
\centering
\includegraphics[width=0.4\textwidth]{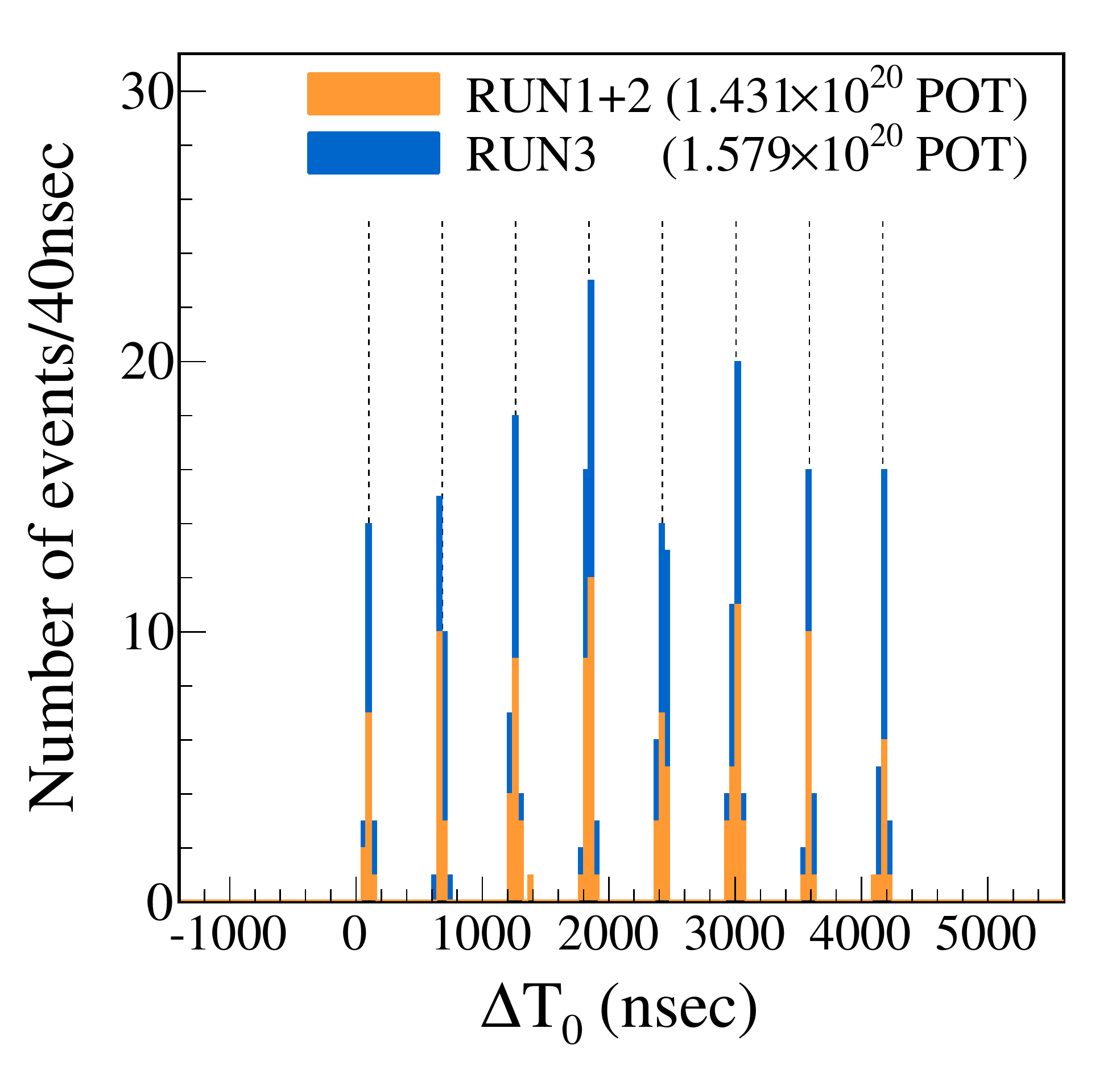}
\caption{
$\Delta T_0$ distribution of FC events zoomed in on the spill time
observed during T2K Run 1+2 and Run 3.
The eight dotted vertical lines represent the 581\,ns interval
bunch center position fitted to the observed FC event times.
}
\label{fig:sk_dt0fc}
\end{figure}

We define the quantity $\Delta T_0$, which is the timing of the event
relative to the leading edge of the spill, accounting for the travel time of the neutrino from production to detection. 
Fig.~\ref{fig:sk_dt0} shows the $\Delta T_0$ distribution of all
FC, OD and LE events within $\pm500\,\mu$s of the beam arrival time; the spill duration is about $5\,\mu$s.
A clear peak at $\Delta T_0=0$ is seen for the FC sample.
We observe five FC events outside of the $5\,\mu$s spill window.
The expected number of such out-of-time FC events,
mainly low energy events
and atmospheric neutrino events,
is estimated to be 3.3 from data collected when the beam is not present.
Fig.~\ref{fig:sk_dt0fc} shows the $\Delta T_0$ distribution of FC events within the spill window. 
We correct the $\Delta T_0$ of each event to account for the position of the neutrino interaction 
vertex and  the photon propagation time from the interaction vertex to the PMTs.
The far detector event timing clearly exhibits the eight bunch beam timing structure.
The eight dotted vertical lines in the figure represent the 8 bunch
centers at intervals of 581\,ns from a fit to the observed FC event timing.
The RMS value of the residual time distribution
between each FC event and the closest of the fitted bunch center times
is about 25\,ns.

We require the $\Delta T_0$ for selected FC events
to be between $-0.2$~$\mu$s to $10$~$\mu$s. We observe 240 such in-time
fully contained events.
We extract a fully contained sample within the fiducial volume (FCFV) 
by further requiring
$E_{\rm vis}$ to be above 30~MeV and
the reconstructed vertex be 2~m away from the ID wall.
We observe 174 such FCFV events, while the 
expected accidental contamination
from events unrelated to the beam,
mostly atmospheric neutrino interactions,
is calculated to be 0.005 events.

\begin{figure}
\centering
\includegraphics[width=0.4\textwidth]{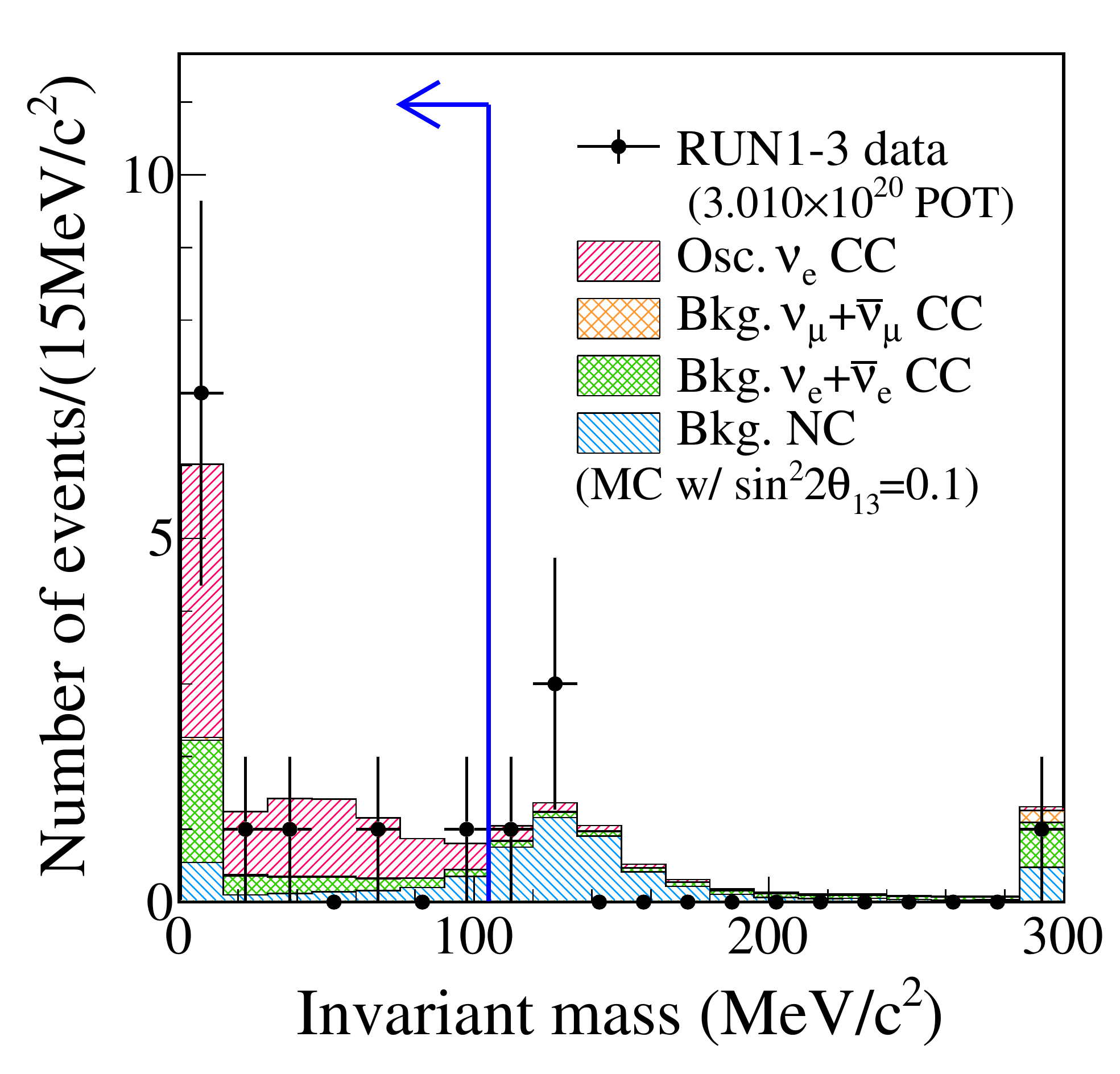}
\caption{
Distribution of invariant mass $M_{\rm inv}$
when each event is forced to be reconstructed as two photon rings.
The data are shown as points with error bars (statistical only)
and the MC predictions are in shaded histograms.
The last bin shows overflow entries.
The arrow shows the selection criterion $M_{\rm inv}<105$~MeV/$c^2$.
}
\label{fig:polfitm}
\end{figure}

\begin{figure}
\centering
\includegraphics[width=0.4\textwidth]{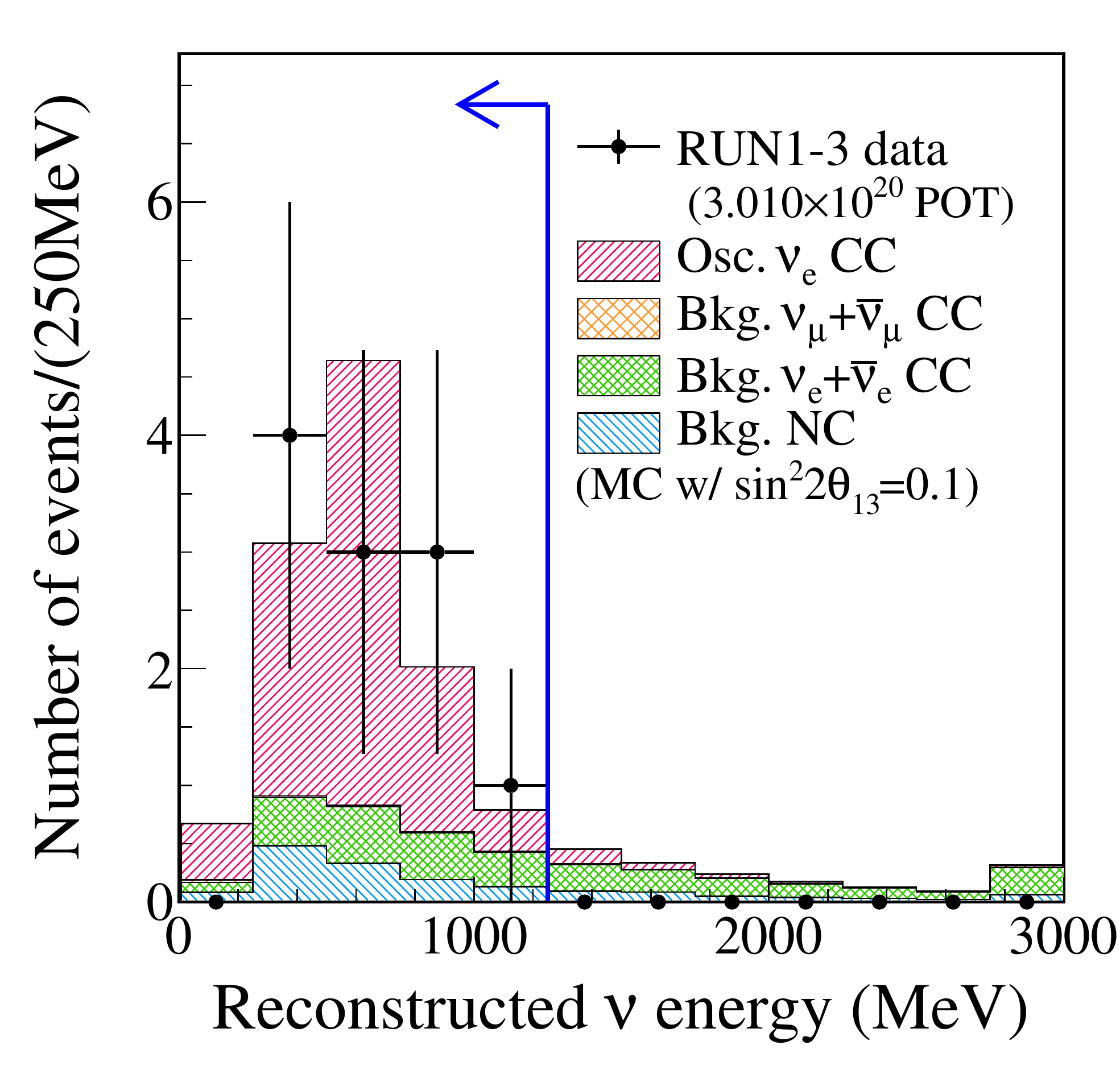}
\caption{
Distribution of the reconstructed
neutrino energy spectrum of the events which pass all
\nue appearance signal selection criteria with the exception
of the energy cut.
The data are shown as points with error bars (statistical only)
and the MC predictions are in shaded histograms.
The arrow shows the selection criterion $E_\nu^{\rm rec}<1250$~MeV.
}
\label{fig:enurec_cut}
\end{figure}

CC \nue interactions (\ccnue) are identified in SK by detecting a single, electron-like ring; at the energy of the T2K neutrino beam, most of the produced particles other than the electron are below Cherenkov threshold or do not exit the nucleus.
The main backgrounds are intrinsic \nue contamination in the beam
and NC interactions with a misidentified \piz.
The analysis relies on the well-established reconstruction techniques
developed for other data samples in SK~\cite{Ashie:2005ik}.
The single, electron-like ring selection criteria are unchanged from our previous measurement of
electron neutrino appearance~\cite{PhysRevLett.107.041801},
and were determined from MC studies before data-taking commenced.  We select CC \nue candidate events which satisfy the following criteria:
\begin{enumerate}
\item[(1)] The event is fully contained in the ID and the reconstructed vertex is within the fiducial volume (FCFV)
\item[(2)] There is only one reconstructed ring
\item[(3)] The ring is electron-like
\item[(4)] The visible energy, $E_{\rm vis}$, is greater than 100~MeV
\item[(5)] There is no Michel electron 
\item[(6)] The event's invariant mass is not consistent with a \piz mass 
\item[(7)] The reconstructed neutrino energy, $E_\nu^{\rm rec}$, is less than 1250~MeV
\end{enumerate}
The $E_{\rm vis}$ cut removes low energy NC interactions
and electrons from the decay of unseen muons and pions, such as cosmic muons outside the beam time window or muons below Cherenkov threshold.
A Michel electron is an electron from muon decay which is identified by looking 
for a time-delayed ID-PMT hit peak after the primary neutrino interaction.
In order to reduce NC \piz events, we utilize a special
fitter which reconstructs each event with a two photon ring
hypothesis. It searches for the direction and energy
of the second ring which maximizes the likelihood based on
the light pattern of the event~\cite{Barszczak:2005sf}.
Fig.~\ref{fig:polfitm} shows the invariant mass $M_{\rm inv}$
distribution of the two photon rings for the data and simulation.
As shown in the figure, the NC background component peaks
around the \piz invariant mass, hence events with $M_{\rm inv}>105$~MeV/$c^2$ are cut.
Finally, the energy of the parent neutrino is computed
assuming CCQE kinematics and neglecting Fermi motion as follows:
\begin{linenomath*}
\begin{equation}
\label{eqn:enurec}
E_\nu^{\rm rec}=\frac{m_p^2-(m_n-E_b)^2-m_e^2+2(m_n-E_b)E_e}
{2(m_n-E_b-E_e+p_e\cos\theta_e)},
\end{equation}
\end{linenomath*}
where $m_p$ is the proton mass, $m_n$ the neutron mass, and $E_b=27$~MeV
is the binding energy of a nucleon inside a $^{16}$O nucleus.
$E_e$, $p_e$, and $\theta_e$ are the reconstructed electron energy,
momentum, and angle with respect to the beam direction, respectively.
We select $E_\nu^{\rm rec}<1250$~MeV since the signal at high energy
is expected to be small for the atmospheric mass splitting, and the
intrinsic \nue background is dominant in this region, as shown in Fig.~\ref{fig:enurec_cut}.


\begin{table*}
\centering
\caption{
Event reduction for the \nue appearance search at the far detector.
After each selection criterion is applied, the numbers of observed 
and MC expected events of CC \num, intrinsic CC \nue, NC,
and the CC \nue signal, are given. All MC samples include three-neutrino oscillations for $\sin^22\theta_{13}=0.1$, $\delta_{CP}=0$,
and normal mass hierarchy.}
\begin{tabular}{lcccccc}
\hline \hline
& Data & MC total & CC \num & CC \nue & NC & CC $\num\to\nue$ \\
\hline
(0) interaction in FV           &  n/a  & 311.4  & 158.3  & 8.3 & 131.6 & 13.2 \\ 
(1) fully contained in FV       &  174  & 180.5  & 119.6  & 8.0 &  40.2 & 12.7 \\ 
(2) single ring                 &   88  &  95.7  &  68.4  & 5.1 &  11.4 & 10.8 \\ 
(3) $e$-like                    &   22  &  26.4  &   2.7  & 5.0 &   8.0 & 10.7 \\ 
(4) $E_{\rm vis}>100$~MeV      &   21  &  24.1  &   1.8  & 5.0 &   6.9 & 10.4 \\ 
(5) no delayed electron         &   16  &  19.3  &   0.3  & 4.0 &   5.9 &  9.1 \\ 
(6) not \piz-like               &   11  &  13.0 &   0.09 & 2.8 &   1.6 &  8.5 \\ 
(7) $E_\nu^{\rm rec}<1250$~MeV &   11  &  11.2 &   0.06 & 1.7 &   1.2 &  8.2 \\ 
\hline \hline
\end{tabular}
\label{tab:sknue0.1}
\end{table*}
\begin{table*}
\centering
\caption{
Same as Table~\ref{tab:sknue0.1} but with MC prediction for
$\sin^22\theta_{13}=0$.}
\begin{tabular}{lcccccc}
\hline \hline
& Data & MC total & CC \num & CC \nue & NC & CC $\num\to\nue$ \\
\hline
(0) interaction in FV           &  n/a  & 299.0  & 158.5  & 8.6 & 131.6 & 0.3 \\ 
(1) fully contained in FV          &  174  & 168.5  & 119.8  & 8.2 &  40.2 & 0.3 \\ 
(2) single ring                 &   88  &  85.4  &  68.5  & 5.3 &  11.4 & 0.2 \\ 
(3) $e$-like                    &   22  &  16.1  &   2.7  & 5.2 &   8.0 & 0.2 \\ 
(4) $E_{\rm vis}>100$~MeV      &   21  &  14.1  &   1.8  & 5.2 &   6.9 & 0.2 \\ 
(5) no delayed electron         &   16  &  10.6  &   0.3  & 4.2 &   5.9 & 0.2 \\ 
(6) not \piz-like               &   11  &   4.8 &   0.09 & 2.9 &   1.6 & 0.2 \\ 
(7) $E_\nu^{\rm rec}<1250$~MeV &   11  &   3.3 &   0.06 & 1.8 &   1.2 & 0.2 \\ 
\hline \hline
\end{tabular}
\label{tab:sknue0.0}
\end{table*}
The numbers of observed events after each selection criterion,
and the MC predictions for $\sin^22\theta_{13}=0.1$ and
$\sin^22\theta_{13}=0$, are shown
in Tables~\ref{tab:sknue0.1} and \ref{tab:sknue0.0}, respectively.
Eleven events remain in the data
after all \nue appearance signal selection criteria are applied.
Using the MC simulation, we estimate the \nue appearance signal efficiency in the SK FV to 
be 62\%, while the rejection rates for CC \num+\numb,
intrinsic CC \nue+\nueb, and NC are
$>99.9$\%, 80\%, and 99\%, respectively.
More than half  of the remaining background
is due to intrinsic CC \nue interactions (57\% for $\sin^22\theta_{13}=0.1$).
The fraction of CCQE events in the CC \nue signal and background 
are 80\% and 65\%, respectively.
NC interactions constitute
41\% of the total surviving background, 80\% of which are due to \piz mesons and 6\% of which originate from NC single photon ($\Delta \rightarrow N\gamma$) production.

\begin{figure}
\centering
\includegraphics[width=0.48\textwidth]{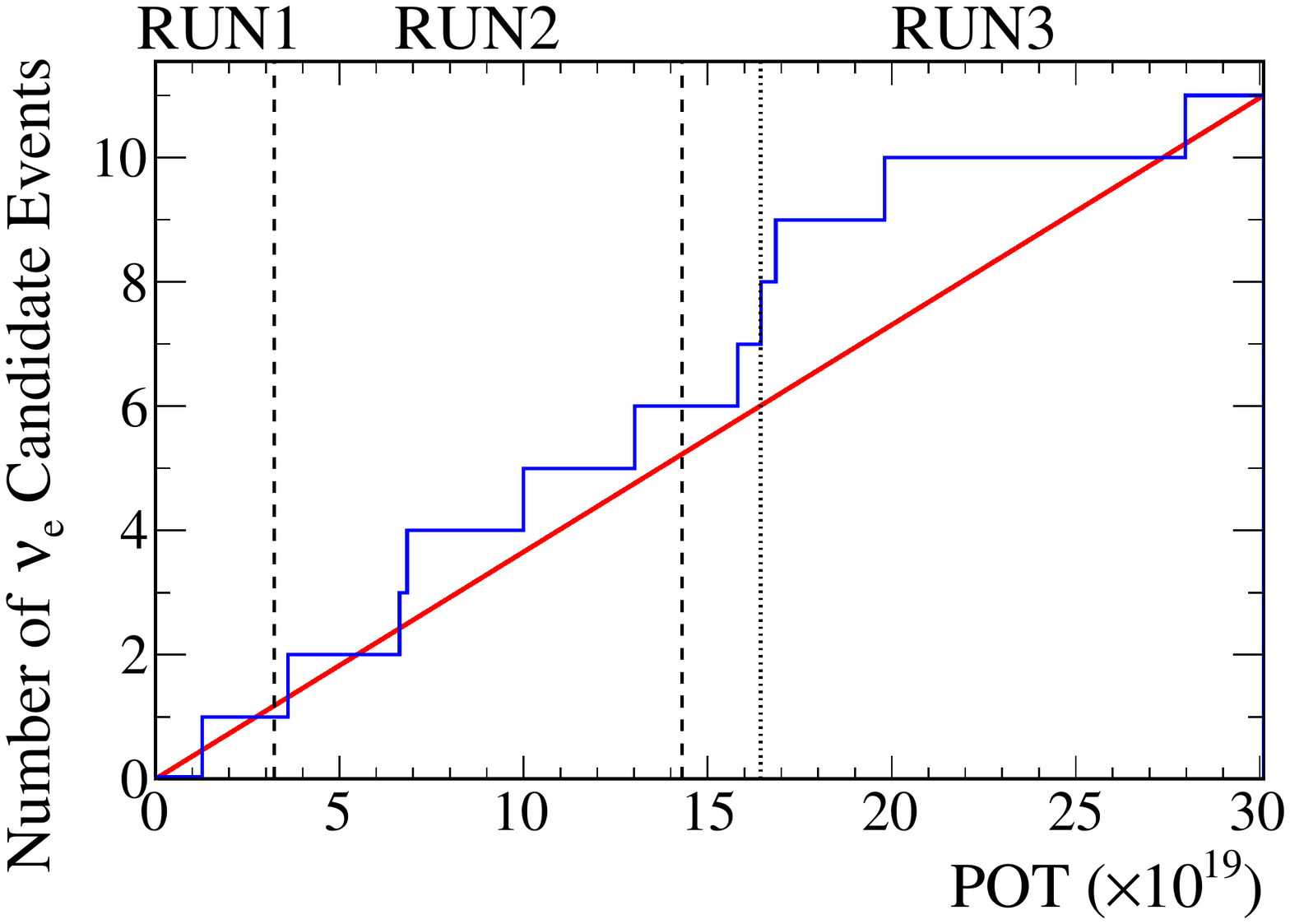}
\caption{
The cumulative number of observed \nue candidate events
as a function of accumulated POT.
The vertical dashed lines separate the three running periods,
and the dotted line indicates the horn current change during Run 3.
The solid line indicates a hypothesis of constant event rate.
}
\label{fig:nue_kstest}
\end{figure}

\begin{figure}
\includegraphics[width=0.4\textwidth]{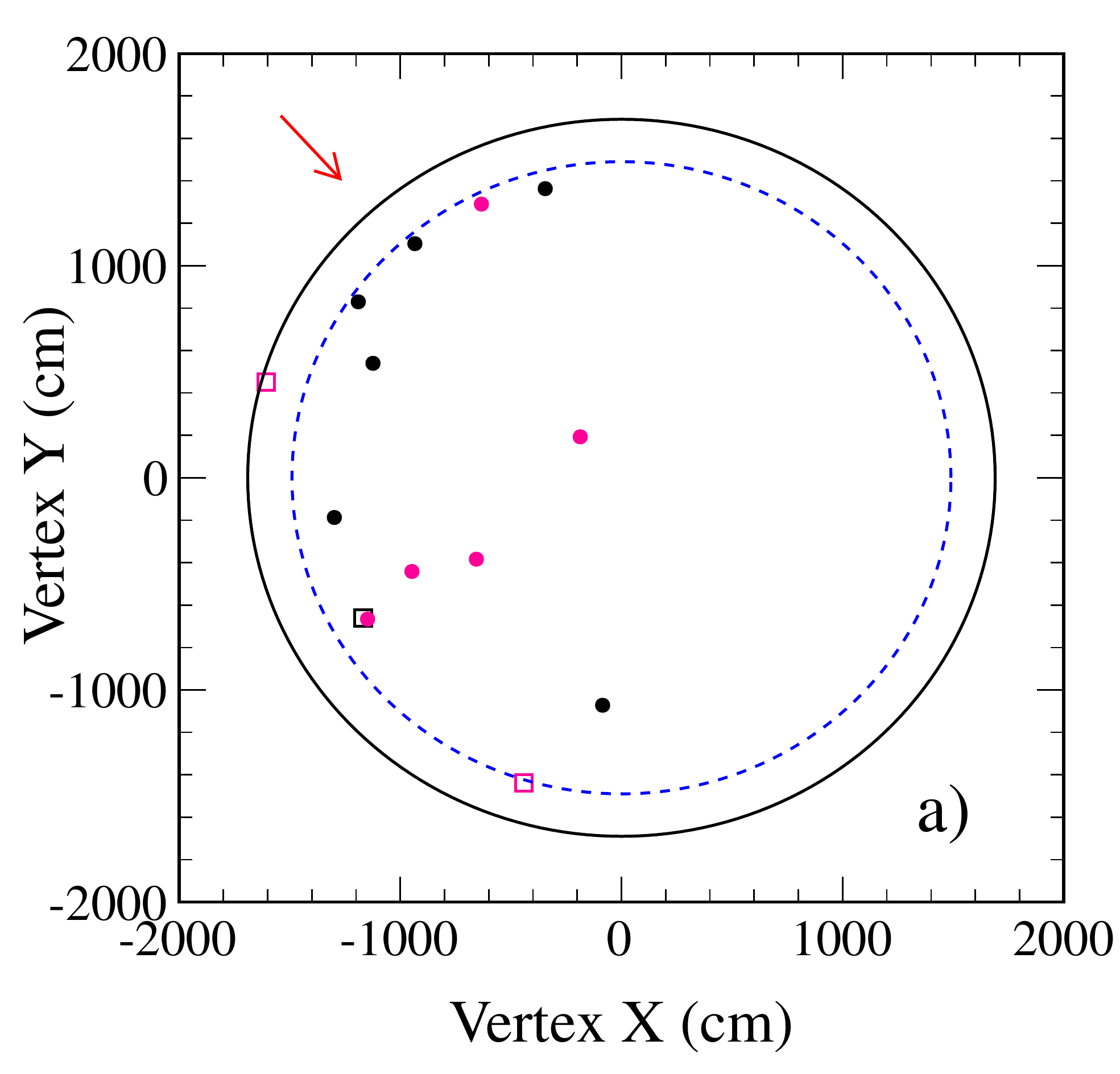} \\
\includegraphics[width=0.4\textwidth]{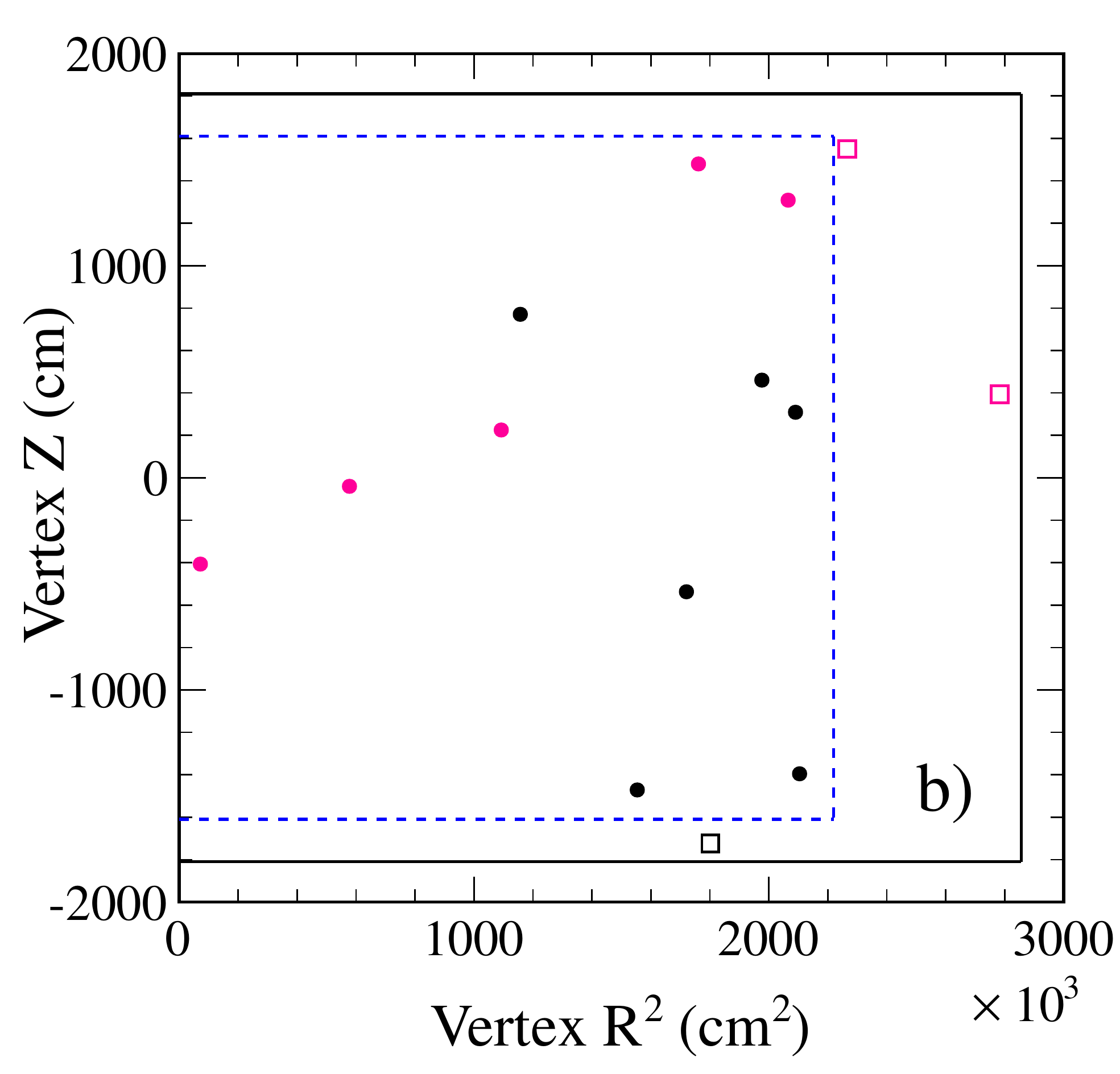} 
\caption{
a) Two-dimensional $(x,y)$ distribution of the reconstructed vertex positions
of observed \nue candidate events.
b) Two-dimensional $(r^2=x^2+y^2,z)$ distribution of the reconstructed vertex positions of the observed \nue candidate events.
The arrow indicates the neutrino beam direction and the
dashed line indicates the fiducial volume boundary.
Black markers are events observed during Run 1+2,
and pink markers are events from Run 3.
Open squares represent events which passed all the \nue selection cuts except for the fiducial volume cut.
}
\label{fig:nue_vtxxy}
\end{figure}

\begin{figure}
\centering
\includegraphics[trim=1cm 0cm 1cm 0cm, clip=true, width=0.45\textwidth]{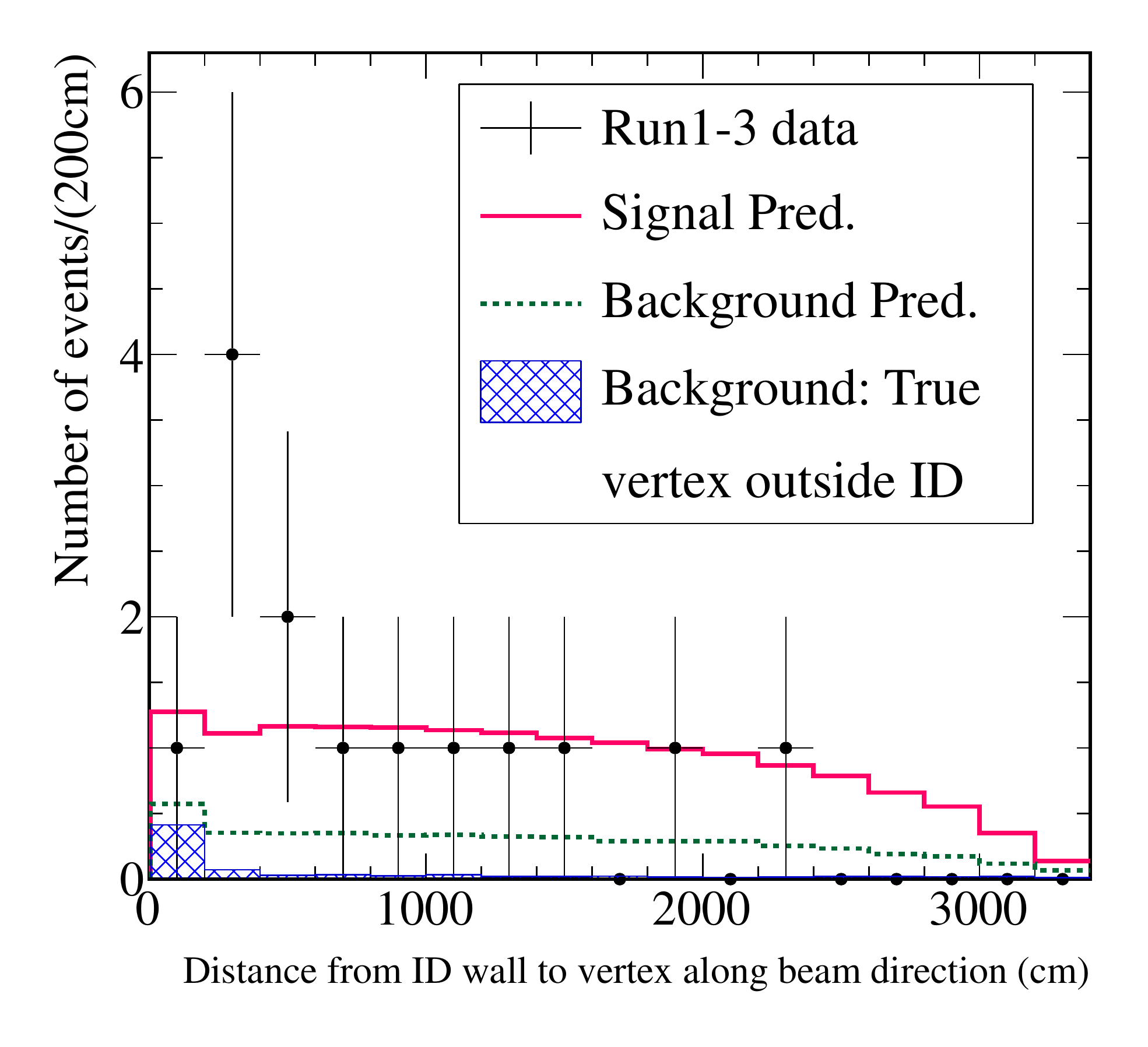}
\caption{
Distribution of events which pass \nue selection cuts except for the FV cut as a function of the distance from the ID wall 
to the vertex, calculated along the beam direction.  The solid line indicates the expected distribution for signal (\stot=0.1) and background, and the background prediction is shown with the dashed line. The hatch-filled histogram shows the subset of background whose true vertex is outside the ID.
}
\label{fig:nue_entering}
\end{figure}

Additional checks of the eleven data events are performed.
From visual inspection, it appears that all events have only
a single, electron-like Cherenkov ring.
A KS test of the observed number of
\nue candidate events as a function of accumulated POT
is compatible with the normalized event rate being constant
($p$-value = 0.48) as shown in Fig.~\ref{fig:nue_kstest}.
Fig.~\ref{fig:nue_vtxxy} shows the $(x,y)$ and $(r^2,z)$ distributions of
the reconstructed vertices of observed \nue candidate events.
As we previously reported, the first 6 candidate events were clustered near the edge of the FV in 
the upstream beam direction.
We observe no such clustering in the newly observed 5 events (pink points
in the figure). All event vertices are $x<0$ in the SK coordinate system which is not related to the beam direction. Other T2K neutrino selections with larger event samples, such as the CC \num selection, populate the entire $x$ and $y$ region.
Figure~\ref{fig:nue_entering} shows the distribution of distance from the ID wall to the vertex along the beam direction for events passing all \nue selection cuts except the FV cut. 
A KS test to this distribution yields a $p$-value of 0.06. 
In addition, a dedicated selection of penetrating particles produced in upstream, out-of-FV neutrino interactions shows no indication of an excess.


\subsection{SK efficiency and reconstruction uncertainties}

We have studied the systematic uncertainties on the simulation of the SK event selection efficiency and reconstruction  using comparisons of data and MC control samples.
The error on the FC event selection is estimated to be 1\%, with
a dominant contribution from the flasher event rejection.
We evaluate the flasher rejection uncertainty from the difference in
the cut efficiency between the atmospheric neutrino data and MC simulation.
We estimate the uncertainty on the fiducial volume definition 
to be 1\% by comparing the reconstructed
vertex distributions of observed and simulated cosmic-ray muons
which have been independently determined to have stopped inside the ID.
We estimate an energy scale uncertainty of 2.3\% from comparisons of distributions between 
cosmic-ray data and simulated samples.  These samples include
the reconstructed momentum spectrum of electrons from the decay of cosmic ray 
muons, cosmic-ray muons which stop in SK and have similar energies to the T2K neutrino events, and 
the reconstructed mass of neutral pions from atmospheric neutrino interactions.
The error on the number of \nue candidate events due to
the uncertainty on the delayed, decay-electron tagging efficiency
is 0.2\%. We evaluate this uncertainty from a comparison of
the tagging efficiency between cosmic-ray stopped muon data and MC samples.

The remaining uncertainties on the detection efficiency are evaluated in categories corresponding to the 
particles exiting the target nucleus.  The ``CC \nue single electron'' category is comprised of interactions 
where a single electron is emitted and is the only detectable particle in the final state. 
The ``CC \nue other'' category includes  all other CC \nue interactions not in the CC \nue single electron category.
NC events are also classified  based on the particle type which exits the nucleus. 
The ``NC single \piz'' category includes events with only one \piz in the detector.

The topological light pattern of the rings provides the information 
needed to construct quantities used in the selection: the number of
rings (cut 2), particle identification (cut 3) and the invariant mass (cut 6). We evaluate the systematic error on the efficiency of each of the three topological cuts 
on the selection  with a fit to SK atmospheric neutrino data using MC simulation-based templates.  We create two control samples in the SK atmospheric neutrino data set which are sensitive to CC \nue single electron and CC \nue other event types. The \nue enriched control samples pass the FCFV, $E_{\rm vis}>100$~MeV criteria; however the number of decay electrons in the event is used to separate QE-like (single ring) from nonQE-like (multiple rings) instead of the ring-counting algorithm.
Each control sample is divided into one ``core'' sub-sample, which passes the three topological cuts, and three ``tail'' sub-samples, where events have failed one of the three topological cuts.
The sub-samples are further divided into 17 bins (labeled with index $i$) in $p_e$ and $\theta_e$, the reconstructed electron momentum and angle with respect to the beam direction, so that we can evaluate the dependence of the systematic errors on these kinematic variables.
The expected number of events in all sub-samples depends 
on the efficiency of each topological cut, $\vec{\epsilon}=\{\epsilon_{1ring},\epsilon_{PID},\epsilon_{inv.mass} \}$, and parameters which represent systematic uncertainties 
on the event rate, $\vec{\alpha}$. 
The $\vec{\alpha}$ parameters include uncertainties on
the atmospheric neutrino flux normalization,
the absolute cross section of CC non-QE and NC interactions,
the \nue/\num relative cross section,
and the energy dependence of the CCQE cross section.
We perform a $\chi^2$ fit to the atmospheric control samples, allowing the $\vec{\epsilon}$ and $\vec{\alpha}$ parameters to vary.

We extract the uncertainties on the CC \nue single electron and CC \nue other event categories 
based on the effect of the selection cuts on the efficiency $\vec{\epsilon}$ within the fit to the control samples. 
We estimate the bias as the difference between the fitted value and the nominal value of the event rate for two categories (CC \nue single electron and CC \nue other) over  17 reconstructed \pthetae bins.  The correlations between bins are considered.
We also include uncertainties on the event categories determined from the
fit; the fit uncertainties are treated as uncorrelated between bins.
For the CC \nue single electron category, the bias is estimated to be 1-9\% across all bins,
while the fit uncertainty is 4-8\% across all bins.
The bias and fit uncertainty for the CC \nue other category 
are 27\% and 14\%, respectively; this component is a small contribution to the signal and background prediction, and so the momentum and angular dependence of the uncertainty is ignored.
As described later, we use these errors and their correlations
as inputs for deriving the total SK systematic error
on the T2K \nue appearance candidate events.

NC interactions producing a single exclusive photon 
via radiative decays of $\Delta$ resonances (NC1$\gamma$) are a background
to the \nue appearance signal, as the photon ring is very similar to an electron ring.
We evaluated the difference in the selection efficiency between the single photon MC sample and the single electron MC sample to estimate the uncertainty on the selection efficiency of  NC1$\gamma$ events.
The difference in relative efficiencies is no larger than 1\%, so 
we assign an additional 1\% uncertainty, added in quadrature to the uncertainty on single electron rings estimated from the CC \nue single electron sample efficiency, as the uncertainty on the selection efficiency for NC1$\gamma$ background events.

We evaluate the systematic uncertainty for events where the muon decays in flight with a MC study.
The Cherenkov ring of the electron from a muon which decays in flight tends to be in the same direction as the parent muon, and therefore these events look similar to CC \nue interactions.
We estimate the uncertainty on the expected number of of muon-decay-in-flight background events to be 16\%, with the largest contribution from the uncertainty on the muon polarization.
The fraction of muons which decay in flight in the selected \nue candidate event sample is estimated to be smaller than 1\%, and so this uncertainty does not contribute substantially to the total uncertainty on the \nue candidates.

The efficiency of NC$1\pi^{0}$ events for the $\nu_e$ selection criteria is
determined to be 6\% from the MC simulation.
To evaluate the systematic uncertainty for events with a 
\piz in the final state, we construct ``hybrid-\piz'' control samples.
The ``hybrid-\piz'' samples contain events where a \piz is constructed using one simulated photon ring and a second electron-like ring from the SK atmospheric or cosmic-ray samples.
The simulated photon ring kinematics are chosen such that the two rings
follow the decay kinematics of a \piz.  The hybrid samples are
constructed with electron rings from data (hybrid-\piz data) and the simulation (hybrid-\piz MC), 
and the comparison of the two is used to evaluate the systematic uncertainties.

We investigate the systematic error coming from the higher-energy ring
and the lower-energy ring separately.
The ``primary'' sample uses
electron rings from the SK atmospheric samples, with the electron ring having higher energy than
the simulated photon ring.  In the ``secondary'' sample the electron ring has a lower energy 
than the photon ring.  Below 60~MeV, electrons from cosmic-ray muons are used; otherwise the
electrons from the SK atmospheric samples are used.

We compare the efficiency of the \nue selection criteria on \piz events
in the hybrid-\piz data and hybrid-\piz MC samples in each of the
 17 ($p_e$,$\theta_e$) bins.
We apply the efficiency differences as correlated systematic errors
among bins, while the statistical errors on the efficiency differences
are applied as uncorrelated systematic errors.
For the NC single \piz component,
we estimate correlated errors in each ($p_e$,$\theta_e$) bin
to be between 2-60\%,
and uncorrelated errors are between 15-50\%. 
The assigned errors are larger in the lower momentum bins, where the \piz selection efficiency is lower. 
We evaluate the systematic uncertainties on events with one or more charged particles above Cherenkov threshold and a \piz by using hybrid-\piz control samples 
with additional simulated rings for the extra particles.

\begin{figure}
\begin{center}
\includegraphics[width=0.50\textwidth]{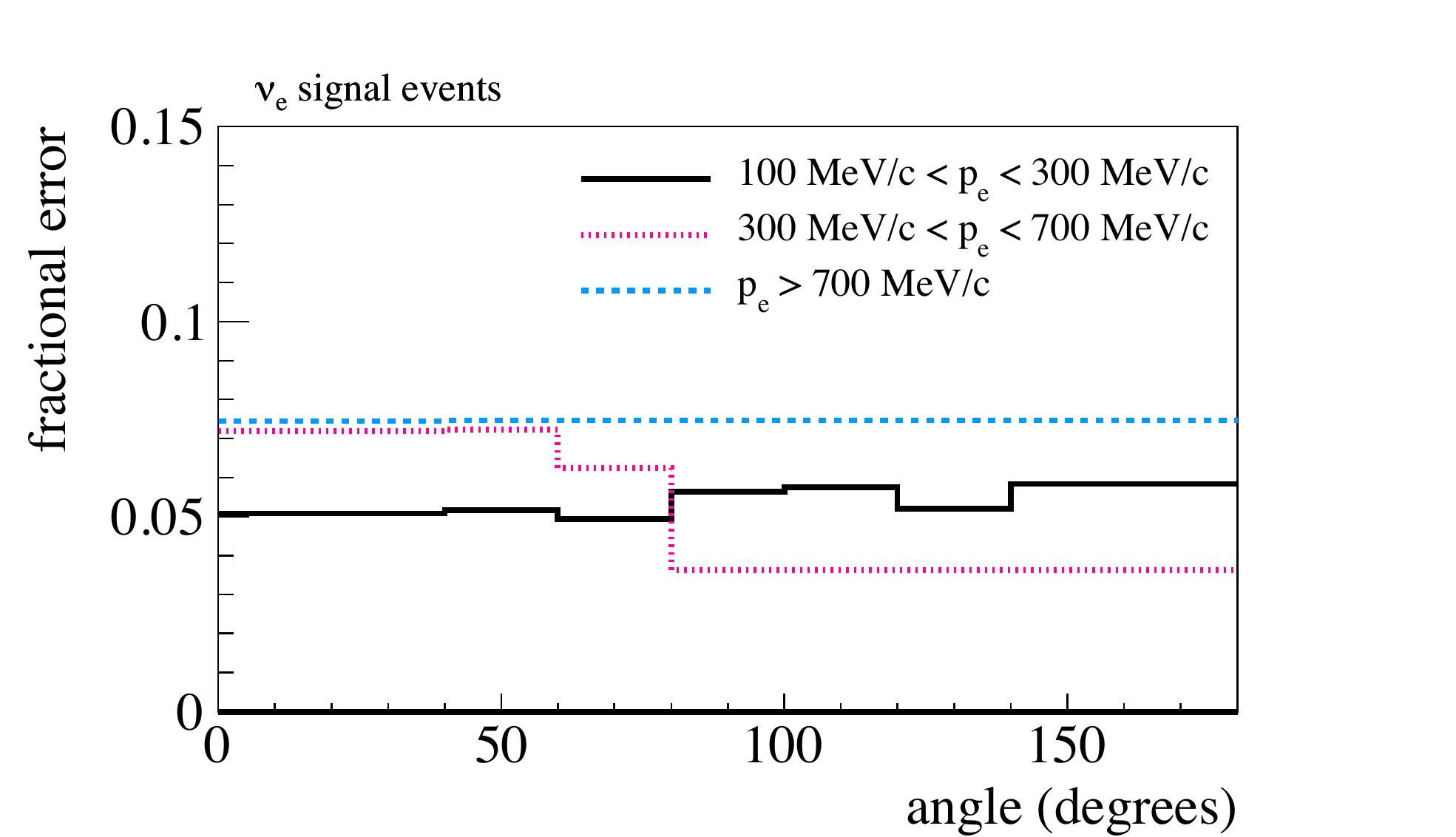}
\includegraphics[width=0.50\textwidth]{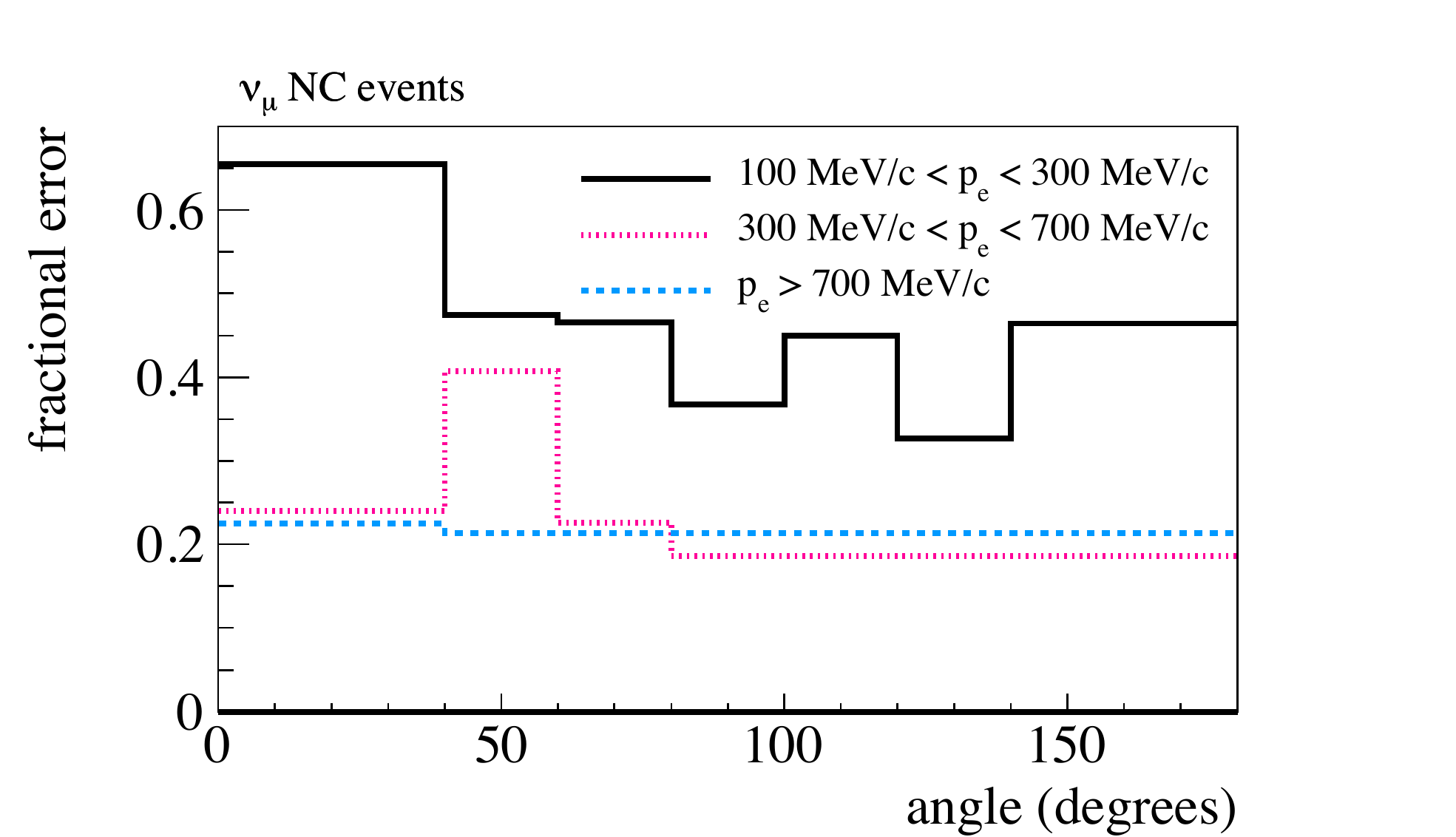}
\end{center}
\caption{The fractional errors on SK $\nu_e$ signal (top) and NC background (bottom) predictions from the SK detector
response uncertainty as a function of the electron candidate momentum and angle.
}
\label{fig:sk_frac_error}
\end{figure}

Finally, we combine all systematic uncertainties on the \nue appearance
signal selection at SK into a single covariance matrix.
The covariance matrix has bins in the observable kinematic variables, ($p_e$,$\theta_e$) or $E_{\nu}^{rec}$,
for the four event categories: signal CC \nue, background CC \num, CC \nue, and NC.
We use this covariance matrix to model the systematic uncertainties on the simulated detector efficiency
and reconstruction in the oscillation fits described in Section~\ref{sec:sk_fit_method}.
The fractional errors as a function of the both the electron momentum and angle are shown in 
Fig.~\ref{fig:sk_frac_error}.

\section{\label{sec:sk_fit_method} Oscillation Fit Method and Results}

The \nue appearance oscillation signal is an excess of \nue candidates over background. Table~\ref{tab:sknue0.1} and Table~\ref{tab:sknue0.0} show the predicted number of \nue candidate events after we apply the tuned neutrino flux and cross section parameters discussed in Sec.~\ref{sec:extrapolation}. If \stot=0.1, we expect 11.2 events, and if \stot=0, we  expect 3.3. We evaluate the systematic uncertainties on the expected signal and background event rates due to the uncertainties on the flux model, neutrino interaction cross section model and SK reconstruction efficiencies, as summarized in Section~\ref{sec:nuepred}.

The probability to observe 11 or more events based on the predicted
background of $3.3\pm0.4$~(syst.) events is  $9\times 10^{-4}$,   
equivalent to an exclusion significance of $3.1\sigma$. This rate-only hypothesis test
makes no assumptions about the energy spectrum of the candidate events
or their consistency with the neutrino oscillation hypothesis; it is a
statement that we observe an excess of electron-like events over background. The background model includes expected \num $\rightarrow$ \nue oscillation through the solar term shown in Eq.~\ref{eq:subleading}, which corresponds to 0.2 events. 
The reported $p$-value corresponds to the probability to observe 11 or more events from background 
sources and oscillations that depend on the $\theta_{12}$ mixing angle.  If instead we consider the 
probability to observe 11 or more events from background sources only, the $p$-value is $6\times10^{-4}$.

\begin{figure*}[tb]
\begin{center}
\includegraphics[width=0.8\textwidth]{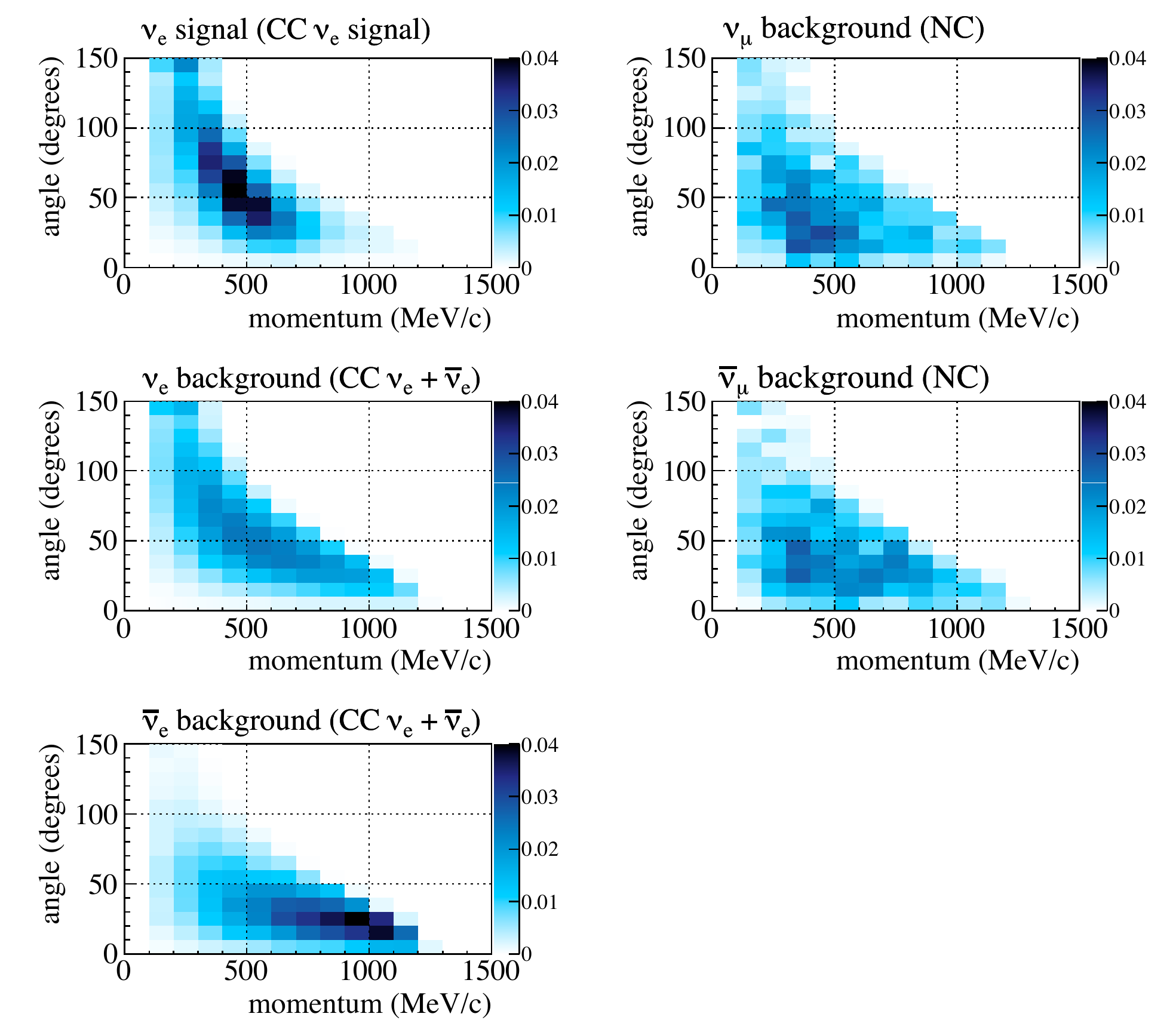}
\end{center}
\caption{ \pthetae distribution for $\nue$ signal (top left), $\num$ background (top right),  $\nue$ background (middle left), $\numb$ background (middle right) and $\nueb$ background (bottom left).  
Each distribution is normalized to unit area. 
}
\label{fig:ptheta_comp}
\end{figure*}

\begin{figure}
\centering
\includegraphics[width=0.5\textwidth]{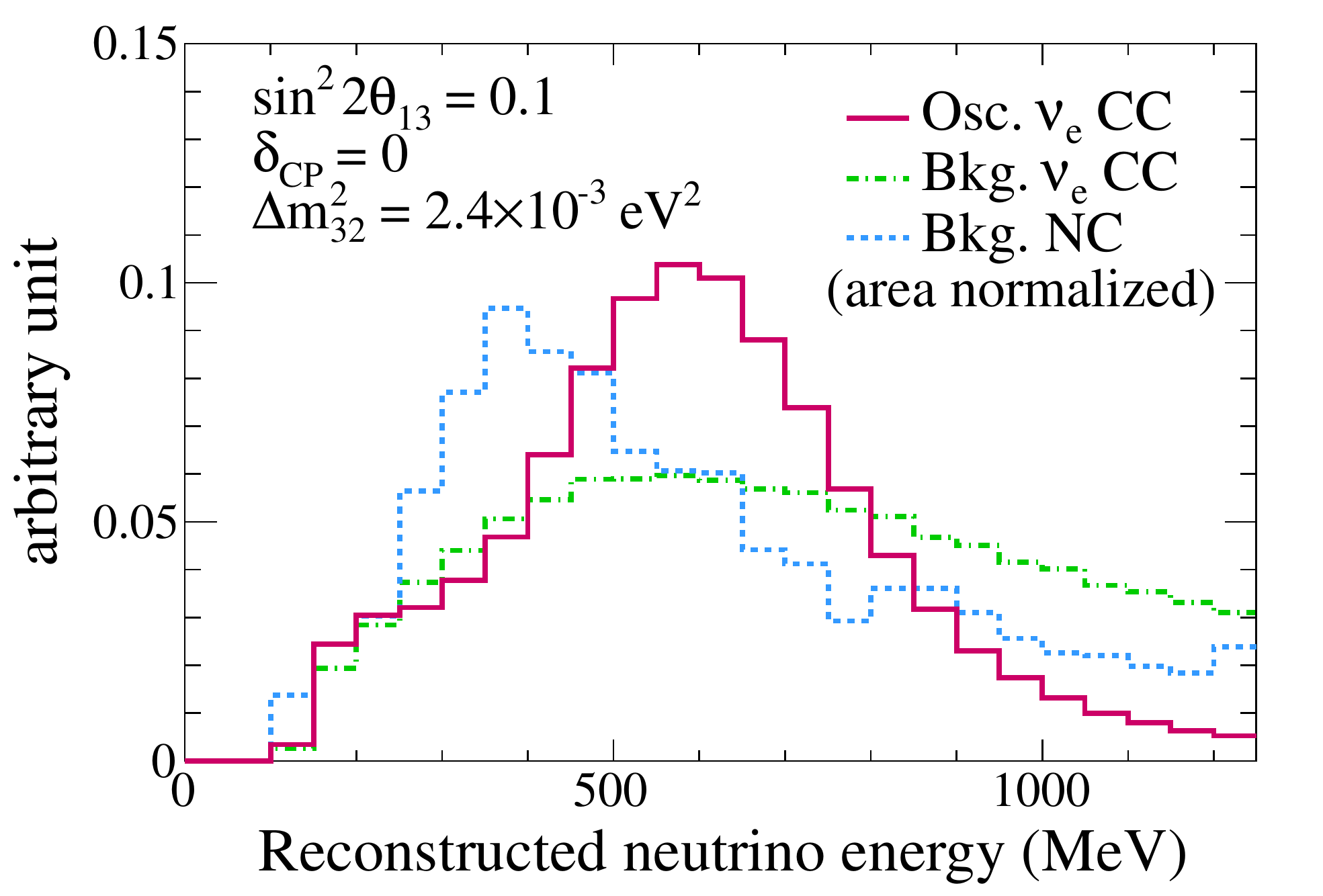}
\caption{
The MC reconstructed neutrino energy distributions of
$\num\to\nue$ CC signal, intrinsic \nue CC background and NC
background components in the \nue candidate event sample.
The histograms are normalized to the same area. 
}
\label{fig:spec_comp}
\end{figure}

We fit the \nue candidate sample in the three-neutrino mixing paradigm to estimate \stot.
The dominant effect of a non-zero \stot~is to increase the overall rate of \nue events. 
However, spectral information,  {\it e.g.} 
electron momentum and angle with respect to the T2K beam direction, \pthetae, or reconstructed neutrino energy, \erec, 
can be used to 
further separate the signal from background.
Fig.~\ref{fig:ptheta_comp} shows the area-normalized \pthetae distribution for the $\nue$ candidate events predicted by the SK simulation. The signal CC \nue are predominantly CCQE, and peaked at $E_\nu \approx 0.6$~GeV, near the first oscillation maximum and neutrino flux peak. This results in a clear kinematic correlation across the \pthetae distribution for signal events.
This peak is also visible in the \erec distribution for signal events, shown in Fig.~\ref{fig:spec_comp}. Conversely, the backgrounds to the \nue signal populate a wider range of kinematic space. 
The NC backgrounds are predominantly photons misidentified as electron neutrino candidates, when one photon from \piz decay is not reconstructed, or when the two photons are co-linear. 
This background predominantly populates the low momentum and forward angle region as well as the signal region.  
The intrinsic beam $\nue$ ($\nueb$) backgrounds have a larger contribution of events at higher energy than the oscillated \nue, and so more often produce electrons with high momentum in the forward direction.

We find that based on studies of the \erec and \pthetae kinematic distributions that the
\pthetae distribution has the best power to discriminate signal and background with the minimal cross section model dependence, hence we perform a two-dimensional extended maximum likelihood fit to the
\pthetae data distribution. 
Section~\ref{sec:pthetafit} describes the \pthetae likelihood fit to estimate \stot, and Section~\ref{sec:altanalysis} describes two additional fits using \erec and rate-only information for comparison to the \pthetae fit.

\subsection{\nue predicted event rate and systematic uncertainties}
\label{sec:nuepred}

The predicted number of \nue candidates and the event shape distribution depend upon the flux, cross section parameters, oscillation probability, and the efficiency and resolution of the SK detector.
We calculate the predicted number of events in a given momentum and angular bin ($i$) as
\begin{linenomath*}
\begin{eqnarray}
\lefteqn{N^{p}_{i}(\vec{o},\vec{f}) \qquad} \nonumber \\ 
& = & \sum_{j}^{flux~type}
\bigg[\sum_{k=1}^{E_{\nu}bins} \; b_{j,k} 
\cdot  \Big\{\sum_{l=1}^{Int.modes} \; P_{j,k,l}^{osc}(\vec{o}) \nonumber \\
& & \times x^{norm}_{k,l}w_{i,j,k,l}(\vec{x}) 
\cdot d_{i,j,k}  \cdot  T_{i,j,k,l}^{p}
\Big\} \bigg]. \nonumber \\
\; \label{eq:nexp_ptheta}
\end{eqnarray}
\end{linenomath*}
Here, 
$T_{i,j,k,l}^{p}$ 
are the nominal Monte Carlo templates  that predict the event rate as a function of:
\begin{itemize}
\item momentum/angular bins ($i$). The momentum bins are 100~MeV/$c$ wide from 0~MeV/$c$ to 1500~MeV/$c$ (15 in total), and the angular bins are 10$^{\circ}$ wide from 0$^{\circ}$ to 140$^{\circ}$ with one bin for $\te>140^{\circ}$ (15 in total). The bins are ordered by increasing $\te$ in groups of increasing momentum.
\item flux type ($j$) with categories for $\nue$ signal, $\num$ background, $\nue$ background, $\numb$ background and $\nueb$ background.
\item true neutrino energy ($k$) with 200 bins (50~MeV wide) from 0~GeV to 10~GeV and one bin from 10~GeV to 30~GeV.
\item interaction mode ($l$) with categories for CCQE, CC1$\pi$, CC coherent, CC other, NC1$\pi^0$, NC coherent and NC other.
\end{itemize}
The systematic parameters are $\vec{f}=(b_{j,k},x_{k,l}^{norm},\vec{x},d_{i,j,k},f^{s})$.
The $b_{j,k}$ vary the flux normalization, and the $x_{k,l}^{norm}$ are cross section normalization parameters.
The $\vec{x}$ are cross section parameters such as $M_{A}^{QE}$ and $p_{F}$ where the effect on the prediction
is modeled with response functions,
$w_{i,j,k,l}$, evaluated for each combination of observable bin, flux type, neutrino energy bin and interaction mode.
The $d_{i,j,k}$ are systematic parameters that vary the normalization of the prediction for each
combination of observable bin, flux type and interaction mode.  These parameters are used to model variations 
due to final state interactions (FSI) and SK efficiency uncertainties.  The momentum scale variation according to the parameter $f^s$
is not shown in Eq.~\ref{eq:nexp_ptheta}.  The parameter $f^s$ scales the momentum range of the bins and the bin
contents are recalculated assuming a flat momentum dependence in each bin.

We compute three-neutrino oscillation probabilities, $P_{k,l,m}^{osc}(\vec{o})$, which include matter effects, according to the numerical technique  defined in Ref~\cite{Barger:1980tf}, for a given set of the oscillation parameters, $\vec{o}$. 
The \dcp~dependence is evaluated
by scanning the value of \dcp~and fitting for \stot~with \dcp~fixed at each scan point.  The remaining oscillation
parameters are always held fixed to the values listed in Table~\ref{tab:oscparam}.

Based on Eq.~\ref{eq:nexp_ptheta},  we predict both the total 
number of events and the normalized \pthetae shape distribution 
(probability density function, PDF).
The predicted number of events and the predicted \pthetae distribution are used 
in the likelihood function of the oscillation fit. 
The effect of the systematic uncertainties on the predicted number of events and  \pthetae PDF are studied by recalculating the rate and PDF under variations of the systematic parameters according to the prior probability distribution of the parameters. Table~\ref{tbl:nexp_error_summary} summarizes the uncertainty on the predicted number of events for each systematic error source
assuming \stot=0 and \stot=0.1.  

Uncertainties related to the nuclear model are applied independently for the
SK prediction and are not constrained by the fit to ND280 data since the primary target nuclei are different in
the ND280 ($ ^{12}C$) and SK ($ ^{16}O$) detectors.  These uncertainties include: the nuclear model uncertainty 
($x_{SF}$), the uncertainty on the Fermi momentum in the relativistic Fermi gas model ($p_{F}$), the uncertainty
on the $N\pi$ invariant mass for resonant production in the nuclear medium ($W_{\textrm{eff}}$), the uncertainty on the
rate of non-pionic decays of $\Delta$ resonances in the nuclear medium ($x_{\pi-less}$), and uncertainties on 
the final state 
interactions of pions in the nucleus.  The nuclear model related uncertainties contribute errors on the 
event rate prediction of 4.8\% for \stot=0 and 7.0\% for \stot=0.1.

The uncertainty on background only predicted number of events (\stot=0) is larger than that of signal+background due to the larger uncertainties on the NC backgrounds (32\%); the uncertainty on CC background events (14\%) is comparable to that of the CC signal events.
The inclusion of the ND280 measurements reduces the uncertainty on the total predicted event rate due to the flux and CCQE, \ccpip cross section model from 18.3\% to 8.5\% (22.6\% to 5.0\%), assuming \stot=0. (\stot=0.1). 
The far detector efficiency uncertainty has been reduced from 14.7\% (9.4\%) in the previous analysis~\cite{PhysRevLett.107.041801} 
to 6.8\% (3.0\%) assuming \stot=0.0 (\stot=0.1) due to new CC \nue and $\pi^0$ SK atmospheric control samples; 
the FSI uncertainty has also been reduced from 10.1\% (5.4\%) in the previous results 
to 2.9\% (2.3\%) in this analysis, as correlations between reconstructed bins are now taken into account (Sec.~\ref{sec:neut_fsi_tune}).

\begin{table}[tb]
\begin{center}
\caption{%
Summary of the contributions to the total uncertainty on the predicted number of events, assuming \stot=0 and \stot=0.1, separated by sources of systematic uncertainty. Each error is given in units of percent.
}%
\label{tbl:nexp_error_summary}
\begin{tabular}{lcc}
\hline
                              & \multicolumn{2}{c}{\stot$=$} \\
Error source                  & 0 & 0.1 \\
\hline
Beam flux \& $\nu$ int. (ND280 meas.)         & 8.5   & 5.0 \\
$\nu$ int. (from other exp.)                  &       & \\
\phantom{M}$x_{CCother}$                      & 0.2   & 0.1 \\
\phantom{M}$x_{SF}$                           & 3.3   & 5.7 \\
\phantom{M}$p_F$                              & 0.3   & 0.0 \\
\phantom{M}$x^{CCcoh}$                        & 0.2   & 0.2 \\
\phantom{M}$x^{NCcoh}$                        & 2.0   & 0.6 \\
\phantom{M}$x^{NCother}$                      & 2.6   & 0.8 \\
\phantom{M}$x_{\nue/\num}$                    & 1.8   & 2.6 \\
\phantom{M}$W_{\textrm{eff}}$                          & 1.9   & 0.8 \\
\phantom{M}$x_{\pi-less}$                     & 0.5   & 3.2 \\
\phantom{M}$x_{1\pi E_{\nu}}$                 & 2.4   & 2.0 \\
Final state interactions                       & 2.9   & 2.3 \\ 
Far detector                                  & 6.8   & 3.0 \\
\hline
Total                                         & 13.0  & 9.9 \\
\hline
\end{tabular}

\end{center}
\end{table}

\begin{figure}[htb]
\begin{center}
\begin{minipage}{0.5\textwidth}
 \includegraphics[width=1.0\textwidth,clip]{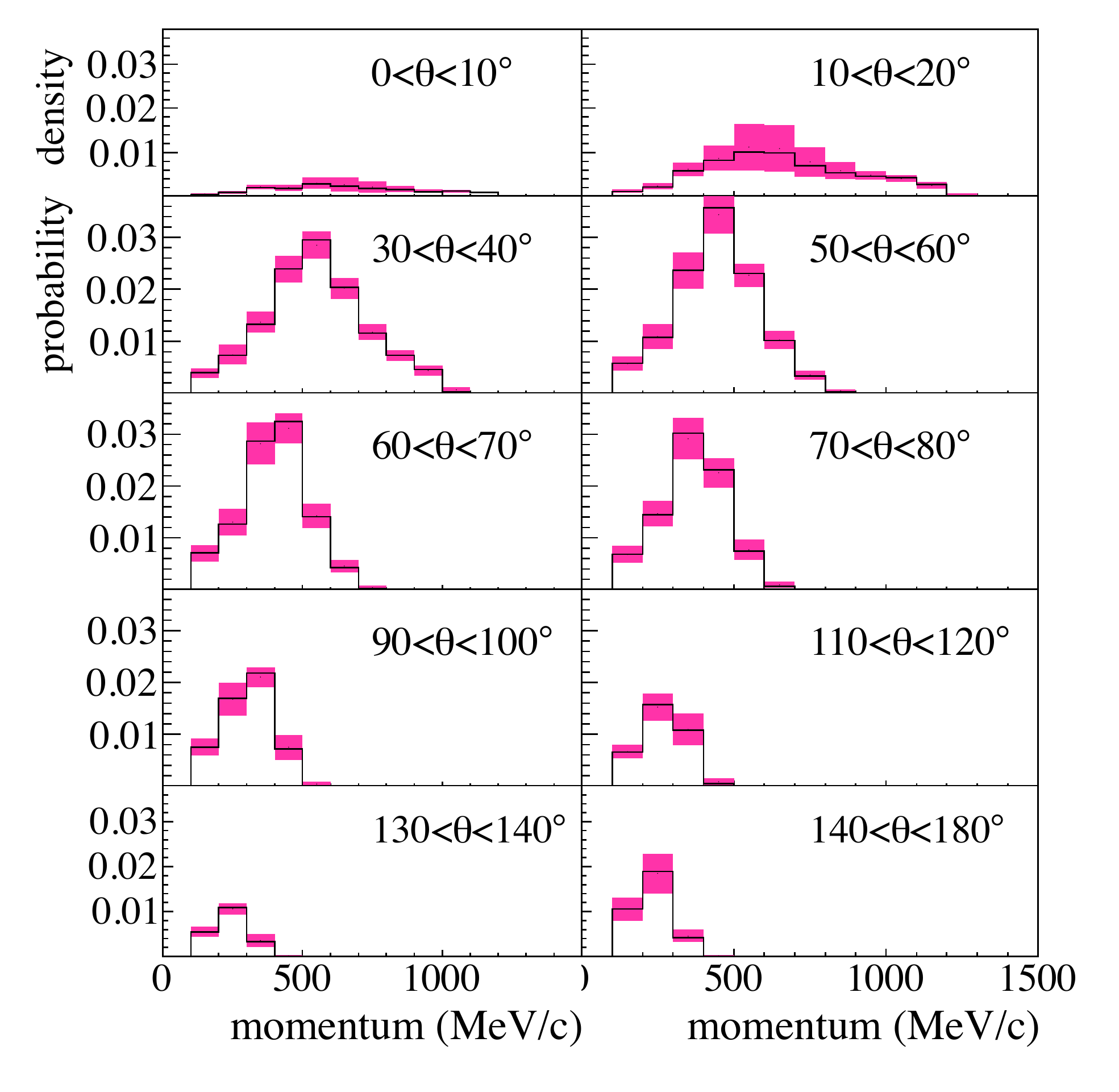}%
 \caption{%
The PDF as a function of momentum for 
different angular bins (10 of 15 \pthetae bins are shown) and
\stot=0.1. 
The shaded areas represent one sigma deviations that are evaluated by fluctuating all of the systematic parameters 
according to a multivariate normal distribution using their prior values and covariance matrix. 
} \label{fig:pl_err_st01}%
\end{minipage}
\end{center}
\end{figure}

\begin{figure}[htb]
\begin{center}
\begin{minipage}{0.5\textwidth}
 \includegraphics[width=1.0\textwidth,clip]{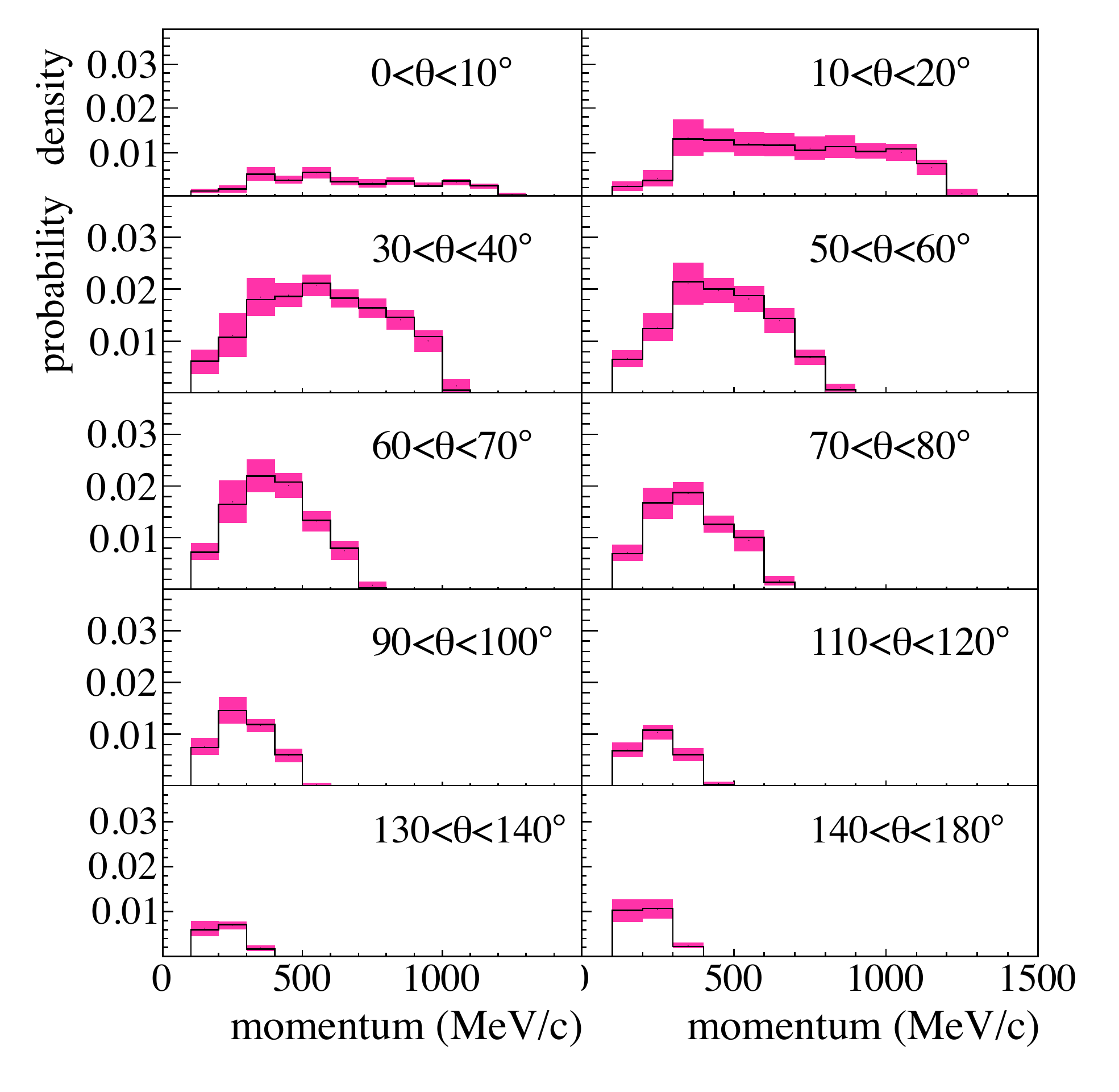}%
 \caption{%
The PDF as a function of momentum for 
for the same bins as in Fig.~\ref{fig:pl_err_st01} and
\stot=0.
The shaded areas  represent one sigma deviations that are evaluated by fluctuating all of the systematic parameters 
according to a multivariate normal distribution using their prior values and covariance matrix. 
} \label{fig:pl_err_st00}%
\end{minipage}
\end{center}
\end{figure}

We also consider the effect on the \pthetae PDF, or ``shape'' of \pthetae, as the systematic parameters are changed.
Fig.~\ref{fig:pl_err_st01} (Fig~\ref{fig:pl_err_st00}) shows the variation of the one-dimensional angular slices of the total signal+background 
as a function of momentum for \stot=0.1 (\stot=0).
The main contributions to the shape systematic uncertainties for \stot=0 are the SK detector efficiency and $W_{\textrm{eff}}$ parameters  in the neutrino interaction models which introduce uncertainties on the \pthetae distribution of $\num$ (NC) background. For \stot=0.1, the dominant contributions to the shape systematic uncertainties are the \num flux, CCQE and CC1$\pi$ cross section parameters, $x_{SF}$, and the SK detector uncertainties.

\subsection{\nue likelihood }
\label{sec:pthetafit}

We define an extended likelihood as the product of the likelihoods for the observed number of $\nue$ candidate events (\LLnorm), 
the shape of \pthetae distribution of those events (\LLshape) and
the constraint term for the nuisance parameters (\LLsyst). 
The normalization term, $\LL_{norm}$, is defined by the Poisson probability
to observe the number of $\nue$ candidate events, $N_{obs}$, given a predicted number of events, $n=\sum_{i,j}^{N_{p\theta}} N^p_{i,j}(\vec{o},\vec{f})$:
\begin{linenomath*}
\begin{equation}
{\cal L}_{norm}(\vec{o},\vec{f}) = \frac{(n^{N_{obs}}) e^{-n}}{N_{obs}!}
\label{eq:lnorm}
\end{equation}
\end{linenomath*} 
The shape term, $\LL_{shape}$ is defined by the product of the probabilities
that each event has a particular value of the momentum and angle \pthetae.
We use a Bayesian marginalization technique in order to incorporate the systematic uncertainties, 
by integrating over all systematic parameters.
Then, the only free parameter in the marginalized likelihood is \stot:
\begin{linenomath*}
\begin{equation}
\LL'(\vec{o}) = \int \LLnorm (\vec{o},\vec{f}) \times \LLshape(\vec{o},\vec{f}) \times \LLsyst(\vec{f})  d\vec{f} \;.
\label{eq:lmarg}
\end{equation}
\end{linenomath*}
Here we assume \LLsyst~is a multivariate normal distribution of the systematic parameters 
defined by the parameters' prior values and covariance matrix. 
The oscillation parameters are obtained by maximizing the marginalized likelihood. 

\begin{figure}[th]
\begin{center}
 \includegraphics[width=1.0\linewidth]{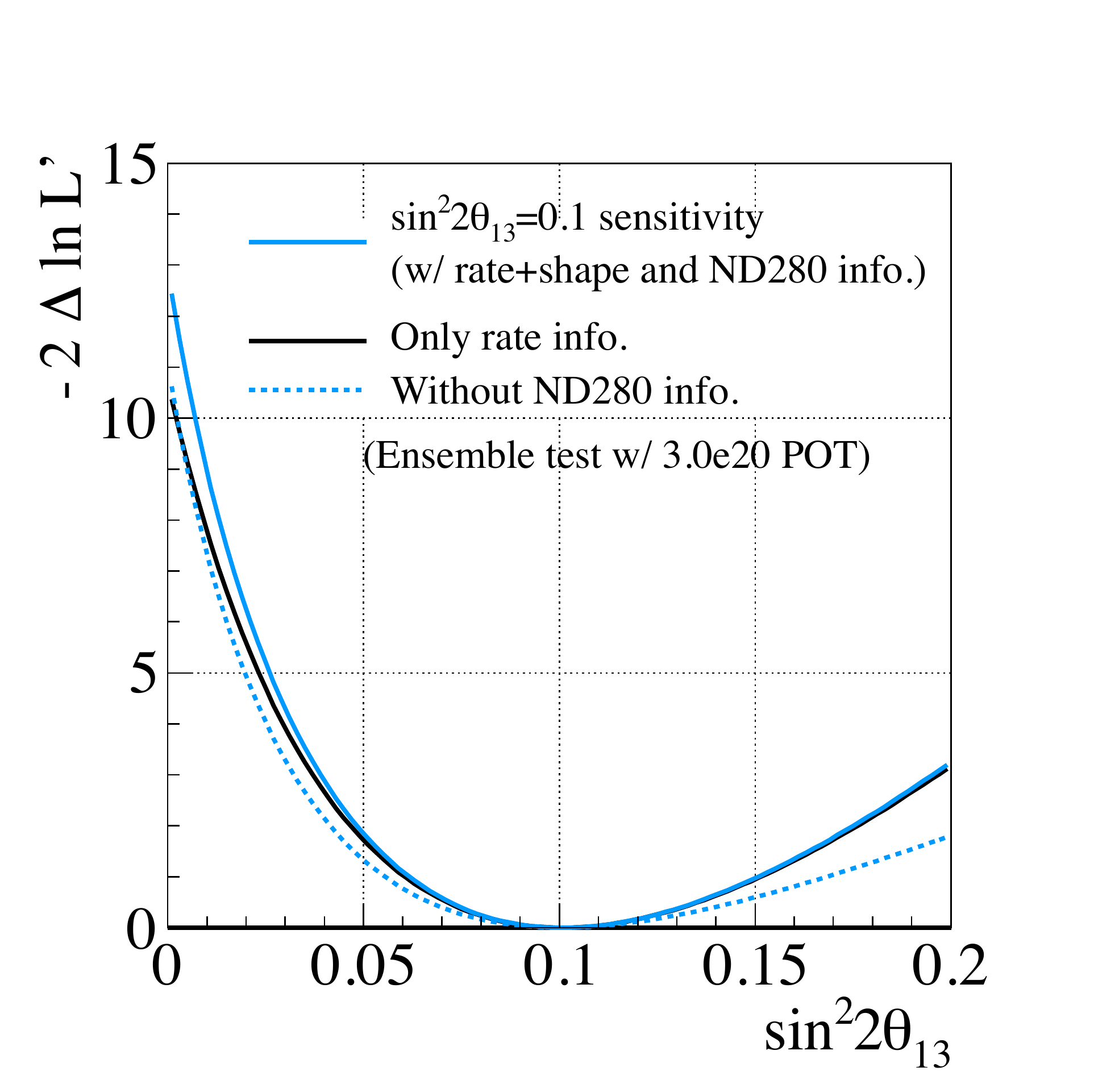}\\%
 \caption{%
The $-2 \Delta\ln{\cal L}$ average sensitivity curve for toy MC data generated at \stot=0.1 with \dcp=0, normal hierarchy and $3.01\times10^{20}$ POT. The likelihood is shown for three cases: where rate, shape and ND280 information is used, where only rate and ND280 information is used, and where rate and shape information is used without ND280 information.
} \label{fig:dlshape}%
\end{center}
\end{figure}

We have studied the increase in sensitivity of the analysis from the use of kinematic \pthetae information and from the ND280 fit. The difference of the log likelihood at the best-fit and at another value of \stot~ is calculated as:
\begin{linenomath*}
\begin{eqnarray}
-2\Delta\ln{\cal L}= & -2[\ln \LL'(\sin^22\theta_{13}) \nonumber \\
&  - \ln \LL'(\sin^22\theta_{13}^{best})]
 \label{eq:dll}
\end{eqnarray}
\end{linenomath*}
The likelihood in $-2 \Delta\ln{\cal L}$ can include just the normalization term,  or the normalization and shape term, and the systematic term in the likelihood can include the ND280 measurements or not. Fig.~\ref{fig:dlshape} shows the average $-2 \Delta\ln{\cal L}$ curves for these three cases, for toy MC data generated at \stot=0.1. 
We obtained a 20\% improvement to  $-2 \Delta\ln{\cal L}$ at \stot=0 when kinematic information is included; this is equivalent to a 20\% increased beam exposure. Similar studies show a comparable increase of 19\%
for the use of ND280 information in the likelihood to reduce the systematic errors.

\subsection{Results for \stot}
\label{sec:nueresults}

\begin{figure}
\centering
 \includegraphics[trim=0.5cm 0cm 0.5cm 1cm, clip=true, width=0.47\textwidth]{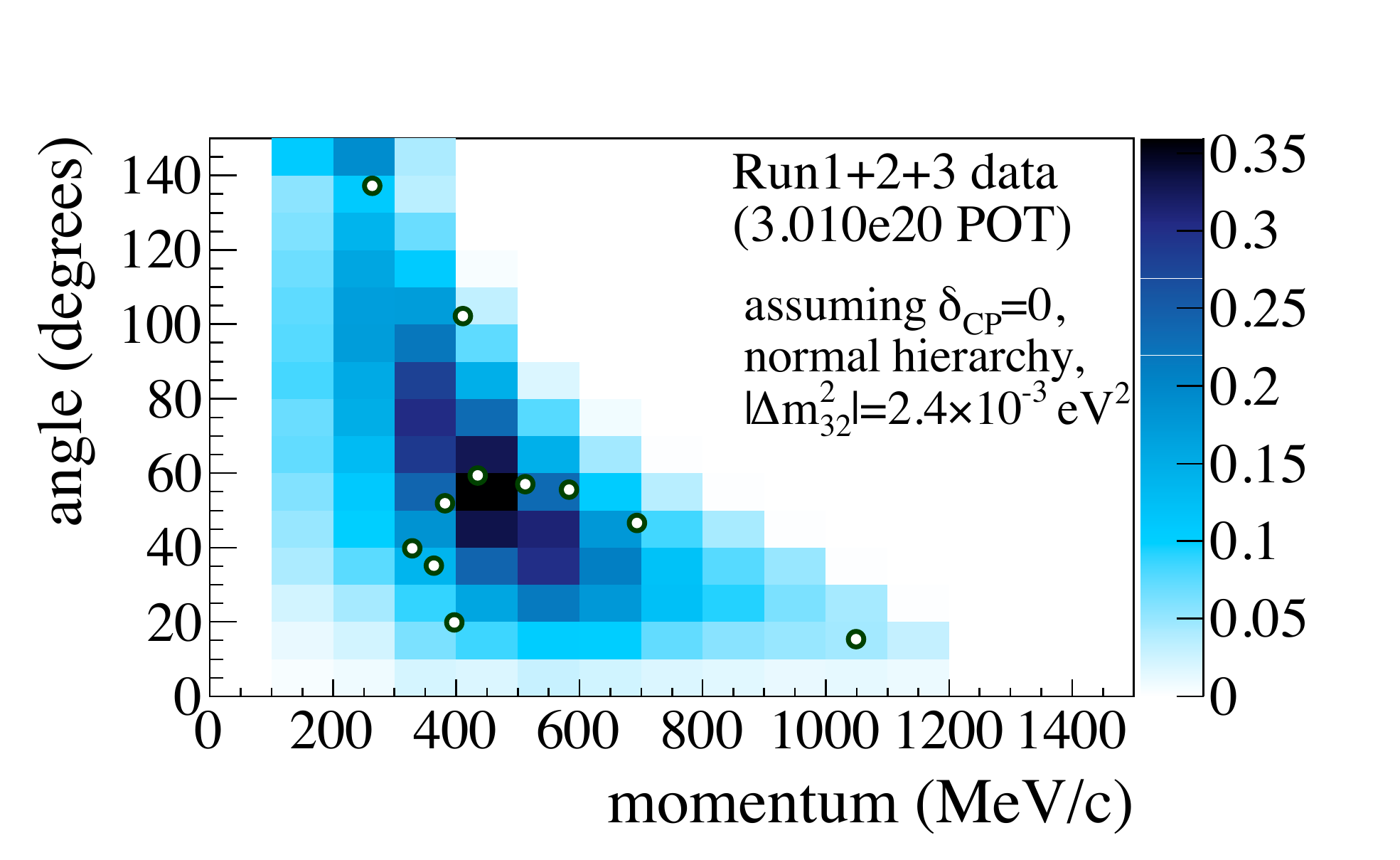}\\
 \includegraphics[trim=0.5cm 0cm 0.5cm 1cm, clip=true, width=0.47\textwidth]{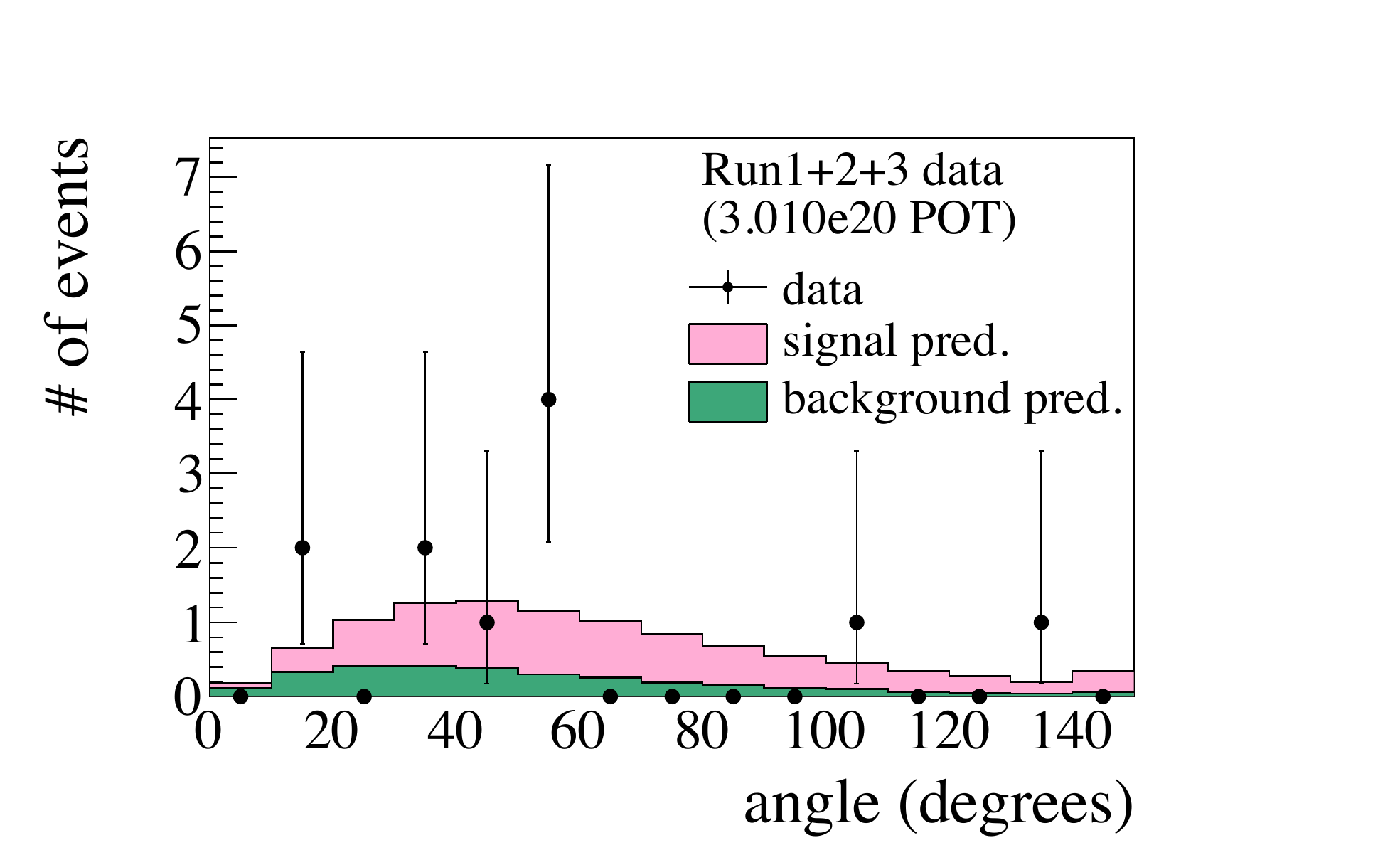}\\ 
 \includegraphics[trim=0.5cm 0cm 0.5cm 1cm, clip=true, width=0.47\textwidth]{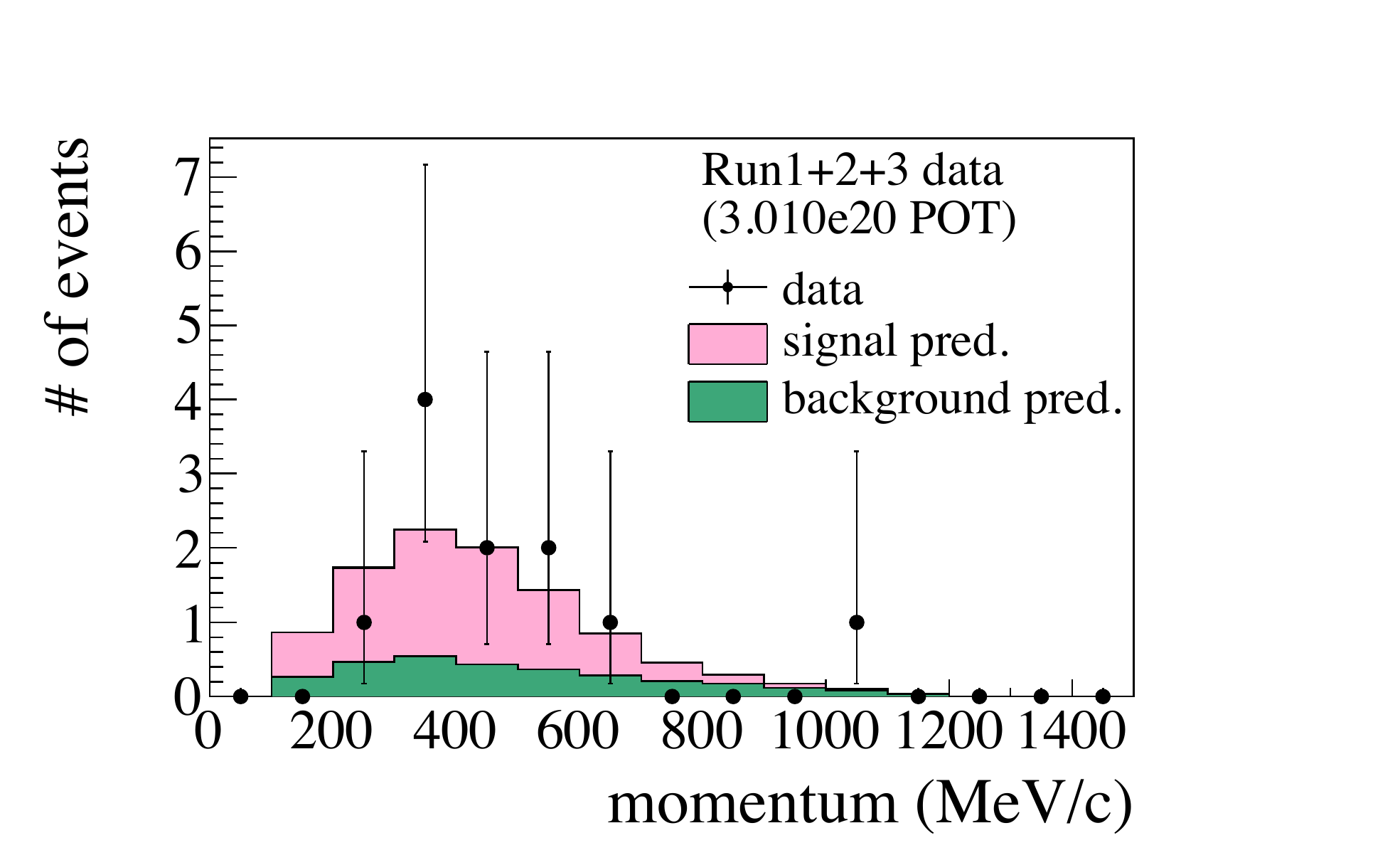}\\
 \caption{%
The \pthetae distribution of the \nue events (dots) (top)  overlaid with the prediction. The prediction includes the rate tuning determined from the fit to near detector information and a signal assuming the best-fit value of \stot=0.088. 
The angular distribution (middle) of the \nue events in data overlaid with prediction,
and the momentum distribution (bottom) with the same convention as above.
} \label{fig:result_run123c}%
\end{figure}

We performed the fit to the observed 11 \nue candidate events by allowing \stot~to vary and scanning the value of \dcp.
Fig.~\ref{fig:result_run123c} compares the \pthetae kinematic distributions observed in data with the prediction at the best-fit value of \stot.

\begin{figure}
\centering
 \includegraphics[trim=1cm 0cm 1.2cm 1.2cm, clip=true, width=0.43\textwidth]{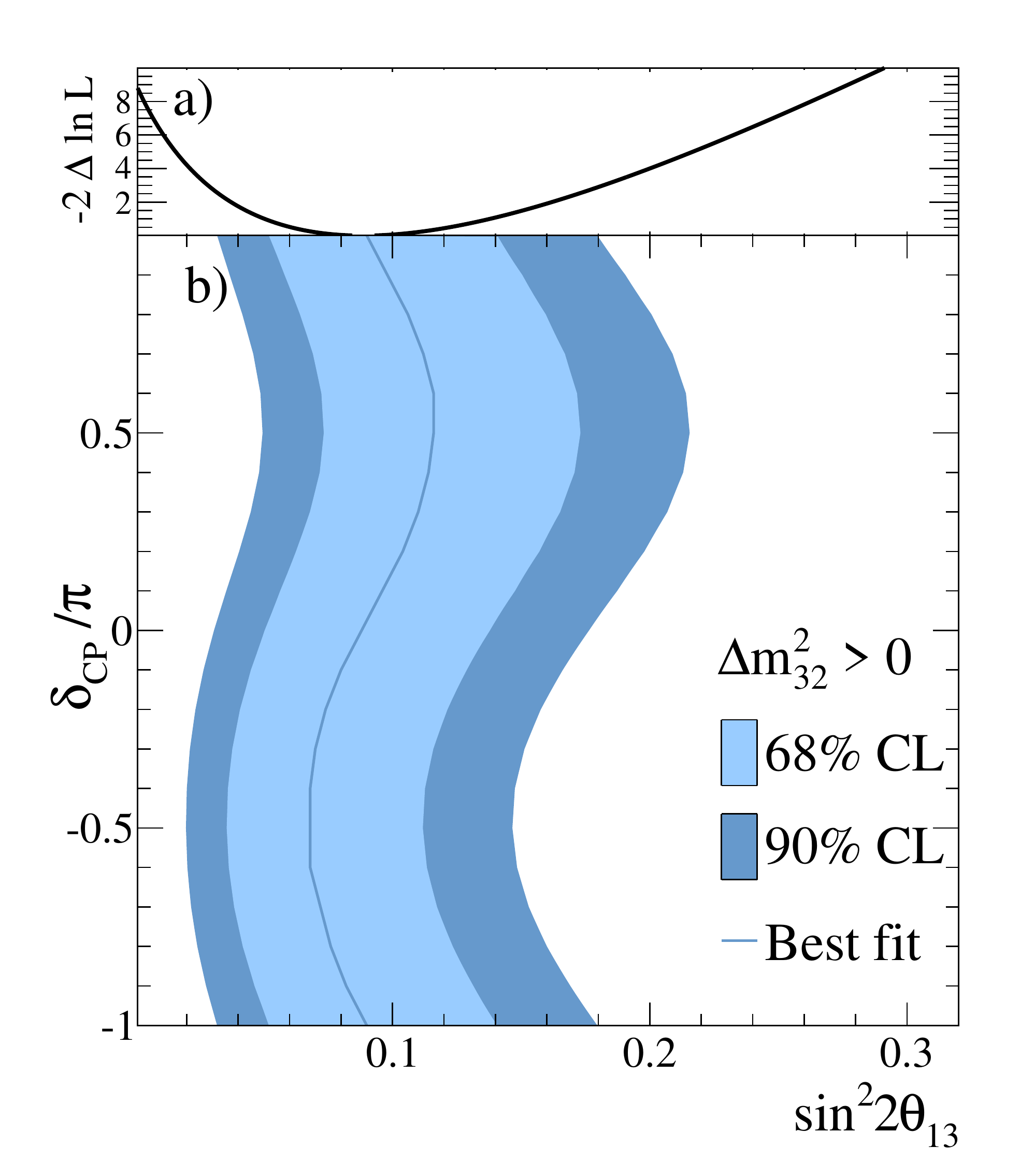}\\%
 \includegraphics[trim=1cm 0cm 1.2cm 1.2cm, clip=true, width=0.43\textwidth]{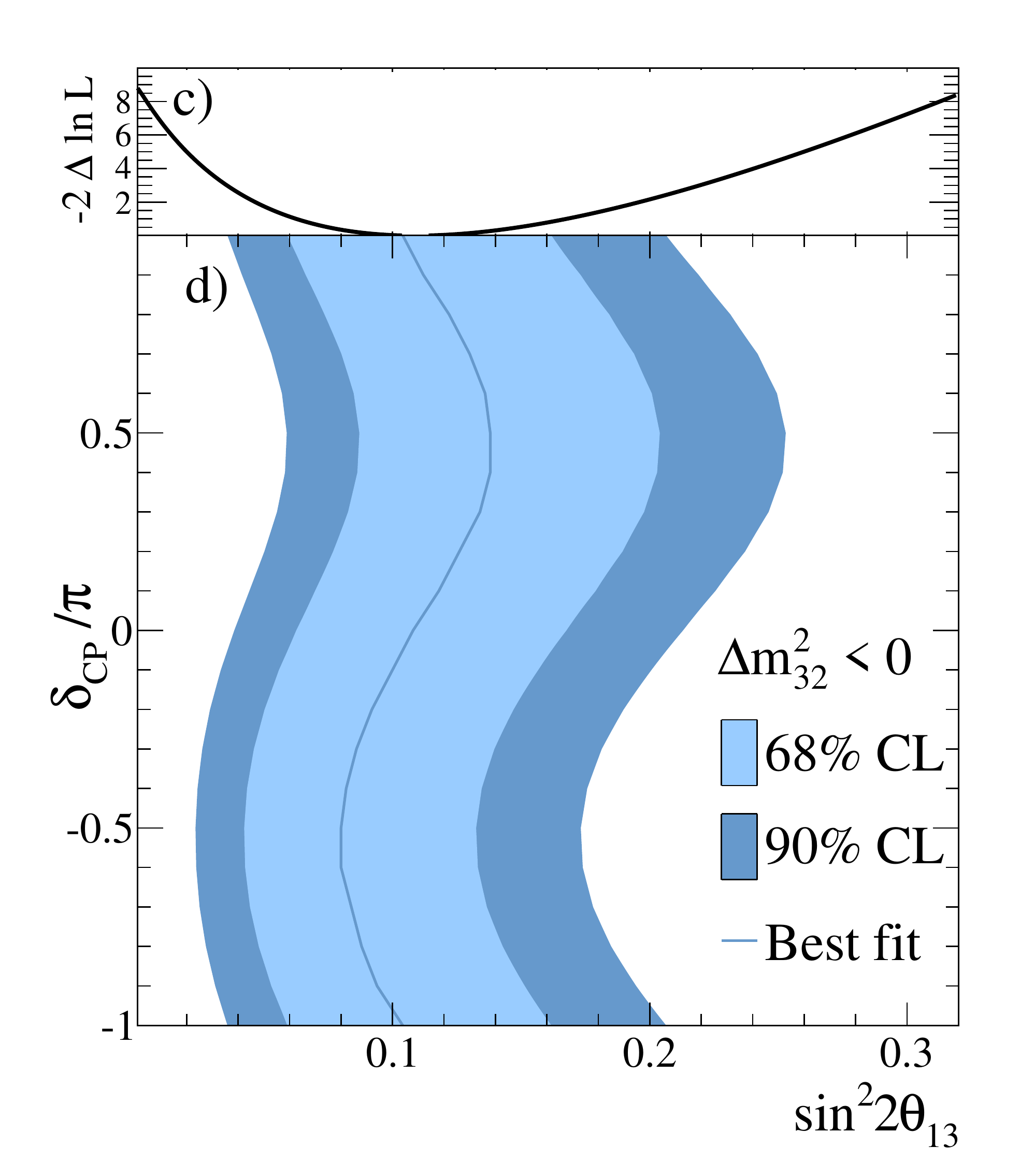}%

 \caption{%
The  68\% and 90\% confidence intervals for \stot~ scanned over values of \dcp~assuming  normal hierarchy (top, b) and inverted hierarchy (bottom, d) with all other oscillation parameters fixed at the values in Table~\ref{tab:oscparam}. The best-fit value of \stot~ for each value of \dcp~is also shown for the \pthetae analysis. The $-2 \Delta\ln{\cal L}$ curve for  normal hierarchy (top, a) and inverted hierarchy (bottom, c) at \dcp=0 are also shown vs. \stot. 
} \label{fig:2d_run123c}%
\end{figure}

Because of the potential bias in the determination of \stot~ near the physical boundary of \stot=0,
we calculate the confidence intervals following the Feldman-Cousins (FC) method~\cite{PhysRevD.57.3873}.
The 68\% and 90\% confidence intervals calculated using the FC method and constant $-2 \Delta\ln{\cal L}$
method are found to be equivalent. 
Assuming \dcp=0, the best-fit values of $\sin^22\theta_{13}$ with the 68\% confidence intervals are:
\begin{linenomath*}
\begin{eqnarray*}
\stot = 0.088{}^{+0.049}_{-0.039} &\;\;\;& \textrm{(normal hierarchy)}\\
\stot = 0.108{}^{+0.059}_{-0.046} &\;\;\;& \textrm{(inverted hierarchy)}
\end{eqnarray*}
\end{linenomath*}
The 90\% confidence intervals are: 
\begin{linenomath*}
\begin{eqnarray*}
0.030 < \sin^22\theta_{13} < 0.175 &\;\;\;& \textrm{(normal hierarchy)}\\
0.038 < \sin^22\theta_{13} < 0.212 &\;\;\;& \textrm{(inverted hierarchy).}
\end{eqnarray*}
\end{linenomath*}
Fig.~\ref{fig:2d_run123c} shows the 68\% and 90\% confidence intervals for $\sin^22\theta_{13}$  and the best-fit \stot~for each value of $\delta_{CP}$.

To compare the data with the best-fit \pthetae distribution, assuming normal hierarchy and \dcp=0,
we perform the Kolmogorov-Smirnov (KS) test.
We reorder the 2D \pthetae distribution into a 1D histogram, and generate 4000 toy MC experiments with the input value of $\sin^22\theta_{13} = 0.088$ (best-fit value) and where the observed number of events is 11. 
We then calculate the maximum distance for each toy experiment and determine the fraction of toy experiments for which the maximum distance is equal to or more than $0.22$, the value obtained for a KS test done on data.
The $p$-value is 0.54 and therefore the \pthetae distribution of data is consistent with the best-fit distribution.

Fig.~\ref{fig:2d_run123c}  shows the $-2 \Delta\ln{\cal L}$ 
curve as a function of \stot, for \dcp=0. 
We consider an alternate test of the background hypothesis using the value of  $-2 \Delta\ln{\cal L}$ at \stot=0.
The probability of obtaining a $-2 \Delta\ln{\cal L}$ at \stot=0 equal to or greater than
the value observed in data, 8.8, is calculated using the distribution
of $-2 \Delta\ln{\cal L}$ from pseudo-experiments generated with \stot=0, \dcp=0, normal hierarchy and fitted with the signal+background model. 
This test makes use of the different \pthetae distributions of signal compared to background, assuming three active neutrino mixing, and yields a similar probability of $1\times 10^{-3}$ to the rate-only test presented earlier.

\subsection{Alternate analysis methods}
\label{sec:altanalysis}
				   
In addition to the \pthetae analysis, we performed an analysis using the reconstructed neutrino energy spectrum, and a rate-only analysis.
Since \erec is closely correlated to the true neutrino energy for QE interactions, it provides the simplest projection for observing the energy dependence of the oscillation probability.
This analysis also provides a consistency check of the use of spectral information in the fit. We also provide an update to the previous \nue appearance analysis~\cite{PhysRevLett.107.041801}, where only rate information was used.

The likelihood including neutrino energy spectrum information is defined as:
\begin{linenomath*}
\begin{eqnarray}
\label{eqn:ereclikelihood}
{\cal L}(
\vec{o},\vec{f})=
{\cal L}_{norm}(\vec{o},\vec{f})
\times{\cal L}_{shape}(\vec{o},\vec{f})
\times{\cal L}_{syst}(\vec{f})
\end{eqnarray}
\end{linenomath*}
In this analysis, we perform a one dimensional scan of \stot~ for each value of \dcp~while the other oscillation parameters 
are fixed.
At each \stot~point, the negative log likelihood $-2\ln{\cal L}(\vec{o},\vec{f})$ is minimized by allowing the nuisance parameters, $\vec{f}$, to vary.
The best-fit value of $\sin^22\theta_{13}$ is the point where $-2\ln{\cal L}(\vec{o})$ is minimized and $-2\Delta\ln{\cal L}$
is used to constructing a confidence interval for $\sin^22\theta_{13}$ 
according to the FC method.

\begin{figure}
\centering
\includegraphics[width=0.5\textwidth]{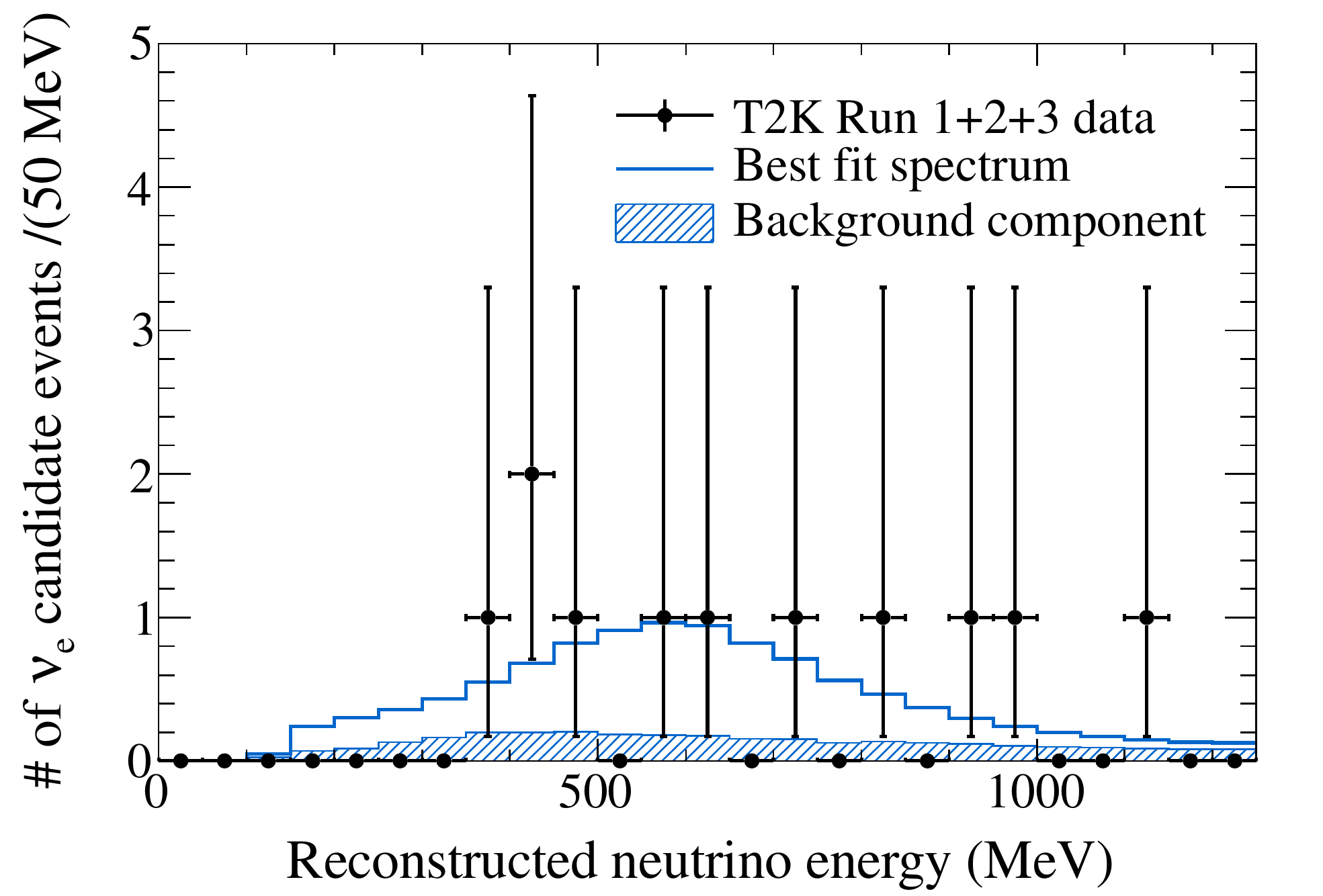}
\caption{
The observed \erec distribution and prediction,
assuming \stot=0.092, $\delta_{CP}=0$, and normal hierarchy.
The background component is also shown.
}
\label{fig:run123bestspec}
\end{figure}

Fig.~\ref{fig:run123bestspec} shows the observed \erec distribution for the \nue events with the best-fit of the \erec analysis applied. 
The observed spectrum agrees with the best-fit expectation, confirmed by a KS test with a $p$-value of  0.7. 
The best-fit values of $\sin^22\theta_{13}$, assuming $\delta_{CP}=0$, are:
\begin{linenomath*}
\begin{eqnarray*}
\stot = 0.092^{\,+0.049}_{\,-0.039} &\;\;\;& \textrm{(normal hierarchy)}\\
\stot = 0.112^{\,+0.058}_{\,-0.047} &\;\;\;& \textrm{(inverted hierarchy)}
\end{eqnarray*}
\end{linenomath*}
The 90\% confidence intervals are: 
\begin{linenomath*}
\begin{eqnarray*}
0.033 < \sin^22\theta_{13} < 0.179 &\;\;\;& \textrm{(normal hierarchy)}\\
0.040 < \sin^22\theta_{13} < 0.215 &\;\;\;& \textrm{(inverted hierarchy).}
\end{eqnarray*}
\end{linenomath*}

The rate-only measurement only uses the number of \nue events at SK to determine \stot.  This analysis uses the normalization likelihood ratio:

\begin{linenomath*}
\begin{equation}
\Delta\chi^{2}=-2\log  \frac{{\cal L}_{norm}(\vec{o},\vec{f})}{{{\cal L}^{\mathrm{best}}_{norm}}(\vec{o},\vec{f})}
\end{equation}
\end{linenomath*}
where ${\cal L}_{norm}$ is defined in Eq.~\ref{eq:lnorm}.
The value of $\Delta\chi^{2}$ is calculated for the 11 observed \nue candidates, in a one dimensional scan of \stot~ for each point of \dcp~ with all other oscillation parameters fixed. The confidence intervals are determined using the FC method. 

The best-fit values of $\sin^22\theta_{13}$, assuming $\delta_{CP}=0$, are:
\begin{linenomath*}
\begin{eqnarray*}
\stot = 0.097_{-0.041}^{+0.053} &\;\;\;& \textrm{(normal hierarchy)}\\
\stot = 0.123_{-0.051}^{+0.065} &\;\;\;& \textrm{(inverted hierarchy).}
\end{eqnarray*}
\end{linenomath*}
The 90\% confidence intervals are:
\begin{linenomath*}
\begin{eqnarray*}
0.034 < \sin^22\theta_{13} < 0.190 &\;\;\;& \textrm{(normal hierarchy)}\\
0.044 < \sin^22\theta_{13} < 0.236 &\;\;\;& \textrm{(inverted hierarchy).}
\end{eqnarray*}
\end{linenomath*}

\begin{figure}
\centering
\includegraphics[trim=0.5cm 0cm 0.7cm 1cm, clip=true,width=0.42\textwidth]{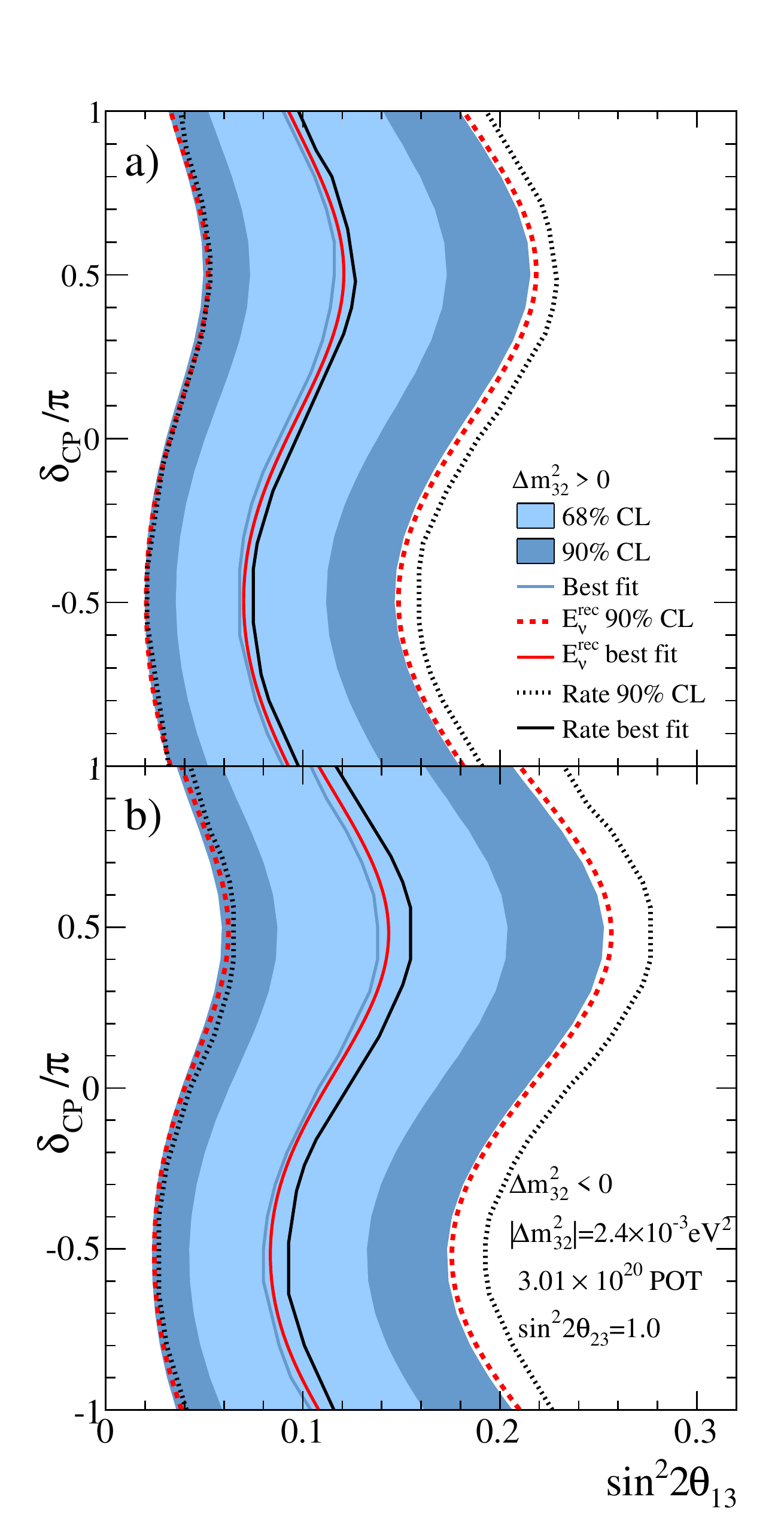}
\caption{
The  68\% and 90\% confidence interval regions for \stot~ scanned over values of \dcp~assuming normal hierarchy (a) and inverted hierarchy (b) with the best-fit value of \stot~ shown for the \pthetae analysis.
The 90\% confidence interval region for the \erec analysis and rate-only analysis are overlaid.  The best-fit values of \stot~ for the \erec analysis  and the rate-only analysis are also shown.  All other oscillation parameters are fixed at the values in Table~\ref{tab:oscparam}.
}
\label{fig:2d_run123c_overlay}
\end{figure}

Fig.~\ref{fig:2d_run123c_overlay} shows the three analyses are consistent with each other. The rate-only analysis has a higher best-fit value of \stot~ than the \erec, \pthetae analyses. This results from the additional discriminatory power of the kinematic information to identify events as 
slightly more similar to the background distribution than  the predicted oscillation signal. In addition, the difference between the best-fit and the 90\% upper confidence interval for the rate-only analysis is larger than the other two analyses. This is due to a slight (2\%) over-coverage of the rate-only analysis.

\section{\label{sec:conclusion} Conclusion}

In summary, we have reported the first evidence of electron neutrino appearance in a muon neutrino beam with a baseline
and neutrino spectrum optimized for the atmospheric mass splitting.
We observed 11 candidate \nue events at the SK detector when $3.3 \pm 0.4$(syst.) background events are expected, and 
rejected the background-only hypothesis 
with a $p$-value of 0.0009, equivalent to a 3.1$\sigma$ significance.
We have employed a fit to the ND280 near detector data that constrains the parametrized neutrino flux and interaction
models used to predict the event rates at SK.
The ND280 constraint on the \nue candidates reduced the overall systematic uncertainty to 10--13\% depending on the value of \stot, an important step towards precision measurements of \nue appearance.
The excess of events at SK corresponds to a best-fit value of $\stot=0.088{}^{+0.049}_{-0.039}$ at 68\%~C.L., 
assuming \dcp=0, \sttt=1.0 and normal hierarchy.

This result represents an important step towards constraining the unknown parameters in the 
three-neutrino oscillation model.  The evidence of electron neutrino appearance opens the door for a rich program
of experimental physics in this oscillation channel. 
T2K measurements of this channel will be an important input to global fits which also combine muon neutrino disappearance measurements and reactor-based measurements of $\theta_{13}$ via $\bar{\nu}_{e}$ disappearance to begin to constrain \dcp~and the octant of $\theta_{23}$.  
Future measurements of the appearance probability for antineutrinos will provide a further constraint on \dcp~and the mass hierarchy.

\begin{acknowledgments}
We thank the J-PARC accelerator team for the superb accelerator performance and
CERN NA61 colleagues for providing essential particle production data and for their fruitful collaboration.
We acknowledge the support of MEXT, Japan; 
NSERC, NRC and CFI, Canada;
CEA and CNRS/IN2P3, France;
DFG, Germany; 
INFN, Italy;
Ministry of Science and Higher Education, Poland;
RAS, RFBR and the Ministry of Education and Science
of the Russian Federation; 
MEST and NRF, South Korea;
MICINN and CPAN, Spain;
SNSF and SER, Switzerland;
STFC, U.K.; NSF and 
DOE, U.S.A.
We also thank CERN for their donation of the UA1/NOMAD magnet 
and DESY for the HERA-B magnet mover system.
In addition, participation of individual researchers
and institutions in T2K has been further supported by funds from: ERC (FP7), EU; JSPS, Japan; Royal Society, UK; 
DOE Early Career program, and the A. P. Sloan Foundation, U.S.A.
Computations were performed on the supercomputers at the SciNet~\cite{scinet}   HPC Consortium. SciNet is funded by: the Canada Foundation for Innovation under the auspices of Compute Canada; the Government of Ontario; Ontario Research Fund - Research Excellence; and the University of Toronto.

\end{acknowledgments}

\bibliographystyle{apsrev4-1}
\bibliography{NueApp2012_PRD_flat}

\providecommand{\noopsort}[1]{}\providecommand{\singleletter}[1]{#1}%
\begin{thebibliography}{115}%
\makeatletter
\providecommand \@ifxundefined [1]{%
 \@ifx{#1\undefined}
}%
\providecommand \@ifnum [1]{%
 \ifnum #1\expandafter \@firstoftwo
 \else \expandafter \@secondoftwo
 \fi
}%
\providecommand \@ifx [1]{%
 \ifx #1\expandafter \@firstoftwo
 \else \expandafter \@secondoftwo
 \fi
}%
\providecommand \natexlab [1]{#1}%
\providecommand \enquote  [1]{``#1''}%
\providecommand \bibnamefont  [1]{#1}%
\providecommand \bibfnamefont [1]{#1}%
\providecommand \citenamefont [1]{#1}%
\providecommand \href@noop [0]{\@secondoftwo}%
\providecommand \href [0]{\begingroup \@sanitize@url \@href}%
\providecommand \@href[1]{\@@startlink{#1}\@@href}%
\providecommand \@@href[1]{\endgroup#1\@@endlink}%
\providecommand \@sanitize@url [0]{\catcode `\\12\catcode `\$12\catcode
  `\&12\catcode `\#12\catcode `\^12\catcode `\_12\catcode `\%12\relax}%
\providecommand \@@startlink[1]{}%
\providecommand \@@endlink[0]{}%
\providecommand \url  [0]{\begingroup\@sanitize@url \@url }%
\providecommand \@url [1]{\endgroup\@href {#1}{\urlprefix }}%
\providecommand \urlprefix  [0]{URL }%
\providecommand \Eprint [0]{\href }%
\providecommand \doibase [0]{http://dx.doi.org/}%
\providecommand \selectlanguage [0]{\@gobble}%
\providecommand \bibinfo  [0]{\@secondoftwo}%
\providecommand \bibfield  [0]{\@secondoftwo}%
\providecommand \translation [1]{[#1]}%
\providecommand \BibitemOpen [0]{}%
\providecommand \bibitemStop [0]{}%
\providecommand \bibitemNoStop [0]{.\EOS\space}%
\providecommand \EOS [0]{\spacefactor3000\relax}%
\providecommand \BibitemShut  [1]{\csname bibitem#1\endcsname}%
\let\auto@bib@innerbib\@empty
\bibitem [{\citenamefont {Cleveland}\ \emph {et~al.}(1998)\citenamefont
  {Cleveland} \emph {et~al.}}]{Cleveland:1998nv}%
  \BibitemOpen
  \bibfield  {author} {\bibinfo {author} {\bibfnamefont {B.}~\bibnamefont
  {Cleveland}} \emph {et~al.},\ }\href {\doibase 10.1086/305343} {\bibfield
  {journal} {\bibinfo  {journal} {Astrophys. J.}\ }\textbf {\bibinfo {volume}
  {496}},\ \bibinfo {pages} {505} (\bibinfo {year} {1998})}\BibitemShut
  {NoStop}%
\bibitem [{\citenamefont {Hirata}\ \emph {et~al.}(1989)\citenamefont {Hirata}
  \emph {et~al.}}]{PhysRevLett.63.16}%
  \BibitemOpen
  \bibfield  {author} {\bibinfo {author} {\bibfnamefont {K.~S.}\ \bibnamefont
  {Hirata}} \emph {et~al.} (\bibinfo {collaboration} {Kamiokande-II
  Collaboration}),\ }\href {\doibase 10.1103/PhysRevLett.63.16} {\bibfield
  {journal} {\bibinfo  {journal} {Phys. Rev. Lett.}\ }\textbf {\bibinfo
  {volume} {63}},\ \bibinfo {pages} {16} (\bibinfo {year} {1989})}\BibitemShut
  {NoStop}%
\bibitem [{\citenamefont {Abdurashitov}\ \emph {et~al.}(1994)\citenamefont
  {Abdurashitov} \emph {et~al.}}]{Abdurashitov1994234}%
  \BibitemOpen
  \bibfield  {author} {\bibinfo {author} {\bibfnamefont {J.}~\bibnamefont
  {Abdurashitov}} \emph {et~al.} (\bibinfo {collaboration} {SAGE
  Collaboration}),\ }\href {\doibase 10.1016/0370-2693(94)90454-5} {\bibfield
  {journal} {\bibinfo  {journal} {Phys. Lett. B}\ }\textbf {\bibinfo {volume}
  {328}},\ \bibinfo {pages} {234 } (\bibinfo {year} {1994})}\BibitemShut
  {NoStop}%
\bibitem [{\citenamefont {Anselmann}\ \emph {et~al.}(1994)\citenamefont
  {Anselmann} \emph {et~al.}}]{Anselmann1994377}%
  \BibitemOpen
  \bibfield  {author} {\bibinfo {author} {\bibfnamefont {P.}~\bibnamefont
  {Anselmann}} \emph {et~al.} (\bibinfo {collaboration} {GALLEX
  Collaboration}),\ }\href {\doibase 10.1016/0370-2693(94)90744-7} {\bibfield
  {journal} {\bibinfo  {journal} {Phys. Lett. B}\ }\textbf {\bibinfo {volume}
  {327}},\ \bibinfo {pages} {377 } (\bibinfo {year} {1994})}\BibitemShut
  {NoStop}%
\bibitem [{\citenamefont {Fukuda}\ \emph {et~al.}(2001)\citenamefont {Fukuda}
  \emph {et~al.}}]{PhysRevLett.86.5651}%
  \BibitemOpen
  \bibfield  {author} {\bibinfo {author} {\bibfnamefont {S.}~\bibnamefont
  {Fukuda}} \emph {et~al.} (\bibinfo {collaboration} {Super-Kamiokande
  Collaboration}),\ }\href {\doibase 10.1103/PhysRevLett.86.5651} {\bibfield
  {journal} {\bibinfo  {journal} {Phys. Rev. Lett.}\ }\textbf {\bibinfo
  {volume} {86}},\ \bibinfo {pages} {5651} (\bibinfo {year}
  {2001})}\BibitemShut {NoStop}%
\bibitem [{\citenamefont {Ahmad}\ \emph {et~al.}(2002)\citenamefont {Ahmad}
  \emph {et~al.}}]{PhysRevLett.89.011301}%
  \BibitemOpen
  \bibfield  {author} {\bibinfo {author} {\bibfnamefont {Q.~R.}\ \bibnamefont
  {Ahmad}} \emph {et~al.} (\bibinfo {collaboration} {SNO Collaboration}),\
  }\href {\doibase 10.1103/PhysRevLett.89.011301} {\bibfield  {journal}
  {\bibinfo  {journal} {Phys. Rev. Lett.}\ }\textbf {\bibinfo {volume} {89}},\
  \bibinfo {pages} {011301} (\bibinfo {year} {2002})}\BibitemShut {NoStop}%
\bibitem [{\citenamefont {Arpesella}\ \emph {et~al.}(2008)\citenamefont
  {Arpesella} \emph {et~al.}}]{PhysRevLett.101.091302}%
  \BibitemOpen
  \bibfield  {author} {\bibinfo {author} {\bibfnamefont {C.}~\bibnamefont
  {Arpesella}} \emph {et~al.} (\bibinfo {collaboration} {Borexino
  Collaboration}),\ }\href {\doibase 10.1103/PhysRevLett.101.091302} {\bibfield
   {journal} {\bibinfo  {journal} {Phys. Rev. Lett.}\ }\textbf {\bibinfo
  {volume} {101}},\ \bibinfo {pages} {091302} (\bibinfo {year}
  {2008})}\BibitemShut {NoStop}%
\bibitem [{\citenamefont {Fukuda}\ \emph {et~al.}(1998)\citenamefont {Fukuda}
  \emph {et~al.}}]{Fukuda:1998ub}%
  \BibitemOpen
  \bibfield  {author} {\bibinfo {author} {\bibfnamefont {Y.}~\bibnamefont
  {Fukuda}} \emph {et~al.} (\bibinfo {collaboration} {Super-Kamiokande
  Collaboration}),\ }\href {\doibase 10.1016/S0370-2693(98)00876-4} {\bibfield
  {journal} {\bibinfo  {journal} {Phys. Lett. B}\ }\textbf {\bibinfo {volume}
  {436}},\ \bibinfo {pages} {33} (\bibinfo {year} {1998})},\ \Eprint
  {http://arxiv.org/abs/hep-ex/9805006} {arXiv:hep-ex/9805006 [hep-ex]}
  \BibitemShut {NoStop}%
\bibitem [{\citenamefont {Hirata}\ \emph {et~al.}(1988)\citenamefont {Hirata}
  \emph {et~al.}}]{Hirata1988416}%
  \BibitemOpen
  \bibfield  {author} {\bibinfo {author} {\bibfnamefont {K.}~\bibnamefont
  {Hirata}} \emph {et~al.},\ }\href {\doibase 10.1016/0370-2693(88)91690-5}
  {\bibfield  {journal} {\bibinfo  {journal} {Phys. Lett. B}\ }\textbf
  {\bibinfo {volume} {205}},\ \bibinfo {pages} {416 } (\bibinfo {year}
  {1988})}\BibitemShut {NoStop}%
\bibitem [{\citenamefont {Becker-Szendy}\ \emph {et~al.}(1992)\citenamefont
  {Becker-Szendy} \emph {et~al.}}]{PhysRevD.46.3720}%
  \BibitemOpen
  \bibfield  {author} {\bibinfo {author} {\bibfnamefont {R.}~\bibnamefont
  {Becker-Szendy}} \emph {et~al.},\ }\href {\doibase 10.1103/PhysRevD.46.3720}
  {\bibfield  {journal} {\bibinfo  {journal} {Phys. Rev. D}\ }\textbf {\bibinfo
  {volume} {46}},\ \bibinfo {pages} {3720} (\bibinfo {year}
  {1992})}\BibitemShut {NoStop}%
\bibitem [{\citenamefont {Ahlen}\ \emph {et~al.}(1995)\citenamefont {Ahlen}
  \emph {et~al.}}]{Ahlen1995481}%
  \BibitemOpen
  \bibfield  {author} {\bibinfo {author} {\bibfnamefont {S.}~\bibnamefont
  {Ahlen}} \emph {et~al.},\ }\href {\doibase 10.1016/0370-2693(95)00958-N}
  {\bibfield  {journal} {\bibinfo  {journal} {Phys. Lett. B}\ }\textbf
  {\bibinfo {volume} {357}},\ \bibinfo {pages} {481 } (\bibinfo {year}
  {1995})}\BibitemShut {NoStop}%
\bibitem [{\citenamefont {Allison}\ \emph {et~al.}(1997)\citenamefont {Allison}
  \emph {et~al.}}]{Allison:1996yb}%
  \BibitemOpen
  \bibfield  {author} {\bibinfo {author} {\bibfnamefont {W.}~\bibnamefont
  {Allison}} \emph {et~al.},\ }\href {\doibase 10.1016/S0370-2693(96)01609-7}
  {\bibfield  {journal} {\bibinfo  {journal} {Phys.Lett.}\ }\textbf {\bibinfo
  {volume} {B391}},\ \bibinfo {pages} {491} (\bibinfo {year} {1997})},\ \Eprint
  {http://arxiv.org/abs/hep-ex/9611007} {arXiv:hep-ex/9611007 [hep-ex]}
  \BibitemShut {NoStop}%
\bibitem [{\citenamefont {Wendell}\ \emph {et~al.}(2010)\citenamefont {Wendell}
  \emph {et~al.}}]{PhysRevD.81.092004}%
  \BibitemOpen
  \bibfield  {author} {\bibinfo {author} {\bibfnamefont {R.}~\bibnamefont
  {Wendell}} \emph {et~al.} (\bibinfo {collaboration} {Super-Kamiokande
  Collaboration}),\ }\href {\doibase 10.1103/PhysRevD.81.092004} {\bibfield
  {journal} {\bibinfo  {journal} {Phys. Rev. D}\ }\textbf {\bibinfo {volume}
  {81}},\ \bibinfo {pages} {092004} (\bibinfo {year} {2010})}\BibitemShut
  {NoStop}%
\bibitem [{\citenamefont {Abe}\ \emph {et~al.}(2008)\citenamefont {Abe} \emph
  {et~al.}}]{PhysRevLett.100.221803}%
  \BibitemOpen
  \bibfield  {author} {\bibinfo {author} {\bibfnamefont {S.}~\bibnamefont
  {Abe}} \emph {et~al.} (\bibinfo {collaboration} {KamLAND Collaboration}),\
  }\href {\doibase 10.1103/PhysRevLett.100.221803} {\bibfield  {journal}
  {\bibinfo  {journal} {Phys. Rev. Lett.}\ }\textbf {\bibinfo {volume} {100}},\
  \bibinfo {pages} {221803} (\bibinfo {year} {2008})}\BibitemShut {NoStop}%
\bibitem [{\citenamefont {Ahn}\ \emph {et~al.}(2006)\citenamefont {Ahn} \emph
  {et~al.}}]{PhysRevD.74.072003}%
  \BibitemOpen
  \bibfield  {author} {\bibinfo {author} {\bibfnamefont {M.~H.}\ \bibnamefont
  {Ahn}} \emph {et~al.} (\bibinfo {collaboration} {K2K Collaboration}),\ }\href
  {\doibase 10.1103/PhysRevD.74.072003} {\bibfield  {journal} {\bibinfo
  {journal} {Phys. Rev. D}\ }\textbf {\bibinfo {volume} {74}},\ \bibinfo
  {pages} {072003} (\bibinfo {year} {2006})}\BibitemShut {NoStop}%
\bibitem [{\citenamefont {Adamson}\ \emph {et~al.}(2012)\citenamefont {Adamson}
  \emph {et~al.}}]{Adamson:2012rm}%
  \BibitemOpen
  \bibfield  {author} {\bibinfo {author} {\bibfnamefont {P.}~\bibnamefont
  {Adamson}} \emph {et~al.} (\bibinfo {collaboration} {MINOS Collaboration}),\
  }\href {\doibase 10.1103/PhysRevLett.108.191801} {\bibfield  {journal}
  {\bibinfo  {journal} {Phys. Rev. Lett.}\ }\textbf {\bibinfo {volume} {108}},\
  \bibinfo {pages} {191801} (\bibinfo {year} {2012})}\BibitemShut {NoStop}%
\bibitem [{\citenamefont {Adamson}\ \emph
  {et~al.}(2011{\natexlab{a}})\citenamefont {Adamson} \emph
  {et~al.}}]{PhysRevLett.106.181801}%
  \BibitemOpen
  \bibfield  {author} {\bibinfo {author} {\bibfnamefont {P.}~\bibnamefont
  {Adamson}} \emph {et~al.} (\bibinfo {collaboration} {MINOS Collaboration}),\
  }\href {\doibase 10.1103/PhysRevLett.106.181801} {\bibfield  {journal}
  {\bibinfo  {journal} {Phys. Rev. Lett.}\ }\textbf {\bibinfo {volume} {106}},\
  \bibinfo {pages} {181801} (\bibinfo {year} {2011}{\natexlab{a}})}\BibitemShut
  {NoStop}%
\bibitem [{\citenamefont {Abe}\ \emph {et~al.}(2012{\natexlab{a}})\citenamefont
  {Abe} \emph {et~al.}}]{Abe:2012gx}%
  \BibitemOpen
  \bibfield  {author} {\bibinfo {author} {\bibfnamefont {K.}~\bibnamefont
  {Abe}} \emph {et~al.} (\bibinfo {collaboration} {T2K Collaboration}),\ }\href
  {\doibase 10.1103/PhysRevD.85.031103} {\bibfield  {journal} {\bibinfo
  {journal} {Phys. Rev. D}\ }\textbf {\bibinfo {volume} {85}},\ \bibinfo
  {pages} {031103} (\bibinfo {year} {2012}{\natexlab{a}})}\BibitemShut
  {NoStop}%
\bibitem [{\citenamefont {Aguilar-Arevalo}\ \emph {et~al.}(2001)\citenamefont
  {Aguilar-Arevalo} \emph {et~al.}}]{Aguilar:2001ty}%
  \BibitemOpen
  \bibfield  {author} {\bibinfo {author} {\bibfnamefont {A.~A.}\ \bibnamefont
  {Aguilar-Arevalo}} \emph {et~al.} (\bibinfo {collaboration} {LSND
  Collaboration}),\ }\href {\doibase 10.1103/PhysRevD.64.112007} {\bibfield
  {journal} {\bibinfo  {journal} {Phys. Rev. D}\ }\textbf {\bibinfo {volume}
  {64}},\ \bibinfo {pages} {112007} (\bibinfo {year} {2001})}\BibitemShut
  {NoStop}%
\bibitem [{\citenamefont {Aguilar-Arevalo}\ \emph
  {et~al.}(2010{\natexlab{a}})\citenamefont {Aguilar-Arevalo} \emph
  {et~al.}}]{PhysRevLett.105.181801}%
  \BibitemOpen
  \bibfield  {author} {\bibinfo {author} {\bibfnamefont {A.}~\bibnamefont
  {Aguilar-Arevalo}} \emph {et~al.} (\bibinfo {collaboration} {MiniBooNE
  Collaboration}),\ }\href {\doibase 10.1103/PhysRevLett.105.181801} {\bibfield
   {journal} {\bibinfo  {journal} {Phys. Rev. Lett.}\ }\textbf {\bibinfo
  {volume} {105}},\ \bibinfo {pages} {181801} (\bibinfo {year}
  {2010}{\natexlab{a}})}\BibitemShut {NoStop}%
\bibitem [{\citenamefont {Abe}\ \emph {et~al.}(2011{\natexlab{a}})\citenamefont
  {Abe} \emph {et~al.}}]{PhysRevLett.107.041801}%
  \BibitemOpen
  \bibfield  {author} {\bibinfo {author} {\bibfnamefont {K.}~\bibnamefont
  {Abe}} \emph {et~al.} (\bibinfo {collaboration} {T2K Collaboration}),\ }\href
  {\doibase 10.1103/PhysRevLett.107.041801} {\bibfield  {journal} {\bibinfo
  {journal} {Phys. Rev. Lett.}\ }\textbf {\bibinfo {volume} {107}},\ \bibinfo
  {pages} {041801} (\bibinfo {year} {2011}{\natexlab{a}})}\BibitemShut
  {NoStop}%
\bibitem [{\citenamefont {Agafonova}\ \emph {et~al.}(2010)\citenamefont
  {Agafonova} \emph {et~al.}}]{Agafonova2010138}%
  \BibitemOpen
  \bibfield  {author} {\bibinfo {author} {\bibfnamefont {N.}~\bibnamefont
  {Agafonova}} \emph {et~al.} (\bibinfo {collaboration} {OPERA
  Collaboration}),\ }\href {\doibase 10.1016/j.physletb.2010.06.022} {\bibfield
   {journal} {\bibinfo  {journal} {Phys. Lett. B}\ }\textbf {\bibinfo {volume}
  {691}},\ \bibinfo {pages} {138 } (\bibinfo {year} {2010})}\BibitemShut
  {NoStop}%
\bibitem [{\citenamefont {Abe}\ \emph {et~al.}(2013{\natexlab{a}})\citenamefont
  {Abe} \emph {et~al.}}]{PhysRevLett.110.181802}%
  \BibitemOpen
  \bibfield  {author} {\bibinfo {author} {\bibfnamefont {K.}~\bibnamefont
  {Abe}} \emph {et~al.} (\bibinfo {collaboration} {Super-Kamiokande
  Collaboration}),\ }\href {\doibase 10.1103/PhysRevLett.110.181802} {\bibfield
   {journal} {\bibinfo  {journal} {Phys. Rev. Lett.}\ }\textbf {\bibinfo
  {volume} {110}},\ \bibinfo {pages} {181802} (\bibinfo {year}
  {2013}{\natexlab{a}})}\BibitemShut {NoStop}%
\bibitem [{\citenamefont {Apollonio}\ \emph {et~al.}(2003)\citenamefont
  {Apollonio} \emph {et~al.}}]{Apollonio:2002gd}%
  \BibitemOpen
  \bibfield  {author} {\bibinfo {author} {\bibfnamefont {M.}~\bibnamefont
  {Apollonio}} \emph {et~al.} (\bibinfo {collaboration} {CHOOZ
  Collaboration}),\ }\href {\doibase 10.1140/epjc/s2002-01127-9} {\bibfield
  {journal} {\bibinfo  {journal} {Eur. Phys. J. C}\ }\textbf {\bibinfo {volume}
  {27}},\ \bibinfo {pages} {331} (\bibinfo {year} {2003})}\BibitemShut
  {NoStop}%
\bibitem [{\citenamefont {Adamson}\ \emph
  {et~al.}(2011{\natexlab{b}})\citenamefont {Adamson} \emph
  {et~al.}}]{Adamson:2011qu}%
  \BibitemOpen
  \bibfield  {author} {\bibinfo {author} {\bibfnamefont {P.}~\bibnamefont
  {Adamson}} \emph {et~al.} (\bibinfo {collaboration} {MINOS Collaboration}),\
  }\href {\doibase 10.1103/PhysRevLett.107.181802} {\bibfield  {journal}
  {\bibinfo  {journal} {Phys. Rev. Lett.}\ }\textbf {\bibinfo {volume} {107}},\
  \bibinfo {pages} {181802} (\bibinfo {year} {2011}{\natexlab{b}})}\BibitemShut
  {NoStop}%
\bibitem [{\citenamefont {Adamson}\ \emph {et~al.}(2013)\citenamefont {Adamson}
  \emph {et~al.}}]{PhysRevLett.110.171801}%
  \BibitemOpen
  \bibfield  {author} {\bibinfo {author} {\bibfnamefont {P.}~\bibnamefont
  {Adamson}} \emph {et~al.} (\bibinfo {collaboration} {MINOS Collaboration}),\
  }\href {\doibase 10.1103/PhysRevLett.110.171801} {\bibfield  {journal}
  {\bibinfo  {journal} {Phys. Rev. Lett.}\ }\textbf {\bibinfo {volume} {110}},\
  \bibinfo {pages} {171801} (\bibinfo {year} {2013})}\BibitemShut {NoStop}%
\bibitem [{\citenamefont {Beringer}\ \emph {et~al.}(2012)\citenamefont
  {Beringer} \emph {et~al.}}]{PhysRevD.86.010001}%
  \BibitemOpen
  \bibfield  {author} {\bibinfo {author} {\bibfnamefont {J.}~\bibnamefont
  {Beringer}} \emph {et~al.} (\bibinfo {collaboration} {Particle Data Group}),\
  }\href {\doibase 10.1103/PhysRevD.86.010001} {\bibfield  {journal} {\bibinfo
  {journal} {Phys. Rev. D}\ }\textbf {\bibinfo {volume} {86}},\ \bibinfo
  {pages} {010001} (\bibinfo {year} {2012})}\BibitemShut {NoStop}%
\bibitem [{\citenamefont {An}\ \emph {et~al.}(2012)\citenamefont {An} \emph
  {et~al.}}]{PhysRevLett.108.171803}%
  \BibitemOpen
  \bibfield  {author} {\bibinfo {author} {\bibfnamefont {F.~P.}\ \bibnamefont
  {An}} \emph {et~al.} (\bibinfo {collaboration} {Daya Bay Collaboration}),\
  }\href {\doibase 10.1103/PhysRevLett.108.171803} {\bibfield  {journal}
  {\bibinfo  {journal} {Phys. Rev. Lett.}\ }\textbf {\bibinfo {volume} {108}},\
  \bibinfo {pages} {171803} (\bibinfo {year} {2012})}\BibitemShut {NoStop}%
\bibitem [{\citenamefont {Ahn}\ \emph {et~al.}(2012)\citenamefont {Ahn} \emph
  {et~al.}}]{PhysRevLett.108.191802}%
  \BibitemOpen
  \bibfield  {author} {\bibinfo {author} {\bibfnamefont {J.~K.}\ \bibnamefont
  {Ahn}} \emph {et~al.} (\bibinfo {collaboration} {RENO Collaboration}),\
  }\href {\doibase 10.1103/PhysRevLett.108.191802} {\bibfield  {journal}
  {\bibinfo  {journal} {Phys. Rev. Lett.}\ }\textbf {\bibinfo {volume} {108}},\
  \bibinfo {pages} {191802} (\bibinfo {year} {2012})}\BibitemShut {NoStop}%
\bibitem [{\citenamefont {Abe}\ \emph {et~al.}(2012{\natexlab{b}})\citenamefont
  {Abe} \emph {et~al.}}]{PhysRevLett.108.131801}%
  \BibitemOpen
  \bibfield  {author} {\bibinfo {author} {\bibfnamefont {Y.}~\bibnamefont
  {Abe}} \emph {et~al.} (\bibinfo {collaboration} {Double Chooz
  Collaboration}),\ }\href {\doibase 10.1103/PhysRevLett.108.131801} {\bibfield
   {journal} {\bibinfo  {journal} {Phys. Rev. Lett.}\ }\textbf {\bibinfo
  {volume} {108}},\ \bibinfo {pages} {131801} (\bibinfo {year}
  {2012}{\natexlab{b}})}\BibitemShut {NoStop}%
\bibitem [{\citenamefont {Freund}(2001)}]{Freund:2001pn}%
  \BibitemOpen
  \bibfield  {author} {\bibinfo {author} {\bibfnamefont {M.}~\bibnamefont
  {Freund}},\ }\href {\doibase 10.1103/PhysRevD.64.053003} {\bibfield
  {journal} {\bibinfo  {journal} {Phys.Rev.}\ }\textbf {\bibinfo {volume}
  {D64}},\ \bibinfo {pages} {053003} (\bibinfo {year} {2001})},\ \Eprint
  {http://arxiv.org/abs/hep-ph/0103300} {arXiv:hep-ph/0103300 [hep-ph]}
  \BibitemShut {NoStop}%
\bibitem [{\citenamefont {Abe}\ \emph {et~al.}(2011{\natexlab{b}})\citenamefont
  {Abe} \emph {et~al.}}]{Abe:2011ks}%
  \BibitemOpen
  \bibfield  {author} {\bibinfo {author} {\bibfnamefont {K.}~\bibnamefont
  {Abe}} \emph {et~al.} (\bibinfo {collaboration} {T2K Collaboration}),\ }\href
  {\doibase 10.1016/j.nima.2011.06.067} {\bibfield  {journal} {\bibinfo
  {journal} {Nucl. Instrum. Methods Phys. Res., Sect. A}\ }\textbf {\bibinfo
  {volume} {659}},\ \bibinfo {pages} {106} (\bibinfo {year}
  {2011}{\natexlab{b}})}\BibitemShut {NoStop}%
\bibitem [{\citenamefont {Beavis}\ \emph {et~al.}(1995)\citenamefont {Beavis}
  \emph {et~al.}}]{bnl_offaxis}%
  \BibitemOpen
  \bibfield  {author} {\bibinfo {author} {\bibfnamefont {D.}~\bibnamefont
  {Beavis}} \emph {et~al.},\ }\href@noop {} {}\bibinfo {type} {Tech. Rep.}\
  \bibinfo {number} {52459}\ (\bibinfo  {institution} {BNL},\ \bibinfo {year}
  {1995})\BibitemShut {NoStop}%
\bibitem [{\citenamefont {Mann}()}]{Mann:1993zk}%
  \BibitemOpen
  \bibfield  {author} {\bibinfo {author} {\bibfnamefont {A.~K.}\ \bibnamefont
  {Mann}},\ }\href@noop {} {\ }\bibinfo {note} {{Prepared for 3rd NESTOR
  Workshop, Pylos, Greece, 19-21 Oct., 1993}}\BibitemShut {NoStop}%
\bibitem [{\citenamefont {Helmer}()}]{Helmer:1994ac}%
  \BibitemOpen
  \bibfield  {author} {\bibinfo {author} {\bibfnamefont {R.~L.}\ \bibnamefont
  {Helmer}},\ }\href@noop {} {\ }\bibinfo {note} {{Proc.\ 9th Lake Louise
  Winter Institute, Lake Louise, Canada, 1994, (World Scientific, 1995; eds. A.
  Astbury et al.), p. 291}}\BibitemShut {NoStop}%
\bibitem [{\citenamefont {Otani}\ \emph {et~al.}(2010)\citenamefont {Otani}
  \emph {et~al.}}]{Otani2010368}%
  \BibitemOpen
  \bibfield  {author} {\bibinfo {author} {\bibfnamefont {M.}~\bibnamefont
  {Otani}} \emph {et~al.},\ }\href {\doibase 10.1016/j.nima.2010.02.251}
  {\bibfield  {journal} {\bibinfo  {journal} {Nucl. Instrum. Methods Phys. Res.
  Sect. A}\ }\textbf {\bibinfo {volume} {623}},\ \bibinfo {pages} {368 }
  (\bibinfo {year} {2010})}\BibitemShut {NoStop}%
\bibitem [{\citenamefont {Abe}\ \emph {et~al.}(2012{\natexlab{c}})\citenamefont
  {Abe} \emph {et~al.}}]{Abe2012}%
  \BibitemOpen
  \bibfield  {author} {\bibinfo {author} {\bibfnamefont {K.}~\bibnamefont
  {Abe}} \emph {et~al.} (\bibinfo {collaboration} {T2K Collaboration}),\ }\href
  {\doibase 10.1016/j.nima.2012.03.023} {\bibfield  {journal} {\bibinfo
  {journal} {Nucl. Instrum. Methods Phys. Res., Sect. A}\ }\textbf {\bibinfo
  {volume} {694}},\ \bibinfo {pages} {211} (\bibinfo {year}
  {2012}{\natexlab{c}})}\BibitemShut {NoStop}%
\bibitem [{\citenamefont {Aoki}\ \emph {et~al.}(2013)\citenamefont {Aoki},
  \citenamefont {Barr}, \citenamefont {Batkiewicz}, \citenamefont {Blocki},
  \citenamefont {Brinson} \emph {et~al.}}]{Aoki:2012mf}%
  \BibitemOpen
  \bibfield  {author} {\bibinfo {author} {\bibfnamefont {S.}~\bibnamefont
  {Aoki}}, \bibinfo {author} {\bibfnamefont {G.}~\bibnamefont {Barr}}, \bibinfo
  {author} {\bibfnamefont {M.}~\bibnamefont {Batkiewicz}}, \bibinfo {author}
  {\bibfnamefont {J.}~\bibnamefont {Blocki}}, \bibinfo {author} {\bibfnamefont
  {J.}~\bibnamefont {Brinson}},  \emph {et~al.},\ }\href {\doibase
  10.1016/j.nima.2012.10.001} {\bibfield  {journal} {\bibinfo  {journal}
  {Nucl.Instrum.Meth.}\ }\textbf {\bibinfo {volume} {A698}},\ \bibinfo {pages}
  {135} (\bibinfo {year} {2013})},\ \Eprint {http://arxiv.org/abs/1206.3553}
  {arXiv:1206.3553 [physics.ins-det]} \BibitemShut {NoStop}%
\bibitem [{\citenamefont {Assylbekov}\ \emph {et~al.}(2012)\citenamefont
  {Assylbekov} \emph {et~al.}}]{Assylbekov201248}%
  \BibitemOpen
  \bibfield  {author} {\bibinfo {author} {\bibfnamefont {S.}~\bibnamefont
  {Assylbekov}} \emph {et~al.},\ }\href {\doibase 10.1016/j.nima.2012.05.028}
  {\bibfield  {journal} {\bibinfo  {journal} {Nucl.Instrum.Meth.}\ }\textbf
  {\bibinfo {volume} {A686}},\ \bibinfo {pages} {48 } (\bibinfo {year}
  {2012})}\BibitemShut {NoStop}%
\bibitem [{\citenamefont {Amaudruz}\ \emph {et~al.}(2012)\citenamefont
  {Amaudruz} \emph {et~al.}}]{Amaudruz:2012pe}%
  \BibitemOpen
  \bibfield  {author} {\bibinfo {author} {\bibfnamefont {P.}~\bibnamefont
  {Amaudruz}} \emph {et~al.} (\bibinfo {collaboration} {T2K ND280 FGD
  Collaboration}),\ }\href {\doibase 10.1016/j.nima.2012.08.020} {\bibfield
  {journal} {\bibinfo  {journal} {Nucl. Instrum. Methods Phys. Res., Sect. A}\
  }\textbf {\bibinfo {volume} {696}},\ \bibinfo {pages} {1} (\bibinfo {year}
  {2012})}\BibitemShut {NoStop}%
\bibitem [{\citenamefont {Abgrall}\ \emph
  {et~al.}(2011{\natexlab{a}})\citenamefont {Abgrall} \emph
  {et~al.}}]{Abgrall:2010hi}%
  \BibitemOpen
  \bibfield  {author} {\bibinfo {author} {\bibfnamefont {N.}~\bibnamefont
  {Abgrall}} \emph {et~al.} (\bibinfo {collaboration} {T2K ND280 TPC
  Collaboration}),\ }\href {\doibase 10.1016/j.nima.2011.02.036} {\bibfield
  {journal} {\bibinfo  {journal} {Nucl. Instrum. Methods Phys. Res., Sect. A}\
  }\textbf {\bibinfo {volume} {637}},\ \bibinfo {pages} {25} (\bibinfo {year}
  {2011}{\natexlab{a}})}\BibitemShut {NoStop}%
\bibitem [{\citenamefont {Fukuda}\ \emph {et~al.}(2003)\citenamefont {Fukuda}
  \emph {et~al.}}]{Fukuda:2002uc}%
  \BibitemOpen
  \bibfield  {author} {\bibinfo {author} {\bibfnamefont {Y.}~\bibnamefont
  {Fukuda}} \emph {et~al.} (\bibinfo {collaboration} {Super-Kamiokande
  Collaboration}),\ }\href {\doibase 10.1016/S0168-9002(03)00425-X} {\bibfield
  {journal} {\bibinfo  {journal} {Nucl. Instrum. Methods Phys. Res., Sect. A}\
  }\textbf {\bibinfo {volume} {501}},\ \bibinfo {pages} {418} (\bibinfo {year}
  {2003})}\BibitemShut {NoStop}%
\bibitem [{\citenamefont {Allan}\ and\ \citenamefont
  {Weiss}(1980)}]{CommonView}%
  \BibitemOpen
  \bibfield  {author} {\bibinfo {author} {\bibfnamefont {D.~W.}\ \bibnamefont
  {Allan}}\ and\ \bibinfo {author} {\bibfnamefont {M.~A.}\ \bibnamefont
  {Weiss}},\ }\href@noop {} {\bibfield  {journal} {\bibinfo  {journal}
  {Proceedings of the 34th Annual Symposium on Frequency Control}\ ,\ \bibinfo
  {pages} {334}} (\bibinfo {year} {1980})}\BibitemShut {NoStop}%
\bibitem [{\citenamefont {Abe}\ \emph {et~al.}(2013{\natexlab{b}})\citenamefont
  {Abe} \emph {et~al.}}]{PhysRevD.87.012001}%
  \BibitemOpen
  \bibfield  {author} {\bibinfo {author} {\bibfnamefont {K.}~\bibnamefont
  {Abe}} \emph {et~al.} (\bibinfo {collaboration} {T2K Collaboration}),\ }\href
  {\doibase 10.1103/PhysRevD.87.012001} {\bibfield  {journal} {\bibinfo
  {journal} {Phys. Rev. D}\ }\textbf {\bibinfo {volume} {87}},\ \bibinfo
  {pages} {012001} (\bibinfo {year} {2013}{\natexlab{b}})}\BibitemShut
  {NoStop}%
\bibitem [{\citenamefont {Abgrall}\ \emph
  {et~al.}(2011{\natexlab{b}})\citenamefont {Abgrall} \emph
  {et~al.}}]{Abgrall:2011ae}%
  \BibitemOpen
  \bibfield  {author} {\bibinfo {author} {\bibfnamefont {N.}~\bibnamefont
  {Abgrall}} \emph {et~al.} (\bibinfo {collaboration} {NA61/SHINE
  Collaboration}),\ }\href {\doibase 10.1103/PhysRevC.84.034604} {\bibfield
  {journal} {\bibinfo  {journal} {Phys. Rev. C}\ }\textbf {\bibinfo {volume}
  {84}},\ \bibinfo {pages} {034604} (\bibinfo {year}
  {2011}{\natexlab{b}})}\BibitemShut {NoStop}%
\bibitem [{\citenamefont {Abgrall}\ \emph {et~al.}(2012)\citenamefont {Abgrall}
  \emph {et~al.}}]{Abgrall:2011ts}%
  \BibitemOpen
  \bibfield  {author} {\bibinfo {author} {\bibfnamefont {N.}~\bibnamefont
  {Abgrall}} \emph {et~al.} (\bibinfo {collaboration} {NA61/SHINE
  Collaboration}),\ }\href {\doibase 10.1103/PhysRevC.85.035210} {\bibfield
  {journal} {\bibinfo  {journal} {Phys. Rev. C}\ }\textbf {\bibinfo {volume}
  {85}},\ \bibinfo {pages} {035210} (\bibinfo {year} {2012})}\BibitemShut
  {NoStop}%
\bibitem [{\citenamefont {Ferrari}\ \emph {et~al.}()\citenamefont {Ferrari},
  \citenamefont {Sala}, \citenamefont {Fasso},\ and\ \citenamefont
  {Ranft}}]{Ferrari:2005zk}%
  \BibitemOpen
  \bibfield  {author} {\bibinfo {author} {\bibfnamefont {A.}~\bibnamefont
  {Ferrari}}, \bibinfo {author} {\bibfnamefont {P.~R.}\ \bibnamefont {Sala}},
  \bibinfo {author} {\bibfnamefont {A.}~\bibnamefont {Fasso}}, \ and\ \bibinfo
  {author} {\bibfnamefont {J.}~\bibnamefont {Ranft}},\ }\href@noop {} {\bibinfo
   {journal} {CERN-2005-010, SLAC-R-773, INFN-TC-05-11}\ }\BibitemShut
  {NoStop}%
\bibitem [{\citenamefont {Battistoni}\ \emph {et~al.}(2007)\citenamefont
  {Battistoni}, \citenamefont {Muraro}, \citenamefont {Sala}, \citenamefont
  {Cerutti}, \citenamefont {Ferrari} \emph {et~al.}}]{Battistoni:2007zzb}%
  \BibitemOpen
\bibfield  {journal} {  }\bibfield  {author} {\bibinfo {author} {\bibfnamefont
  {G.}~\bibnamefont {Battistoni}}, \bibinfo {author} {\bibfnamefont
  {S.}~\bibnamefont {Muraro}}, \bibinfo {author} {\bibfnamefont {P.~R.}\
  \bibnamefont {Sala}}, \bibinfo {author} {\bibfnamefont {F.}~\bibnamefont
  {Cerutti}}, \bibinfo {author} {\bibfnamefont {A.}~\bibnamefont {Ferrari}},
  \emph {et~al.},\ }\href {\doibase 10.1063/1.2720455} {\bibfield  {journal}
  {\bibinfo  {journal} {AIP Conf.Proc.}\ }\textbf {\bibinfo {volume} {896}},\
  \bibinfo {pages} {31} (\bibinfo {year} {2007})},\ \bibinfo {note} {we used
  FLUKA2008, which was the latest version at the time of this study. A new
  version, FLUKA2011, has been already released now and the comparison with
  data would be different.}\BibitemShut {Stop}%
\bibitem [{\citenamefont {Brun}\ \emph {et~al.}(1994)\citenamefont {Brun},
  \citenamefont {Carminati},\ and\ \citenamefont {Giani}}]{GEANT3}%
  \BibitemOpen
  \bibfield  {author} {\bibinfo {author} {\bibfnamefont {R.}~\bibnamefont
  {Brun}}, \bibinfo {author} {\bibfnamefont {F.}~\bibnamefont {Carminati}}, \
  and\ \bibinfo {author} {\bibfnamefont {S.}~\bibnamefont {Giani}},\
  }\href@noop {} {\bibfield  {journal} {\bibinfo  {journal} {CERN-W5013}\ }
  (\bibinfo {year} {1994})}\BibitemShut {NoStop}%
\bibitem [{\citenamefont {Zeitnitz}\ and\ \citenamefont
  {Gabriel}(1993)}]{GCALOR}%
  \BibitemOpen
  \bibfield  {author} {\bibinfo {author} {\bibfnamefont {C.}~\bibnamefont
  {Zeitnitz}}\ and\ \bibinfo {author} {\bibfnamefont {T.~A.}\ \bibnamefont
  {Gabriel}},\ }\href@noop {} {\bibfield  {journal} {\bibinfo  {journal} {Proc.
  of International Conference on Calorimetry in High Energy Physics}\ }
  (\bibinfo {year} {1993})}\BibitemShut {NoStop}%
\bibitem [{\citenamefont {Bhadra}\ \emph {et~al.}(2013)\citenamefont {Bhadra}
  \emph {et~al.}}]{Bhadra201345}%
  \BibitemOpen
  \bibfield  {author} {\bibinfo {author} {\bibfnamefont {S.}~\bibnamefont
  {Bhadra}} \emph {et~al.},\ }\href {\doibase 10.1016/j.nima.2012.11.044}
  {\bibfield  {journal} {\bibinfo  {journal} {Nucl. Instrum. Methods Phys.
  Res., Sect. A}\ }\textbf {\bibinfo {volume} {703}},\ \bibinfo {pages} {45 }
  (\bibinfo {year} {2013})}\BibitemShut {NoStop}%
\bibitem [{\citenamefont {McDonald}\ and\ \citenamefont
  {Russell}(1989)}]{McDonald:1988nv}%
  \BibitemOpen
  \bibfield  {author} {\bibinfo {author} {\bibfnamefont {K.}~\bibnamefont
  {McDonald}}\ and\ \bibinfo {author} {\bibfnamefont {D.}~\bibnamefont
  {Russell}},\ }\href {\doibase 10.1007/BFb0018284} {\bibfield  {journal}
  {\bibinfo  {journal} {Lect. Notes Phys.}\ }\textbf {\bibinfo {volume}
  {343}},\ \bibinfo {pages} {122} (\bibinfo {year} {1989})}\BibitemShut
  {NoStop}%
\bibitem [{\citenamefont {Eichten}\ \emph {et~al.}(1972)\citenamefont {Eichten}
  \emph {et~al.}}]{eichten}%
  \BibitemOpen
  \bibfield  {author} {\bibinfo {author} {\bibfnamefont {T.}~\bibnamefont
  {Eichten}} \emph {et~al.},\ }\href@noop {} {\bibfield  {journal} {\bibinfo
  {journal} {Nucl. Phys. B}\ }\textbf {\bibinfo {volume} {44}} (\bibinfo {year}
  {1972})}\BibitemShut {NoStop}%
\bibitem [{\citenamefont {Allaby}\ \emph {et~al.}(1970)\citenamefont {Allaby}
  \emph {et~al.}}]{allaby}%
  \BibitemOpen
  \bibfield  {author} {\bibinfo {author} {\bibfnamefont {J.~V.}\ \bibnamefont
  {Allaby}} \emph {et~al.},\ }\href@noop {} {}\bibinfo {type} {Tech. Rep.}\
  \bibinfo {number} {70-12}\ (\bibinfo  {institution} {CERN},\ \bibinfo {year}
  {1970})\BibitemShut {NoStop}%
\bibitem [{\citenamefont {Chemakin}\ \emph {et~al.}(2008)\citenamefont
  {Chemakin} \emph {et~al.}}]{e910}%
  \BibitemOpen
  \bibfield  {author} {\bibinfo {author} {\bibfnamefont {I.}~\bibnamefont
  {Chemakin}} \emph {et~al.} (\bibinfo {collaboration} {E910 Collaboration}),\
  }\href {\doibase 10.1103/PhysRevC.77.049903, 10.1103/PhysRevC.77.015209}
  {\bibfield  {journal} {\bibinfo  {journal} {Phys. Rev. C}\ }\textbf {\bibinfo
  {volume} {77}},\ \bibinfo {pages} {015209} (\bibinfo {year}
  {2008})}\BibitemShut {NoStop}%
\bibitem [{\citenamefont {Abrams}\ \emph {et~al.}(1970)\citenamefont {Abrams}
  \emph {et~al.}}]{Abrams}%
  \BibitemOpen
  \bibfield  {author} {\bibinfo {author} {\bibfnamefont {R.~J.}\ \bibnamefont
  {Abrams}} \emph {et~al.},\ }\href {\doibase 10.1103/PhysRevD.1.1917}
  {\bibfield  {journal} {\bibinfo  {journal} {Phys. Rev. D}\ }\textbf {\bibinfo
  {volume} {1}},\ \bibinfo {pages} {1917} (\bibinfo {year} {1970})}\BibitemShut
  {NoStop}%
\bibitem [{\citenamefont {Allaby}\ \emph {et~al.}(1969)\citenamefont {Allaby}
  \emph {et~al.}}]{Allaby1969500}%
  \BibitemOpen
  \bibfield  {author} {\bibinfo {author} {\bibfnamefont {J.}~\bibnamefont
  {Allaby}} \emph {et~al.},\ }\href {\doibase 10.1016/0370-2693(69)90184-1}
  {\bibfield  {journal} {\bibinfo  {journal} {Phys. Lett. B}\ }\textbf
  {\bibinfo {volume} {30}},\ \bibinfo {pages} {500 } (\bibinfo {year}
  {1969})}\BibitemShut {NoStop}%
\bibitem [{\citenamefont {Allardyce}\ \emph {et~al.}(1973)\citenamefont
  {Allardyce} \emph {et~al.}}]{Allardyce:1973ce}%
  \BibitemOpen
  \bibfield  {author} {\bibinfo {author} {\bibfnamefont {B.}~\bibnamefont
  {Allardyce}} \emph {et~al.},\ }\href {\doibase 10.1016/0375-9474(73)90049-3}
  {\bibfield  {journal} {\bibinfo  {journal} {Nuclear Physics A}\ }\textbf
  {\bibinfo {volume} {209}},\ \bibinfo {pages} {1 } (\bibinfo {year}
  {1973})}\BibitemShut {NoStop}%
\bibitem [{\citenamefont {Bellettini}\ \emph {et~al.}(1966)\citenamefont
  {Bellettini} \emph {et~al.}}]{Bellettini:1966zz}%
  \BibitemOpen
  \bibfield  {author} {\bibinfo {author} {\bibfnamefont {G.}~\bibnamefont
  {Bellettini}} \emph {et~al.},\ }\href {\doibase 10.1016/0029-5582(66)90267-7}
  {\bibfield  {journal} {\bibinfo  {journal} {Nucl.Phys.}\ }\textbf {\bibinfo
  {volume} {79}},\ \bibinfo {pages} {609} (\bibinfo {year} {1966})}\BibitemShut
  {NoStop}%
\bibitem [{\citenamefont {Bobchenko}\ \emph {et~al.}(1979)\citenamefont
  {Bobchenko} \emph {et~al.}}]{Bobchenko}%
  \BibitemOpen
  \bibfield  {author} {\bibinfo {author} {\bibfnamefont {B.~M.}\ \bibnamefont
  {Bobchenko}} \emph {et~al.},\ }\href@noop {} {\bibfield  {journal} {\bibinfo
  {journal} {Sov. J. Nucl. Phys.}\ }\textbf {\bibinfo {volume} {30}},\ \bibinfo
  {pages} {805} (\bibinfo {year} {1979})}\BibitemShut {NoStop}%
\bibitem [{\citenamefont {Carroll}\ \emph {et~al.}(1979)\citenamefont {Carroll}
  \emph {et~al.}}]{Carroll}%
  \BibitemOpen
  \bibfield  {author} {\bibinfo {author} {\bibfnamefont {A.}~\bibnamefont
  {Carroll}} \emph {et~al.},\ }\href {\doibase 10.1016/0370-2693(79)90226-0}
  {\bibfield  {journal} {\bibinfo  {journal} {Phys. Lett. B}\ }\textbf
  {\bibinfo {volume} {80}},\ \bibinfo {pages} {319 } (\bibinfo {year}
  {1979})}\BibitemShut {NoStop}%
\bibitem [{\citenamefont {Cronin}\ \emph {et~al.}(1957)\citenamefont {Cronin},
  \citenamefont {Cool},\ and\ \citenamefont {Abashian}}]{Cronin}%
  \BibitemOpen
  \bibfield  {author} {\bibinfo {author} {\bibfnamefont {J.~W.}\ \bibnamefont
  {Cronin}}, \bibinfo {author} {\bibfnamefont {R.}~\bibnamefont {Cool}}, \ and\
  \bibinfo {author} {\bibfnamefont {A.}~\bibnamefont {Abashian}},\ }\href
  {\doibase 10.1103/PhysRev.107.1121} {\bibfield  {journal} {\bibinfo
  {journal} {Phys. Rev.}\ }\textbf {\bibinfo {volume} {107}},\ \bibinfo {pages}
  {1121} (\bibinfo {year} {1957})}\BibitemShut {NoStop}%
\bibitem [{\citenamefont {Chen}\ \emph {et~al.}(1955)\citenamefont {Chen},
  \citenamefont {Leavitt},\ and\ \citenamefont {Shapiro}}]{Chen}%
  \BibitemOpen
  \bibfield  {author} {\bibinfo {author} {\bibfnamefont {F.~F.}\ \bibnamefont
  {Chen}}, \bibinfo {author} {\bibfnamefont {C.~P.}\ \bibnamefont {Leavitt}}, \
  and\ \bibinfo {author} {\bibfnamefont {A.~M.}\ \bibnamefont {Shapiro}},\
  }\href {\doibase 10.1103/PhysRev.99.857} {\bibfield  {journal} {\bibinfo
  {journal} {Phys. Rev.}\ }\textbf {\bibinfo {volume} {99}},\ \bibinfo {pages}
  {857} (\bibinfo {year} {1955})}\BibitemShut {NoStop}%
\bibitem [{\citenamefont {Denisov}\ \emph {et~al.}(1973)\citenamefont {Denisov}
  \emph {et~al.}}]{Denisov}%
  \BibitemOpen
  \bibfield  {author} {\bibinfo {author} {\bibfnamefont {S.}~\bibnamefont
  {Denisov}} \emph {et~al.},\ }\href {\doibase 10.1016/0550-3213(73)90351-9}
  {\bibfield  {journal} {\bibinfo  {journal} {Nuclear Physics B}\ }\textbf
  {\bibinfo {volume} {61}},\ \bibinfo {pages} {62 } (\bibinfo {year}
  {1973})}\BibitemShut {NoStop}%
\bibitem [{\citenamefont {Longo}\ and\ \citenamefont {Moyer}(1962)}]{Longo}%
  \BibitemOpen
  \bibfield  {author} {\bibinfo {author} {\bibfnamefont {M.~J.}\ \bibnamefont
  {Longo}}\ and\ \bibinfo {author} {\bibfnamefont {B.~J.}\ \bibnamefont
  {Moyer}},\ }\href {\doibase 10.1103/PhysRev.125.701} {\bibfield  {journal}
  {\bibinfo  {journal} {Phys. Rev.}\ }\textbf {\bibinfo {volume} {125}},\
  \bibinfo {pages} {701} (\bibinfo {year} {1962})}\BibitemShut {NoStop}%
\bibitem [{\citenamefont {Vlasov}\ \emph {et~al.}(1978)\citenamefont {Vlasov}
  \emph {et~al.}}]{Vlasov}%
  \BibitemOpen
  \bibfield  {author} {\bibinfo {author} {\bibfnamefont {A.~V.}\ \bibnamefont
  {Vlasov}} \emph {et~al.},\ }\href@noop {} {\bibfield  {journal} {\bibinfo
  {journal} {Sov. J. Nucl. Phys.}\ }\textbf {\bibinfo {volume} {27}},\ \bibinfo
  {pages} {222} (\bibinfo {year} {1978})}\BibitemShut {NoStop}%
\bibitem [{\citenamefont {Feynman}(1969)}]{Feynman69}%
  \BibitemOpen
  \bibfield  {author} {\bibinfo {author} {\bibfnamefont {R.~P.}\ \bibnamefont
  {Feynman}},\ }\href {\doibase 10.1103/PhysRevLett.23.1415} {\bibfield
  {journal} {\bibinfo  {journal} {Phys. Rev. Lett.}\ }\textbf {\bibinfo
  {volume} {23}},\ \bibinfo {pages} {1415} (\bibinfo {year}
  {1969})}\BibitemShut {NoStop}%
\bibitem [{\citenamefont {Hayato}(2009)}]{Hayato:2009}%
  \BibitemOpen
  \bibfield  {author} {\bibinfo {author} {\bibfnamefont {Y.}~\bibnamefont
  {Hayato}},\ }\href@noop {} {\bibfield  {journal} {\bibinfo  {journal} {Acta
  Phys. Pol. B}\ }\textbf {\bibinfo {volume} {40}},\ \bibinfo {pages} {2477}
  (\bibinfo {year} {2009})}\BibitemShut {NoStop}%
\bibitem [{\citenamefont {Llewellyn~Smith}(1972)}]{LlewellynSmith:1972}%
  \BibitemOpen
  \bibfield  {author} {\bibinfo {author} {\bibfnamefont {C.}~\bibnamefont
  {Llewellyn~Smith}},\ }\href {\doibase 10.1016/0370-1573(72)90010-5}
  {\bibfield  {journal} {\bibinfo  {journal} {Phys. Rept.}\ }\textbf {\bibinfo
  {volume} {3}},\ \bibinfo {pages} {261} (\bibinfo {year} {1972})}\BibitemShut
  {NoStop}%
\bibitem [{\citenamefont {Smith}\ and\ \citenamefont
  {Moniz}(1972)}]{SmithMoniz:1972}%
  \BibitemOpen
  \bibfield  {author} {\bibinfo {author} {\bibfnamefont {R.}~\bibnamefont
  {Smith}}\ and\ \bibinfo {author} {\bibfnamefont {E.}~\bibnamefont {Moniz}},\
  }\href {\doibase 10.1016/0550-3213(72)90040-5} {\bibfield  {journal}
  {\bibinfo  {journal} {Nucl. Phys. B}\ }\textbf {\bibinfo {volume} {43}},\
  \bibinfo {pages} {605} (\bibinfo {year} {1972})}\BibitemShut {NoStop}%
\bibitem [{\citenamefont {Smith}\ and\ \citenamefont
  {Moniz}(1975)}]{SmithMonizErratum}%
  \BibitemOpen
  \bibfield  {author} {\bibinfo {author} {\bibfnamefont {R.}~\bibnamefont
  {Smith}}\ and\ \bibinfo {author} {\bibfnamefont {E.}~\bibnamefont {Moniz}},\
  }\href {\doibase 10.1016/0550-3213(75)90612-4} {\bibfield  {journal}
  {\bibinfo  {journal} {Nucl. Phys. B}\ }\textbf {\bibinfo {volume} {101}},\
  \bibinfo {pages} {547 } (\bibinfo {year} {1975})}\BibitemShut {NoStop}%
\bibitem [{\citenamefont {Rein}\ and\ \citenamefont
  {Sehgal}(1981)}]{ReinSehgal:1981}%
  \BibitemOpen
  \bibfield  {author} {\bibinfo {author} {\bibfnamefont {D.}~\bibnamefont
  {Rein}}\ and\ \bibinfo {author} {\bibfnamefont {L.~M.}\ \bibnamefont
  {Sehgal}},\ }\href {\doibase 10.1016/0003-4916(81)90242-6} {\bibfield
  {journal} {\bibinfo  {journal} {Annals Phys.}\ }\textbf {\bibinfo {volume}
  {133}},\ \bibinfo {pages} {79} (\bibinfo {year} {1981})}\BibitemShut
  {NoStop}%
\bibitem [{\citenamefont {Gluck}\ \emph {et~al.}(1998)\citenamefont {Gluck},
  \citenamefont {Reya},\ and\ \citenamefont {Vogt}}]{Gluck:1998xa}%
  \BibitemOpen
  \bibfield  {author} {\bibinfo {author} {\bibfnamefont {M.}~\bibnamefont
  {Gluck}}, \bibinfo {author} {\bibfnamefont {E.}~\bibnamefont {Reya}}, \ and\
  \bibinfo {author} {\bibfnamefont {A.}~\bibnamefont {Vogt}},\ }\href {\doibase
  10.1007/s100520050289} {\bibfield  {journal} {\bibinfo  {journal} {Eur. Phys.
  J. C}\ }\textbf {\bibinfo {volume} {5}},\ \bibinfo {pages} {461} (\bibinfo
  {year} {1998})}\BibitemShut {NoStop}%
\bibitem [{\citenamefont {Nakahata}\ \emph {et~al.}(1986)\citenamefont
  {Nakahata} \emph {et~al.}}]{Nakahata:1986zp}%
  \BibitemOpen
  \bibfield  {author} {\bibinfo {author} {\bibfnamefont {M.}~\bibnamefont
  {Nakahata}} \emph {et~al.} (\bibinfo {collaboration} {KAMIOKANDE
  Collaboration}),\ }\href {\doibase 10.1143/JPSJ.55.3786} {\bibfield
  {journal} {\bibinfo  {journal} {J. Phys. Soc. Jap.}\ }\textbf {\bibinfo
  {volume} {55}},\ \bibinfo {pages} {3786} (\bibinfo {year}
  {1986})}\BibitemShut {NoStop}%
\bibitem [{\citenamefont {Sjostrand}(1994)}]{Sjostrand:1993yb}%
  \BibitemOpen
  \bibfield  {author} {\bibinfo {author} {\bibfnamefont {T.}~\bibnamefont
  {Sjostrand}},\ }\href {\doibase 10.1016/0010-4655(94)90132-5} {\bibfield
  {journal} {\bibinfo  {journal} {Comput. Phys. Commun.}\ }\textbf {\bibinfo
  {volume} {82}},\ \bibinfo {pages} {74} (\bibinfo {year} {1994})}\BibitemShut
  {NoStop}%
\bibitem [{\citenamefont {Bodek}\ and\ \citenamefont
  {Yang}(2003)}]{Bodek:2003wd}%
  \BibitemOpen
  \bibfield  {author} {\bibinfo {author} {\bibfnamefont {A.}~\bibnamefont
  {Bodek}}\ and\ \bibinfo {author} {\bibfnamefont {U.}~\bibnamefont {Yang}},\
  }\href@noop {} {\bibfield  {journal} {\bibinfo  {journal} {Nucl. Phys. Proc.
  Suppl.}\ } (\bibinfo {year} {2003})}\BibitemShut {NoStop}%
\bibitem [{\citenamefont {Salcedo}\ \emph {et~al.}(1988)\citenamefont
  {Salcedo}, \citenamefont {Oset}, \citenamefont {Vicente-Vacas},\ and\
  \citenamefont {Garcia-Recio}}]{Salcedo:1988}%
  \BibitemOpen
  \bibfield  {author} {\bibinfo {author} {\bibfnamefont {L.}~\bibnamefont
  {Salcedo}}, \bibinfo {author} {\bibfnamefont {E.}~\bibnamefont {Oset}},
  \bibinfo {author} {\bibfnamefont {M.}~\bibnamefont {Vicente-Vacas}}, \ and\
  \bibinfo {author} {\bibfnamefont {C.}~\bibnamefont {Garcia-Recio}},\ }\href
  {\doibase 10.1016/0375-9474(88)90310-7} {\bibfield  {journal} {\bibinfo
  {journal} {Nucl. Phys. A}\ }\textbf {\bibinfo {volume} {484}},\ \bibinfo
  {pages} {557} (\bibinfo {year} {1988})}\BibitemShut {NoStop}%
\bibitem [{\citenamefont {de~Perio}(2011)}]{dePerio:2011zz}%
  \BibitemOpen
  \bibfield  {author} {\bibinfo {author} {\bibfnamefont {P.}~\bibnamefont
  {de~Perio}},\ }\href {\doibase 10.1063/1.3661590} {\bibfield  {journal}
  {\bibinfo  {journal} {AIP Conf.Proc.}\ }\textbf {\bibinfo {volume} {1405}},\
  \bibinfo {pages} {223} (\bibinfo {year} {2011})}\BibitemShut {NoStop}%
\bibitem [{\citenamefont {Ashery}\ \emph {et~al.}(1981)\citenamefont {Ashery}
  \emph {et~al.}}]{Ashery:1981tq}%
  \BibitemOpen
  \bibfield  {author} {\bibinfo {author} {\bibfnamefont {D.}~\bibnamefont
  {Ashery}} \emph {et~al.},\ }\href {\doibase 10.1103/PhysRevC.23.2173}
  {\bibfield  {journal} {\bibinfo  {journal} {Phys. Rev. C}\ }\textbf {\bibinfo
  {volume} {23}},\ \bibinfo {pages} {2173} (\bibinfo {year}
  {1981})}\BibitemShut {NoStop}%
\bibitem [{\citenamefont {Jones}\ \emph {et~al.}(1993)\citenamefont {Jones}
  \emph {et~al.}}]{PhysRevC.48.2800}%
  \BibitemOpen
  \bibfield  {author} {\bibinfo {author} {\bibfnamefont {M.~K.}\ \bibnamefont
  {Jones}} \emph {et~al.},\ }\href {\doibase 10.1103/PhysRevC.48.2800}
  {\bibfield  {journal} {\bibinfo  {journal} {Phys. Rev. C}\ }\textbf {\bibinfo
  {volume} {48}},\ \bibinfo {pages} {2800} (\bibinfo {year}
  {1993})}\BibitemShut {NoStop}%
\bibitem [{\citenamefont {Giannelli}\ \emph {et~al.}(2000)\citenamefont
  {Giannelli} \emph {et~al.}}]{PhysRevC.61.054615}%
  \BibitemOpen
  \bibfield  {author} {\bibinfo {author} {\bibfnamefont {R.~A.}\ \bibnamefont
  {Giannelli}} \emph {et~al.},\ }\href {\doibase 10.1103/PhysRevC.61.054615}
  {\bibfield  {journal} {\bibinfo  {journal} {Phys. Rev. C}\ }\textbf {\bibinfo
  {volume} {61}},\ \bibinfo {pages} {054615} (\bibinfo {year}
  {2000})}\BibitemShut {NoStop}%
\bibitem [{\citenamefont {Aguilar-Arevalo}\ \emph
  {et~al.}(2010{\natexlab{b}})\citenamefont {Aguilar-Arevalo} \emph
  {et~al.}}]{mb-ccqe}%
  \BibitemOpen
  \bibfield  {author} {\bibinfo {author} {\bibfnamefont {A.~A.}\ \bibnamefont
  {Aguilar-Arevalo}} \emph {et~al.} (\bibinfo {collaboration} {MiniBooNE
  Collaboration}),\ }\href {\doibase 10.1103/PhysRevD.81.092005} {\bibfield
  {journal} {\bibinfo  {journal} {Phys. Rev. D}\ }\textbf {\bibinfo {volume}
  {81}},\ \bibinfo {pages} {092005} (\bibinfo {year}
  {2010}{\natexlab{b}})}\BibitemShut {NoStop}%
\bibitem [{\citenamefont {Lyubushkin}\ \emph {et~al.}(2009)\citenamefont
  {Lyubushkin} \emph {et~al.}}]{Lyubushkin:2008pe}%
  \BibitemOpen
  \bibfield  {author} {\bibinfo {author} {\bibfnamefont {V.}~\bibnamefont
  {Lyubushkin}} \emph {et~al.} (\bibinfo {collaboration} {NOMAD
  Collaboration}),\ }\href {\doibase 10.1140/epjc/s10052-009-1113-0} {\bibfield
   {journal} {\bibinfo  {journal} {Eur. Phys. J. C}\ }\textbf {\bibinfo
  {volume} {63}},\ \bibinfo {pages} {355} (\bibinfo {year} {2009})}\BibitemShut
  {NoStop}%
\bibitem [{\citenamefont {Martini}\ \emph {et~al.}(2009)\citenamefont
  {Martini}, \citenamefont {Ericson}, \citenamefont {Chanfray},\ and\
  \citenamefont {Marteau}}]{Martini:2009}%
  \BibitemOpen
  \bibfield  {author} {\bibinfo {author} {\bibfnamefont {M.}~\bibnamefont
  {Martini}}, \bibinfo {author} {\bibfnamefont {M.}~\bibnamefont {Ericson}},
  \bibinfo {author} {\bibfnamefont {G.}~\bibnamefont {Chanfray}}, \ and\
  \bibinfo {author} {\bibfnamefont {J.}~\bibnamefont {Marteau}},\ }\href
  {\doibase 10.1103/PhysRevC.80.065501} {\bibfield  {journal} {\bibinfo
  {journal} {Phys. Rev. C}\ }\textbf {\bibinfo {volume} {80}},\ \bibinfo
  {pages} {065501} (\bibinfo {year} {2009})}\BibitemShut {NoStop}%
\bibitem [{\citenamefont {Martini}\ \emph {et~al.}(2010)\citenamefont
  {Martini}, \citenamefont {Ericson}, \citenamefont {Chanfray},\ and\
  \citenamefont {Marteau}}]{Martini:2010}%
  \BibitemOpen
  \bibfield  {author} {\bibinfo {author} {\bibfnamefont {M.}~\bibnamefont
  {Martini}}, \bibinfo {author} {\bibfnamefont {M.}~\bibnamefont {Ericson}},
  \bibinfo {author} {\bibfnamefont {G.}~\bibnamefont {Chanfray}}, \ and\
  \bibinfo {author} {\bibfnamefont {J.}~\bibnamefont {Marteau}},\ }\href
  {\doibase 10.1103/PhysRevC.81.045502} {\bibfield  {journal} {\bibinfo
  {journal} {Phys. Rev. C}\ }\textbf {\bibinfo {volume} {81}},\ \bibinfo
  {pages} {045502} (\bibinfo {year} {2010})}\BibitemShut {NoStop}%
\bibitem [{\citenamefont {Amaro}\ \emph {et~al.}(2011)\citenamefont {Amaro},
  \citenamefont {Barbaro}, \citenamefont {Caballero}, \citenamefont
  {Donnelly},\ and\ \citenamefont {Williamson}}]{Amaro:2011}%
  \BibitemOpen
  \bibfield  {author} {\bibinfo {author} {\bibfnamefont {J.}~\bibnamefont
  {Amaro}}, \bibinfo {author} {\bibfnamefont {M.}~\bibnamefont {Barbaro}},
  \bibinfo {author} {\bibfnamefont {J.}~\bibnamefont {Caballero}}, \bibinfo
  {author} {\bibfnamefont {T.}~\bibnamefont {Donnelly}}, \ and\ \bibinfo
  {author} {\bibfnamefont {C.}~\bibnamefont {Williamson}},\ }\href {\doibase
  10.1016/j.physletb.2010.12.007} {\bibfield  {journal} {\bibinfo  {journal}
  {Phys. Lett. B}\ }\textbf {\bibinfo {volume} {696}},\ \bibinfo {pages} {151 }
  (\bibinfo {year} {2011})}\BibitemShut {NoStop}%
\bibitem [{\citenamefont {Bodek}\ \emph {et~al.}(2011)\citenamefont {Bodek},
  \citenamefont {Budd},\ and\ \citenamefont {Christy}}]{Bodek:2011}%
  \BibitemOpen
  \bibfield  {author} {\bibinfo {author} {\bibfnamefont {A.}~\bibnamefont
  {Bodek}}, \bibinfo {author} {\bibfnamefont {H.}~\bibnamefont {Budd}}, \ and\
  \bibinfo {author} {\bibfnamefont {M.}~\bibnamefont {Christy}},\ }\href
  {http://dx.doi.org/10.1140/epjc/s10052-011-1726-y} {\bibfield  {journal}
  {\bibinfo  {journal} {Eur. Phys. J. C}\ }\textbf {\bibinfo {volume} {71}},\
  \bibinfo {pages} {1} (\bibinfo {year} {2011})}\BibitemShut {NoStop}%
\bibitem [{\citenamefont {Nieves}\ \emph {et~al.}(2012)\citenamefont {Nieves},
  \citenamefont {Simo},\ and\ \citenamefont {Vacas}}]{Nieves:2012}%
  \BibitemOpen
  \bibfield  {author} {\bibinfo {author} {\bibfnamefont {J.}~\bibnamefont
  {Nieves}}, \bibinfo {author} {\bibfnamefont {I.~R.}\ \bibnamefont {Simo}}, \
  and\ \bibinfo {author} {\bibfnamefont {M.~V.}\ \bibnamefont {Vacas}},\ }\href
  {\doibase 10.1016/j.physletb.2011.11.061} {\bibfield  {journal} {\bibinfo
  {journal} {Phys. Lett. B}\ }\textbf {\bibinfo {volume} {707}},\ \bibinfo
  {pages} {72 } (\bibinfo {year} {2012})}\BibitemShut {NoStop}%
\bibitem [{\citenamefont {Meloni}\ and\ \citenamefont
  {Martini}(2012)}]{Meloni:2012fq}%
  \BibitemOpen
  \bibfield  {author} {\bibinfo {author} {\bibfnamefont {D.}~\bibnamefont
  {Meloni}}\ and\ \bibinfo {author} {\bibfnamefont {M.}~\bibnamefont
  {Martini}},\ }\href {\doibase 10.1016/j.physletb.2012.08.007} {\bibfield
  {journal} {\bibinfo  {journal} {Phys. Lett. B}\ }\textbf {\bibinfo {volume}
  {716}},\ \bibinfo {pages} {186} (\bibinfo {year} {2012})}\BibitemShut
  {NoStop}%
\bibitem [{\citenamefont {Lalakulich}\ and\ \citenamefont
  {Mosel}(2012)}]{Lalakulich:2012hs}%
  \BibitemOpen
  \bibfield  {author} {\bibinfo {author} {\bibfnamefont {O.}~\bibnamefont
  {Lalakulich}}\ and\ \bibinfo {author} {\bibfnamefont {U.}~\bibnamefont
  {Mosel}},\ }\href {\doibase 10.1103/PhysRevC.86.054606} {\bibfield  {journal}
  {\bibinfo  {journal} {Phys. Rev. C}\ }\textbf {\bibinfo {volume} {86}},\
  \bibinfo {pages} {054606} (\bibinfo {year} {2012})}\BibitemShut {NoStop}%
\bibitem [{\citenamefont {Aguilar-Arevalo}\ \emph
  {et~al.}(2011{\natexlab{a}})\citenamefont {Aguilar-Arevalo} \emph
  {et~al.}}]{mb-cc1pip}%
  \BibitemOpen
  \bibfield  {author} {\bibinfo {author} {\bibfnamefont {A.~A.}\ \bibnamefont
  {Aguilar-Arevalo}} \emph {et~al.} (\bibinfo {collaboration} {MiniBooNE
  Collaboration}),\ }\href {\doibase 10.1103/PhysRevD.83.052007} {\bibfield
  {journal} {\bibinfo  {journal} {Phys. Rev. D}\ }\textbf {\bibinfo {volume}
  {83}},\ \bibinfo {pages} {052007} (\bibinfo {year}
  {2011}{\natexlab{a}})}\BibitemShut {NoStop}%
\bibitem [{\citenamefont {Aguilar-Arevalo}\ \emph
  {et~al.}(2011{\natexlab{b}})\citenamefont {Aguilar-Arevalo} \emph
  {et~al.}}]{mb-cc1pi0}%
  \BibitemOpen
  \bibfield  {author} {\bibinfo {author} {\bibfnamefont {A.~A.}\ \bibnamefont
  {Aguilar-Arevalo}} \emph {et~al.} (\bibinfo {collaboration} {MiniBooNE
  Collaboration}),\ }\href {\doibase 10.1103/PhysRevD.83.052009} {\bibfield
  {journal} {\bibinfo  {journal} {Phys. Rev. D}\ }\textbf {\bibinfo {volume}
  {83}},\ \bibinfo {pages} {052009} (\bibinfo {year}
  {2011}{\natexlab{b}})}\BibitemShut {NoStop}%
\bibitem [{\citenamefont {Aguilar-Arevalo}\ \emph
  {et~al.}(2010{\natexlab{c}})\citenamefont {Aguilar-Arevalo} \emph
  {et~al.}}]{mb-nc1pi0}%
  \BibitemOpen
  \bibfield  {author} {\bibinfo {author} {\bibfnamefont {A.~A.}\ \bibnamefont
  {Aguilar-Arevalo}} \emph {et~al.} (\bibinfo {collaboration} {MiniBooNE
  Collaboration}),\ }\href {\doibase 10.1103/PhysRevD.81.013005} {\bibfield
  {journal} {\bibinfo  {journal} {Phys. Rev. D}\ }\textbf {\bibinfo {volume}
  {81}},\ \bibinfo {pages} {013005} (\bibinfo {year}
  {2010}{\natexlab{c}})}\BibitemShut {NoStop}%
\bibitem [{\citenamefont {Nakayama}\ \emph {et~al.}(2005)\citenamefont
  {Nakayama} \emph {et~al.}}]{k2k-nc1pi0}%
  \BibitemOpen
  \bibfield  {author} {\bibinfo {author} {\bibfnamefont {S.}~\bibnamefont
  {Nakayama}} \emph {et~al.} (\bibinfo {collaboration} {K2K Collaboration}),\
  }\href {\doibase 10.1016/j.physletb.2005.05.044} {\bibfield  {journal}
  {\bibinfo  {journal} {Phys. Lett. B}\ }\textbf {\bibinfo {volume} {619}},\
  \bibinfo {pages} {255 } (\bibinfo {year} {2005})}\BibitemShut {NoStop}%
\bibitem [{\citenamefont {Golan}\ \emph {et~al.}(2012)\citenamefont {Golan},
  \citenamefont {Juszczak},\ and\ \citenamefont {Sobczyk}}]{Golan:2012}%
  \BibitemOpen
  \bibfield  {author} {\bibinfo {author} {\bibfnamefont {T.}~\bibnamefont
  {Golan}}, \bibinfo {author} {\bibfnamefont {C.}~\bibnamefont {Juszczak}}, \
  and\ \bibinfo {author} {\bibfnamefont {J.~T.}\ \bibnamefont {Sobczyk}},\
  }\href {\doibase 10.1103/PhysRevC.86.015505} {\bibfield  {journal} {\bibinfo
  {journal} {Phys. Rev. C}\ }\textbf {\bibinfo {volume} {86}},\ \bibinfo
  {pages} {015505} (\bibinfo {year} {2012})}\BibitemShut {NoStop}%
\bibitem [{\citenamefont {Wu}\ \emph {et~al.}(2008)\citenamefont {Wu} \emph
  {et~al.}}]{Wu:2007ab}%
  \BibitemOpen
  \bibfield  {author} {\bibinfo {author} {\bibfnamefont {Q.}~\bibnamefont {Wu}}
  \emph {et~al.} (\bibinfo {collaboration} {NOMAD Collaboration}),\ }\href
  {\doibase 10.1016/j.physletb.2007.12.027} {\bibfield  {journal} {\bibinfo
  {journal} {Phys. Lett. B}\ }\textbf {\bibinfo {volume} {660}},\ \bibinfo
  {pages} {19} (\bibinfo {year} {2008})}\BibitemShut {NoStop}%
\bibitem [{\citenamefont {Hasegawa}\ \emph {et~al.}(2005)\citenamefont
  {Hasegawa} \emph {et~al.}}]{Hasegawa:2005}%
  \BibitemOpen
  \bibfield  {author} {\bibinfo {author} {\bibfnamefont {M.}~\bibnamefont
  {Hasegawa}} \emph {et~al.} (\bibinfo {collaboration} {K2K Collaboration}),\
  }\href {\doibase 10.1103/PhysRevLett.95.252301} {\bibfield  {journal}
  {\bibinfo  {journal} {Phys. Rev. Lett.}\ }\textbf {\bibinfo {volume} {95}},\
  \bibinfo {pages} {252301} (\bibinfo {year} {2005})}\BibitemShut {NoStop}%
\bibitem [{\citenamefont {Hiraide}\ \emph {et~al.}(2008)\citenamefont {Hiraide}
  \emph {et~al.}}]{Hiraide:2008}%
  \BibitemOpen
  \bibfield  {author} {\bibinfo {author} {\bibfnamefont {K.}~\bibnamefont
  {Hiraide}} \emph {et~al.} (\bibinfo {collaboration} {SciBooNE
  Collaboration}),\ }\href {\doibase 10.1103/PhysRevD.78.112004} {\bibfield
  {journal} {\bibinfo  {journal} {Phys. Rev. D}\ }\textbf {\bibinfo {volume}
  {78}},\ \bibinfo {pages} {112004} (\bibinfo {year} {2008})}\BibitemShut
  {NoStop}%
\bibitem [{\citenamefont {Kurimoto}\ \emph {et~al.}(2010)\citenamefont
  {Kurimoto} \emph {et~al.}}]{PhysRevD.81.111102}%
  \BibitemOpen
  \bibfield  {author} {\bibinfo {author} {\bibfnamefont {Y.}~\bibnamefont
  {Kurimoto}} \emph {et~al.} (\bibinfo {collaboration} {SciBooNE
  Collaboration}),\ }\href {\doibase 10.1103/PhysRevD.81.111102} {\bibfield
  {journal} {\bibinfo  {journal} {Phys. Rev. D}\ }\textbf {\bibinfo {volume}
  {81}},\ \bibinfo {pages} {111102} (\bibinfo {year} {2010})}\BibitemShut
  {NoStop}%
\bibitem [{\citenamefont {Musset}\ and\ \citenamefont
  {Vialle}(1978)}]{Musset19781}%
  \BibitemOpen
  \bibfield  {author} {\bibinfo {author} {\bibfnamefont {P.}~\bibnamefont
  {Musset}}\ and\ \bibinfo {author} {\bibfnamefont {J.-P.}\ \bibnamefont
  {Vialle}},\ }\href {\doibase 10.1016/0370-1573(78)90051-0} {\bibfield
  {journal} {\bibinfo  {journal} {Physics Reports}\ }\textbf {\bibinfo {volume}
  {39}},\ \bibinfo {pages} {1 } (\bibinfo {year} {1978})}\BibitemShut {NoStop}%
\bibitem [{\citenamefont {Kim}\ \emph {et~al.}(1981)\citenamefont {Kim},
  \citenamefont {Langacker}, \citenamefont {Levine},\ and\ \citenamefont
  {Williams}}]{RevModPhys.53.211}%
  \BibitemOpen
  \bibfield  {author} {\bibinfo {author} {\bibfnamefont {J.~E.}\ \bibnamefont
  {Kim}}, \bibinfo {author} {\bibfnamefont {P.}~\bibnamefont {Langacker}},
  \bibinfo {author} {\bibfnamefont {M.}~\bibnamefont {Levine}}, \ and\ \bibinfo
  {author} {\bibfnamefont {H.~H.}\ \bibnamefont {Williams}},\ }\href {\doibase
  10.1103/RevModPhys.53.211} {\bibfield  {journal} {\bibinfo  {journal} {Rev.
  Mod. Phys.}\ }\textbf {\bibinfo {volume} {53}},\ \bibinfo {pages} {211}
  (\bibinfo {year} {1981})}\BibitemShut {NoStop}%
\bibitem [{\citenamefont {Adamson}\ \emph {et~al.}(2010)\citenamefont {Adamson}
  \emph {et~al.}}]{Adamson:2009ju}%
  \BibitemOpen
  \bibfield  {author} {\bibinfo {author} {\bibfnamefont {P.}~\bibnamefont
  {Adamson}} \emph {et~al.} (\bibinfo {collaboration} {MINOS Collaboration}),\
  }\href {\doibase 10.1103/PhysRevD.81.072002} {\bibfield  {journal} {\bibinfo
  {journal} {Phys. Rev. D}\ }\textbf {\bibinfo {volume} {81}},\ \bibinfo
  {pages} {072002} (\bibinfo {year} {2010})}\BibitemShut {NoStop}%
\bibitem [{\citenamefont {Moniz}\ \emph {et~al.}(1971)\citenamefont {Moniz},
  \citenamefont {Sick}, \citenamefont {Whitney}, \citenamefont {Ficenec},
  \citenamefont {Kephart} \emph {et~al.}}]{Moniz:1971mt}%
  \BibitemOpen
  \bibfield  {author} {\bibinfo {author} {\bibfnamefont {E.}~\bibnamefont
  {Moniz}}, \bibinfo {author} {\bibfnamefont {I.}~\bibnamefont {Sick}},
  \bibinfo {author} {\bibfnamefont {R.}~\bibnamefont {Whitney}}, \bibinfo
  {author} {\bibfnamefont {J.}~\bibnamefont {Ficenec}}, \bibinfo {author}
  {\bibfnamefont {R.~D.}\ \bibnamefont {Kephart}},  \emph {et~al.},\ }\href
  {\doibase 10.1103/PhysRevLett.26.445} {\bibfield  {journal} {\bibinfo
  {journal} {Phys. Rev. Lett.}\ }\textbf {\bibinfo {volume} {26}},\ \bibinfo
  {pages} {445} (\bibinfo {year} {1971})}\BibitemShut {NoStop}%
\bibitem [{\citenamefont {Ankowski}\ and\ \citenamefont
  {Sobczyk}(2006)}]{Ankowski:2005wi}%
  \BibitemOpen
  \bibfield  {author} {\bibinfo {author} {\bibfnamefont {A.~M.}\ \bibnamefont
  {Ankowski}}\ and\ \bibinfo {author} {\bibfnamefont {J.~T.}\ \bibnamefont
  {Sobczyk}},\ }\href {\doibase 10.1103/PhysRevC.74.054316} {\bibfield
  {journal} {\bibinfo  {journal} {Phys. Rev. C}\ }\textbf {\bibinfo {volume}
  {74}},\ \bibinfo {pages} {054316} (\bibinfo {year} {2006})}\BibitemShut
  {NoStop}%
\bibitem [{\citenamefont {Day}\ and\ \citenamefont
  {McFarland}(2012)}]{Day:2012gb}%
  \BibitemOpen
  \bibfield  {author} {\bibinfo {author} {\bibfnamefont {M.}~\bibnamefont
  {Day}}\ and\ \bibinfo {author} {\bibfnamefont {K.~S.}\ \bibnamefont
  {McFarland}},\ }\href {\doibase 10.1103/PhysRevD.86.053003} {\bibfield
  {journal} {\bibinfo  {journal} {Phys. Rev. D}\ }\textbf {\bibinfo {volume}
  {86}},\ \bibinfo {pages} {053003} (\bibinfo {year} {2012})}\BibitemShut
  {NoStop}%
\bibitem [{\citenamefont {Agostinelli}\ \emph {et~al.}(2003)\citenamefont
  {Agostinelli} \emph {et~al.}}]{Agostinelli2003250}%
  \BibitemOpen
  \bibfield  {author} {\bibinfo {author} {\bibfnamefont {S.}~\bibnamefont
  {Agostinelli}} \emph {et~al.},\ }\href {\doibase
  10.1016/S0168-9002(03)01368-8} {\bibfield  {journal} {\bibinfo  {journal}
  {Nucl. Instrum. Methods Phys. Res. Sect. A}\ }\textbf {\bibinfo {volume}
  {506}},\ \bibinfo {pages} {250 } (\bibinfo {year} {2003})}\BibitemShut
  {NoStop}%
\bibitem [{\citenamefont {Allison}\ \emph {et~al.}(Feb)\citenamefont {Allison}
  \emph {et~al.}}]{1610988}%
  \BibitemOpen
  \bibfield  {author} {\bibinfo {author} {\bibfnamefont {J.}~\bibnamefont
  {Allison}} \emph {et~al.},\ }\href {\doibase 10.1109/TNS.2006.869826}
  {\bibfield  {journal} {\bibinfo  {journal} {Nuclear Science, IEEE
  Transactions on}\ }\textbf {\bibinfo {volume} {53}},\ \bibinfo {pages} {270}
  (\bibinfo {year} {Feb.})}\BibitemShut {NoStop}%
\bibitem [{\citenamefont {Baker}\ and\ \citenamefont
  {Cousins}(1984)}]{Baker1984437}%
  \BibitemOpen
  \bibfield  {author} {\bibinfo {author} {\bibfnamefont {S.}~\bibnamefont
  {Baker}}\ and\ \bibinfo {author} {\bibfnamefont {R.~D.}\ \bibnamefont
  {Cousins}},\ }\href {\doibase 10.1016/0167-5087(84)90016-4} {\bibfield
  {journal} {\bibinfo  {journal} {Nuclear Instruments and Methods in Physics
  Research}\ }\textbf {\bibinfo {volume} {221}},\ \bibinfo {pages} {437 }
  (\bibinfo {year} {1984})}\BibitemShut {NoStop}%
\bibitem [{\citenamefont {Ashie}\ \emph {et~al.}(2005)\citenamefont {Ashie}
  \emph {et~al.}}]{Ashie:2005ik}%
  \BibitemOpen
  \bibfield  {author} {\bibinfo {author} {\bibfnamefont {Y.}~\bibnamefont
  {Ashie}} \emph {et~al.} (\bibinfo {collaboration} {Super-Kamiokande
  Collaboration}),\ }\href {\doibase 10.1103/PhysRevD.71.112005} {\bibfield
  {journal} {\bibinfo  {journal} {Phys. Rev. D}\ }\textbf {\bibinfo {volume}
  {71}},\ \bibinfo {pages} {112005} (\bibinfo {year} {2005})}\BibitemShut
  {NoStop}%
\bibitem [{\citenamefont {Barszczak}(2005)}]{Barszczak:2005sf}%
  \BibitemOpen
  \bibfield  {author} {\bibinfo {author} {\bibfnamefont {T.}~\bibnamefont
  {Barszczak}},\ }\emph {\bibinfo {title} {{The Efficient discrimination of
  electron and pi-zero events in a water Cherenkov detector and the applicati
  on to neutrino oscillation experiments}}},\ \href@noop {} {Ph.D. thesis},\
  \bibinfo  {school} {University of California, Irvine} (\bibinfo {year}
  {2005})\BibitemShut {NoStop}%
\bibitem [{\citenamefont {Barger}\ \emph {et~al.}(1980)\citenamefont {Barger},
  \citenamefont {Whisnant}, \citenamefont {Pakvasa},\ and\ \citenamefont
  {Phillips}}]{Barger:1980tf}%
  \BibitemOpen
  \bibfield  {author} {\bibinfo {author} {\bibfnamefont {V.~D.}\ \bibnamefont
  {Barger}}, \bibinfo {author} {\bibfnamefont {K.}~\bibnamefont {Whisnant}},
  \bibinfo {author} {\bibfnamefont {S.}~\bibnamefont {Pakvasa}}, \ and\
  \bibinfo {author} {\bibfnamefont {R.}~\bibnamefont {Phillips}},\ }\href
  {\doibase 10.1103/PhysRevD.22.2718} {\bibfield  {journal} {\bibinfo
  {journal} {Phys. Rev. D}\ }\textbf {\bibinfo {volume} {22}},\ \bibinfo
  {pages} {2718} (\bibinfo {year} {1980})}\BibitemShut {NoStop}%
\bibitem [{\citenamefont {Feldman}\ and\ \citenamefont
  {Cousins}(1998)}]{PhysRevD.57.3873}%
  \BibitemOpen
  \bibfield  {author} {\bibinfo {author} {\bibfnamefont {G.~J.}\ \bibnamefont
  {Feldman}}\ and\ \bibinfo {author} {\bibfnamefont {R.~D.}\ \bibnamefont
  {Cousins}},\ }\href {\doibase 10.1103/PhysRevD.57.3873} {\bibfield  {journal}
  {\bibinfo  {journal} {Phys. Rev. D}\ }\textbf {\bibinfo {volume} {57}},\
  \bibinfo {pages} {3873} (\bibinfo {year} {1998})}\BibitemShut {NoStop}%
\bibitem [{\citenamefont {C.~Loken}\ \emph {et~al.}(2010)\citenamefont
  {C.~Loken} \emph {et~al.}}]{scinet}%
  \BibitemOpen
  \bibfield  {author} {\bibinfo {author} {\bibfnamefont {D.~G.}\ \bibnamefont
  {C.~Loken}} \emph {et~al.},\ }\href {\doibase 10.1088/1742-6596/256/1/012026}
  {\bibfield  {journal} {\bibinfo  {journal} {J. Phys.: Conf. Ser.}\ }\textbf
  {\bibinfo {volume} {256}},\ \bibinfo {pages} {012026} (\bibinfo {year}
  {2010})}\BibitemShut {NoStop}%
\bibitem [{\citenamefont {Aguilar-Arevalo}\ \emph {et~al.}(2009)\citenamefont
  {Aguilar-Arevalo} \emph {et~al.}}]{AguilarArevalo:2008yp}%
  \BibitemOpen
  \bibfield  {author} {\bibinfo {author} {\bibfnamefont {A.~A.}\ \bibnamefont
  {Aguilar-Arevalo}} \emph {et~al.} (\bibinfo {collaboration} {MiniBooNE
  Collaboration}),\ }\href {\doibase 10.1103/PhysRevD.79.072002} {\bibfield
  {journal} {\bibinfo  {journal} {Phys. Rev. D}\ }\textbf {\bibinfo {volume}
  {79}},\ \bibinfo {pages} {072002} (\bibinfo {year} {2009})}\BibitemShut
  {NoStop}%
\bibitem [{\citenamefont {D'Agostini}(1994)}]{covariance}%
  \BibitemOpen
  \bibfield  {author} {\bibinfo {author} {\bibfnamefont {G.}~\bibnamefont
  {D'Agostini}},\ }\href {\doibase 10.1016/0168-9002(94)90719-6} {\bibfield
  {journal} {\bibinfo  {journal} {Nucl. Instrum. Methods Phys. Res., Sect. A}\
  }\textbf {\bibinfo {volume} {346}},\ \bibinfo {pages} {306 } (\bibinfo {year}
  {1994})}\BibitemShut {NoStop}%
\end{thebibliography}%

\appendix
\newpage

\section{\label{sec:neut_data} \neut model external data comparisons and tuning }
We fit external pion scattering data and neutrino scattering data with the \neut model
while allowing subsets of the systematic parameters described in 
Table~\ref{tab:xsec_params} to vary.  These fits constrain the \neut FSI, CCQE and resonant
pion production models.  The details of these fits are described here.

\subsubsection{\label{sec:neut_fsi_fit} FSI model}
\begin{figure}[tp]
  \centering
 \includegraphics[width=1.0\linewidth]{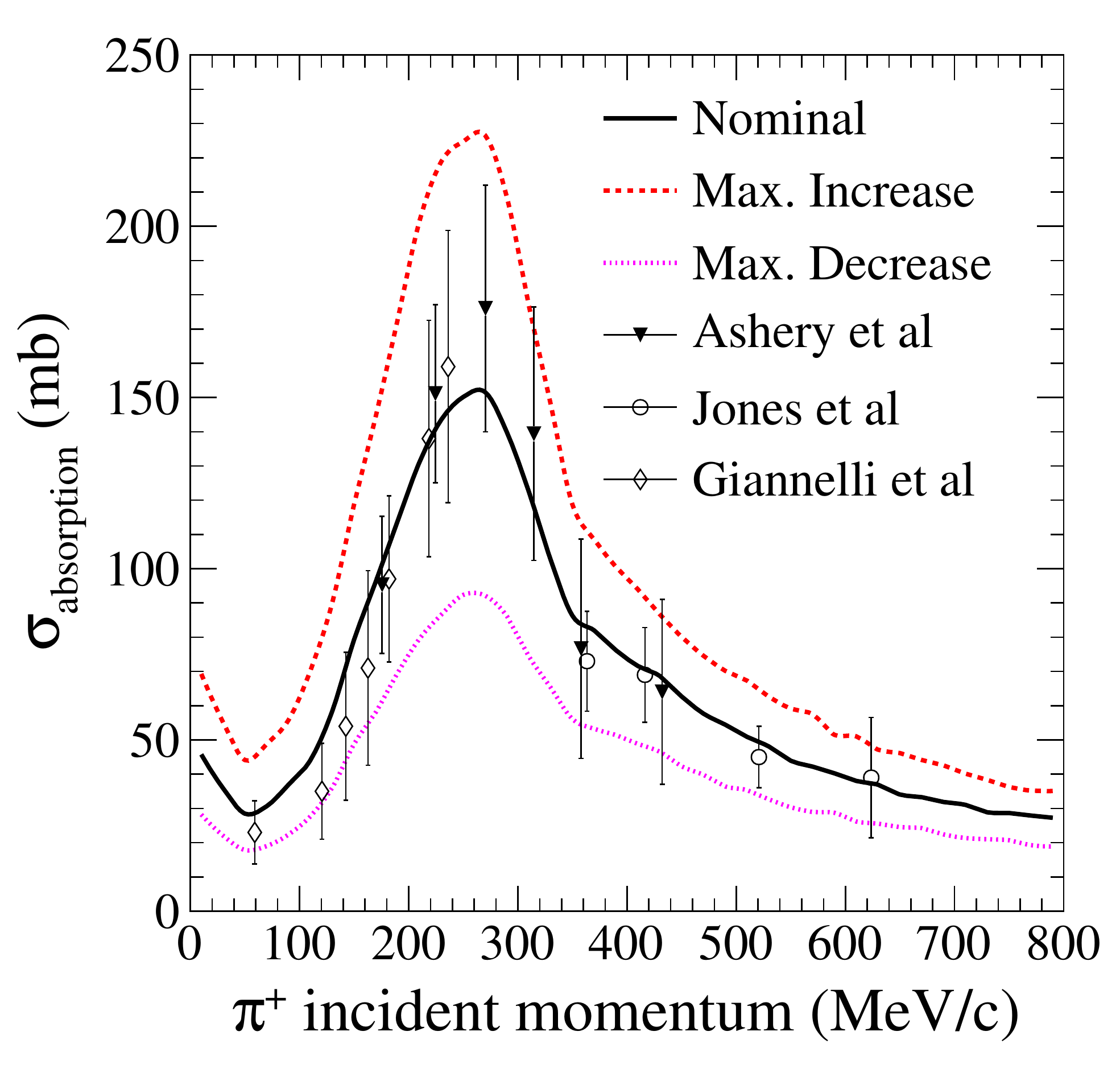} \\
  \caption{Pion absorption cross section as a function of pion momentum overlaid with   $\pi^+$-${}^{12}$C scattering data, Ashery~{\it et~al.}~\cite{Ashery:1981tq}, Jones~{\it et~al.}~\cite{PhysRevC.48.2800}, and Giannelli~{\it et~al.}~\cite{PhysRevC.61.054615}.
}
  \label{fig:fsi-parameter-sets}
\end{figure}

The \neut FSI model includes parameters which alter the pion interaction probabilities for absorption,
charge exchange, and quasi-elastic scattering~\cite{dePerio:2011zz}.
The values of these parameters and their uncertainties are determined from fits to pion scattering data. 
Fig.~\ref{fig:fsi-parameter-sets} shows
the tuned cascade model compared to macroscopic measurements of the pion absorption cross section and the maximum variation of the model parameters chosen to cover the uncertainties on the data.

In total, we consider 16 variations of the FSI model parameters to cover the uncertainties
on macroscopic pion scattering data.  
For each of the modified FSI parameter sets and the nominal \neut model,  we evaluate with weights the effect on ND280, SK or external predicted observables by calculating the covariance matrix of
the predicted observables. FSI covariance matrices are generated for MiniBooNE, ND280 and SK predictions. The external data covariance matrices use observable bins from 
external data, such as reconstructed $(T_\mu, \cos \theta_\mu)$ bins for MiniBooNE. The ND280 covariance matrix corresponds to 
the two ND280 selections' reconstructed \pt bins and the SK covariance matrices correspond to the 
\nue selection with either reconstructed \pthetae or \erec bins:

\begin{linenomath*}
\begin{equation}
  \label{eq:fsi-covariance}
  V_{ij} = \frac{1}{16} \sum_{k=1}^{k=16} (p_i^\mathrm{nom} - p_i^k)(p_j^\mathrm{nom} - p_j^k),
\end{equation}
\end{linenomath*}
where $p_i^k$ is the expected event rate in the $i$th observable bin assuming
the $k$th FSI parameter set, and $p_i^\mathrm{nom}$ is the expected event rate in the same
bin assuming the nominal FSI parameter set.  For the oscillation analysis, we add these FSI covariance matrices to the 
detector efficiency and reconstruction covariance matrices evaluated for ND280 
(Section~\ref{sec:nd280_numusystematics}) and SK  (Section~\ref{sec:sk_selection}) selections.

\subsubsection{\label{sec:neut_ccqe_fit}CCQE model }
As discussed in Section~\ref{sec:neut_ccqe}, we fit the \mb measurement of 
the CCQE double-differential cross sections in bins of muon
kinetic energy and angle, $(T_\mu, \cos
\theta_\mu)$~\cite{mb-ccqe} with the \neut model.  While the CCQE model
can be directly constrained with T2K ND280 data, we also fit the \mb 
measurement since the \mb detector's 4$\pi$ acceptance provides coverage for backwards
produced muons that are currently excluded in the ND280 selection.

To compare the \neut model of CCQE interactions with MiniBooNE data,
we use the MiniBooNE flux prediction~\cite{AguilarArevalo:2008yp} to generate CCQE interactions. 
We fit the MiniBooNE double-differential cross section data with the \neut prediction, allowing
 $M^{QE}_A$ and the overall cross section normalization to vary, by 
minimizing the $\chi^2$ defined as:
\begin{linenomath*}
\begin{equation}
  \label{eq:mb-ccqe-chi2}
  \chi^2(M^{QE}_A,\lambda) = \sum_{i=0}^N \left( \frac{D_i -
      \lambda M_i(M^{QE}_A)}{\sigma_i} \right)^2 + \left(\frac{\lambda
      - 1}{\sigma_\lambda} \right)^2.
\end{equation}
\end{linenomath*}

\begin{figure}
  \centering
 \includegraphics[width=1.0\linewidth]{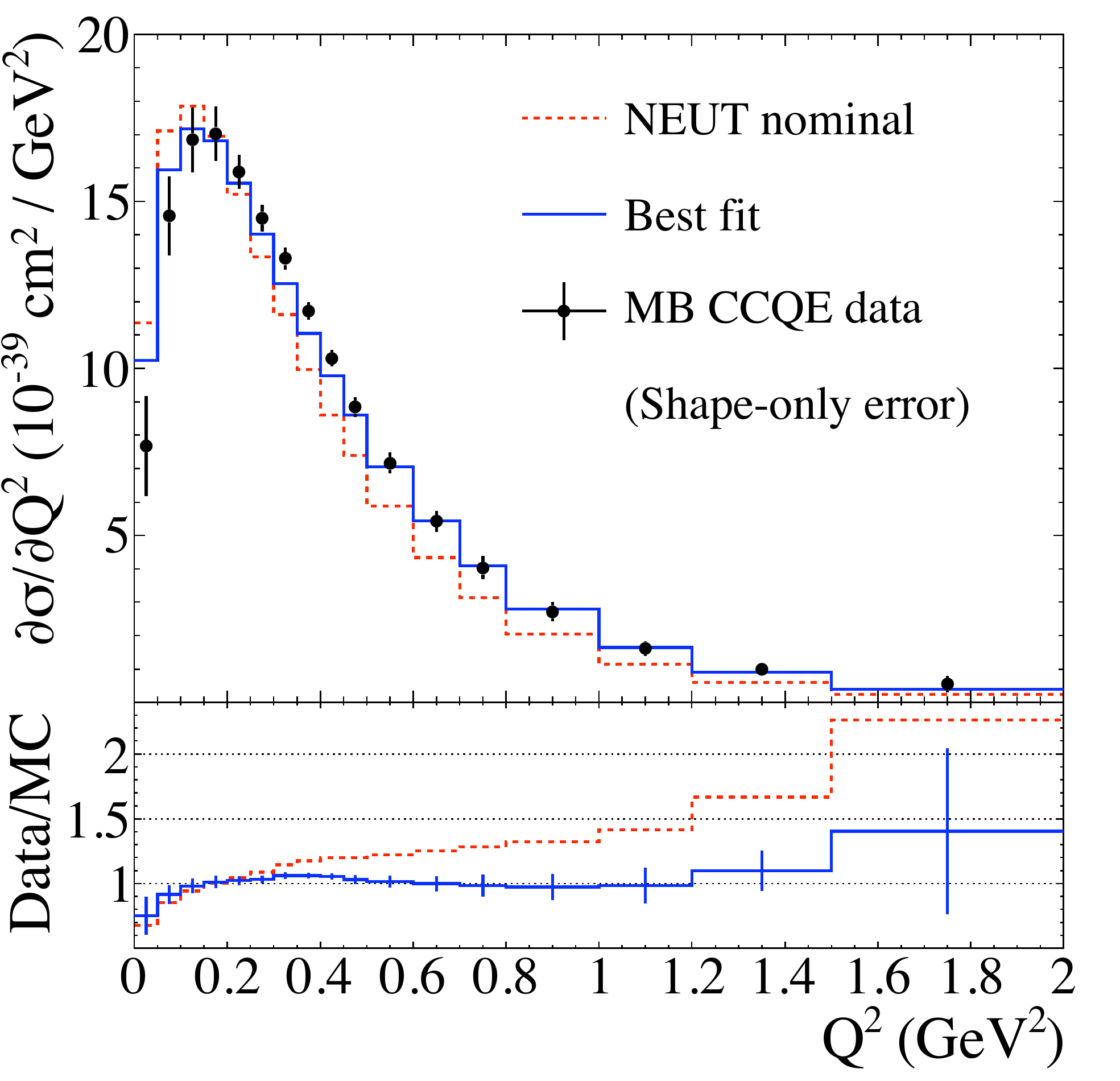} \\
  \caption{
The CCQE  cross section as a function of $Q^2$ (top) as measured by \mb (points), with the \neut nominal and \neut at the best-fit of the \mb CCQE $(T_\mu, \cos \theta_\mu)$ spectrum.
Ratio of data to \neut (bottom) for nominal (dashed) and best-fit (solid). 
}
  \label{fig:ccqe_qsq}
\end{figure}

\begin{figure}
  \centering
 \includegraphics[width=1.0\linewidth]{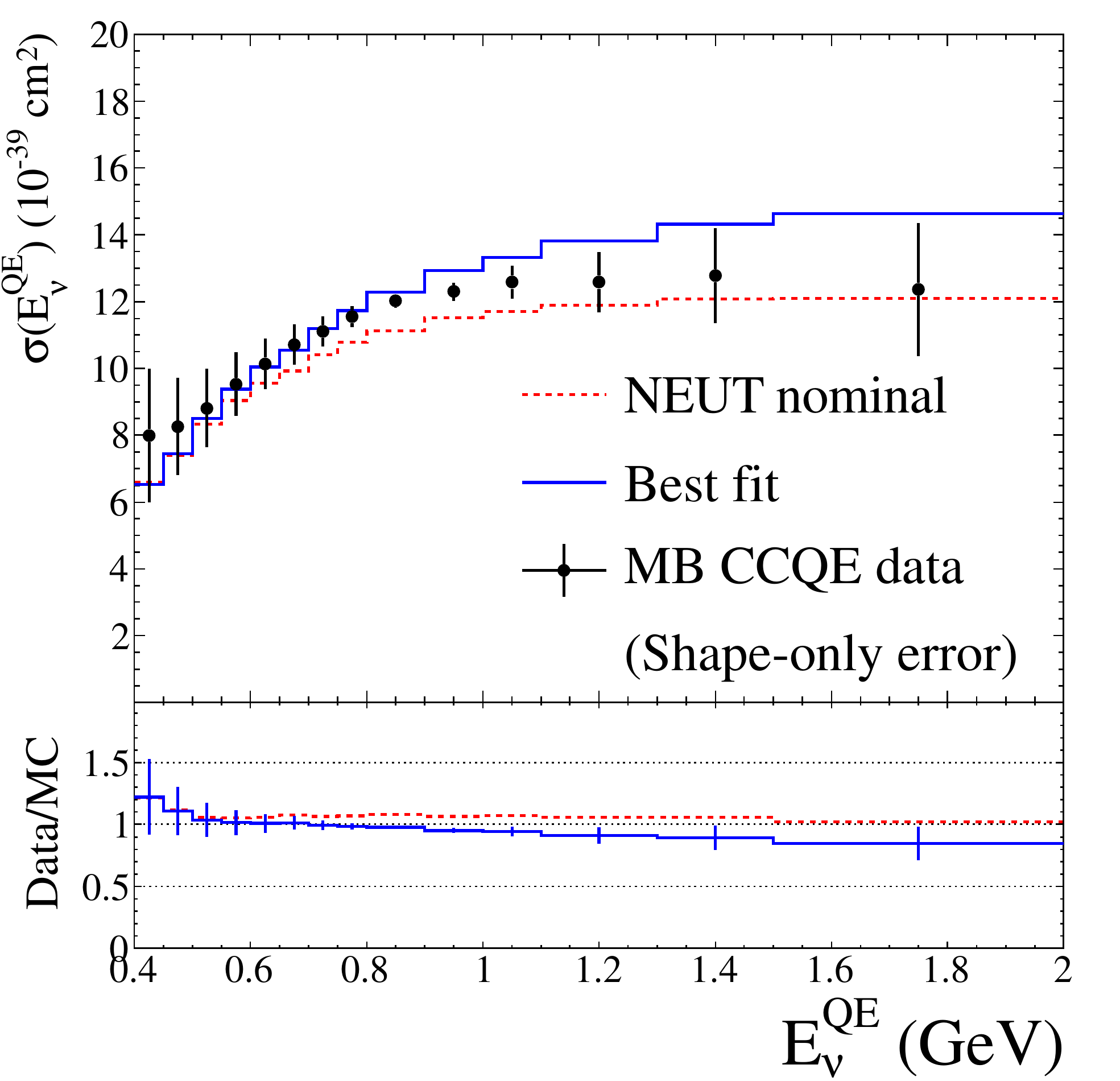} \\
  \caption{
The CCQE  cross section as a function of neutrino energy (top) as measured by \mb (points), with the \neut nominal and \neut at the best-fit of the \mb CCQE $(T_\mu, \cos \theta_\mu)$ spectrum.
Ratio of data to \neut (bottom) for nominal (dashed) and best-fit (solid). 
}
  \label{fig:ccqe_enu}
\end{figure}

Here, the sum runs over the $N$ bins in the $(T_\mu, \cos \theta_\mu)$
differential cross section, $D_i$ is the cross section measured by
\mb in the $i$th bin, $M_i$ is the \neut prediction in that bin
and $\sigma_i$ is the reported shape-only component of the error on the measured cross section. 
The second term adds a penalty to the normalization parameter $\lambda$, which is 
constrained within the \mb flux uncertainty, $\sigma_\lambda=10.7\%$.
The best-fit parameter
values are $M^{QE}_A =
1.64 \pm 0.04 \, \mathrm{GeV}$ and $\lambda = 0.88 \pm 0.02$. 
Fig~\ref{fig:ccqe_qsq} shows the measured MiniBooNE cross section as a function of $Q^2$ for the nominal and best-fit value of $M^{QE}_A$, which is well reproduced except at lowest values of $Q^2$. However, this value of $M^{QE}_A$ is 
significantly larger than the value of $M^{QE}_A =
1.35 \pm 0.17 \, \mathrm{GeV}$ obtained by the \mb collaboration in a fit to the
single-differential $\mathrm{d}\sigma/\mathrm{d}Q^2$ spectrum, with an
uncertainty that is smaller by a factor of 4.  
We postulate that the difference in central values is
due to deficiencies in the nuclear model at low $Q^2$, which
\mb addressed by adding an empirical parameter $\kappa$ to modify
Pauli blocking, and the lack of full correlations between the measured $(T_\mu, \cos \theta_\mu)$ bins which are not included in the provided uncertainties. We assume the lack of bin correlations also causes the discrepancy in the fitted uncertainty, and this is supported by  the relatively small $\chi^2=26.9$ that is observed for 137 degrees of freedom.
Furthermore, the fitted prediction for the total CCQE cross section as a function of energy is
poor, as illustrated in Fig.~\ref{fig:ccqe_enu}.  The fitted model is systematically higher than 
the \mb data above 1~GeV, although agreement is improved near the T2K peak energy of 600~MeV.

As is discussed in Section~\ref{sec:nd280}, a CCQE-like selection of interactions
in ND280 has power to constrain the CCQE cross section model.  Since the fit to 
\mb data poorly reproduces the energy dependent cross section and lacks the full
correlation of data points, we do not directly tune the \neut model with the fitted value
for $M^{QE}_A$.  Instead, we set large prior uncertainties
on the CCQE model parameters and allow the ND280 data to constrain the model.
We set $M^{QE}_A$ to the \neut nominal value (1.21~GeV), with the prior uncertainty
set to the difference between the nominal value and 
best-fit value from the \mb fit, \emph{viz.}\ $(1.64-1.21 = 0.43)\, \mathrm{GeV}$. 
We set the uncertainty on the low energy CCQE normalization, $x_{1}^{QE}$, to the size 
of the MiniBooNE flux uncertainty (11\%). 

\subsubsection{\label{sec:neut_singlepi_fit}Single pion production model}

As discussed in Section~\ref{sec:neut_singlepi} we consider measurements of single 
pion production cross sections on light 
nuclei in the T2K energy range by \mb~\cite{mb-cc1pip,mb-cc1pi0,mb-nc1pi0},
and K2K~\cite{k2k-nc1pi0}.
We perform a joint fit to the \mb measurements of charged current
single $\pi^+$ production (\ccpip), charged current single $\pi^0$
production (\ccpi) and neutral current single $\pi^0$ production
(\ncpi), and we check the fit results with the K2K measurement.

An important feature of the \mb single pion measurements is that they are defined by the particles exiting the target nucleus, not the particles produced at the neutrino interaction vertex. The measurements do not include corrections for FSI, but do include uncertainties of interactions of the pions in the detector.
To derive the \neut predictions for these selections, we generate interactions according
to the \mb flux as was done for the CCQE fits. Instead of selecting generated events based on the true
neutrino interaction mode, such as \ccpip, we select the events based on the
presence of a single pion exiting the nucleus.
Hence, multiple interaction types are present in the prediction for each of the
\mb measurements.
For example, \ccpip interactions chiefly result in a single charged pion exiting the nucleus, but these events may instead pass the \ccpi selection if $\pi^+$ undergoes single charge exchange within the nucleus. 
This interdependence within the \mb
selections, as well as the fact that all three are predicted by the same
model in \neut, justifies the use of a joint fit to the three
measurements. 

We fit to the measured $\mathrm{d}\sigma/\mathrm{d}Q^2$ spectra from
\ccpip and \ccpi samples and the $\mathrm{d}\sigma /
\mathrm{d}p_{\pi^0}$ spectrum from the \ncpi samples. 
\mb provides uncertainties for each of the measurement.  In the case of
the  \ccpi and \ncpi measurements, covariance matrices account for correlations
between the measured points in the spectra arising from the \mb flux model and detector response.
\mb only provides diagonal errors for the \ccpip measurement.  We construct a covariance
matrix for the \ccpip by assuming a 10\% flux uncertainty correlated across all bins
and by adding an additional uncorrelated uncertainty to the diagonal terms to recover
the diagonal errors provided by \mb.  While the flux is shared for the three measurements, 
at this time no correlation between the three measurements was considered.

For each of the three measured distributions ($k$) we construct the $\chi^2$ based on the data and \neut prediction:
\begin{linenomath*}
 \begin{equation}
   \label{eq:chi2-ccpi0}
   \chi^2_{k} = \sum_i \sum_j [D^k_i - M^k_i(\vec{x})] (C^k_{ij})^{-1} [D^k_j - M^k_j(\vec{x})].
 \end{equation}
\end{linenomath*}
Here, $i$ and $j$ sum over the bins in the $k$th measurement, $D^k_{i}$ are the measured differential
cross sections, $C^k_{ij}$ is the covariance matrix describing the uncertainty on the measurement and 
$M^k_i(\vec{x})$ are the \neut predictions for each measurement.  

The cross section parameters that are allowed to vary in the fit, $\vec{x}$, along with their prior values
and prior uncertainties for penalty terms are listed in Table~\ref{tab:singlepi-fitparams}.
Contributions to the predictions from CC multi-pion/DIS ($x_{CCother}$)
interactions, NC coherent interactions, NC1$\pi^{\pm}$ interactions and NC multi-pion/DIS interactions 
are relatively small, so penalty terms are used for the associated parameters according to the prior uncertainties.

We minimize the total $\chi^2$ that includes the $\chi^2$ for each of the measurements and the penalty terms:
\begin{linenomath*}
 \begin{equation}
   \label{eq:chi2-1pi}
   \chi^2_{total} = \chi^{2}_{CC1\pi^{+}}+\chi^{2}_{CC1\pi^0}+\chi^2_{NC1\pi^0}+\sum_k \frac{(s_k - s_k^{nom})^2}{\sigma_k^2},
 \end{equation}
\end{linenomath*}
where, for each penalized parameter $k$, $s_k$ is the value of the parameter, $s_k^{nom}$ is the nominal value, and $\sigma_k$ is the prior uncertainty assigned to the penalty parameter.

In practice, the inclusion of the \ncpi covariance matrix in the fit results in a
best-fit which lies outside the range of the data points. This
behavior results from strongly-correlated measurements combined with a model
which does not correctly describe the data~\cite{covariance}.
To achieve a fit that better reproduces the central values of the data points,  we only use the diagonal terms of the \ncpi covariance matrix in our fit.  
The missing correlations also result in uncertainties on the fit parameters which do not cover the uncertainties in the data points. To remedy this, we multiply the fit parameter uncertainties by a scale factor of 2 (2.5) for CC (NC) parameters, while keeping their correlations the same. These scale factor ensure that the flux-integrated cross section uncertainty matches that given by \mb (16\% for each measurement).

The results of the fit are discussed in Section~\ref{sec:neut_singlepi}.  We propagate the fitted values 
and uncertainties for $M_{A}^{RES}$, $x^{CC1\pi}_{1}$ and $x^{NC1\pi^0}$ to model the cross section in 
the fit to ND280 data described in Section~\ref{sec:extrapolation}.  In addition, we keep parameter
$W_{\textrm{eff}}$ at its nominal value, but apply an uncertainty equal to the amount it is pulled in the fit
to the \mb data.

\begin{figure}
  \centering
  \includegraphics[width=0.95\linewidth]{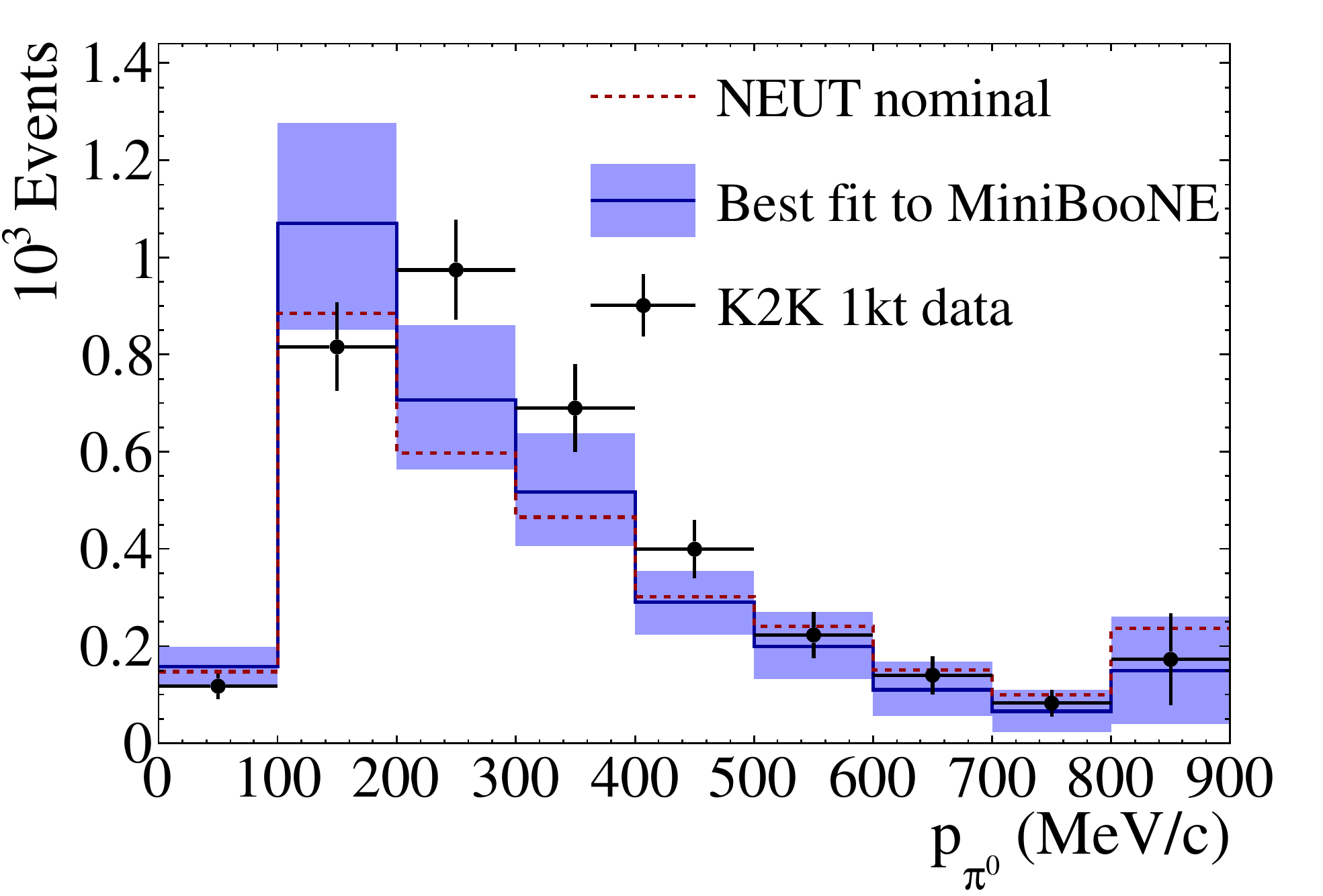}
  \caption{Differential $\mathrm{d}\sigma /\mathrm{d}p_{\pi^0}$ cross
    section measured by K2K and the nominal and best-fit from the \mb single pion fits \neut predictions, with error band showing the uncertainties after
    the fit to \mb data.}
  \label{fig:k2k_ncpi0}
\end{figure}

We compare the results of the fitted \neut pion production model to the NC K2K measurement.
The $\mathrm{d}\sigma /\mathrm{d}p_{\pi^0}$ distribution measured by K2K in the 1000 ton water Cherenkov
detector is shown with the nominal and tuned \neut model in Fig.~\ref{fig:k2k_ncpi0}.  As with the \mb data,
the data prefer a peak at higher momentum and fewer events in the high momentum tail compared to the 
nominal \neut prediction.  The use of \neut assuming the best-fit parameters from the \mb single pion production fits  does not significantly improve the agreement between \neut and the K2K data.
However, the discrepancy is covered by the uncertainties on the single pion production and FSI model.

\end{document}